\documentclass[10pt]{iopart}
\usepackage{epsfig}  
\usepackage{iopams}
\def\qbar {\bar{q}}
\def\ubar {\bar{u}}
\def\dbar {\bar{d}}
\def\sbar {\bar{s}}
\def\cbar {\bar{c}}
\def\bbar {\bar{b}}
\def\tbar {\bar{t}}
\def\pbar {\bar{p}}
\def\ebar {\bar{e}}
\def\Phibf {\mathbf{\Phi}}

\def\msbar {\mathrm{\overline{MS}}} 
\def\mtop  {M_\mathrm{top}}

\def\BR   {\mathrm{BR}}
\def\DZero {D\O\ }

\def\EtMiss  {E_\mathrm{T}\!\!\!\!\!\!\!/ \ \; }
\def\EtMissVec {\bi{E}_\mathrm{T}\!\!\!\!\!\!\!/ \;\;}
\def\OneDig {$\ \,$}

\begin{document}
\hfill Published in: Rep. Prog. Phys. {\bf 68} (2005) 2409--2494

\hfill IEKP-KA/2005-8 

\hfill Corrections added: August 3, 2007

\title{Top quark physics in hadron collisions}
\author{Wolfgang Wagner}
\address{Institut f\"ur Experimentelle Kernphysik, Universit\"at Karlsruhe,
76128 Karlsruhe, Germany}
\ead{wagner@ekp.uni-karlsruhe.de}
\begin{abstract}
The top quark is the heaviest elementary particle observed to date.
Its large mass makes the top quark an ideal laboratory to test predictions of 
perturbation theory concerning heavy quark production at hadron colliders.
The top quark is also a powerful probe for new phenomena beyond the Standard Model 
of particle physics. 
In addition, the top quark mass is a crucial parameter for scrutinizing the 
Standard Model in electroweak precision tests and for predicting
the mass of the yet unobserved Higgs boson.
Ten years after the discovery of the top quark at the Fermilab Tevatron
top quark physics has entered an era where detailed measurements of 
top quark properties are undertaken. In this review article 
an introduction to the phenomenology of top quark production in hadron
collisions is given, the lessons learned in Tevatron Run I are 
summarized, and first Run II results are discussed. A brief outlook to 
the possibilities of top quark research a the Large Hadron Collider,
currently under construction at CERN, is included. 

\end{abstract}
\maketitle

\pagestyle{myheadings}
\tableofcontents
\newpage
\pagestyle{headings}
\markboth{Top quark physics in hadron collisions}{Wolfgang Wagner}

\section{Introduction}
The top quark is the by far heaviest of the six fundamental fermions in the 
Standard Model (SM) of particle physics. Its large mass made the search
for the top quark a long and tedious process, since accelerators with 
high centre-of-mass energies are needed. 
In 1977 the discovery of the bottom quark indicated the existence of a 
third quark generation, and shortly thereafter the quest for the top quark 
began.   
Searches were conducted in electron-positron ($e^+e^-$) and proton-antiproton
($p\pbar$) collisions during the 1980s and early 1990s. 
Finally, in 1995 the top quark was discovered at the
Fermilab Tevatron $p\pbar$ collider. Subsequently, its mass was precisely measured
to be $M_\mathrm{top} = (178.0\pm4.3)\,\mathrm{GeV}/c^2$~\cite{mtopCombined2004}. 
The relative precision of this measurement (2.4\%) is better than our 
knowledge of any other quark mass. 
The top quark is about 40 times heavier than the second-heaviest quark,
the bottom quark. 
Its huge mass makes the top quark an ideal probe for new physics 
beyond the SM.
It remains an open question to particle physics research whether
the observed mass hierarchy is a result of unknown fundamental particle dynamics.
It has been argued that the top quark could be the key to 
understand the dynamical origin  
of how particle masses are generated by the mechanism of 
electroweak symmetry breaking, 
since its mass is close to the energy scale at which the break down occurs
(vacuum expectation value of the Higgs field = 246 GeV)~\cite{peccei1991}.
The most favoured framework to describe electroweak symmetry breaking
is the Higgs mechanism. The masses of the Higgs boson, the $W$ boson and the
top quark are closely related through higher order corrections to various
physics processes. A precise knowledge of the top quark mass together with
other electroweak precision measurements can therefore be used to predict
the Higgs boson mass. 

At present, top quarks can only be directly produced at the Tevatron.
The physics results of the Tevatron experiments CDF and D\O \ will therefore
be the focus of this article.
We review the experimental status of top quark physics at the
beginning of the Tevatron Run II which will yield considerably improved
measurements in the top sector.
We discuss the first Run II analysis results and 
summarise the lessons learned from Run I data taken between 1990 -- 1995.

The outline of this article is as follows:
In chapter~\ref{sec:topInSM} we give a
brief introduction to the SM of particle physics and stress the importance
of the top quark for higher order corrections to electroweak perturbation
theory. In particular, we discuss electroweak precision measurements used 
to predict the top quark and Higgs boson mass within the SM. 
Chapter~\ref{sec:topProd} covers the theoretical description of SM top quark 
production at hadron colliders and the top quark decay.
In chapter~\ref{sec:topdetect} we elaborate on analysis techniques used to 
detect top quarks in particle detectors. 
A short description of the Tevatron experiments CDF and D\O \  is provided. 
In chapter~\ref{sec:discover} we recall the early
searches for the top quark in the 1980s and the discovery at the Tevatron
in 1994/95. 
In chapter~\ref{sec:ttbarxs} we present various techniques to measure
the top-antitop pair production cross section.
Chapter~\ref{sec:mass} covers the top quark mass measurements and
chapter~\ref{sec:decayProperties} summarises various advanced tests of top quark 
properties. In chapter~\ref{sec:anotop} we discuss
the search for anomalous (non SM) top quark production.
Previous reviews of top quark physics can be found in
references~\cite{manganoTrippe2004,konigsberg2003,topLHC2000,Bhat:1998cd,CampagnariTop,kuehnSLAC1997,teubner99}.

\section{The top quark in the Standard Model}
\label{sec:topInSM}
\subsection{The standard model of particle physics}  
  \renewcommand\arraystretch{1.5}
  The Standard Model (SM) of particle physics postulates that all matter be 
  composed of a few basic, point-like and structureless constituents: 
  elementary particles. 
  One distinguishes two groups: quarks and leptons. 
  Both of them are fermions and carry spin 1/2. 
  The quarks come in six different flavours: up, down, charm, strange, 
  top and bottom; formally described by assigning flavour quantum
  numbers.  
  The SM incorporates six leptons: the electron ($e^-$) and the 
  electron-neutrino ($\nu_e$), the muon ($\mu^-$) and the 
  muon-neutrino ($\nu_\mu$), the tau ($\tau^-$) and the tau-neutrino 
  ($\nu_\tau$). They carry electron, muon and tau quantum numbers.
  Quarks and leptons can be grouped into three generations (or families)
  as shown in table~\ref{tab:QuarksAndLeptons} which also contains the 
  charges and masses of the particles. 
  \begin{table}
  \caption{\label{tab:QuarksAndLeptons}Three generations of quarks and leptons, 
    their charges and masses are given~\cite{PDG2004}.}
    \begin{math}\begin{array}{cccrcrr}
      \br
      &\multicolumn{3}{c}{\mathrm{Quarks}}&\multicolumn{3}{c}{\mathrm{Leptons}} \\ 
      \mathrm{Generation} &        
        \mathrm{Symbol} & \mathrm{Charge} & \mathrm{Mass\;[ MeV/}c^2 ] \ \ \ \  & 
        \mathrm{Symbol} & \mathrm{Charge} & \mathrm{Mass\;[ MeV/}c^2 ] \ \ \ \  \\ \mr
        \mathbf{1} &         
        u & +\frac{2}{3}  & 1.5 \ \mathrm{to} \ 4 & \nu_e & 0 & < 3\cdot 10^{-6} \\ 
        \mathbf{1} & d & -\frac{1}{3}  & 4 \ \mathrm{to} \ 8 & 
        e^- & -1 & 0.51 \\ \hline
        \mathbf{2} & c & +\frac{2}{3} & (1.15 \ \mathrm{to} \ 1.35)\cdot 10^3 &
        \nu_\mu & 0 & < 0.19 \\ 
        \mathbf{2} & s & -\frac{1}{3} & 80 \ \mathrm{to} \ 130 & 
        \mu^- & -1 & 106\\ \hline
        \mathbf{3} &
        t & +\frac{2}{3} & (178.0\pm4.3)\cdot 10^3 & 
        \nu_\tau & 0 & < 18.2 \\ 
        \mathbf{3} & 
        b & -\frac{1}{3} & (4.1 \ \mathrm{to} \ 4.4)\cdot 10^3 & 
        \tau^- & -1 & 1777 \\ \br
      \end{array}
    \end{math}
  \end{table}
  The three generations exhibit a striking mass hierarchy,  
  the top quark having by far the highest mass. Understanding the deeper 
  reason behind the hierarchy and generation structure is one of the open 
  questions of particle physics.
  Each quark and each lepton has an associated antiparticle with the
  same mass but opposite charge. 
  The antiquarks are denoted $\ubar$, $\dbar$, etc. The antiparticle of 
  the electron is the positron ($e^+$).  

  The forces of nature acting between quarks and leptons are described by 
  quantized fields. The interactions between elementary particles are due
  to the
  exchange of field quanta which are said to mediate the forces. 
  The SM incorporates the electromagnetic force, responsible for example for
  the emission of light from excited atoms, the weak force, which for 
  instance causes nuclear beta decay, and the strong force which keeps nuclei 
  stable. Gravitation is not included in the framework of the SM but rather
  described by the theory of general relativity. 
  All particles with mass or energy 
  feel the gravitational force. However, due to the weakness of 
  gravitation with respect to the other forces acting in elementary particle 
  reactions it is not further considered in this article. 
  The electromagnetic, weak and strong forces are described by so called
  quantum gauge field theories (see explanation below). 
  The quanta of these fields carry spin 1 and are 
  therefore called gauge bosons.
  The electromagnetic force is mediated by the
  massless photon ($\gamma$), the weak force by the massive $W^\pm$, 
  $M_W = (80.425\pm0.038)$~GeV/$c^2$~\cite{PDG2004}, and the $Z^0$, 
  $M_Z = (91.1876\pm0.0021)$~GeV/$c^2$~\cite{PDG2004},
  and the strong force by eight massless gluons (g). 
  Quarks participate in electromagnetic, weak and strong interactions. 
  All leptons experience the weak force, the charged ones also feel the 
  electromagnetic force.
  But leptons do not take part in strong interactions.
  A thorough introduction to the SM can be found in various text books
  of particle 
  physics~\cite{bargerPhillips,halzenMartin,griffiths,perkins,lohrmann}.

  \subsubsection{Electroweak interactions.}
  \renewcommand\arraystretch{1.0}
  In quantum field theory quarks and leptons are represented by spinor fields
  $\bPsi$ which are functions of the continuous space-time coordinates $x_\mu$.
  To take into account that the weak interaction only couples to the left-handed
  particles, left- and right-handed fields 
  $\bPsi_\mathbf{L} = \frac{1}{2} (1 - \gamma_5 )\,\bPsi$ and 
  $\bPsi_\mathbf{R} = \frac{1}{2} (1 + \gamma_5 )\,\bPsi$ 
  are introduced. The left-handed states of one generation are grouped into
  weak-isospin 
  doublets, the right-handed states form singlets:
  \[ \begin{array}{cccccccc}
       \left( \begin{array}{c} u \\ d \\ \end{array} \right)_L &
       \left( \begin{array}{c} c \\ s \\ \end{array} \right)_L &
       \left( \begin{array}{c} t \\ b \\ \end{array} \right)_L &
         & &
       \left( \begin{array}{c} \nu_e \\ e \\ \end{array} \right)_L &
       \left( \begin{array}{c} \nu_\mu \\ \mu \\ \end{array} \right)_L &
       \left( \begin{array}{c} \nu_\tau \\ \tau \\ \end{array} \right)_L \\
        & & \\
       u_R & c_R & t_R & & & & & \\
       d_R & s_R & b_R & & & e_R & \mu_R & \tau_R \\   
     \end{array}
  \]
  The weak-isospin assignment for the doublet is: up-type quarks (u,c,t) 
  and neutrinos carry $T_3 = +\frac{1}{2}$; down-type quarks (d,s,b),
  electron, muon and tau lepton have $T_3 = - \frac{1}{2}$.  
  In the original SM the right-handed neutrino states are omitted, since 
  neutrinos are assumed to be massless.
  Recent experimental evidence~\cite{superK,SNO2001,SNO2002}, however,
  strongly indicates that neutrinos have mass and the SM needs to be 
  extended in this respect. 

  The dynamics of the electromagnetic and weak forces follow from the free 
  particle Lagrangian density
  \begin{equation} 
    \mathcal{L}_0 = \rmi \; \overline{\bPsi} \; \gamma^\mu \partial_\mu \;\bPsi
  \end{equation}
  by demanding the invariance of $\mathcal{L}_0$ under local phase 
  transformations:
  \begin{equation}
    \bPsi_\mathbf{L}\longrightarrow\rme^{\rmi g \balpha(x) 
    \mathbf{\cdot T} + \rmi g' \beta(x) Y}\;\bPsi_\mathbf{L} 
    \ \ \ \ \ \mathrm{and} \ \ \ \ \  
    \bPsi_\mathbf{R}\longrightarrow\rme^{\rmi g' \beta(x) Y}\;\bPsi_\mathbf{R}
    \;. 
    \label{eq:gauge}
  \end{equation}
  For historical reasons these transformations are also referred to as 
  gauge transformations.
  In (\ref{eq:gauge}) the parameter $\balpha(x)$ is an arbitrary 
  three-component vector and
  $\mathbf{T} = (T_1,\, T_2,\, T_3 )^t$ is the weak-isospin operator whose 
  components
  $T_i$ are the generators of $SU(2)_L$ symmetry transformations. The index
  $L$ indicates that the phase transformations act only on left-handed states.
  The matrix representations are given by $T_i = \frac{1}{2}\; \tau_i$ where
  the $\tau_i$ are the Pauli matrices. The $T_i$ do not commute:
  $[ T_i, T_j ]=\rmi\,\epsilon_{ijk}\, T_k$. That is why the $SU(2)_L$ gauge group
  is said to be non-Abelian. 
  $\beta(x)$ is a one-dimensional function of $x$.
  $Y$ is the weak 
  hypercharge which satisfies the relation $Q = T_3 + Y/2$, where $Q$ is the
  electromagnetic charge. $Y$  is the generator of the symmetry group $U(1)_Y$.
  Demanding the Lagrangian $\mathcal{L}_0$ to be invariant under the combined 
  gauge transformations of $SU(2)_L \times U(1)_Y$, see (\ref{eq:gauge}), 
  requires the addition of terms to the free Lagrangian 
  which involve four additional vector (spin 1) fields: the isotriplet
  $\mathbf{W_\mu} = (W_{1\,\mu}\, , W_{2\,\mu}\, , W_{3\,\mu})^t$ for $SU(2)_L$ and 
  the singlet $B_\mu$ for $U(1)_Y$. This is technically done by replacing the
  derivative $\partial_\mu$ in $\mathcal{L}_0$ by the covariant derivative
  \begin{equation}
    D_\mu = \partial_\mu + \rmi\,g\,\mathbf{W_\mu \cdot T} + \rmi\, g' \frac{1}{2}
    B_\mu Y
  \end{equation}
  and adding the kinetic energy terms of the gauge fields:
  $-\frac{1}{4} \mathbf{W_{\mu\nu} \cdot W^{\mu\nu}} 
   -\frac{1}{4} B_{\mu\nu} \mathbf{\cdot} B^{\mu\nu}$. 
  The field tensors $\mathbf{W_{\mu\nu}}$ and $B_{\mu\nu}$ are given by 
  $\mathbf{W_{\mu\nu}} = \partial_\mu \mathbf{W_\nu} - \partial_\nu \mathbf{W_\mu}
   - g \cdot \mathbf{W_\mu \times W_\nu}$ and 
  $B_{\mu\nu} = \partial_\mu B_\nu - \partial_\nu B_\mu$.
  Since the vector fields $\mathbf{W_\mu}$ and $B_\mu$ are introduced via
  gauge transformations they are called gauge fields and the quanta of
  these fields are named gauge bosons.
  For an electron-neutrino pair, for example, the resulting Lagrangian
  is: 
  \begin{eqnarray}
    \mathcal{L}_1 & = & \rmi\;\overline{\left( 
      \begin{array}{c} \nu_e \\ e \end{array} \right)}_L \; \gamma^\mu 
      \left[ \partial_\mu + \rmi\, g\, \mathbf{W_\mu \cdot T} +
      \rmi\,g'\,Y_L\,\frac{1}{2}\,B_\mu \right] 
      \left( \begin{array}{c} \nu_e \\ e \end{array} \right)_L + \nonumber \\ 
      \nonumber \\
      & & \rmi \; \ebar_R\; \gamma^\mu 
      \left[ \partial_\mu - g'\,Y_R\,\frac{1}{2}\,B_\mu \right] e_R
      -\frac{1}{4} \; \mathbf{W_{\mu\nu} \cdot W^{\mu\nu}} 
      -\frac{1}{4} B_{\mu\nu} \mathbf{\cdot} B^{\mu\nu}
      \label{eq:exampleLagrangian}
  \end{eqnarray} 
  This model developed by Glashow~\cite{Glashow}, 
  Weinberg and Salam~\cite{Weinberg,Salam} in 
  the 1960s
  allows to describe electromagnetic and weak interactions in one
  framework. One therefore refers to it as unified electroweak theory. 

  \subsubsection{The Higgs mechanism.}      
  One has to note, however, that $\mathcal{L}_1$ describes only massless
  gauge bosons and massless fermions. Mass-terms such as 
  $\frac{1}{2} M^2 B_\mu B^\mu$ or $-m \overline{\bPsi}\bPsi$ are not gauge invariant
  and therefore cannot be added. To include massive particles into the model
  in a gauge invariant way the Higgs mechanism is used. Four scalar fields
  are added to the theory in form of the isospin doublet 
  $\Phibf = \left( \phi^+ , \; \phi^0 \right)^t$
  where $\phi^+$ and $\phi^0$ are complex fields. This is the minimal choice. 
  The term
  $\mathcal{L}_H = \left| D_\mu \bPhi \right|^2 - V({\bPhi}^\dag \bPhi)$
  is added to $\mathcal{L}_1$. The scalar potential takes the form
  $V({\bPhi}^\dag \bPhi)=\mu^2\;{\Phibf}^\dag \Phibf + \lambda\;
   ({\bPhi}^\dag \bPhi )^2$.
  \par
  In most cases particle reactions cannot be calculated from first principles. 
  One rather has to use perturbation theory and expand a solution starting
  from the ground state of the system which is in particle physics called
  the vacuum expectation value. The parameters $\mu$ and 
  $\lambda$ can be chosen such that the vacuum expectation value of the Higgs 
  potential $V$ 
  is different from zero: $|\bPhi_\mathrm{vac}| = \sqrt{-\frac{1}{2}\,\mu^2/\lambda}$
  and thus does not share the symmetry of $V$. 
  The scalar Higgs fields inside
  $\bPhi$ are redefined such that the new fields, 
  $\bxi (x) = (\xi_1(x), \xi_2(x), \xi_3(x) )^t$ and $H(x)$, have zero vacuum
  expectation value. When the new parameterization of $\Phibf$ is inserted
  into the Lagrangian, the
  symmetry of the Lagrangian is broken, that is, the Lagrangian is not
  an even function of the Higgs fields anymore.
  This mechanism where the ground states do not share the symmetry of
  the Lagrangian is called spontaneous symmetry breaking.
  As a result, one of the Higgs fields, the $H(x)$ field, has acquired
  mass, while the other three fields, $\bxi$, remain 
  massless~\cite{goldstone,goldstoneSalam}.

  Applying spontaneous symmetry breaking as described above to the combined
  Lagrangian $\mathcal{L}_2 = \mathcal{L}_1 + \mathcal{L}_H$ and enforcing 
  local gauge invariance
  of $\mathcal{L}_2$, makes three electroweak gauge bosons acquire mass.
  This is the aim of the whole formalism. 
  One field, the photon field, remains massless.
  The resulting boson fields after spontaneous symmetry breaking are, however, 
  not the original fields 
  $\mathbf{W_\mu}$ and $B_\mu$ but rather mixtures of those: 
  the $W^\pm_\mu = ( W^1_\mu \mp \rmi\;W^2_\mu ) / \sqrt{2}$, the $Z^0$ and the photon
  field $A_\mu$:
  \begin{equation}
    \left( \begin{array}{c} A_\mu \\ Z_\mu \end{array} \right) = 
    \left( \begin{array}{cc} \cos \theta_W & \sin \theta_W \\
                       - \sin \theta_W & \cos \theta_W   \end{array} \right)
    \ \ 
    \left( \begin{array}{c} B_\mu \\ W^{3}_\mu \end{array}  \right)
  \end{equation}      
  The mixing angle $\theta_W$ is the Weinberg angle defined by the coupling
  constants $g'/g = \tan \theta_W$. 
  While the original massless vector fields have two degrees of freedom
  (transverse polarizations), the new massive fields acquired a third degree of 
  freedom, their longitudinal polarization. The longitudinal modes of the
  $W^\pm$ and the $Z^0$ are due to the disappearance of the $\bxi$ states
  from the theory. The total number of degrees of freedom is thus conserved.

  Spontaneous symmetry breaking also generates lepton masses if Yukawa 
  in\-ter\-action terms of the lepton and Higgs fields are added to the 
  Lagrangian:
  \begin{equation}
    \mathcal{L}_\mathrm{Yukawa}^\mathrm{lepton} = - G_e \left[ \ebar_R \left(
    \mathbf{\Phi}^\dag 
    \left( \begin{array}{c} \nu_e \\ e \end{array} \right)_L \right) +
    \left( \overline{\left( \nu_e, e \right)}^t_L
    \mathbf{\Phi} \right) \ e_R 
    \right] 
  \end{equation} 
  Here the Yukawa terms for the electron-neutrino doublet are given as an
  example. $G_e$ is a further coupling constant describing the coupling
  of the electron and electron-neutrino to the Higgs field. In this 
  formalism neutrinos are assumed to be massless.

  Quark masses are also generated by adding Yukawa terms to the 
  Lagrangian. However, for the quarks both, the upper and the lower member of
  the weak-isospin doublet, need to acquire mass. For this to happen an additional
  conjugate Higgs multiplet has to be constructed: 
  $\bPhi_c = \rmi \tau_2 \bPhi^* = ( \phi^{0^*}, - \phi^-)^t$.
  The Yukawa terms for the quarks are given by:
  \begin{equation}
    \mathcal{L}_\mathrm{Yukawa}^\mathrm{quark} = 
    \sum_{i=1}^3 \sum_{j=1}^3 \ \tilde{G}_{ij} \ 
    \ubar_{iR} \tilde{\bPhi}^\dag \ 
    \left( \begin{array}{c} u_j \\ d_j \end{array} \right)_L +
    G_{ij} \  \dbar_{iR} \ \bPhi^\dag \ 
    \left( \begin{array}{c} u_j \\ d_j \end{array} \right)_L + 
    \ \mathrm{h.c.}
  \end{equation}
  The $u_j$ and $d_j$ are the weak eigenstates of the up-type ($u,\,c,\,t$)
  and the down-type ($d,\,s,\,b$) quarks, respectively.
  Couplings between quarks of different generations are allowed by this ansatz.   
  After spontaneous symmetry breaking the Yukawa terms produce mass terms for
  the quarks which can be described by mass matrices in generation space:
  $\overline{ ( u_1, u_2, u_3) }_R \;\mathcal{M}^u \;( u_1, u_2, u_3)^t_L$
  and
  $\overline{ ( d_1, d_2, d_3) }_R \;\mathcal{M}^d \;( d_1, d_2, d_3)^t_L$
  with $\mathcal{M}_{ij}^u = |\bPhi_\mathrm{vac}|\cdot\tilde{G}_{ij}$
  and  $\mathcal{M}_{ij}^d = |\bPhi_\mathrm{vac}|\cdot G_{ij}$.
  The mass matrices are non-diagonal but can be diagonalized by unitary
  transformations, which essentially means to change basis from
  weak eigenstates to mass eigenstates, which are identical to the flavour 
  eigenstates
  $u$, $c$, $t$ and $d$, $s$, $b$. 
  In charged-current interactions ($W^\pm$ exchange) this leads to
  transitions between mass eigenstates of different generations
  referred to as generation mixing.
  It is possible to set weak and mass eigenstates equal for the 
  up-type quarks and ascribe the mixing entirely to the down-type quarks:
  \begin{equation}
    \left( \begin{array}{ccc} d' \\ s' \\ b' \end{array} \right)_L =
    \mathbf{V} \ 
    \left( \begin{array}{ccc} d \\ s \\ b \end{array} \right)_L 
    = \left( \begin{array}{ccc} V_{ud} & V_{us} & V_{ub}\\ V_{cd} & V_{cs}&V_{cb} \\
      V_{td} & V_{ts} & V_{tb} \\ \end{array} \right)
    \left( \begin{array}{ccc} d \\ s \\ b \end{array} \right)_L          
  \end{equation}
  where $d'$, $s'$ and $b'$ are the weak eigenstates. The mixing matrix
  $\mathbf{V}$ is called the Cabbibo-Kobayashi-Maskawa (CKM) 
  matrix~\cite{KM}.

  \subsubsection{Strong interactions.}
  The theory of strong interactions is called quantum chromodynamics (QCD)
  since it attributes a colour charge to the quarks. 
  There are three different types of strong charges (colours):
  ``red'', ``green'' and ``blue''.
  Strong interactions conserve the flavour of quarks.
  Leptons do not carry colour at all, they are inert with respect to strong
  interactions.
  QCD is a quantum field theory based on the non-Abelian gauge group $SU(3)_C$
  of phase transformations on the quark colour fields.
  Invoking local gauge invariance of the Lagrangian yields eight massless
  gauge bosons: the gluons. 
  The gauge symmetry is exact and not broken as in the case of weak 
  interactions.
  Each gluon carries one unit of colour and one unit of anticolour.
  The strong force binds quarks together to form bound-states called 
  hadrons. There are two groups of hadrons: mesons consisting of a quark
  and an antiquark and baryons built of either three quarks or three
  antiquarks. All hadrons are colour-singlet states. Quarks cannot exist
  as free particles. This experimental fact is summarised in the notion
  of quark confinement: quarks are confined to exist in hadrons.
  
  \subsection{Model predictions of top quark properties}
  In 1975 the tau lepton was the first particle of the third generation to be 
  discovered~\cite{perl1975}. Only two years later, in 1977, a new heavy meson,
  the $\Upsilon$, was discovered~\cite{herb1977}. 
  It was quickly recognised to contain a new
  fifth quark, the $b$ quark ($\Upsilon = b\bbar$).
  After these discoveries the doublet structure of the SM strongly suggested
  the existence of a third neutrino associated with the tau lepton and the
  existence of a sixth quark, called the top quark.
  The properties of the $b$ quark and $b$ hadrons were subject to extensive
  experimental research in the past 25 years. As a result, charge and 
  weak isospin of the $b$ quark are well established quantities: 
  $Q_b = -\frac{1}{3}$ and $T_3 = - \frac{1}{2}$.
  The charge $Q_b$ was first deduced from measurements of the cross section 
  $\sigma_\mathrm{had}$ for hadron production on the 
  $\Upsilon$ resonance at the DORIS $e^+e^-$ storage 
  ring~\cite{berger1978,bienlein1978,darden1978}.
  The integral over $\sigma_\mathrm{had}$ is related to the leptonic width 
  $\Gamma_{ee}$ of the $\Upsilon$:
  $\int \sigma_\mathrm{had}\, \rmd M = 6\,\pi^2 \Gamma_{ee} / M_\Upsilon^2$
  (using the approximation that $\Gamma_\mathrm{had}\approx\Gamma_\mathrm{total}$).
  Nonrelativistic potential models of the $\Upsilon$ relate $\Gamma_{ee}$
  to the charge of the constituent $b$ quarks. 
  \par
  The weak isospin of the $b$ quark was obtained from the measurement of the
  forward-backward asymmetry $A_\mathrm{FB}$ in $e^+e^-\rightarrow b\bar{b}$ 
  reactions. The asymmetry is defined as the difference in rate of $b$ quarks
  produced in the forward hemisphere (polar angle $\theta > 90^\circ$) 
  minus the rate of $b$ quarks produced in the backward hemisphere 
  ($\theta < 90^\circ$) over the sum of the two rates.
  In 1984 the JADE experiment measured 
  $A_\mathrm{FB}=(-22.8\pm6.0\pm2.5)\%$~\cite{jadeAFB}
  in excellent agreement with the value of -25.2\% predicted for a $b$ 
  quark that is the lower member of a SM weak isospin doublet.
  In case of a weak isospin singlet the asymmetry would be zero.    
  The assignment of quantum numbers for the $b$ quark 
  allows to predict the charge and the weak isospin of
  the top quark to be: $Q_t = + \frac{2}{3}$ and $T_3 = + \frac{1}{2}$.

  Another argument which supported the existence of a complete third quark 
  generation comes from perturbation theory. In particle physics the terms of 
  a perturbation series are depicted in Feynman diagrams. The first order 
  terms are pictured as tree level diagrams. Higher order terms correspond
  to loop diagrams. Certain loop diagrams are divergent. These divergencies
  can only be overcome by summing up over several divergent terms in a
  consistent manner and have the divergencies cancel each other. This formalism
  is called renormalization. For electroweak theory to be renormalizable,
  the sum over fermion triangle diagrams, such as the one shown in 
  figure~\ref{fig:radCorrections}a, has to vanish~\cite{thooft71a,thooft71b,leader1982}. 
  With a third lepton doublet in place the cancellation only occurs if also   
  a complete third quark doublet is there.
  While the SM predicts the charge and the weak isospin of the top quark,
  its mass remains a free parameter. 

\subsection{Top quark mass and electroweak precision measurements} 
  \label{sec:ewktopdetermine}
  Even though the top quark mass is not predicted by the SM it enters as a parameter
  in the calculation of radiative corrections to electroweak processes.
  With highly precise measurements at hand it is therefore possible to 
  indirectly determine the top quark mass from those 
  processes~\cite{Langacker:1989sm,Hollik:1990ii}.
  In this context radiative corrections denote higher order contributions to a
  perturbation series, e.g. to calculate cross sections of electroweak
  processes. 

  The most precise electroweak measurements are available from $e^+e^-$
  colliders operating at the $Z^0$ pole, where $\sqrt{s} = M_Z$. Here  
  $\sqrt{s}$ is the centre-of-mass energy of the colliding electrons and
  positrons. $M_Z$ is the mass of the $Z^0$ boson. 
  In the 1990s there were two particle colliders operating on the $Z^0$ pole: 
  the Large Electron Positron Collider (LEP)~\cite{myers1990} 
  at CERN with four experiments 
  (ALEPH~\cite{Decamp:1990jr,Buskulic1995}, 
   DELPHI~\cite{delphiDet1991,delphiDet1996}, L3~\cite{l3Det1994a,l3Det1996a} 
   and OPAL~\cite{Ahmet:1991eg}) 
  and the Stanford Linear Collider (SLC)~\cite{bRichter} 
  with the SLD~\cite{sldDesign,sldVXD3} experiment. 
  The LEP experiments collected about
  4 million $Z^0$ events each, SLD on the order of 0.5 million events.
  Although the SLD sample is considerably smaller, the experiment reached 
  a competitive sensitivity for many measurements, since at SLC it was possible
  to polarize the colliding electrons and positrons.

\begin{figure}[!t]
  (a) \hspace*{45mm} (b) \hspace*{45mm} (c)\\
  \includegraphics[width=0.28\textwidth]{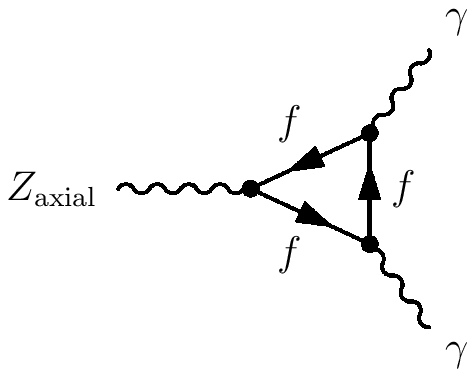}
  \hspace{0.05\textwidth}
  \includegraphics[width=0.28\textwidth]{feynEpsi/ZpropattbarLoopPict.epsi} 
  \hspace*{0.04\textwidth}
  \includegraphics[width=0.28\textwidth]{feynEpsi/ZttWbbtrianglePict.epsi} 
  \caption{\label{fig:radCorrections}
    (a) An example of a fermion triangle diagram.
    (b,c) Two example Feynman diagrams of radiative corrections 
    involving the top quark. (b) is a radiative correction 
    to the $Z^0$ propagator. (c) shows a vertex 
    correction to the decay $Z^0 \rightarrow b\bbar$.}    
\end{figure}

  Two examples of radiative corrections to the $Z^0$ propagator and to the $Z^0$
  vertex involving the top quark are given in figure~\ref{fig:radCorrections}b 
  and~\ref{fig:radCorrections}c.
  The top mass plays a particular large role in the corrections represented
  by these diagrams due to the large mass difference between the top quark 
  and its weak-isospin partner, the $b$ quark. The correction terms introduce
  a quadratic dependence on the top mass, whereas the dependence on the
  Higgs mass is only logarithmic.
  The LEP electroweak working group (LEPEWWG) has combined 
  the measurements of the four LEP experiments to obtain a common data set.
  To test SM predictions and check the overall consistency of the data with the
  model the SLD data and Tevatron measurements in $p\pbar$ collisions
  are included~\cite{lepewwg2004}. 
  Among the results of the fits is an indirect determination 
  of the top mass and indirect limits on the Higgs mass. 
  The quantities used for a constraint fit to the SM are discussed below:
  \begin{enumerate}
    \item The mass $M_Z$ and the total width $\Gamma_Z$ of the $Z^0$. 
      The definition of $M_Z$ and
      $\Gamma_Z$ is based on the Breit-Wigner denominator of the electroweak
      cross section for fermion pair ($f\bar{f}$) production due to $Z^0$
      exchange: 
      $\sigma_{f\bar{f}}(s)=\sigma^0_f\cdot s \Gamma_Z^2/
       ((s-M_Z^2)^2+ (s \Gamma_Z/M_Z)^2)$
      where 
      $\sigma_f^0$ is the pole cross section at $\sqrt{s}=M_Z$.
    \item The hadronic pole cross section: 
      \[ \sigma_h^0 = \frac{12\;\pi}{M_Z^2}\cdot
         \frac{\Gamma_{ee}\,\Gamma_\mathrm{had}}{\Gamma_Z^2} \]
      where $\Gamma_{ee}$ and $\Gamma_\mathrm{had}$ are the partial widths for the 
      $Z^0$ decaying into electrons or hadrons, respectively.
    \item The ratio of the hadronic and leptonic widths at the $Z^0$ pole:
      $R_\ell^0 \equiv \Gamma_\mathrm{had}/\Gamma_{\ell\ell}$.
      $\Gamma_{\ell\ell}$ is obtained from $Z^0$ decays into $e^+\,e^-$, 
      $\mu^+\,\mu^-$
      or $\tau^+\,\tau^-$ assuming lepton universality 
      ($\Gamma_{ee} = \Gamma_{\mu\mu} = \Gamma_{\tau\tau}\equiv\Gamma_{\ell\ell}$)
      which is only correct for massless leptons. To account for the tau mass
      $\Gamma_{\tau\tau}$ is corrected, in average by $0.2\%$.
    \item The pole forward-backward asymmetry $A_{\mathrm{FB}}^{0,\ell}$ 
      ($\ell \in \{e, \mu, \tau \}$) for $Z^0$ decays 
      into charged leptons. 
      The asymmetry is defined as the total cross section for a negative
      lepton to be scattered into the forward direction 
      ($90^\circ < \theta \leq 180^\circ$) minus the total cross section for 
      a negative lepton to be scattered into the backward direction
      ($0^\circ \leq \theta \leq 90^\circ$) divided by the sum.
      \[ A_{\mathrm{FB}}^{0,l} = \frac{\sigma_{\mathrm{F}}(\ell^-, 90^\circ < \theta
         \leq 180^\circ) - \sigma_{\mathrm{B}}(\ell^-, 0^\circ \leq \theta 
         \leq 90^\circ)}{\sigma_{\mathrm{F}}(\ell^-, 90^\circ < \theta
         \leq 180^\circ) + \sigma_{\mathrm{B}}(\ell^-, 0^\circ \leq \theta 
         \leq 90^\circ)} \]
      Here $\theta$ is the angle between the produced lepton and the incoming
      electron. The forward-backward asymmetry originates from parity 
      violating terms in the total cross section.
      $A_{\mathrm{FB}}^{0,\ell}$ can be written as a combination of effective 
      couplings 
      $A_{\mathrm{FB}}^{0,\ell} = \frac{3}{4}\cdot\mathcal{A}_e\,\mathcal{A}_\ell$
      with 
      $\mathcal{A}_\ell=(2\;g_{V,\ell}\,g_{A,\ell})/(g_{V,\ell}^2 + g_{A,\ell}^2)$
      where $g_{V,\ell}$ and $g_{A,\ell}$ are the vector and axial couplings,
      respectively. 
    \item The effective leptonic coupling $A_\ell$ derived from the 
      tau polarization
      measurement. Parity violation in the weak interaction leads to 
      longitudinally polarized final state fermions from the $Z^0$ decay.
      In case of the decay $Z^0 \rightarrow \tau^+ \tau^-$ this polarization
      can be measured via the subsequent parity violating decays of the
      tau as polarimeter. The polarization $P_\tau$ is defined in terms of
      the cross section for the production of right-handed and left-handed
      $\tau^-$, respectively: 
      $P_\tau = (\sigma_R - \sigma_L)/(\sigma_R + \sigma_L)$.
      Averaging $P_\tau$ over all production angles $\theta$ gives a measure
      of the effective coupling: $\langle P_\tau \rangle  =  - \mathcal{A}_\tau$.
    \item The effective leptonic weak mixing angle 
      $\sin^2 \theta^\mathrm{lept}_{\mathrm{eff}}$ from
      the inclusive hadronic charge asymmetry $\langle Q_{\mathrm{FB}} \rangle$ 
      which
      is measured from $Z^0\rightarrow q\qbar$ decays. An estimator for
      the quark charge is derived from the sum of momentum weighted track 
      charges in the quark and antiquark hemispheres, respectively.
    \item The effective leptonic coupling $\mathcal{A}_\ell$ derived from
      left-right asymmetries measured with longitudinally polarized 
      electron and positron beams at SLD. The left-right asymmetry 
      $A_\mathrm{LR}$ is formed as the number of $Z^0$ bosons $N_L$
      produced by left-handed electrons minus the 
      number of $Z^0$ bosons $N_R$ produced by right-handed electrons:
      \[ A_\mathrm{LR} = \frac{N_L - N_R}{N_L+N_R} \cdot \frac{1}{P_e} \]
      Additionally, one divides by the luminosity weighted polarization
      $P_{e}$ of the electron beam. $A_\mathrm{LR}$ is measured for all
      three  charged lepton final states. The pole values $A_\mathrm{LR}^0$
      are equivalent to the effective couplings $\mathcal{A}_e$, 
      $\mathcal{A}_\mu$ and $\mathcal{A}_\tau$, respectively. Assuming lepton
      universality the results are combined to form a single value $A_\ell$.
    \item The measurement of the top mass by CDF and 
      \DZero~\cite{mtopCombined2004}.
    \item The ratios of $b$ and $c$ quark partial widths of the $Z^0$ with 
      respect to its total hadronic partial width: 
      $R_b^0 \equiv \Gamma_{b\bbar} / \Gamma_\mathrm{had}$ and 
      $R_c^0 \equiv \Gamma_{c\cbar} / \Gamma_\mathrm{had}$.
    \item The forward-backward asymmetries for the decays 
      $Z^0 \rightarrow b\bbar$ and $Z^0 \rightarrow c\cbar$:
      $A_\mathrm{FB}^{0,b}$ and $A_\mathrm{FB}^{0,c}$, defined analogously
      to the leptonic asymmetries.
    \item The effective couplings for $b$ and $c$ quarks: 
      $\mathcal{A}_b$ and $\mathcal{A}_c$, defined analogously to 
      $\mathcal{A}_\ell$.
    \item The mass and width of the $W$ boson from the combination of measurements 
      in $p\pbar$ collisions by UA2, CDF and \DZero and measurements at 
      LEP 2. 
  \end{enumerate} 
  The quantities describing the decay $Z^0 \rightarrow b\bbar$ ($R_b^0$, 
  $A_\mathrm{FB}^{0,b}$ and $\mathcal{A}_b$) exhibit a
  particular strong dependence on $\mtop$ because they are sensitive to weak
  vertex corrections as the one shown in figure~\ref{fig:radCorrections}c.
  In many cases vertex corrections involving a $W$  boson are suppressed
  due to small CKM matrix elements. The vertices 
  in figure~\ref{fig:radCorrections}, however, contain the matrix element 
  $V_{tb}$ which is close to 1. Therefore, the graphs lead to significant
  corrections depending on the top quark mass. 

  The measured values of the above listed quantities are given in 
  \renewcommand\arraystretch{1.3}
  table.~\ref{tab:fitinputs}. 
  \begin{table}[tb]
    \begin{tabular}{lrrlrr}
      \br
      \begin{minipage}{20mm} Physical \\ quantity \end{minipage} & 
        \begin{minipage}{29mm} Measurement \\ with total error \end{minipage} & 
        SM fit &
        \begin{minipage}{15mm} Physical \\ quantity \end{minipage} & 
        \begin{minipage}{29mm} \vspace*{1mm} Measurement \\ with 
          total error \vspace*{2mm} 
          \end{minipage} & 
        SM fit \\ \mr
      \multicolumn{3}{c}{LEP measurements} & 
        \multicolumn{3}{c}{LEP and SLD heavy flavour} \\
      $M_Z$ [GeV/$c^2$] & 91.1875 $\pm$ 0.0021 & 91.1875 & 
        $R_b^0$ & 0.21630 $\pm$ 0.00065 & 0.21562 \\
      $\Gamma_Z$ [GeV/$c^2$] & 2.4952 $\pm$ 0.0023 & 2.4966 &
        $R_c^0$ & 0.1723 $\pm$ 0.0031\OneDig & 0.1723\OneDig \\
      $\sigma_h^0$ [nb] & 41.540 $\pm$ 0.037\OneDig & 41.481\OneDig &
        $A_\mathrm{FB}^\mathrm{0,\,b}$ & 0.0998 $\pm$ 0.0017\OneDig & 
        0.1040\OneDig \\
      $R_\ell^0$ & 20.767 $\pm$ 0.025\OneDig & 20.739\OneDig & 
        $A_\mathrm{FB}^\mathrm{0,\,c}$ & 0.0706 $\pm$ 0.0035\OneDig & 
        0.0744\OneDig \\
      $A_{\mathrm{FB}}^{0, \ell}$ & 0.0171 $\pm$ 0.0010 & 0.0165 & 
        $\mathcal{A}_b$ & 0.923 $\pm$ 0.020\OneDig\OneDig & 
        0.935\OneDig\OneDig \\
      $\mathcal{A}_\ell \ ( \langle P_\tau \rangle)$ & 0.1465 $\pm$ 0.0033 & 
        0.1483 & 
        $\mathcal{A}_c$ & 0.670 $\pm$ 0.026\OneDig\OneDig & 
        0.668\OneDig\OneDig \\
      $\sin^2 \theta_{eff}^{\mathrm{lep}} \ (\langle Q_\mathrm{FB} \rangle)$ & 
        0.2324 $\pm$ 0.0012 & 0.2314 & & & \\ \hline 
      \multicolumn{3}{c}{SLD only measurements} & 
      \multicolumn{3}{c}{UA2, Tevatron and LEP2 measurements} \\
      $\mathcal{A}_\ell$ & 0.1513 $\pm$ 0.0021 & 0.1483 &
      $M_W$ [GeV/$c^2$] & 80.425 $\pm$ 0.034\OneDig\OneDig &  80.394\OneDig\OneDig \\
      \cline{1-3}
      \multicolumn{3}{c}{Tevatron only measurements} &
      $\Gamma_W$ [GeV/$c^2$] & 2.133 $\pm$ 0.069\OneDig\OneDig & 
      2.093\OneDig\OneDig \\
      $M_\mathrm{top}$ [GeV/$c^2$] & 178.0 $\pm$ 4.3\OneDig\OneDig\OneDig\OneDig  & 
      178.1\OneDig\OneDig\OneDig\OneDig & & & \\ \br
    \end{tabular}
    \caption{\label{tab:fitinputs} Experimental input to overall fits to the 
       SM by the LEP 
       electro-weak working group~\cite{lepewwg2004}. The experimental 
       values of the physical quantities, which are described in detail in the 
       text, are given with the total error. 
       The results of the constraint fit to all data is given in the columns
       named ``SM fit''. In the fit the Higgs mass is treated as a free 
       parameter.}    
  \end{table} 
  Other input quantities not shown in the table
  are the Fermi constant $G_F$, the electromagnetic coupling constant
  at the $Z^0$ mass scale, $\alpha(M_Z^2)$, and the fermion masses. 
  The Higgs mass and the strong
  coupling constant at the $Z^0$ mass scale, $\alpha_s(M_Z^2)$, are treated 
  as free parameters in the fits. Detailed reviews of
  electroweak physics at LEP are, for instance, given in 
  references~\cite{dSchaile,gQuast}.

\begin{figure}[tb]
    (a) \hspace*{70mm} (b) \\
      \resizebox{0.48\textwidth}{!}
      {\includegraphics[0pt,35pt][566pt,559pt]{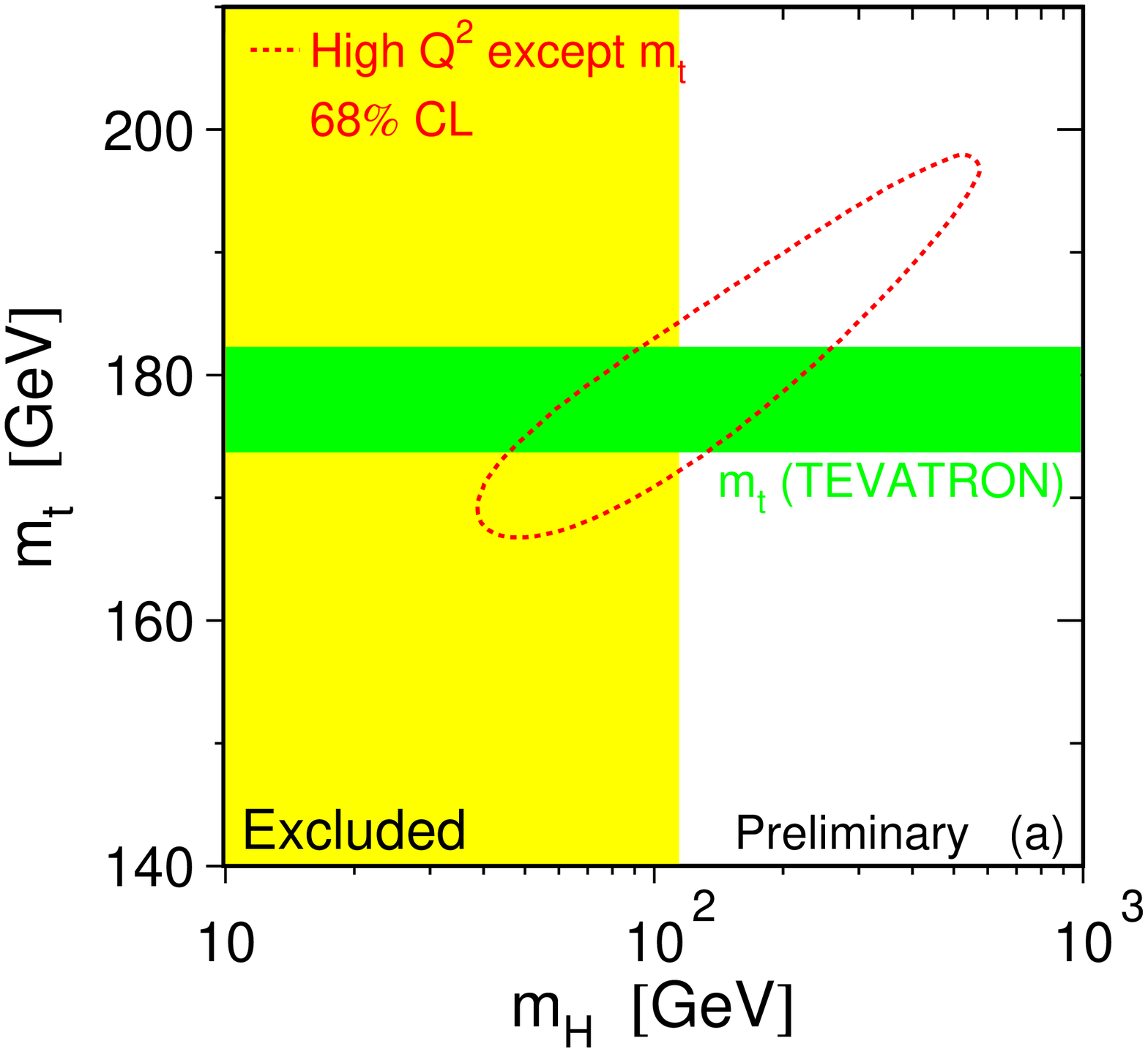}}
      \hspace*{0.02\textwidth}
      \resizebox{0.48\textwidth}{!}
      {\includegraphics[0pt,35pt][566pt,559pt]{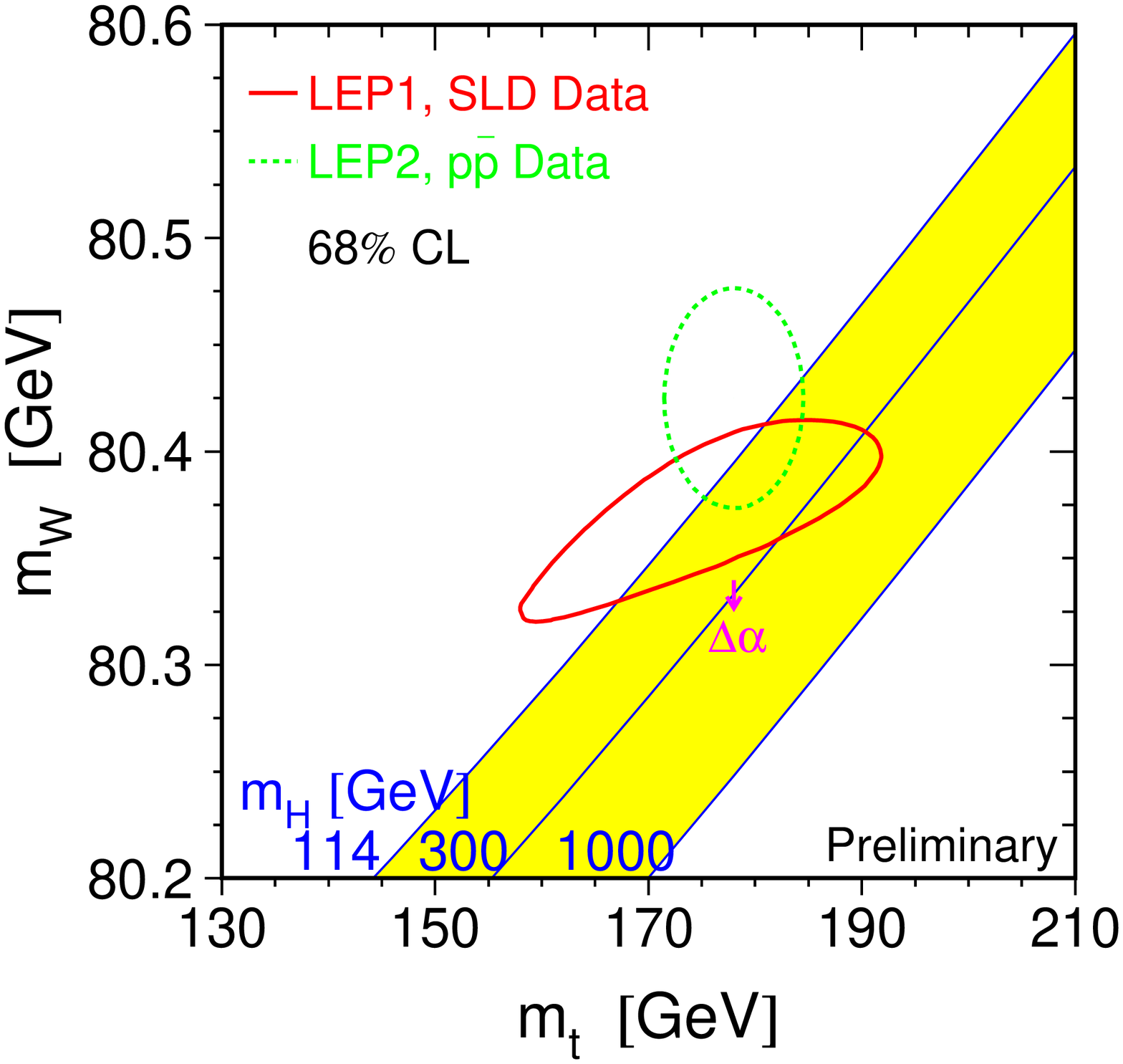}}
    \caption{Results of fits to electroweak data~\cite{lepewwg2004}. Direct
      Higgs boson searches at LEP2 and the top mass measurement at the 
      Tevatron are summarised in these plots. 
      (a) shows the plane ``top mass versus Higgs mass''. 
      The light shaded area
      is excluded by direct searches for the Higgs boson at LEP2. 
      The dark shaded area
      is given by the central value of the direct top mass measurement at the
      Tevatron and its errors. The dashed line encloses the region preferred by 
      the electroweak data at 68\% confidence level.
      (b) shows the ``$M_W$ versus $\mtop$'' plane. 
      The full line encloses the area preferred by the SM fit to data from 
      LEP1 and SLD.
      The dashed contour indicates the result of the LEP2, UA2 and Tevatron 
      $W$ mass measurements and the direct top quark mass measurement.
      The plot also shows the SM relationship of the masses as function
      of the Higgs boson mass.        
      }     
    \label{fig:MtopMHiggs}
  \end{figure}

  If the top mass is left floating in the fit it is possible to obtain a prediction
  for $M_\mathrm{top}$ from the model based on all other measurements. Comparing this
  prediction with the direct measurement from CDF and \DZero is an important
  cross-check of the SM. The fit yields a value of 
  $\mtop = 179^{+12}_{-9}$~GeV/$c^2$ which is in very good agreement with the 
  measured value of $\mtop = (178.0\pm4.3)$~GeV/$c^2$. The result of the fit is
  depicted in figure~\ref{fig:MtopMHiggs}a where $\mtop$
  is plotted versus $M_{H}$. 
  The plot shows that the overlap region of the
  direct measurement with the indirect determination prefers a low value for 
  the Higgs mass. It is obvious from the diagram that there is a large
  positive correlation, about 70\%, between the top quark and the Higgs boson mass.

  The constraint on the Higgs mass becomes significantly stronger if 
  the measured value for $\mtop$ 
  is included
  in the SM fit. Only $M_H$ and $\alpha_S(M_Z^2)$ are free parameters in this
  case. The resulting value for the Higgs mass is:
  $M_H = 114^{+69}_{-45}$~GeV/$c^2$, a value close to the exclusion limit 
  obtained by direct searches at LEP2 which yield 
  $M_H > 114.4$~GeV/$c^2$~\cite{lepHiggsFinal}. 
  The SM fit 
  (not using the direct search result)
  yields an upper limit of $M_H < 260$~GeV/$c^2$ at the 95\% confidence level.
  The SM prediction for all observables resulting from the constrained
  fit is given in table~\ref{tab:fitinputs}.

\begin{figure}[tb]
    \begin{center}
      \includegraphics[width=0.2\textwidth]{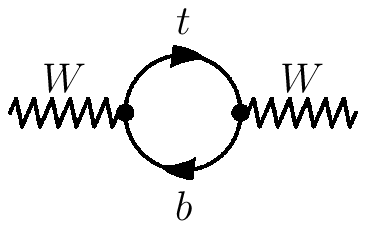}
      \hspace*{15mm}
      \includegraphics[width=0.2\textwidth]{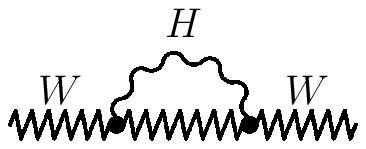}
    \end{center}
    \caption{Examples of loop diagrams of the $W$ propagator introducing
      a dependence of the $W$ mass on the top and Higgs mass.}
    \label{fig:mWloops}  
  \end{figure}

  \renewcommand\arraystretch{1.0} 
  The last fit discussed here is the one made with $M_W$ and $\mtop$ left as
  free parameters. The result is shown in figure~\ref{fig:MtopMHiggs}b which also 
  includes SM predictions for Higgs
  masses from 114 to 1000~GeV/$c^2$. The direct measurement of $M_W$ and
  $\mtop$ are in fair agreement with the indirect determination. A low
  Higgs mass is preferred.
  The dependence of the $W$ mass on the top and Higgs mass is introduced
  via loop diagrams like the ones shown in figure~\ref{fig:mWloops}.  
  As becomes clear from figure~\ref{fig:MtopMHiggs} increasing
  the precision of the direct measurements of $M_W$ and $\mtop$ is a
  precondition to gain higher 
  leverage for SM predictions of the Higgs mass.

  Summarising this section on electroweak precision measurements we point
  out two major issues with respect to the top quark. 
  First, the very good agreement of the predicted value for $\mtop$
  based on higher order electroweak corrections with the direct measurement
  is a significant success of the SM. Second, more precise measurements
  of $\mtop$ and $M_W$ are needed to obtain tighter constraints on the 
  Higgs boson mass. This is one of the major physics motivations for the Run II
  of the Tevatron.
   
\subsection{The top quark in flavour physics}
  Flavour physics describes the transitions between quarks of different 
  flavour. These transitions always involve the exchange of a W boson.
  They are charged current interactions. Higher order transitions of this kind
  involve
  loops in which particles can occur that are much heavier than the hadrons
  involved in the interaction. 
  Diagrams with down-type quarks ($d$, $s$ and $b$) in the loops cancel each 
  other via the GIM
  mechanism very effectively, since they are nearly degenerate in their masses.
  The large degree of mass splitting in the up-quark sector prevents the 
  cancellation and leads to sizeable loop contributions, e.g. in radiative
  $B$ meson decays.
  One other example where the top quark plays a 
  prominent r\^{o}le is the mixing of neutral $B$ mesons.

  The phenomenon of particle-antiparticle mixing has been experimentally
  estab\-lished in the neutral kaon system ($K^0$--$\overline{K^0}$)
  and the $B^0_d$--$\overline{B^0_d}$ system~\cite{bdMixUA1,bdMixArgus}.
  Mixing is also expected to take place in the $D^0$--$\overline{D^0}$
  and $B^0_s$--$\overline{B^0_s}$ system, but was not observed yet.
  Only limits were 
  set~\cite{alephBsMix2003,delphiBsMix2003,opalBsMixing2001,sldBs,cdfBsMixing99,D0FOCUS1,D0FOCUS2,D0CLEO,D0Babar}.   

  Mixing changes the flavour quantum number (strange, charm, bottom) 
  of the mesons by two units ($\Delta S = 2$, $\Delta C = 2$, $\Delta B = 2$).
  A neutral meson is produced in a well defined flavour eigenstate
  $| P^0 \rangle$ or $|\overline{P^0}\rangle$ with 
  $P \in \{ K, D, B_d, B_s \}$. This initial state
  evolves due to second order weak interactions into a time-dependent
  quantum superposition of the two flavour eigenstates:
  $|P(t)\rangle = a(t)\; |P^0\rangle + b(t)\;|\overline{P^0}\rangle$.
  The time-evolution of the coefficients $a(t)$ and $b(t)$ is given
  by the effective Hamiltonian matrix 
  $\mathbf{H} = \mathbf{M} - \rmi/2\;\mathbf{\Gamma}\ $:
  \begin{equation}
    \rmi \ \frac{\partial}{\partial t} \ \left( \begin{array}{c} a(t) \\ b(t)
    \end{array} \right) = \left( \mathbf{M} - \frac{\rmi}{2}\; \mathbf{\Gamma} \right)
    \ \left( \begin{array}{c} a(t) \\ b(t) \end{array} \right)  
  \end{equation}  
  $\mathbf{M}$ and $\mathbf{\Gamma}$ are $2\times2$ Hermitian matrices
  denoted as mass and decay matrix, respectively. CPT invariance requires
  $M_{11}=M_{22}=M$ and $\Gamma_{11}=\Gamma_{22}=\Gamma$, where M and 
  $\Gamma$ are the mass and the decay width of the flavour eigenstates.

\begin{figure}[!t]
  \begin{center}
  \includegraphics[width=0.32\textwidth]{feynEpsi/bmixing1Pict.epsi}
  \hspace*{10mm}
  \includegraphics[width=0.32\textwidth]{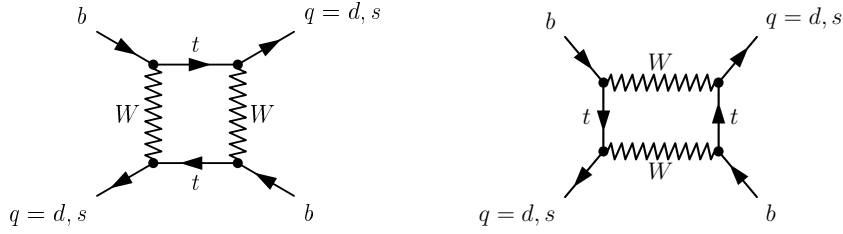}
  \end{center}
    \caption{\label{fig:BmixingBox} Lowest order Feynman diagrams for 
      B meson mixing. A $\overline{B^0}\;(b \dbar)$ or a 
      $\overline{B^0_s}\;(b \sbar)$ oscillates into a $B^0\;(\bbar d)$ or a
      $B^0_s\;(\bbar s)$, respectively.}
  \end{figure}

  In case of the B meson systems ($B^0_d$--$\overline{B^0_d}$ and
  $B^0_s$--$\overline{B^0_s}$) the transition amplitude for the mixing
  matrix is dominated by box diagrams involving the top quark, see
  figure~\ref{fig:BmixingBox}.
  All up-type quarks ($q=u, c, t$) can be exchanged in the box and contribute 
  in general to the mixing amplitude. 
  In the case of $B$ mesons it is, however, found that the mixing is strongly
  dominated by the diagrams with two top quarks in the loop as shown in 
  figure~\ref{fig:BmixingBox}. All other diagrams involving at least one
  u- or c-quark can be neglected with respect to the top-top 
  diagram~\cite{cheng1982,chau1983,buras1984}. 
  As a side remark: In the neutral kaon system the case is different. 
  Charm and top quark box diagrams as well as intermediate virtual pion states 
  contribute significantly to the mixing. For $B$ mixing the off-diagonal elements
  of $\mathbf{M}$ and $\mathbf{\Gamma}$,
  $\langle B^0_q | H_\mathit{eff}^{\Delta B = 2} | \overline{B^0_q}\rangle = 
    M_{12} - i/2\;\Gamma_{12}$ ($q \in \{d, s\}$),
  are in very good approximation given 
  by~\cite{cheng1982,buras1984,PDG2004}:
  \begin{eqnarray}
  M_{12, q} & = & - \frac{G_F^2 M^2_W }{12\;\pi^2} \ m_{B_q} f^2_{B_q} B_{B_q}
    \eta_{B} \cdot \left( V^*_{tq}\,V_{tb} \right)^2 \cdot 
    S_0(x_t) \\
  \Gamma_{12, q} & = & \frac{G_F^2}{8 \;\pi} \ m_b^2 m_{B_q} f^2_{B_q} B_{B_q}
    \eta^\prime_B \cdot \left[ \left( V^*_{tq} V_{tb} \right)^2 +
    V^*_{tq} V_{tb} V^*_{cq} V_{cb} \ \mathcal{O} 
    \left( \frac{m_c^2}{m_b^2} \right) + \right. \nonumber \\
    & & \left. \left( V_{cq}^* V_{cb} \right)^2
    \ \mathcal{O} \left(\frac{m_c^4}{m_b^4} \right) \right]
  \end{eqnarray}
  where $G_F$ is the Fermi constant and $M_W$ the W mass. $m_b$ and $m_c$
  are the $b$ quark and $c$ quark masses, respectively.  $m_{B_q}$ represents the 
  masses of the $B_d^0$ and $B^0_s$ mesons. $f_{B_q}$ is the decay constant 
  and $B_{B_q}$ the bag parameter introduced as a correction factor to 
  hadronic matrix elements.
  The $V_{ij}$ are the CKM matrix elements. $x_t = (\mtop/M_W)^2$ is
  the squared ratio of top quark mass over W mass. The function $S_0(x_t)$
  is an Inami-Lim function~\cite{inami} 
  and can be well approximated by
  $S_0(x_t) = 0.784\; x_t^{0.76}$~\cite{burasFleischer}. 
  The parameters $\eta_B$ and $\eta^\prime_B$ represent QCD corrections to
  the box diagram. 
  The mass eigenstates $B_h$ (h for heavy) and $B_l$ 
  (l for light) diagonalize the effective Hamiltonian $H$ and are given
  by: $|B_h \rangle= p\,| B^0 \rangle - q\, | \overline{B^0} \rangle$
  and $|B_l \rangle= p\,| B^0 \rangle + q\, | \overline{B^0} \rangle$
  with $q^2 = M_{12}^* - \rmi/2\,\Gamma^*_{12}$ and
  $p^2 = M_{12} - \rmi/2\,\Gamma^*_{12}$. Mixing experiments do not determine 
  the masses $m_h$ and $m_l$ but rather the
  mass difference between the two $\Delta m = m_h - m_l$ which is
  in good approximation (about 1\% accuracy) theoretically predicted to be
  $\Delta m = 2\, | M_{12}|$~\cite{buras1984}. This quantity depends
  on the top quark mass via $\eta_B S_0(x_t)$. 
  To give a flavour of the dependency: $\eta_B S_0(x_t)$ changes from
  roughly 1.1 for $\mtop = 150$~GeV to 1.7 at 
  $\mtop = 200$~GeV~\cite{buras1990}.
  Historically, the ARGUS measurement of $B^0_d$--$\overline{B^0_d}$ mixing yielded a
  mass difference $\Delta m_d$ found to be surprisingly large~\cite{bdMixArgus}.
  Using the dependence of $\Delta m_d$ on the top quark mass several authors 
  interpreted this measurement as a hint of a large top quark 
  mass~\cite{ellis1987,bigi1987,barger1987,Datta1987,Du1987}. 
   
  The dependence on $M_\mathrm{top}$ drops out if the ratio
  \begin{equation}
    \frac{\Delta m_s}{\Delta m_d} = 
    \frac{m_{B_s}  f^2_{B_s} B_{B_s}}{m_{B_d}  f^2_{B_d} B_{B_d}}
    \cdot \left| \frac{V_{ts}}{V_{td}} \right|^2
  \end{equation}
  is taken. Once $\Delta m_s$ is measured in $B_s$ mixing this relation 
  will allow to extract the absolute value of the ratio $V_{ts}/V_{td}$
  with good precision.

\section{Top quark production at hadron colliders}
\label{sec:topProd}
In this chapter we present the phenomenology of top quark production 
at hadron colliders.
We limit the discussion to SM processes.
Anomalous top quark production and non-SM decays will be 
covered in chapter~\ref{sec:anotop}.
Specific theoretical cross section predictions refer to the 
Fermilab Tevatron, running at $\sqrt{s} = 1.8\;\mathrm{TeV}$ (Run I) or 
 $\sqrt{s} = 1.96\;\mathrm{TeV}$ (Run II), or to
the future Large Hadron Collider (LHC) at CERN ($\sqrt{s} = 14\;\mathrm{TeV}$).
In the intermediate future the Tevatron and the LHC are the only
two colliders where SM top quark production can be observed.

The two basic production modes of top quarks at hadron colliders are
the production of $t\tbar$ pairs, which is dominated by the strong interaction,
and the production of single top quarks due to electroweak interactions.

\subsection{$t\bar{t}$ production}
We discuss only top quark pair production via the strong interaction.
$t\bar{t}$ pairs can also be produced by electroweak interactions
if a $Z^0$ or a photon are exchanged between the in- and outgoing
quarks. 
However, at a hadron collider the cross sections for theses processes 
are completely negligible
compared to the QCD cross section.
The cross section for the pair production of heavy quarks has been 
calculated in perturbative QCD, i.e. as a perturbation series in the QCD
running coupling constant $\alpha_s(\mu^2)$. The results are applicable to 
the bottom and to the top quark. In the following we will refer only
to $t\tbar$ production, the scope of this article.

\subsubsection{The factorization ansatz}
\label{sec:factor}
The underlying theoretical framework of the calculation is the parton model 
which regards 
a high-energy hadron $A$, in our case a proton or antiproton, as a 
composition of quasi-free partons (quarks and gluons) which share the
longitudinal hadron momentum $p_A$. 
The parton $i$ has the longitudinal momentum $p_i$, i.e. it carries the momentum 
fraction $x_i=p_i/p_A$. 
The cross section calculation is based on the factorization theorem
stating that the cross section is given by the convolution of 
parton distribution functions (PDF) $f_i(x, \mu^2)$ for the colliding
hadrons ($A$, $B$) and the hard parton-parton cross section $\hat{\sigma}_{ij}$:
\begin{equation}
  \sigma ( AB \rightarrow t\tbar ) = \sum_{i, j} \int\, \rmd x_i \rmd x_j \;
  f_{i, A} (x_i, \mu^2) f_{j, B} (x_j, \mu^2) \cdot
  \hat{\sigma}_{ij}(ij \rightarrow t\tbar; \hat{s}, \mu^2)
  \label{eq:factorization}  
\end{equation}
The hadrons $AB$ are either $p\pbar$ (at the Tevatron) or $pp$ (at the LHC).
The parton distribution function $f_{i, A}(x_i, \mu^2)$ describes the 
probability density for finding a parton  $i$ inside the hadron $A$ 
carrying a longitudinal momentum fraction $x_i$.

The parton distribution functions and the parton-parton cross section
$\hat{\sigma}_{ij}$ depend on the factorization
and renormalization scale $\mu$.
For calculating heavy quark production the scale is usually set to be
of the order of the heavy quark mass, here specifically $\mu = \mtop$.
Strictly speaking one has to distinguish between the factorization
scale $\mu_F$ introduced by the factorization ansatz 
and the renormalization
scale $\mu_R$ due to the renormalization procedure invoked to regulate
divergent terms in the perturbation series when calculating the 
parton-parton cross section $\hat{\sigma}_{ij}$.
Since both scales are to some extent arbitrary parameters most authors have
adopted the practice to use only one scale $\mu = \mu_F = \mu_R$ in their
calculations.
If the complete perturbation series could be calculated, the result for the
cross section would be independent of $\mu$. However, since calculations are
performed at finite order in perturbation theory, cross section predictions
do in general depend on the choice of $\mu$. The $\mu$-dependence is
usually tested by varying the scale between $\mu = \mtop/2$ and 
$\mu = 2\,\mtop$. 
The variations in the cross section are quoted as an indicative theoretical 
uncertainty
of the prediction, but should not be mistaken for Gaussian errors.

In (\ref{eq:factorization}) the variable $\hat{s}$ denotes the square of the 
centre-of-mass energy of the colliding partons:
$\hat{s} = x_i x_j (p_A + p_B)^2$.
In symmetric colliders, where $p_A=p_B=p$, 
we have $\hat{s} = 4\,x_i x_j p^2 = x_i x_j s$. 
The sum in (\ref{eq:factorization}) runs over all pairs of light partons $(i, j)$
contributing to the process.
The factorization scheme serves as a method to systematically eliminate
collinear divergencies from the parton cross section $\hat{\sigma}_{ij}$
and absorb them into the parton distribution functions.
A detailed theoretical justification for the applicability of the 
factorization ansatz to heavy quark production can be found in
reference~\cite{collins1986}. According to this analysis effects such as an
intrinsic heavy quark component in the hadron wave function and 
flavour excitation processes like $qt \rightarrow qt$ do not lead
to a breakdown of the conventional factorization formula.   

\subsubsection{Parametrizations of parton distribution functions}
The PDFs are extracted from measurements in deep inelastic scattering 
experiments where either electrons, positrons or neutrinos collide with
nucleons. A variety of experiments of this kind have been conducted since the
late 1960s and provided hard evidence for the parton (quark) model in 
the first place. During the 1990s the experiments ZEUS and 
H1~\cite{h1det} at the $ep$
storage ring HERA at DESY have reached an outstanding precision in
measurements of the proton structure, which meant a big leap forward
for the prediction of cross sections at hadron colliders. 

\begin{figure}[t]
    \begin{center}
      \epsfig{file=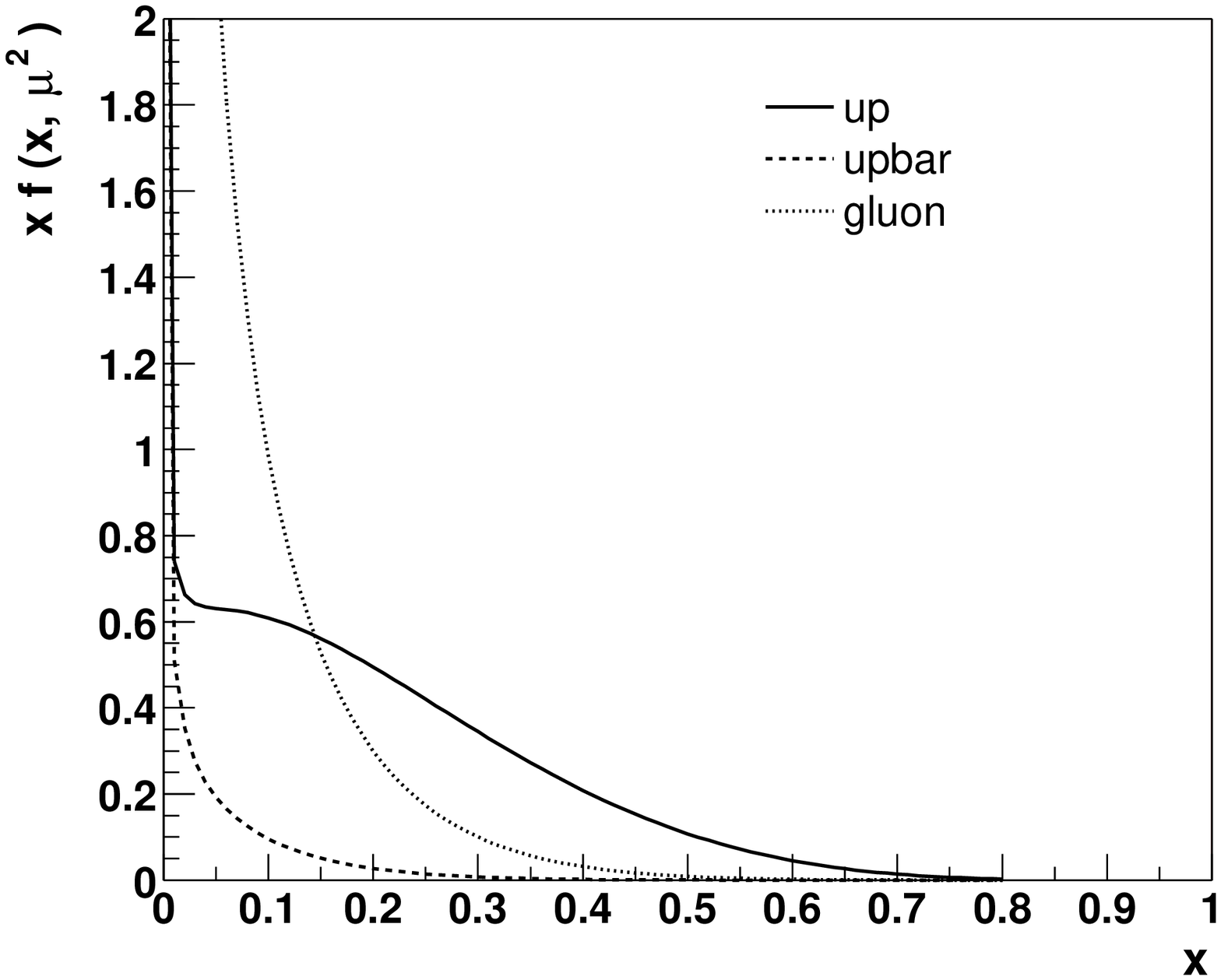, width=7.0cm}
      \epsfig{file=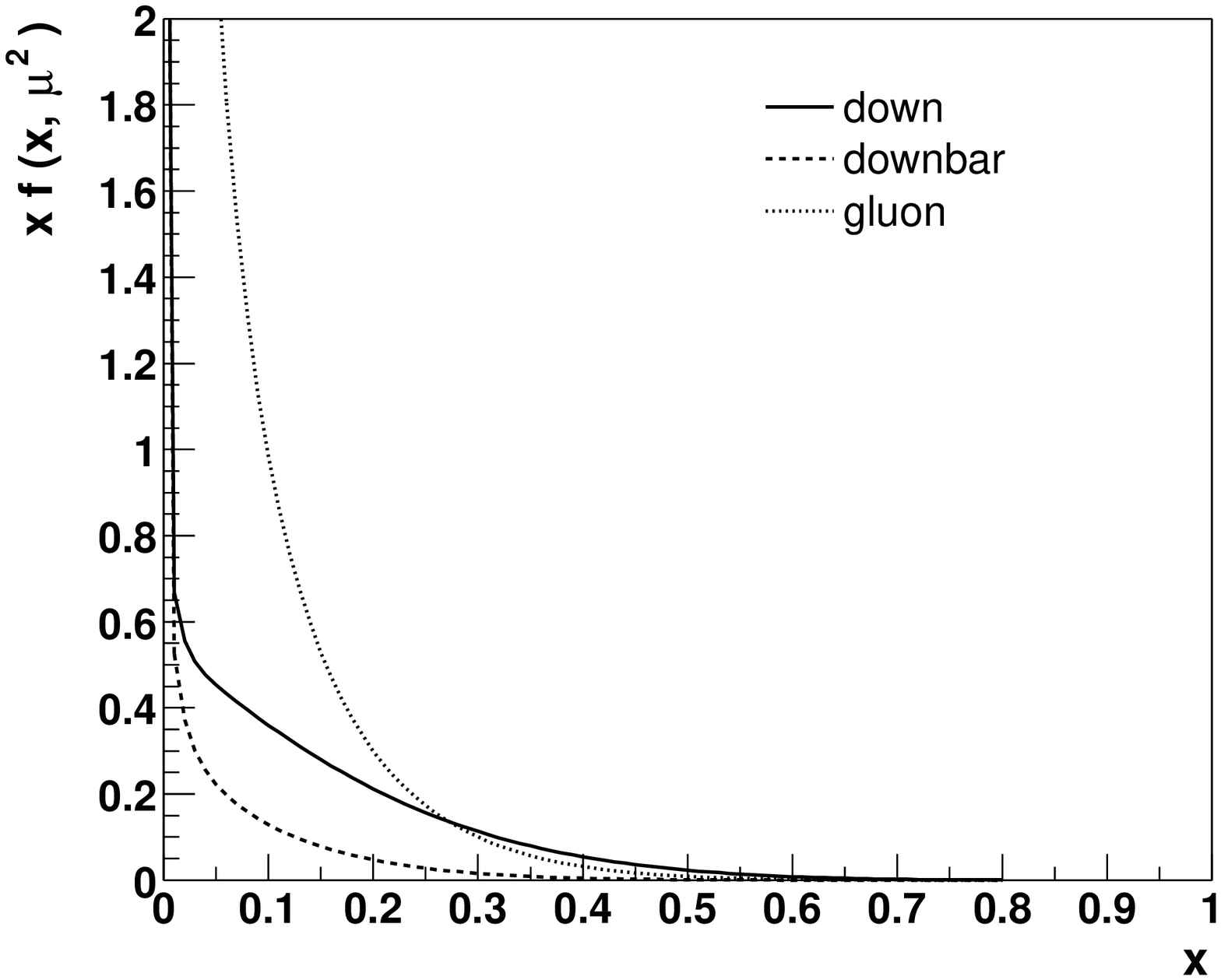, width=7.0cm}
    \end{center}
    \caption{\label{fig:pdfExample}Parton distribution functions (PDFs) of 
     $u$ quarks,
     $\bar{u}$ quarks, $d$ quarks, $\dbar$ quarks and gluons inside the proton. 
     The parametrization is CTEQ3M~\cite{lai1995}.
     The scale at which
     the PDFs are evaluated was chosen to be $\mu=175\;\mathrm{GeV}$ 
     ($\mu^2 = 30625\;\mathrm{GeV^2}$).}
  \end{figure}

Several parametrizations of proton PDFs have been extracted from the 
experimental data by different groups of physicists. As an example 
figure~\ref{fig:pdfExample} shows PDFs of the CTEQ3M parametrization~\cite{lai1995}.
Plotted are the distributions
most relevant for $p\pbar$ and $pp$ collisions, the PDFs for $u$, $\ubar$, 
$d$, $\dbar$ and the gluons. For antiprotons the
distributions in figure~\ref{fig:pdfExample} have to be reversed between
quark and antiquark. The scale $\mu$ was chosen to be 
$\mu = 175\;\mathrm{GeV}$.
One sees that the gluons start to dominate in the $x$ region below 0.15.
The $t\tbar$ production cross section at the Tevatron is dominated
by the large $x$ region, since the top quark mass is relatively large
compared to the Tevatron beam energy ($\mtop/\sqrt{s}= 0.0875$).
At the LHC ($\mtop/\sqrt{s}= 0.0125$) the lower x region becomes 
more important.

\begin{figure}[!bt]
    \begin{center}
      \epsfig{file=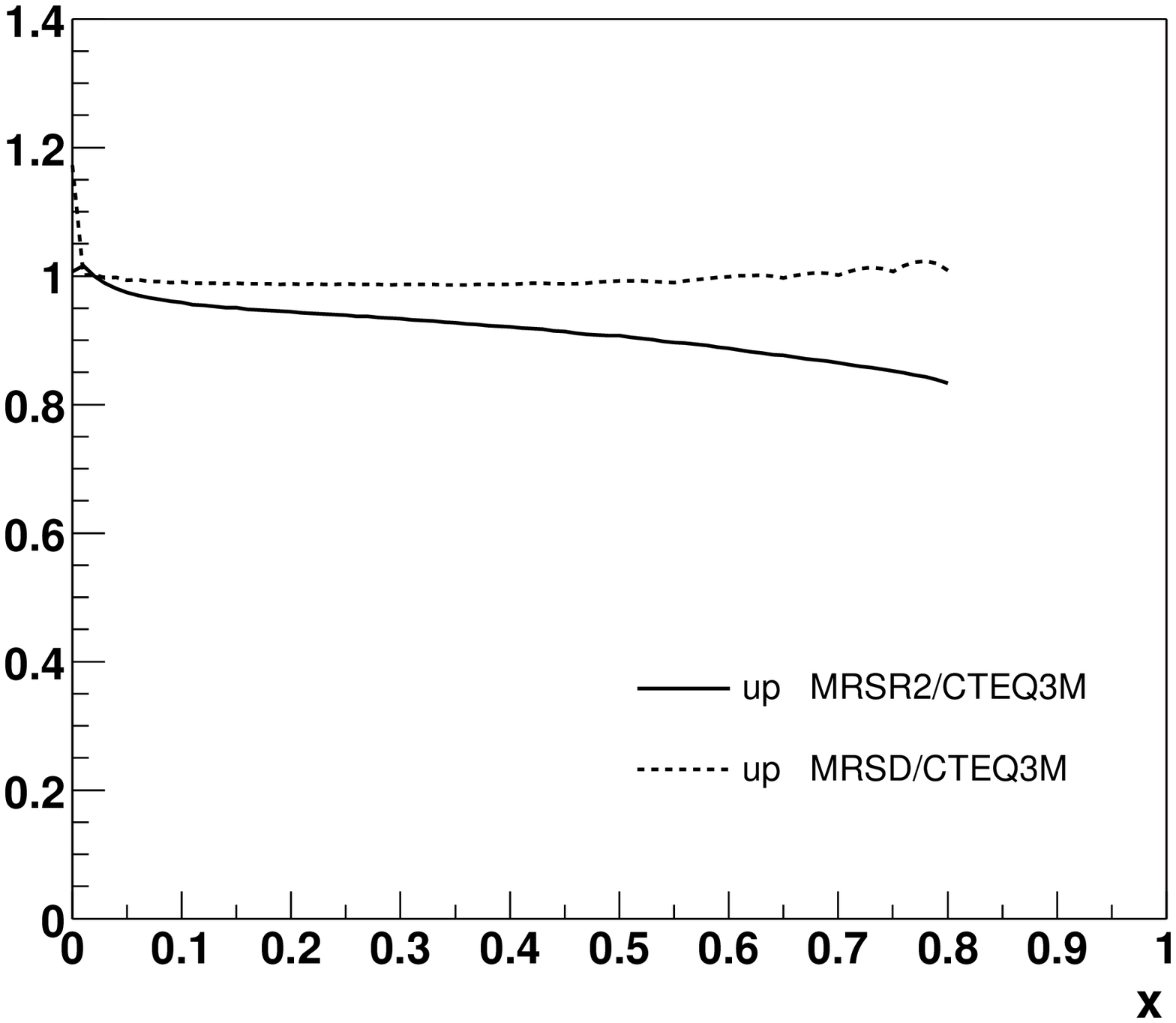, width=7.0cm}
      \epsfig{file=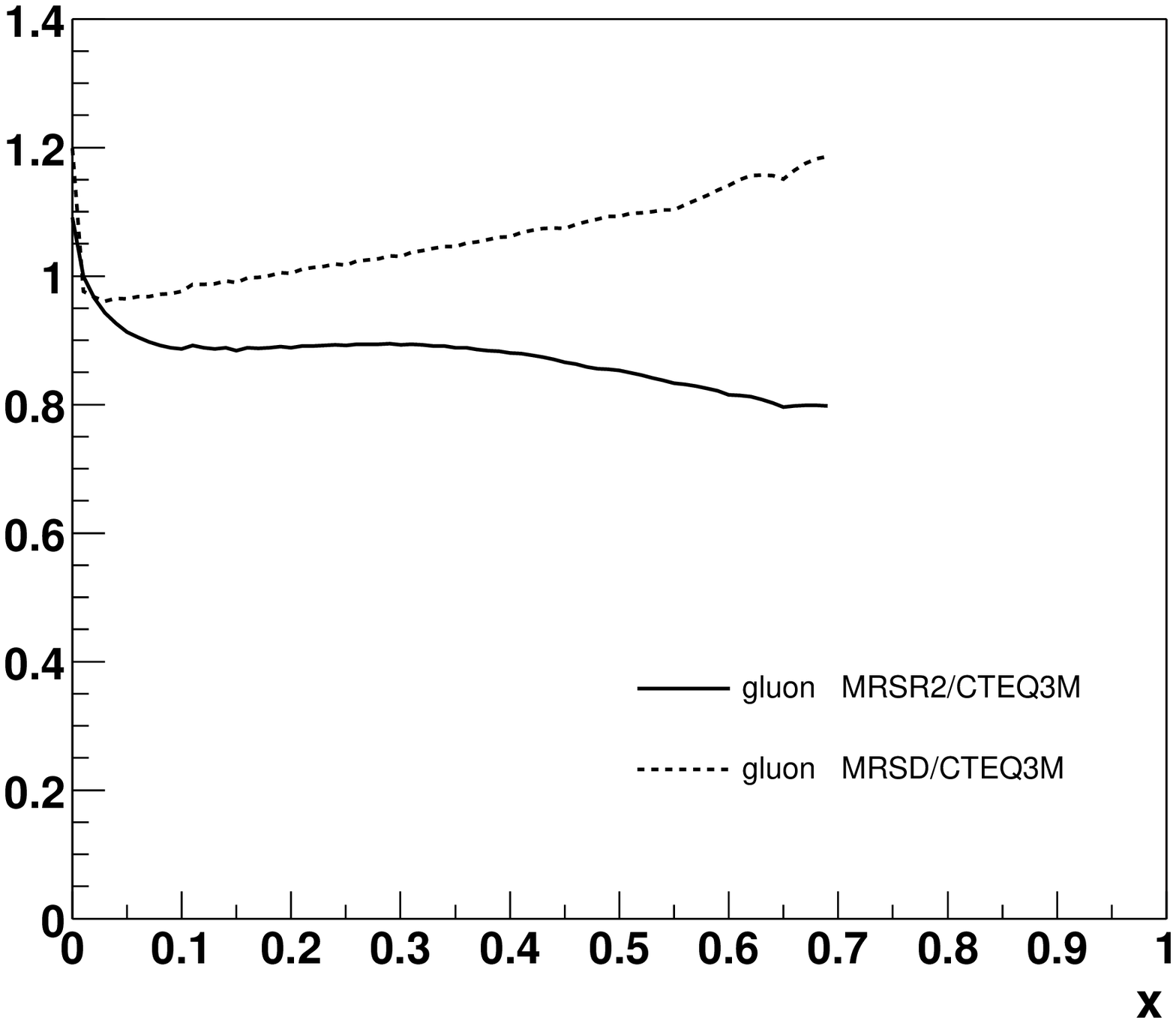, width=7.0cm}
    \end{center}
    \caption{\label{fig:pdfCompare} Comparison of the CTEQ3M, MRSR2 and MRSD 
    parametrizations for the PDF of the up quark and the gluon. 
    The left plot shows the ratios $f_u(x)_\mathrm{MRSR2}/f_u(x)_\mathrm{CTEQ3M}$ 
    (full line) 
    and $f_u(x)_\mathrm{MRSD}/f_u(x)_\mathrm{CTEQ3M}$ (dashed line). On the right
    hand side we plot $f_g(x)_\mathrm{MRSR2}/f_g(x)_\mathrm{CTEQ3M}$ (full line) 
    and 
    $f_g(x)_\mathrm{MRSD}/f_g(x)_\mathrm{CTEQ3M}$ (dashed line).  
    The factorization scale is set to $\mu^2 = 30625\;\mathrm{GeV^2}$. }
  \end{figure}

There are differences between the various sets of PDF parametrizations.
To give the reader a quantitative impression we compare the CTEQ3M, MRSR2 
and MRSD parametrizations in figure~\ref{fig:pdfCompare}. 
As a representative example we plot the ratios 
$f_u(x)_\mathrm{MRSR2}/f_u(x)_\mathrm{CTEQ3M}$ and 
$f_u(x)_\mathrm{MRSD}/f_u(x)_\mathrm{CTEQ3M}$
for up-quarks as well as $f_g(x)_\mathrm{MRSR2}/f_g(x)_\mathrm{CTEQ3M}$ and 
$f_g(x)_\mathrm{MRSD}/f_g(x)_\mathrm{CTEQ3M}$ for gluons. 
These three PDF parametrizations were
chosen because they are used for predictions of the $t\tbar$ cross section
at the Tevatron~\cite{berger1998,bonciani1998,laenen1994}.
Typical differences are on the order of 5 to 10\%.

\subsubsection{The parton cross section} 
\begin{figure}[t]
    \begin{center}
      \includegraphics[width=0.15\textwidth]{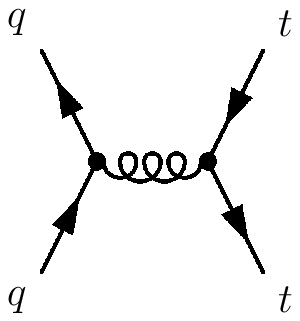}  \hspace*{4mm}
      \includegraphics[width=0.15\textwidth]{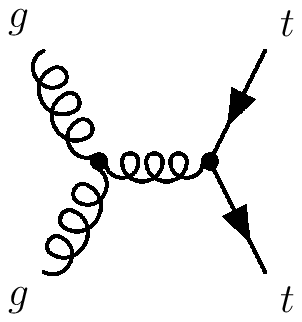}  \hspace*{4mm}
      \includegraphics[width=0.15\textwidth]{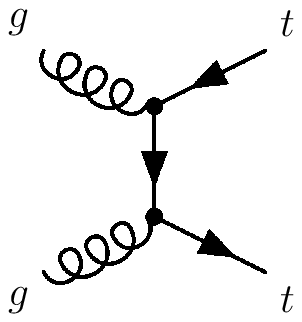}  \hspace*{4mm}
      \includegraphics[width=0.15\textwidth]{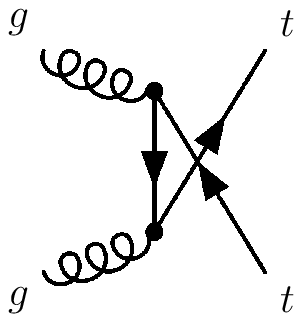}
    \end{center}
    \caption{\label{fig:leadingOrderttbar} Feynman diagrams of the leading
      order processes for $t\tbar$ production: 
      quark-antiquark annihilation ($q\qbar \rightarrow t\tbar$) and 
      gluon fusion ($gg \rightarrow t\tbar$).
     }
  \end{figure}

The cross section $\hat{\sigma}$ of the hard, i.e. short distance, 
parton-parton process 
$ij \rightarrow t \tbar$ can be calculated in perturbative QCD. 
The leading order processes, contributing with $\alpha_s^2$ to the perturbation
series, are quark-antiquark annihilation, 
$q\qbar \rightarrow t\tbar$, and gluon fusion, $gg \rightarrow t\tbar$.
The corresponding Feynman diagrams for these processes are depicted
in figure~\ref{fig:leadingOrderttbar}.
The leading order (Born) cross sections for heavy quark production were calculated
in the late 
1970s~\cite{glueck1978,jones1978,babcock1977,georgi1978,combridge1979,hagiwara1979},
most of them having charm production as a concrete application in mind. 
The differential cross section for quark-antiquark annihilation is given
by
\begin{equation}
  \frac{\rmd\hat{\sigma}}{\rmd\hat{t}} \left( q\qbar \rightarrow t \tbar \right) =
  \frac{4\,\pi\,\alpha_s^2}{9\, \hat{s}^4} \cdot
  \left[ (m^2 - \hat{t} )^2 + (m^2 - \hat{u})^2 + 2m^2\hat{s} \right]
  \label{eq:qqttbar}
\end{equation}
where $\hat{s}$, $\hat{t}$ and $\hat{u}$ are the Lorentz-invariant 
Mandelstam 
variables of the process. They are defined by $\hat{s} = (p_q + p_{\,\qbar})^2$,
$\hat{t} = (p_q - p_{t})^2$ and $\hat{u} = (p_q - p_{\,\tbar})^2$
with $p_i$ being the corresponding momentum 4-vector of the quark $i$.
$m$ denotes the top quark mass.
The differential cross section for the gluon-gluon fusion process is given
by:
\begin{small}
\begin{eqnarray}
  \fl
  \frac{\rmd\hat{\sigma}}{\rmd\hat{t}} \left( g_1 g_2 \rightarrow t \tbar \right) & = &
  \frac{\pi\,\alpha_s^2}{8\, \hat{s}^2} \cdot 
  \left[ 
  \frac{6 (m^2 - \hat{t}) (m^2 - \hat{u})}{\hat{s}^2} -
  \frac{m^2 (\hat{s} - 4\,m^2)}{3\,(m^2-\hat{t})(m^2 - \hat{u})} \right. 
  \nonumber \\ 
  & & \left. + \frac{4}{3} \cdot 
  \frac{(m^2 - \hat{t})(m^2 - \hat{u}) - 2\,m^2 (m^2 + \hat{t})}{(m^2 -\hat{t})^2}
  + \frac{4}{3} \cdot
  \frac{(m^2 - \hat{t})(m^2 - \hat{u}) - 2\,m^2 (m^2 + \hat{u})}{(m^2 -\hat{u})^2}
  \right. \nonumber \\   
  & & \left. - 3 \cdot 
  \frac{(m^2-\hat{t})(m^2-\hat{u})-m^2(\hat{u}-\hat{t})}{\hat{s}\,(m^2 -\hat{t})^2}
  - 3 \cdot 
  \frac{(m^2-\hat{t})(m^2-\hat{u})-m^2(\hat{t}-\hat{u})}{\hat{s}\,(m^2 -\hat{u})^2}
  \right] 
  \label{eq:ggttbar}
\end{eqnarray}
\end{small}
The invariant variables are in this case $\hat{s}=(p_{g1}+p_{g2})^2$,
$\hat{t}=(p_{g1}-p_{t})^2$ and $\hat{u} = (p_{g1}-p_{\,\tbar})^2$.
The cross sections in (\ref{eq:qqttbar}) and (\ref{eq:ggttbar})
are quoted in the form given in reference~\cite{bargerPhillips}.
The invariants $\hat{t}$ and $\hat{u}$ may be expressed in terms of the
cosine of the scattering angle $\hat{\theta}$ in the parton-parton
centre-of-mass system:
\begin{equation}
  \cos{\hat{\theta}} = \sqrt{1-\frac{4\,p_\perp^2}{\hat{s}}}     
  \ \ \ \ \ \ \ \ \ \ \  
  \hat{t} = - \frac{\hat{s}}{2} (1 - \cos{\hat{\theta}}) \ \ \ \ \ \ \ \ \ \ \ 
  \hat{u} = - \frac{\hat{s}}{2} (1 + \cos{\hat{\theta}}) 
  \label{eq:tucostheta}
\end{equation}
where $p_\perp$ is the common transverse momentum of the outgoing top quarks.

A full calculation of next-to-leading order (NLO) corrections contributing
in order $\alpha_s^3$ to the inclusive parton-parton cross section 
for heavy quark pair production was performed independently by two groups:
Nason et al. in 1988~\cite{nason1988} and Beenakker et al. in 
1991~\cite{beenakker1989,beenakker1991}, yielding consistent results.
The NLO calculations involve virtual contributions to the leading order
processes, gluon bremsstrahlung processes ($q\qbar \rightarrow t\tbar + g$
and $gg \rightarrow t\tbar + g$) as well as processes like
$g+q(\qbar) \rightarrow t\tbar + q(\qbar)$. Examples of Feynman diagrams of
NLO processes are given in figure~\ref{fig:NLOgraphs}:
  \begin{figure}[t]
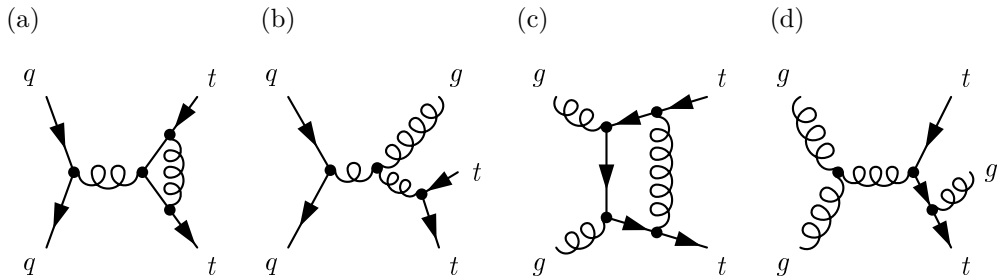

    \begin{center}
       (a)  \hspace{27mm} (b) \hspace{27mm} (c) \hspace{27mm} (d) \hspace{27mm} 
      
      \vspace*{4mm}

      \epsfig{file=feynEpsi/qqbar2ttbarNLOPict.epsi, height=28mm} \hspace*{4mm}
      \epsfig{file=feynEpsi/qqbar2ttbarGBPict.epsi, height=28mm} \hspace*{4mm}
      \epsfig{file=feynEpsi/ggtttbarNLOPict.epsi, height=28mm}  \hspace*{4mm}
      \epsfig{file=feynEpsi/gggttbarGBPict.epsi, height=28mm}
    \end{center}
    \caption{\label{fig:NLOgraphs} Example Feynman diagrams of next-to-leading 
      order (NLO) corrections to quark-antiquark annihilation and
      gluon fusion. a) and c) show virtual corrections, b) and d) gluon
      bremsstrahlung graphs.
     }
  \end{figure}
a) and c) display two virtual graphs, b) and d) two gluon bremsstrahlung graphs.
$\alpha^3_s$ corrections arise when interfering those graphs with 
the leading order graphs of figure~\ref{fig:leadingOrderttbar}.
For the NLO calculation of the hadron-hadron cross section 
$\sigma(AB \rightarrow t\tbar)$ to be consistent one has to use next-to-leading
order determinations of the coupling constant $\alpha_s$ and the PDFs.
All quantities have to be consistently defined in the same renormalization
scheme because different approaches distribute the radiative corrections
differently among the parton-parton cross section, the PDFs and $\alpha_s$.
Most authors use the $\mathrm{\overline{MS}}$ or an extension of the 
$\mathrm{\overline{MS}}$ scheme. 
First NLO cross section predictions for $t\tbar$ production 
at the Tevatron ($\sqrt{s} = 1.8\;\mathrm{TeV}$) 
yielded values of about 
4~pb~\cite{nason1989,beenakker1991,altarelli1988,ellis1991}

At energies close to the kinematic threshold, $\hat{s} = 4\,\mtop^2$,
the quark-antiquark annihilation process is the dominant one, if the
incoming quarks are valence quarks, as is the case of $p\pbar$
collisions. At the Tevatron 80 to 90\% of the $t\tbar$ cross section
is due to quark-antiquark 
annihilation~\cite{laenen1994,bonciani1998,cacciari2004}. 
At higher energies the gluon-gluon
fusion process dominates for both $p\pbar$ and $pp$ collisions.
That is why one can built the LHC as a $pp$ machine without compromising  
the parton-parton cross section.
Technically, it is of course much easier to operate a $pp$ collider,
since one spares the major challenge to produce high antiproton currents
in a storage ring. 
For the Tevatron the ratio of NLO over LO cross sections for gluon-gluon fusion 
is predicted to be 1.8 at $\mtop = 175\;\mathrm{GeV}/c^2$,
for quark-antiquark annihilation the value is only about 1.2~\cite{laenen1994}.
Since the annihilation process is dominating, the overall NLO enhancement
is about 1.25.

\subsubsection{Soft gluon resummation}
\label{sec:softgluonresum}
Contributions to the total cross section due to radiative corrections 
are large in the region near threshold ($\hat{s} = 4\,\mtop^2$) and at high
energies ($\hat{s} > 400\;\mtop^2$). 
Near threshold the cross section is enhanced due to initial state gluon 
bremsstrahlung~(ISGB)~\cite{laenen1992}. This effect is important for
$t\tbar$ production at the Tevatron, but not for the LHC 
where gluon splitting and flavour excitation are increasingly important
effects.
The calculation at fixed next-to-leading order ($\alpha_s^3$) perturbation theory
has been refined to systematically incorporate higher order corrections due 
to soft gluon radiation. Technically, this is done by applying an
integral transform (Mellin transform) to the cross section:
\begin{equation}
  \sigma_N (t\tbar) \equiv \int_0^1 \rmd\rho \ \rho^{N-1} \;\sigma(\rho; t\tbar)
\end{equation}
where $\rho = 4\,\mtop^2/s$ is a dimensionless parameter.
In Mellin moment space the corrections due to soft gluon radiation
are given by a power series of logarithms $\ln N$.
For $\mu = \mtop$ the corrections are positive at all orders. 
Therefore, the resummation of the soft gluon logarithms yields an 
increase of the $t\tbar$ cross section with respect to the NLO value.
\renewcommand\arraystretch{1.8}
Four different groups have presented cross section predictions
based on the resummation of soft gluon contributions:
(1) Laenen et al.~\cite{laenen1992,laenen1994}, 
(2) Berger and Contopanagos~\cite{berger1995,berger1996,berger1998},
(3) Bonciani et al. (BCMN)~\cite{catani1996a,catani1996b,bonciani1998,cacciari2004} and
(4) Kidonakis et al.~\cite{kidonakis2001a,kidonakis2001b,kidonakis2003}.
Their predictions for $\mtop = 175\;\mathrm{GeV}/c^2$ 
are summarised in table~\ref{tab:ttbarXSection}.
\begin{table}[!t]
  \caption{\label{tab:ttbarXSection} Cross section predictions for $t\tbar$
    production at next-to-leading order (NLO) in perturbation theory including
    the resummation of initial state gluon bremsstrahlung~(ISGB).
    The predictions are given for a top mass of $\mtop = 175\;\mathrm{GeV}/c^2$.
    The cross sections are given for $p\pbar$ collisions at the Tevatron
    ($\sqrt{s} = 1.8\;\mathrm{TeV}$ and $\sqrt{s} = 1.96\;\mathrm{TeV}$)
    and $pp$ collisions at the LHC ($\sqrt{s} = 14\;\mathrm{TeV}$).   
    The factorization and renormalization scale is set to $\mu = \mtop$
    to derive the central values.
    It has to be stressed that the authors use different sets of PDF
    parametrizations. Part of the differences can be attributed to this
    fact.
    In reference~\cite{laenen1994} Laenen et al. do not
  provide a direct prediction for $\mtop = 175\;\mathrm{GeV}/c^2$.
  To compare their prediction with those of other authors I choose to
  linearly interpolate the given values for
  $\mtop = 174\;\mathrm{GeV}/c^2$ and $\mtop = 176\;\mathrm{GeV}/c^2$,
  thereby neglecting the functional form of the mass dependence. 
  Within the assigned errors this simplification is acceptable.}
  \begin{center}
  \begin{tabular}{lllcl}
    \br
    Group &  $\sqrt{s}$ & PDF set & $\sigma(t\tbar)$ & Reference\\
    \mr
    (1) Laenen at al.           & 1.8 TeV & MRSD\_' & 
    $4.95^{+0.70}_{-0.42}$ pb &
    \cite{laenen1994} \\
    (2) Berger and Contopanagos & 1.8 TeV & CTEQ3M  & 
    $5.52^{+0.07}_{-0.45}$ pb & 
    \cite{berger1995,berger1996,berger1998} \\
    (3) Bonciani et al.           & 1.8 TeV & CTEQ6M   & 
    $5.19^{+0.52}_{-0.68}$ pb &
    \cite{cacciari2004} \\ 
    (4) Kidonakis et al.      &   1.8 TeV & MRST2002 & 
    $(5.24\pm0.31)$ pb & \cite{kidonakis2003} \\
    \hline
    (2) Berger and Contopanagos & 2.0 TeV & CTEQ3M  & $7.56^{+0.10}_{-0.55}$ pb &
    \cite{berger1996,berger1998} \\
    (3) Bonciani et al.         & 1.96 TeV & CTEQ6M   & $6.70^{+0.71}_{-0.88}$ pb &
    \cite{cacciari2004} \\  
    (4) Kidonakis et al.        & 1.96 TeV & MRST2002 & $(6.77\pm0.42)$ pb &
    \cite{kidonakis2003} \\
    \hline
    (2) Berger and Contopanagos & 14.0 TeV & CTEQ3M & 760 pb &
    \cite{berger1996,berger1998} \\   
    (3) Bonciani et al.         & 14.0 TeV & MRSR2  & $833^{+52}_{-39}$ pb &
    \cite{bonciani1998} \\  
    (4) Kidonakis et al.        & 14.0 TeV & MRST2002 & 870 pb &
    \cite{kidonakis2003} \\
    \br
  \end{tabular}
  \end{center}
\end{table}

In the region very close to the kinematic threshold non-perturbative effects
become dominant and the perturbative approach breaks down.
The resummation can therefore not sensibly be extended into this
region. That is why Laenen et al. introduce a new scale $\mu_0$ with
$\Lambda_\mathrm{QCD} \ll \mu_0 \ll \mtop$ where they stop the resummation.
The concrete choice of scale is to some extent arbitrary. 
The uncertainty quoted by Laenen et al. is derived by varying the
scale $\mu_0$. To calculate the central value of the cross section 
they use $\mu_0 = 0.1\,\mtop$ for $q\qbar$-annihilation
and $\mu_0 = 0.25\,\mtop$ for $gg$-fusion. The values are not 
required to be the same, since the respective perturbation series have
different convergence properties.

Berger and Contopanagos derive the infra-red cut-off $\mu_0$ within their
calculation and thereby define a perturbative region where resummation 
can be applied. Since $\mu_0$ is derived, it is not treated as a source
of error in this approach.  
The theoretical uncertainty quoted by Berger and Contopanagos
is derived by varying the factorization and renormalization
scale $\mu$ between $0.5\,\mtop$ and $2\,\mtop$. 
The central value, given in table~\ref{tab:ttbarXSection}, is calculated 
using the CTEQ3M PDFs.
The uncertainty due to the choice of the PDF parametrization is about 4\%.
Resummation effects are of appreciable size:
The resummed total cross sections 
(for $\sqrt{s} = 1.8\;\mathrm{TeV}$ and $\sqrt{s} = 2.0\;\mathrm{TeV}$, 
respectively) are about 9\% above the NLO cross sections.
Berger and Contopanagos predict an increase of 37\% in cross section,
when going from $\sqrt{s} = 1.8\;\mathrm{TeV}$ to 
$\sqrt{s} = 2.0\;\mathrm{TeV}$ in Run II of the Tevatron.
The cross section value for the LHC, $\sigma = 760\;\mathrm{pb}$, merely 
reflects an estimate and is not accompanied by an uncertainty. 
Due to the much larger centre-of-mass energy at the LHC the
near threshold region is much less important for $t\tbar$ production, 
reducing the significance of ISGB and the
need for resummation of these contributions. 

\renewcommand\arraystretch{1.0}
While the first two groups have only resummed leading logarithmic 
terms (LL), BCMN also include next-to-leading logarithms (NLL). 
They used a different resummation
prescription which does not demand the introduction of an additional
infra-red cut-off $\mu_0$~\cite{catani1996a}. 
The LL result of BCMN shows only an increase of about 1\%
compared to the NLO cross section at the Tevatron~\cite{catani1996b}.
When taking NLL terms into account the increase is 4\%~\cite{bonciani1998}.
In table~\ref{tab:ttbarXSection} we quote the NLL results as updated
in reference~\cite{cacciari2004} with the newest set of PDFs.
The errors quoted by BCMN include PDF uncertainties, which are evaluated
by using sets of PDF parametrizations that provide an estimate of
``1-$\sigma$'' uncertainties. In the case of 
CTEQ~\cite{stump2002,pumplin2002} 40 different sets are available, for
MRST~\cite{martin2002} there are 30 sets. 
Kidonakis {\it et al.} resum leading and subleading logarithms up to order
$\alpha_s^4$ in an attempt to reduce the dependence of the 
cross-section on the renormalization scale $\mu$ compared to the NLO
calculation.
The uncertainty quoted by Kidonakis {\it et al.} is dominated by the choice 
of the kinematic description
of the scattering process, either in one-particle-inclusive or
pair-invariant-mass kinematics. The Tevatron predictions given in 
table~\ref{tab:ttbarXSection} are the average of the two choices.
The LHC prediction is based on the one-particle-inclusive value, since this
is believed to be more appropriate when the cross section is dominated
by gluon-gluon fusion.     

\begin{figure}[t]
  \begin{center}
    \epsfig{file=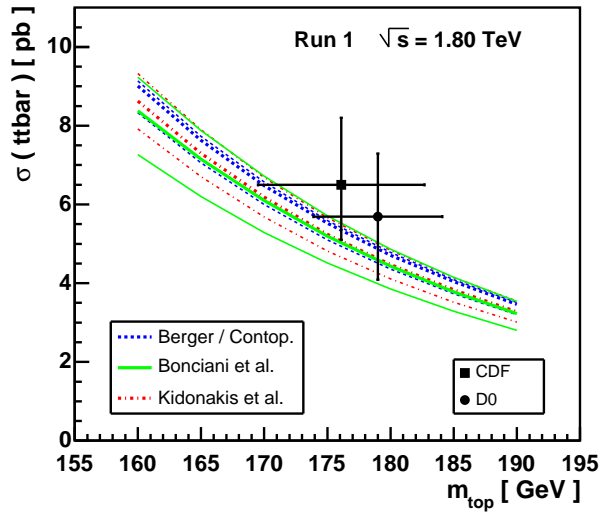, width=8.0cm}
  \end{center}
  \caption{\label{fig:sigmattbar} The $t\tbar$ cross section at 
  $\sqrt{s} = 1.8\;\mathrm{TeV}$ as a function of the top quark mass. 
  The thick full line represents the central value predicted by 
  Catani et al.~\cite{catani1996b,bonciani1998}.
  The thick dashed line shows the prediction by Berger and 
  Contopanagos~\cite{berger1996}. 
  The thick dashed-dotted line is the prediction by Kidonakis et 
  al.~\cite{kidonakis2003}.
  The thin lines indicate the upper and lower uncertainties of the predictions.
  The Run I measurements of 
CDF~\cite{CDFttbarCrossSection2001LeptonPlusJetsRun1,CDFtopMass2001LeptonPlusJetsRun1} 
  and \DZero~\cite{d0Sigmattbar2003Run1,d0mtopnature} are shown for comparison.}
\end{figure}

The cross section predictions are strongly dependent on the top quark mass,
which is illustrated in figure~\ref{fig:sigmattbar}.
The plot shows the predictions for the Tevatron at 
$\sqrt{s} = 1.8\;\mathrm{TeV}$, they 
are in good agreement within the given errors.
For comparison the Run I measurements of 
CDF~\cite{CDFttbarCrossSection2001LeptonPlusJetsRun1,CDFtopMass2001LeptonPlusJetsRun1} 
and D\O ~\cite{d0Sigmattbar2003Run1,d0mtopnature} are shown.
Within the large errors the measurements agree well with the theoretical
predictions.
It is obvious that a precise cross section measurement has to be accompanied
by a precise measurement of the top quark mass to provide a basis for 
a stringent test of the theory.  

\subsection{Single top quark production} 
\label{sec:singleTopTheo}

\begin{figure}[t]
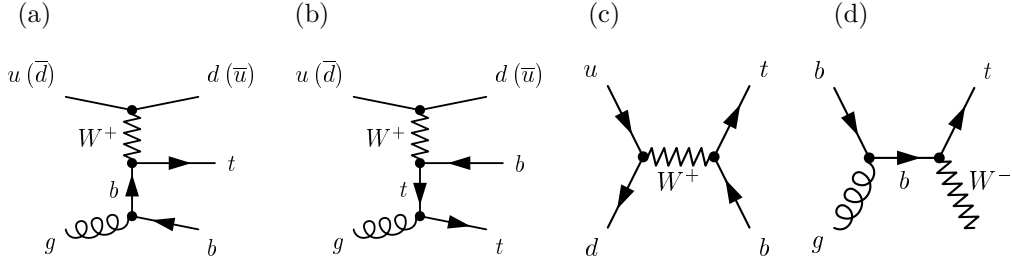

  \hspace*{12mm} (a)  \hspace{30mm} (b) \hspace{32mm} (c) \hspace{26mm} (d)
  \begin{center}    
    \vspace*{2mm}

    \epsfig{file=feynEpsi/singTopWg1Pict.epsi, height=26mm} \hspace*{3mm}
    \epsfig{file=feynEpsi/singTopWg2Pict.epsi, height=26mm} \hspace*{3mm}
    \epsfig{file=feynEpsi/singTopSChanPict.epsi, height=26mm}  \hspace*{3mm}
    \epsfig{file=feynEpsi/singTopAssocPict.epsi, height=26mm}
  \end{center}
  \caption{\label{fig:singleTop} Representative Feynman diagrams for the three
    single top production modes. a) and b) show $W$-gluon fusion graphs, 
    c) the $s$-channel process and d) associated production. We chose to draw
    the graphs in a), b) and c) with the $(u, d)$ weak-isospin doublet 
    coupling to the $W$. This is by far the dominating contribution. 
    In general, also the $(c, s)$ doublet contributes. 
    The graphs show single top
    quark production, the diagrams for single antitop quark production 
    can be obtained by interchanging quarks and antiquarks.
   }
\end{figure}

Top quarks can be produced singly via electroweak interactions involving
the $Wtb$ vertex. There are three production modes which are distinguished by
the virtuality $Q^2$ of the $W$ boson ($Q^2 = - q^2$, where $q$ is the 
four-momentum of the $W$):  
\begin{enumerate}
\item the {\bf t-channel} ($q^2 = \hat{t}\;$): 
  A virtual W strikes a $b$ quark (a sea quark) inside 
  the proton. The W boson is spacelike ($q^2  < 0$). 
  This mode is also known as \emph{W-gluon fusion},
  since the $b$ quark originates from a gluon splitting into a $b\overline{b}$ 
  pair. Feynman diagrams representing this process are shown
  in figure~\ref{fig:singleTop}a and figure~\ref{fig:singleTop}b .
  $W$-gluon fusion is the dominant production mode, both at the Tevatron and 
  at the LHC, as will be shown in the discussion below.
\item the {\bf s-channel} ($q^2 = \hat{s}\;$): 
  This production mode is of Drell-Yan type. 
  A timelike $W$ boson with $q^2 \geq (M_\mathrm{top} + m_b)^2$ 
  is produced by the fusion of two quarks belonging to an SU(2) isospin doublet. 
  See figure~\ref{fig:singleTop}c for the Feynman diagram. 
\item {\bf associated production}: The top quark is produced in association with a
  real (or close to real) $W$ boson ($q^2 = M_W^2$). 
  The initial $b$ quark is a sea quark inside
  the proton. Figure~\ref{fig:singleTop}d shows the Feynman diagram.
  The cross section is negligible at the Tevatron, but of
  considerable size at LHC energies where associated production even supercedes 
  the $s$-channel. 
\end{enumerate}
In $p\pbar$ and $pp$ collisions the cross section is dominated by 
contributions from up and down quarks coupling to the $W$ boson on one
hand side of the Feynman diagrams. That is why the $(u, d\,)$ quark doublet is 
shown in the graphs of figure~\ref{fig:singleTop}. There is of course also a small
contribution from the second weak isospin quark doublet, $(c, s)$; an effect
of about 2\% for $s$- and 6\% for $t$-channel production~\cite{heinson1997}.  
Furthermore, we will only consider single top quark production via a $Wtb$ 
vertex. 
The production channels involving a $Wtd$ or a $Wts$ vertex are strongly 
suppressed due to small CKM matrix elements:
$ 0.0048 < |V_{td}| < 0.014$ and $0.037 < |V_{ts}| < 0.043$~\cite{PDG2004}.
Thus, their contribution to the total cross section is quite small:
$\sim 0.1\%$ and $\sim 1\%$, respectively~\cite{anselmo1992}.   
In the following paragraphs we will review the theoretical cross section
predictions for the three single top processes and the methods with which 
they are obtained.

\subsubsection{W-gluon fusion}
\label{sec:wgfusion}
\begin{figure}[t]
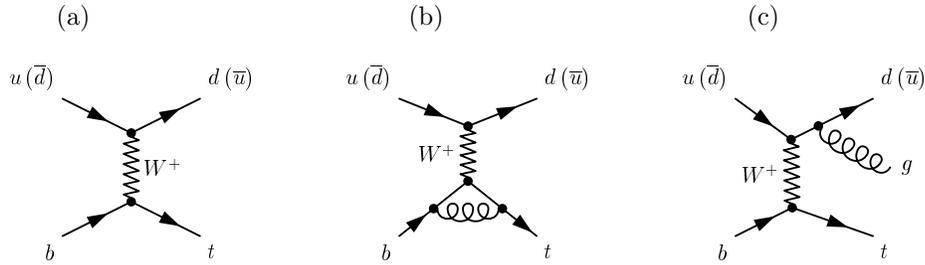

  \begin{center}
     (a)  \hspace{40mm} (b) \hspace{38mm} (c) \hspace{12mm}
    
    \vspace*{4mm}

    \epsfig{file=feynEpsi/singTopWg3Pict.epsi, height=26mm} \hspace*{10mm}
    \epsfig{file=feynEpsi/singTopWg3NLO1Pict.epsi, height=26mm} \hspace*{10mm}
    \epsfig{file=feynEpsi/singTopWg3NLO2Pict.epsi, height=26mm}
  \end{center}
  \caption{\label{fig:Wg} a) shows the leading order Feynman diagram for
    $W$ gluon fusion if the perturbative calculation is performed involving
    a $b$ quark distribution function. b) and c) show two examples of 
    $\alpha_s$ corrections
    to the leading order diagram. b) is a correction to the $b$ vertex, c) 
    a correction to the light quark vertex.
   }
\end{figure}

The $W$-gluon fusion process was already suggested as a potentially interesting source
of top quarks in the mid 1980s~\cite{WillenbrockDicus1986,DawsonWillenbrock1987} 
and early 1990s~\cite{Yuan1990}. If the $b$ quark is taken to be massless, 
a singularity arises when computing the diagram in figure~\ref{fig:singleTop}a 
in case the final $\bar{b}$ quark is collinear with the incoming gluon. 
In reality the non-zero mass of the $b$ quark regulates this collinear divergence. 
When calculating the total cross section the collinear singularity manifests
itself as terms proportional to $\ln ((Q^2 + \mtop^2)/m_b^2)$ with
$Q^2 = -q^2$ being the virtuality of the W boson.
These logarithmic terms cause the perturbation series to converge rather
slowly. 
This difficulty can be obviated by introducing a parton distribution function 
(PDF) for the $b$ quark, $b(x, \mu^2)$, which effectively resums the logarithms
to all orders of perturbation theory and 
implicitly describes the splitting of gluons into $b\bbar$ pairs inside the
colliding hadrons~\cite{bordes1995,stelzerSullivan1997}.
Once a $b$ quark distribution function is introduced into the calculation, the
leading order process is $q_i + b \rightarrow q_j + t$ as shown in 
figure\ref{fig:Wg}a. 
In this formalism the process shown in figure~\ref{fig:singleTop}a
is a higher order correction which is already partially included
in the b-quark distribution function.
The remaining contribution is of order $1/\ln ((Q^2 + \mtop^2)/m_b^2)$ 
with respect to the leading order process in figure~\ref{fig:Wg}a.
Additionally, there are also corrections of order $\alpha_s$:
Two examples of those are shown in figure~\ref{fig:Wg}b and figure~\ref{fig:Wg}c.
The leading order differential cross section calculated from the Feynman graph in 
figure~\ref{fig:Wg}a for quark-quark or antiquark-antiquark collisions
is given by~\cite{WillenbrockDicus1986}:
\begin{equation}
  \frac{\rmd\hat{\sigma}_{ij}}{\rmd\hat{t}} = \frac{\pi\, \alpha_w^2}{4} \cdot
  |V_{ij}|^2\; |V_{tb}|^2 \cdot 
  \frac{\hat{s}-\mtop^2}{(\hat{s}-m_b^2)\,(\hat{t} - M_{W}^2)^2} 
  \label{eq:WgQQXsec}
\end{equation}  
For quark-antiquark collisions the result is:
\begin{equation}
\frac{\rmd\hat{\sigma}_{ij}}{\rmd\hat{t}} = \frac{\pi\, \alpha_w^2}{4} \cdot
  |V_{ij}|^2\; |V_{tb}|^2 \cdot 
  \frac{\hat{u}-m_b^2}{(\hat{s}-m_b^2)^2}\; 
  \frac{\hat{u}-\mtop^2}{(\hat{t}-M_W^2)^2}  
  \label{eq:WgQQbarXec}
\end{equation}
$\alpha_w=g_w^2/(4\,\pi)$ is the weak fine structure constant. 
The Mandelstam variable $\hat{t}$ is given by
\begin{equation}
 \hat{t} = - \frac{\hat{s}}{2}  \left( 1 - \frac{m_b^2}{\hat{s}} \right)
 \left( 1 - \frac{\mtop^2}{\hat{s}} \right) \cdot ( 1 - \cos \hat{\theta}) 
\end{equation} 
and $\hat{s}+\hat{t}+\hat{u} = m_b^2 + \mtop^2$.
Another leading order calculations of the cross section was done
van der Heide et al.~\cite{vanderHeide2000}.

The NLO calculation at first order in $\alpha_s$ comprises the square
of the Born terms, (\ref{eq:WgQQXsec}) and (\ref{eq:WgQQbarXec}),
plus the interference with the virtual graphs plus the square
of the real graphs with one single QCD coupling, 
e.g. figure~\ref{fig:Wg}c. 
Bordes and van Eijk presented an NLO calculation based on the
formalism described above in 1995~\cite{bordes1995}. 
They predict an enhancement of +28\% of the NLO cross section over the 
Born cross section for the Tevatron operating at $\sqrt{s} = 1.8\;\mathrm{TeV}$. 
Bordes and van Eijk used small masses for gluons and quarks to regularize
infrared and collinear divergencies. Mass factorization was performed in the
deep inelastic scattering (DIS) scheme.
Stelzer et al.~\cite{stelzerSullivan1997,stelzer1998} performed an 
NLO calculation entirely based on the $\overline{MS}$ factorization scheme. 
They predict a decrease of the cross section when going from leading order
to NLO by about -8\% to -10\%.

\renewcommand\arraystretch{2.0}
The latest calculation was done by Harris et al. in 2002~\cite{harris2002}
and contains full differential information, such that experimental 
acceptance cuts and jet definitions can be applied.
Their results are summarised in table~\ref{tab:singleTopCrossSection}.
We quote only these latest results for the cross sections, since previous
calculations used different PDFs, which by itself leads to big differences
in the predicted values. Harris et al. compare their results with those
given by Stelzer et al. using the latest PDFs. The agreement is very good,
within 1\%.
\begin{table}[t]
  \caption{\label{tab:singleTopCrossSection} Predicted total cross sections 
    for single top quark production processes.
    The cross sections are given for $p\pbar$ collisions at the Tevatron
    ($\sqrt{s} = 1.8\;\mathrm{TeV}$ or $\sqrt{s} = 1.96\;\mathrm{TeV}$) and
    $pp$ collisions at the LHC ($\sqrt{s} = 14\;\mathrm{TeV}$).
    The cross sections of the $t$- and $s$-channel process are taken from 
    reference~\cite{harris2002} and were evaluated with 
    CTEQ5M1 PDFs. The uncertainties were evaluated in 
    reference~\cite{sullivan2004}.
    The values for associated production are taken from 
    reference~\cite{ttait1999} (Tevatron at $\sqrt{s} = 1.96\;\mathrm{TeV}$ and 
    LHC) and
    reference.~\cite{heinson1997} (Tevatron at $\sqrt{s} = 1.8\;\mathrm{TeV}$).
    All cross sections are given for a top quark mass of 
    $\mtop = 175\;\mathrm{GeV}/c^2$. 
    }
  \begin{center}
  \begin{tabular}{lcccc}
  \br
  Process & $\sqrt{s}$ & $\sigma(\mathrm{t-channel})$ & 
  $\sigma(\mathrm{s-channel})$ & $\sigma(Wt)$ \\ \mr
  $p\pbar \rightarrow t / \tbar$ & 1.80 TeV & $1.45^{+0.20}_{-0.16}\;\mathrm{pb}$ &
  $(0.75\pm0.10)\;\mathrm{pb}$ & $0.14^{+0.05}_{-0.02}\;\mathrm{pb}$ \\
  $p\pbar \rightarrow t / \tbar$ & 1.96 TeV & $1.98^{+0.28}_{-0.22}\;\mathrm{pb}$ & 
  $(0.88\pm0.11)\;\mathrm{pb}$ & 
  $0.094^{+0.015}_{-0.012}\;\mathrm{pb}$ \\ \hline
  $p p \rightarrow t$            & 14.0 TeV & $(156\pm 8)\;\mathrm{pb}$ & 
  $(6.6\pm0.6)\;\mathrm{pb}$ & $14.0^{+3.8}_{-2.8}\;\mathrm{pb}$ \\ 
  $p p \rightarrow \tbar$        & 14.0 TeV & $( 91 \pm 5)\;\mathrm{pb}$ & 
  $(4.1\pm0.4)\;\mathrm{pb}$ & $14.0^{+3.8}_{-2.8}\;\mathrm{pb}$ \\
  \br
  \end{tabular}
  \end{center}
\end{table}
The cross sections given for the Tevatron are the sum of top and antitop quark 
production. 
In $pp$ collisions at the LHC the $W$-gluon fusion cross section differs 
for top and antitop quark production, which are therefore treated
separately in table~\ref{tab:singleTopCrossSection}. 
The increase in the centre-of-mass energy from 1.80$\;$TeV to 1.96$\;$TeV 
in Run II of the Tevatron is predicted to yield a 33\% increase in the total
cross section. The ratio of $\sigma_{Wg}/\sigma_{t\tbar}$ is about 30\%,
for the Tevatron as well as for the LHC. 
The uncertainties quoted in table~\ref{tab:singleTopCrossSection}
are evaluated in reference~\cite{sullivan2004}
and include the uncertainties due to the factorization scale $\mu$,
the choice of PDF parameterization, and the uncertainty in the 
top quark mass.
The factorization scale uncertainty is $\pm 4\%$ at the Tevatron and
$\pm 3\%$ at the LHC.
The central value was calculated with $\mu^2 = Q^2+\mtop^2$ for the
$b$ quark PDF. 
The scale for the light quark PDFs was set to $\mu^2 = Q^2$.
\renewcommand\arraystretch{1.0}

Of course the $t$-channel single top cross section depends on the top 
quark mass. 
The current uncertainty in the top quark mass 
($\Delta m = 4.3\,\mathrm{GeV}/c^2$) corresponds to about 7\% uncertainty
in the cross section at the Tevatron and 3\% at the LHC. 
The dependence is approximately linear in the relevant mass range. 

At first glance it is astonishing that the cross section for $W$-gluon fusion
is of the same order of magnitude as $t\tbar$ production although
it is a weak interaction process. There are several issues that lead
to this relative enhancement~\cite{WillenbrockDicus1986}:
\begin{enumerate}
\item The parton cross section of the $W$-gluon fusion mode scales 
  like $1/M_W^2$ as opposed to the $t\tbar$ cross section which, as a typical
  strong interaction process, scales like $1/\hat{s}$.
  At the Tevatron the subprocess energies are not much greater than
  $M_W$, so one does not gain from the scaling behaviour. However,
  at the LHC the effect is present. 
\item Single top production is kinematically enhanced compared to
  $t\tbar$ production, since only one heavy top quark is produced.
  For single top quark production the parton distribution functions 
  are therefore typically evaluated at half of the value of $x$ 
  needed for the strong process. Since the PDFs are monotonically decreasing
  functions, see figure~\ref{fig:pdfExample}, single top quark production
  is relatively enhanced to $t\bar{t}$ pair production.  
\item The $W$-gluon fusion process is enhanced by logarithmic terms 
  originating from the collinear singularity discussed above.
\end{enumerate}

\begin{figure}[t]
  \begin{center}
  \epsfig{file=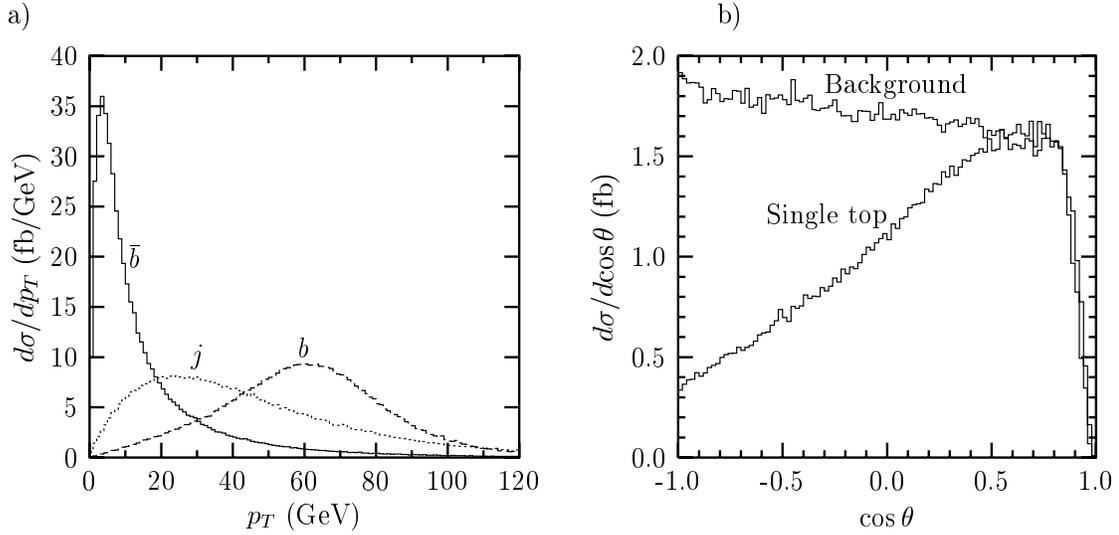, width=0.95\textwidth} 
  \end{center}
  \caption{\label{fig:wgJetPt} (a) Transverse momentum ($p_t$) distribution of 
    the quark jets from $W$-gluon fusion events: (1) the $b$ quark 
    from the top quark decay (solid line), (2) the light quark jet (dotted line),
    and (3) the $\bbar$ quark from the gluon splitting. The distributions
    are calculated for the Tevatron at $\sqrt{s} = 2\,\mathrm{TeV}$.
    (b) Angular distribution of the charged lepton in $W$-gluon fusion
    events at the Tevatron ($\sqrt{s} = 2\,\mathrm{TeV}$). $\theta$ is the
    angle between the lepton momentum and the light quark jet axis. 
    For comparison the same distribution is shown for the
    sum of all background processes ($W + jj$ and $t\tbar$). 
    Both plots, a) and b), are taken from reference~\cite{stelzer1998}.
  }
\end{figure}

In general, $W$-gluon fusion events have three quark jets originating 
from the hard interaction: (1) the $b$ quark jet from the top quark decay,
(2) the light quark jet, and (3) the $\bbar$-quark jet which comes from the
initial gluon splitting. The transverse momentum distribution 
of these jets is shown in figure~\ref{fig:wgJetPt}a. The $b$ quark 
is most of the times the hardest jet, the peak of the distribution is
around 60$\;$GeV. The light quark $p_t$ distribution peaks around
25 GeV, but has a long tail to high values. The $\bbar$ quark $p_t$
distribution peaks at low values. A large share of the $\bbar$ jets will therefore
not be identified, since an experimental analysis will require some
lower $p_t$ cut off, typically at 15~GeV. 

An interesting feature of single top quark production 
(in the t-channel and in the s-channel)
is that in its rest frame the top quark is 100\% polarized along
the direction of the $d$ quark 
($\dbar$ quark)~\cite{carlson1993,mahlon1997,heinson1997,stelzer1998}.
The reason for this is that the $W$ boson couples only to 
fermions with left-handed chirality.
Consequently, the ideal basis to study the top quark spin is the one
which uses the direction of the $d$ quark as the spin axis~\cite{mahlon1997}.
In $p\pbar$ collisions at the Tevatron $W$-gluon fusion proceeds via
$ug \rightarrow dt\bbar$ in 77\% of the cases. The $d$ quark can then be
measured by the light quark jet in the event.
The top quark spin is best analyzed in the spectator basis for which
the spin axis is defined along the light quark jet direction.
However, 23\% of the events proceed via $\dbar g \rightarrow \ubar t \bbar$, 
in which case the $\dbar$-quark is moving along one of the beam directions.
For these events the spectator basis is not ideal, but since the 
light quark jet occurs typically at high rapidity the dilution is small.
In total, the top quark has a net spin polarization of 96\% along the
direction of the light quark jet in $t$-channel single top quark 
production~\cite{mahlon1997}.
In $s$-channel events the best choice is the antiproton beam direction 
as spin basis. In 98\% of the cases the top quark spin is aligned in 
the antiproton direction~\cite{mahlon1997}. 

At the Tevatron top quarks are not produced as ultrarelativistic particles.
Therefore, the chirality eigenstates are not identical to the helicity
eigenstates.
The spin asymmetry $A = (N_\uparrow-N_\downarrow)/(N_\uparrow+N_\downarrow)$
is 0.91 for the $t$-channel and 0.96 for the $s$-channel.
 
Since the top quark does not hadronize, its decay products
carry information about the top quark polarization. 
A suitable variable to investigate the top quark polarization is the 
angular distribution of electrons and muons originating
from the decay chain $t \rightarrow W^+ + b$, $W^+ \rightarrow \ell^+ + \nu_\ell$.
If $\theta_{q\ell}$ is the angle between the charged lepton momentum and the
light quark jet axis in the top quark rest frame, 
the angular distribution is given by
$0.5\, (1+\cos \theta_{q\ell})$. 
A theoretical prediction for this quantity is shown in
figure~\ref{fig:wgJetPt}b~\cite{stelzer1998}. 
Single top quark events show a distinct slope which differs significantly
from the nearly flat background.

\subsubsection{s-channel production}
\begin{figure}[t]
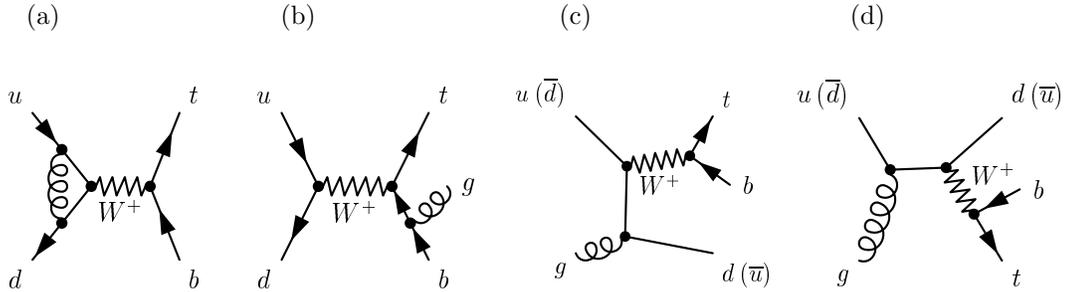

  \begin{center}
     (a)  \hspace{27mm} (b) \hspace{30mm} (c) \hspace{32mm} (d) \hspace{20mm}
    
    \vspace*{6mm}

    \epsfig{file=feynEpsi/sChanNLO3Pict.epsi, height=27mm} \hspace{5mm}
    \epsfig{file=feynEpsi/sChanNLO4Pict.epsi, height=27mm} \hspace{3mm}
    \epsfig{file=feynEpsi/sChanNLO2Pict.epsi, height=28mm} \hspace{1mm}
    \epsfig{file=feynEpsi/sChanNLO1Pict.epsi, height=28mm} 
  \end{center}
  \caption{\label{fig:sChannNLO} Examples of $\alpha_s$-corrections to the
    $s$-channel single top production mode.
  }
\end{figure}

The $s$-channel production mode of single top quarks probes a complementary aspect
of the weak charged current interaction of the top quark, since it is
mediated by a timelike $W$ boson with $q^2 \geq (M_\mathrm{top} + m_b)^2$ 
as opposed
to a spacelike $W$ boson in the $t$-channel process.
The leading order $s$-channel process is depicted in figure~\ref{fig:singleTop}c.
Feynman diagrams yielding corrections of order $\alpha_s$ are shown in
figure~\ref{fig:sChannNLO}. 
The first order corrections depicted in figure~\ref{fig:sChannNLO}c and
figure~\ref{fig:sChannNLO}d have the same initial and final states as
the W-gluon fusion diagrams shown in figure~\ref{fig:singleTop}a 
and figure~\ref{fig:singleTop}b. However, these two classes of diagrams
do not interfere because they have a different colour structure.
The $t\bbar$ pair in the $W$-gluon fusion process is in a colour-octet
state, since it originates from a gluon. In $s$-channel production
the $t\bbar$ pair forms a colour-singlet because it comes from 
a $W$. The different colour structure implies that both groups of
processes must be separately gauge invariant and, therefore, they
cannot interfere~\cite{WillenbrockDicus1986,Yuan1990,heinson1997}.

Several groups have calculated the cross section for the $s$-channel 
production mode in leading or next-to-leading order, 
respectively~\cite{cortese1991,smithWillenbrock96,heinson1997,mrenna1998,harris2002}. 
The leading order result for the partonic cross section
is given by~\cite{cortese1991}
\begin{equation}
 \fl 
 \hat{\sigma}_{ij} (\hat{s}) = \frac{\pi\,\alpha_w}{2}\cdot
 |V_{ij}|^2\,|V_{tb}^2|\cdot
 \frac{\sqrt{q_0^2-\mtop^2}}{\left(\hat{s}-M_W^2\right)^2}\cdot
 \left( q_0 - \frac{\mtop^2+2\,q_0^2}{3\sqrt{\hat{s}}}\right) 
 \ \ \ \mathrm{with} \ \ \  
 q_0 \equiv \frac{\hat{s}+\mtop^2-m_b^2}{2\,\sqrt{\hat{s}}}\;.
\end{equation}

In table~\ref{tab:singleTopCrossSection} we quote the latest 
NLO results by Harris et al.~\cite{harris2002} which are in very good 
agreement with the earlier calculations by 
Smith/Willenbrock~\cite{smithWillenbrock96} and Mrenna/Yuan~\cite{mrenna1998}.
The later authors have resummed soft gluon emission terms in the
$s$-channel cross section. Their resummed cross section is about 3\% above
the NLO value.
The predictions for the $s$-channel cross section
have a smaller uncertainty from the PDFs than for the $t$-channel 
because they do not
depend as strongly on gluon distribution functions as the $t$-channel calculation does. 
The uncertainty in the top quark mass leads to an uncertainty in the cross
section of about 10\% ($\Delta m = 4.3\,\mathrm{GeV}/c^2$, 
$\sqrt{s} = 1.96\;\mathrm{TeV}$))~\cite{sullivan2004}. 

The ratio of cross sections for the $t$-channel and $s$-channel mode
is 2.3 at the Tevatron ($\sqrt{s} = 1.96\;\mathrm{TeV}$) and 23
at the LHC. In $pp$ collisions the gluon initiated processes,
$t\tbar$ production and $W$-gluon fusion, dominate by far 
over $s$-channel single top quark production which is 
a quark-antiquark annihilation process. The $s$-channel signal
will therefore most likely be obscured at the LHC.
Thus, it will be essential to
observe $s$-channel single top quark production at the Tevatron.

\subsubsection{Associated production}
The third single top quark production mode is characterised by the
associated production of a top quark and an on-shell (or close to
on-shell) $W$ boson.
Studying associated production is interesting, since it probes a different
kinematic region of the $Wtb$ interaction vertex than $s$-channel
or $t$-channel production and thereby provides complementary information.
Predictions for the cross sections are given in
table~\ref{tab:singleTopCrossSection}. 
The calculation by T.~Tait~\cite{ttait1999} includes higher order
corrections proportional to $1/\ln(\mtop^2/m_b^2)$. The prescription
is very similar to the one described in section~\ref{sec:wgfusion}
for the $t$-channel mode. However, it is not a full NLO calculation.
Tait provides predictions for the Tevatron  
($\sqrt{s} = 2.00\;\mathrm{TeV}$) and the LHC 
($\sqrt{s} = 14\;\mathrm{TeV}$).
It is obvious from these numbers that associated production is
negligible at the Tevatron, but is quite important at the 
LHC where it even exceeds the $s$-channel production rate.
The errors quoted in table~\ref{tab:singleTopCrossSection} include
the uncertainty due to the choice of the factorization scale
($\pm 15\%$ at the LHC) and the parton 
distribution functions ($\pm 8\%$ at the LHC). 
The uncertainty in the top mass 
($\Delta m = 5\,\mathrm{GeV}/c^2$) causes a spread of the cross section
by $\pm 9\%$ at the LHC.
A second calculation by Belyaev and Boos yields a much higher cross section
for associated production: 
$\sigma(W\,t/\bar{t})= 62.0^{+16.6}_{-3.6}\,\mathrm{pb}$~\cite{boos2001}. 
This result was obtained using the CompHEP program~\cite{boos1995}. 
First studies~\cite{ttait1999} show that associated single top quark 
production will be observable at the LHC with data corresponding to 
about $1\;\mathrm{fb^{-1}}$ of integrated luminosity.

\subsection{Top quark decay}
\label{sec:topdecay}
In the SM top quarks decay predominantly into a $b$ quark and a $W$ boson.
The decays $t\rightarrow d + W^+$ and $t\rightarrow s + W^+$ are 
CKM suppressed relatively to $t \rightarrow b + W^+$ by factors of
$|V_{td}|^2$ and $|V_{ts}|^2$. If we assume the CKM matrix to be unitary
the values of these matrix elements can be inferred from other measured
matrix elements: 
$ 0.0048 < |V_{td}| < 0.014$ and $0.037 < |V_{ts}| < 0.043$~\cite{PDG2004}.
In the discussion of the following paragraphs we will therefore only 
consider the decay $t \rightarrow b + W^+$. 
Potential non-SM decays which would signal new physics will
be discussed in section~\ref{sec:anotop}.

At Born level the amplitude of the decay $t \rightarrow b + W^+$ is given 
by 
\begin{equation}
 \mathcal{M}(t\rightarrow b+W) = \frac{i\;g}{\sqrt{2}}\;\bbar\; 
 \epsilon^{\mu}_W\gamma_\mu \;
 \frac{1 - \gamma_5}{2}\;t \ .
\end{equation}
The decay amplitude is dominated by the contribution from longitudinal $W$ 
bosons because the decay rate of the longitudinal component scales 
with $\mtop^3$. 
In contrast, the top quark decay rate into transverse $W$ bosons increases
only linearly with $\mtop$. In both cases the $W^+$ couples solely to $b$ quarks of
left-handed chirality (a general feature of the SM). 
Since the $b$ quark is effectively massless, compared
to the mass scale set by $\mtop$, left-handed chirality translates
into left-handed helicity for the $b$ quark. 
If the $b$ quark is emitted anti-parallel to the top quark spin axis, 
the $W^+$ must be longitudinally polarized, $h^W$ = 0, to conserve angular
momentum.
If the $b$ quark is emitted parallel to the top quark spin axis, the $W^+$ boson
has helicity $h^W = -1$ and is transversely polarized. 
Thus, elementary angular momentum conservation 
forbids the production of $W$ bosons with positive helicity, $h^W=+1$, in 
top quark decays.
The ratios of decay rates into the three $W$ helicity states are given 
by~\cite{kuehnSLAC1997}:
\begin{equation} 
  \mathcal{A}(h^W=-1):\mathcal{A}(h^W=0):\mathcal{A}(h^W=+1) = 1 : 
  \frac{\mtop^2}{2\,M_W^2}:0.
\end{equation}
Strong next-to-leading order corrections to the decay rate ratios
have been calculated, lowering the fraction of longitudinal $W$ bosons
by 1.1\% and increasing the fraction of left-handed $W$ bosons by 
2.2\%~\cite{Fischer:2000kx,Fischer:2001gp}. Electroweak and finite widths 
effects have even smaller effects on the helicity ratios, inducing
corrections at the per-mille level~\cite{Do:2002ky}.
For the decay of antitop quarks negative helicity is forbidden.
In the SM the top quark decay rate, including first order QCD corrections, 
is given by
\begin{equation}
  \Gamma_{t} = \frac{G_F\;\mtop^3}{8\,\pi\,\sqrt{2}}\;
  \; \left| V_{tb} \right|^2 \;
  \left( 1 - \frac{M_W^2}{\mtop^2}\right)^2\;
  \left( 1 + 2\; \frac{M_W^2}{\mtop^2}\right)\;
  \left[ 1 - \frac{2\,\alpha_s}{3\,\pi} \cdot 
  f\left(y\right)
  \right]
  \label{eq:topdecay}
\end{equation}
with $y=(M_W/\mtop)^2$ and 
$f(y) = 2\,\pi^2/3-2.5-3y+4.5y^2-3y^2\ln{y}$~\cite{kuehnSLAC1997,jezabekKuehn1989,jezabekKuehn1988}.
Using $y=(80.45/174.3)^2$ we find $f(y)=3.85$.
The QCD corrections of order $\alpha_s$ lower the Born decay rate by
$-10\%$. A useful approximation of (\ref{eq:topdecay}) is given by
$\Gamma_\mathrm{top} \simeq 175\;\mathrm{MeV}/c^2\cdot(\mtop/M_W)^3$~\cite{kuehnSLAC1997,bigi1986}. 
The decay width increases from 
1.07 GeV/$c^2$ 
at $\mtop = 160\;\mathrm{GeV}/c^2$ to 
1.53 GeV/$c^2$
at $\mtop = 180\;\mathrm{GeV}/c^2$. 
Expression (\ref{eq:topdecay}) neglects higher order terms proportional to 
$m^2_b/\mtop^2$ and $\alpha^2_s$.
Corrections of order $\alpha_s^2$ were lately calculated, they lower
$\Gamma_\mathrm{top}$ by about -2\%~\cite{czarnecki1999,chetyrkin1999}.
Because the top quark width is small compared to its mass, 
interference between QCD corrections to production and decay
amplitudes has a small effect of order 
$\mathcal{O}(\alpha_s\Gamma_\mathrm{top}/\mtop)$~\cite{fadin1994}.
The decay width for events with hard gluon radiation ($E_g > 20\,\mathrm{GeV}$)
in the final state has been estimated to be 5 -- 10\% of $\Gamma_\mathrm{top}$,
depending on the gluon jet definition 
(cone size $\Delta R = 0.5$ to 1.0)~\cite{mrenna92}.  
Electroweak corrections to $\Gamma_\mathrm{top}$ have also been calculated
and increase the decay width by 
$\delta_\mathrm{EW}=+1.7\%$~\cite{denner1991,migneron1991}.
Taking the finite width of the $W$ boson into account leads to a
negative correction $\delta_\Gamma=-1.5\%$ such that $\delta_\mathrm{EW}$ and
$\delta_\Gamma$ almost cancel each other~\cite{jezabekKuehn1993}.

The large top decay rate implies a very short lifetime of 
$\tau_\mathrm{top} = 1/\Gamma_\mathrm{top} \approx 4\cdot10^{-25}\,\mathrm{s}$ 
which is smaller than the characteristic formation time of hadrons 
$\tau_\mathrm{form} \approx 1/\Lambda_\mathrm{QCD}\approx2\cdot 10^{-24}\;\mathrm{s}$.
In other words top quarks decay before they can couple hadronically 
to light quarks and form hadrons. The lifetime of $t\tbar$ bound states, 
toponium, is too small, $\Gamma_{t\tbar} \sim 2\,\Gamma_\mathrm{top}$, 
to allow for a proper
definition of a bound state with sharp binding energy. 
This feature of a heavy top quark was already pointed out 
in the early and mid 1980s~\cite{kuehn1981,kuehn1982,bigi1986}.

Even though top hadrons cannot be formed, there are other long-distance {QCD}
effects associated with hadronization which have to be considered.
The colour structure of the hard interaction process influences the subsequent
fragmentation and hadronization process.
In the process $e^+e^- \rightarrow t\bar{t}$ the top and antitop quark
are produced in a colour-singlet state. In hadronic collisions, on the
contrary, the production cross section is dominated by configurations
where the $t$ or $\bar{t}$ forms a colour-singlet with the proton or
antiproton remnant, respectively. The colour field -- or in the picture of
string fragmentation --  the string carries the more energy the further
the top quark and the remnant are apart.
If the distance in the top-remnant centre-of-mass system reaches 
about 1 fm before the top quark decays,
the colour string carries enough energy to form light hadrons.
Whether or whether not a significant fraction of top events exhibit 
the described ``early'' fragmentation process, depends strongly on the 
centre-of-mass energy of the hadron collider.
While at Tevatron energies early top quark fragmentation effects are
negligible~\cite{orr1991}, they may well play a r\^{o}le at the LHC,
 where top quarks are produced with a large Lorentz boost. 
If there is no early fragmentation, long-range QCD effects connect
the top quark decay products, the $b$ quarks or the quarks from
hadronic $W$ decays. Even if early fragmentation happens, the fragmentation
of heavy quarks is hard, as seen in $c$ and $b$ quark decays, i.e. 
the fractional energy loss of top quarks as they hadronize is small.
Therefore, it will be quite challenging to observe top quark fragmentation
experimentally, even at the LHC.

Within the constraints discussed above we can assume that top quarks are 
produced and decay like free quarks. The angular distribution
of their decay products follow spin $\frac{1}{2}$ predictions.
The angular distribution of $W$ bosons from top decays is propagated to its
decay products. In case of leptonic $W$ decays the polarization is preserved
and can be measured~\cite{jezabekKuehn1994}.
The angular distribution of charged leptons from $W$ decays originating
from top quarks is given by
\begin{equation}
  \frac{1}{\Gamma}\frac{\rmd\Gamma}{\rmd\cos\theta_\ell} = \frac{3}{4}
  \frac{\mtop^2 \sin^2 \theta_\ell+ M_W^2 (1-\cos \theta_\ell)^2}{\mtop^2 +
  2\,M_W^2}   
\end{equation}
where $\pi-\theta_\ell$ is the angle between the $b$ quark direction and 
the charged lepton in the $W$ boson rest frame~\cite{mahlon1998}.

\section{Experimental techniques}
\label{sec:topdetect}
Advanced experimental techniques are needed to detect and reconstruct
top quark events in hadronic collisions. 
Large scale general-purpose detectors are employed for that task,
their overall structure is quite similar.
Early searches for the top quark were conducted with the UA1 and UA2 experiments 
at the CERN S$p\bar{p}$S. 
The detectors CDF and \DZero are currently in operation at the
Fermilab Tevatron and we will discuss those as typical examples
of collider detectors in more detail in 
sections~\ref{sec:CDF} and~\ref{sec:DZero}, respectively.
In the future the LHC will also feature two general-purpose experiments,
CMS and ATLAS, which are currently under construction at 
CERN.

\renewcommand\arraystretch{1.4}
Usually, general purpose collider detectors feature rotational symmetry with 
respect to the nominal beam axis and forward-backward symmetry with respect 
to the nominal interaction point in the centre of the detector. 
Therefore, a right-handed coordinate system is chosen such that the origin is at the
centre of the detector and the $z$-axis points along the symmetry axis
parallel to the beam (in case of the Tevatron the proton beam). 
The $x$- and $y$-axes define the transverse plane, the $y$-axis points
vertically upwards.
It is often practical to replace the $x$ and $y$ coordinates by 
the azimuth angle $\phi=\arctan (y/x)$, measured in the transverse plane 
with respect to the $x$-axis, and
the polar angle $\theta$, measured with respect to the $z$-axis ($\theta = 0$).
$\phi$ takes values from 0 to $2\pi$, 
$\theta$ from 0 to $\pi$.
Instead of $\theta$ it is often handy to use the pseudorapidity $\eta$, which is
defined as $\eta=-\ln(\tan\theta/2)$. For massless particles $\eta$ is equal 
to the rapidity $y=1/2\,\ln((E+p_z)/(E-p_z))$,
which is an important quantity because the rapidity difference $\Delta y$ 
between two particles is Lorentz-invariant. 
It is also common to calculate the angle between two particles in terms of
the distance in the $\eta$-$\phi$ plane:
$\Delta R \equiv \sqrt{ \Delta \eta^2 + \Delta \phi^2}$.
Table~\ref{tab:kinematicVariables} gives a summary on the definition
of kinematic variables.
\begin{table}[bht]
  \caption{\label{tab:kinematicVariables}Summary of the definition of kinematic 
     variables.}
  \begin{center}
  \begin{tabular}{ll}
    \br
    invariant mass     & $ m^2 = p^\mu p_\mu = E^2 - \vec{p}\,^2 $ \\    
    tranverse momentum & $ p^2_t = p^2_x + p^2_y$ \\
    transverse mass    & 
      $ m^2_{t} \equiv m^2 + p^2_t = E^2 - p^2_z = (E + p_z) \cdot (E - p_z)$ \\
    rapidity           & 
      $ y \equiv \frac{1}{2} \ln \left( \frac{E+p_z}{E-p_z} \right) 
           = \ln{ \left( \frac{E + p_z}{m_t}\right)}$ \\
    pseudo-rapidity    & 
      $ \eta \equiv -\ln{\left(\tan{\frac{\theta}{2}} \right)}$ \\
    distance in the $\eta-\phi$ plane &  
    $\Delta R \equiv \sqrt{ \Delta \eta^2 + \Delta \phi^2}$ \\
    \br
  \end{tabular}
  \end{center}
\end{table}

\subsection{The Tevatron Collider}
As most of the knowledge about the top quark was obtained from measurements
of $p\bar{p}$ collisions at the Tevatron we briefly discuss this 
accelerator here. The Tevatron is located at the Fermi National Accelerator 
Laboratory (Fermilab) near Chicago.
In order to reach energies of 980~GeV per beam, a system of several 
accelerators is needed. 
In the first stage of acceleration, a Cockcroft-Walton pre-accelerator is 
used to generate negatively charged hydrogen ions out of hydrogen gas and then 
accelerate them via electric fields up to an energy of 750~keV. 
Afterwards, the ions enter an approximately 150~m long linear accelerator, 
where they are accelerated up to 400~MeV by oscillating electric fields. 
Before leaving this acceleration stage, the ions pass through a carbon foil, 
which removes their negative charges (electrons).
As a result one obtains a beam of protons that is subsequently bent in a 
circular path by the magnets of a circular accelerator, called the booster. 
On its way out of the booster, the beam has an energy of 8~GeV. 
In the next stage, the protons enter the Main Injector, 
a multitask accelerator completed in~1999. This machine accelerates protons 
up to 150~GeV. 
Some protons are accelerated to 120~GeV and used for antiproton production. 
They are forced to collide with a nickel target, which is installed at the 
antiproton source facility. The interactions with the target produce a variety 
of particles, among them many antiprotons, which are being collected, focused 
and finally stored in the Accumulator Ring. As soon as a sufficient number of 
antiprotons has been produced, they are sent to the Main Injector, which 
accelerates them up to an energy of 150~GeV. In the final stage, the proton and 
antiproton beams with 150~GeV energy are injected in the Tevatron, a circular 
accelerator with a circumference of about 6~km, which is the most powerful 
operational hadron accelerator worldwide. Each beam is accelerated to an energy 
of 0.98~TeV which is equal to (anti-)protons reaching velocities of 
0.9999995~times the speed of light. 
These beams are forced to collide with each other at two interaction regions, 
producing a centre of mass energy of $\sqrt{s}$~=~1.96~TeV. 
The general purpose experiments CDF and D\O \ are placed at these collision 
points. The performance of a collider is described with a quantity called 
{\it luminosity},
$\mathcal{L}$.
The event rate for a certain process with cross section $\sigma$ is given by 
the product $\dot{N}$~=~$\sigma\cdot\cal{L}$.
For a certain time interval the number of events produced by this process 
is given by $N$~=~$\sigma\cdot\int{\cal{L}}\,\rmd t$, where
$\int{\cal{L}}\,\rmd t$ is the {\it integrated luminosity}.
The peak luminosity is reached at the begin of a store after protons and 
antiprotons have been injected. Over the period of a store the luminosity 
slowly decreases as collisions and beam gas interactions lead to a lowering 
of beam currents, that is a loss of protons and antiprotons stored in the 
Tevatron. The maximum value of luminosity that has been achieved until 
May~2005 is~13.0$\cdot$10$^{31}$cm$^{-2}$s$^{-1}$, which is about 60\% above 
the design value of 8$\cdot$10$^{31}$cm$^{-2}$s$^{-1}$.

\subsection{The Collider Detector at Fermilab (CDF)}
\label{sec:CDF}
\begin{figure}[!t]
    \begin{center}
      \epsfig{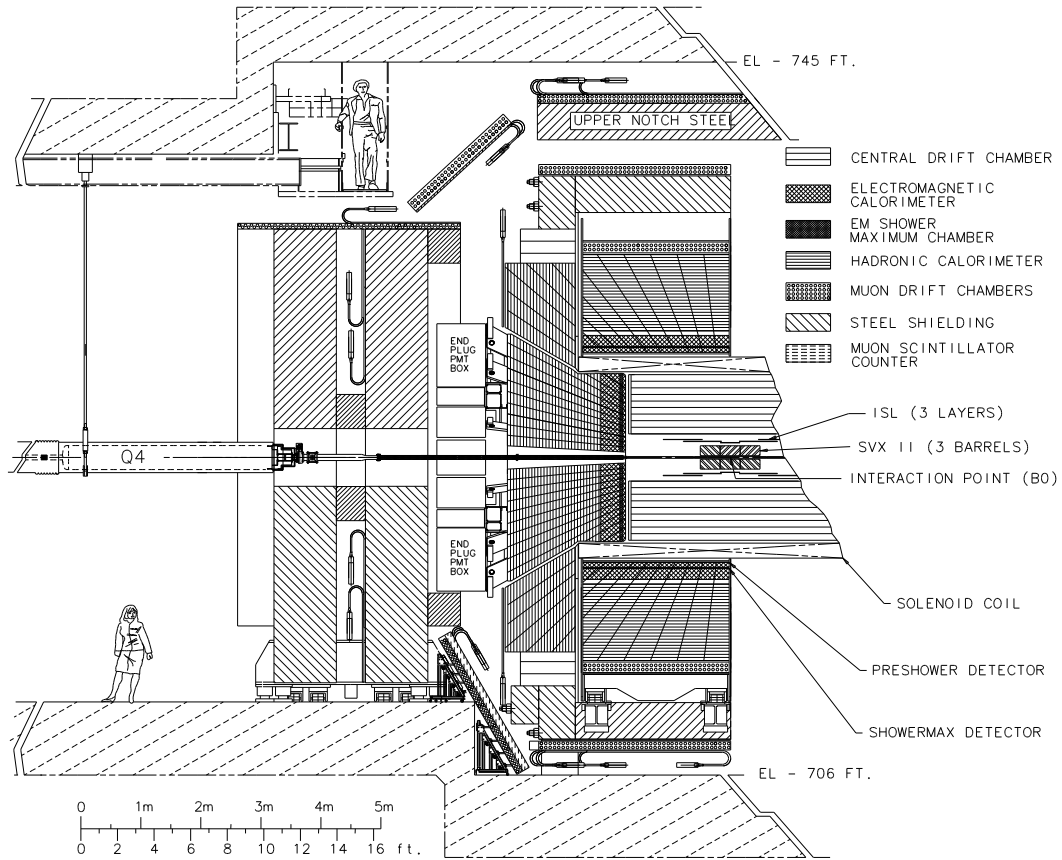}
    \end{center}
    \caption{\label{fig:CDFview} Elevation view of one half of the CDF II detector.}
  \end{figure}

CDF is in operation since 1987. The first Tevatron run (Run I) 
lasted until 1995, when a substantial upgrade program of the 
accelerator complex and the collider detectors began. 
The installation of the renewed experiment 
finished in 2001 and after a commissioning period of about
1 year CDF II is taking physics quality data. 
Run II is expected to last until 2009 and accumulate an integrated 
luminosity corresponding to 4.4 -- 
$8.5\,\mathrm{fb^{-1}}$~\cite{tevatron2004}.
Although we include physics results obtained in Run I,
we describe CDF in its upgraded form after 2001 (CDF II), see 
reference~\cite{CDFJPsiXS2005} for a brief overview.
A detailed description of the Run I detector can be found 
elsewhere~\cite{cdf1detNIM,cdfWmass95}.
Figure~\ref{fig:CDFview} shows an elevation view of one half of the CDF II 
detector.

\paragraph{Tracking System}
In CDF II particle collisions take place within a luminous region which is 
approximately described by a Gaussian distribution with a width of
30 cm. 
The inner part of CDF II is dedicated to reconstruct the trajectories of charged
particles and measure their momenta. The entire tracking volume is
immersed into a solenoidal magnetic field of $|\vec{B}|=1.4\,\mathrm{T}$.
The transverse momentum $p_T$ of charged particles is measured by determining
the curvature of their trajectories in the magnetic field.
The tracking system has two major components, 
the central drift chamber (COT)~\cite{cdfCOT}
and the silicon tracker that itself is composed of three subsystems:
Layer 00, the SVX II~\cite{SVXII} and the Intermediate Silicon Layers 
(ISL)~\cite{ISL}.
Layer 00 consists of one layer of radiation-tolerant single-sided silicon strip
detectors which are directly glued onto the beam pipe.
SVX II extends radially from $r_1=2.4\,\mathrm{cm}$ to $r_2=10.7\,\mathrm{cm}$, 
it is 96 cm long and provides full angular coverage in the pseudorapidity
region of $|\eta|\leq2.0$.
The SVX II is segmented in three cylindrical barrels with beryllium bulkheads 
at each end for mechanical support, cooling the modules and facilitating the 
readout.
The SVX II features five layers of double-sided silicon strip detectors
which provide measurements in the $r$-$\phi$ and $r$-$z$ views.
The ISL consists of one layer of double-sided strip detectors in the 
central region, $|\eta|\leq1.0$, and two layers in the forward and backward 
regions. 
The central ISL is located at a radius of 22 cm and allows to robustly link
tracks between the SVX II and the central drift chamber (COT).
The COT is a cylindrical open-cell drift chamber with inner and outer radii
of 40 and 137 cm. It is designed to find charged particles in the pseudorapidity
region of $|\eta|\leq1.0$ with transverse momenta as low as 400 MeV/$c$.
Four axial and four stereo super-layers with 12 sense wires each provide
a total of 96 measurements. 
The drift gas is a 50:50 Argon-Ethane admixture and the drift field is
1.9 kV/cm, yielding a maximum drift time of 180 ns.
When combining measurements in the silicon tracker and the COT the momentum 
resolution for charged particles is
$\delta p_T/p_T^2 < 0.17\%\,\mathrm{GeV^{-1}}\,c$.

Between the COT and the solenoid a Time-of-Flight (TOF) system has been 
installed to improve particle identification~\cite{cdfTOF}. The TOF consists of 
scintillation panels which provide both timing and amplitude information.
The main purpose of the TOF system is to distinguish between charged 
pions and kaons with momenta $p<1.6\;\mathrm{GeV/c}$,
which is important to reconstruct clean samples of exclusive $b$ hadron decays 
and to
tag the flavor of $b$ hadrons in CP violation and oscillation measurements. 

\paragraph{Calorimetry}
The solenoid of CDF is surrounded by calorimeters, where measurements of the 
electromagnetic and hadronic showers are performed. The calorimeters are 
segmented in azimuth and in pseudorapidity to form a projective tower geometry
which points back to the nominal interaction point.
In addition, the calorimeters are segmented longitudinally to distinguish between
electromagnetic (EM) and hadronic (HAD) showers. All CDF calorimeter systems 
are sampling calorimeters consisting of a sandwich of absorber (lead or iron)
and scintillator material. 
The inner part of the calorimeters is used to measure and identify 
electromagnetic showers. The central electromagnetic calorimeter is called
CEM~\cite{run1CEM}, the forward (plug) part PEM~\cite{CDFPlugUpgrade}.
To improve the spacial resolution a layer of proportional wire chambers is
located at the shower maximum in the electromagnetic calorimeter.
Additional proportional chambers located between the solenoid and the 
CEM sample the early development of electromagnetic showers in the magnet.
The outer parts of the calorimeters measure hadronic showers. There are three
hadron calorimeter subsystems:
the central (CHA)~\cite{run1CHA}, the wall (WHA) and the plug (PHA) calorimeter.
Table~\ref{tab:cdfCalo} summarises information on CDF calorimetry
including the coverage in pseudorapidity, the radiation thickness and the 
energy resolution.
\begin{table}
  \caption{\label{tab:cdfCalo} Summary of CDF calorimeter properties in Run II. The energy
    resolutions for the electromagnetic calorimeters are for incident electrons and 
    photons; in case of the hadron calorimeter for incident isolated pions.
    The $\oplus$ signifies that the constant term is added in quadrature. The 
    transverse energy $E_T$ and the energy $E$ are measured in units of GeV.}
  \begin{center}
  \begin{tabular}{ccrr} 
    \br
    System & Pseudorapidity    & Thickness             & 
      Energy Resolution \\ \mr
    CEM    & $\ \ \ |\eta|<1.1$      & 19 $X_0$, 1 $\lambda$ & 
      13.5\%/$\sqrt{E_T}\,\oplus$ 2\%  \\
    PEM    & $1.1<|\eta|<3.64$ & 21 $X_0$, 1 $\lambda$ & 
      16\%/$\sqrt{E}\,\oplus$ 1\% \\ \hline
    CHA    & $\ \ \ |\eta|<0.9$      & 4.5 $\lambda$ & 75\%/$\sqrt{E_T}\,\oplus$ 3\% \\
    WHA    & $0.7<|\eta|<1.3$  & 4.5 $\lambda$ & 75\%/$\sqrt{E}\,\oplus$ 4\% \\
    PHA    & $1.1<|\eta|<3.64$ & 7   $\lambda$ & 74\%/$\sqrt{E}\,\oplus$ 4\% \\ \br
  \end{tabular}
  \end{center}
\end{table}

\paragraph{Muon System}
In general, high-$p_T$ muons can traverse the calorimeter loosing only a small
fraction of their energy due to ionization. They act as minimum ionizing
particles. Most hadrons, on the other hand, interact strongly within the
calorimeter volume and produce hadronic showers. This effect allows
to identify muons by placing additional scintillation counters and wire chambers 
surrounding the calorimeter.
The CDF muon identification system is composed out of four subdetectors:
the Central Muon Detection System (CMU)~\cite{cdfCMU}, 
the Central Muon Upgrade (CMP),
the Central Muon Extension (CMX) and the Intermediate Muon System (IMU).
The muon system covers the pseudorapidity region of $|\eta|<1.5$. 
To reach the muon system the muons must have a minimum transverse momentum of
1.4 GeV/$c$.

\paragraph{Trigger and Data Acquisition}
The CDF trigger and data acquisition system features three distinct levels.
Level 1 is implemented in customized hardware, it uses input from the muon
system, the calorimeters and COT tracks reconstructed by the extra-fast
tracker (XFT). The maximum accept rate (i.e. output rate) of Level~1 is 50 kHz.
Level~2 is a software trigger implemented on an Alpha processor.
It allows to identify physics objects like electrons, photons, muons and 
jets. In addition, it is possible to select events with tracks which have
large impact parameter; a feature which resembles a revolution in enriching
samples for bottom- and charm-physics at hadron colliders.
The maximum output rate of Level 2 is approximately 300 Hz.
The highest trigger level, Level 3, runs part of the offline reconstruction
code on a PC farm and classifies events in different data streams. 
The output rate to permanent storage is about 75 Hz.
In most analyses top quark events are selected by requiring a high-$p_T$
muon or electron. The respective triggers are quite simple and
have high trigger efficiencies between 95 and 100\%.

\paragraph{Luminosity Measurement}
An important set of measurements at a hadron collider is to determine
the cross sections for the production of heavy particles, such
as $b$ quarks, top quarks or $W$ and $Z$ bosons.
To accomplish these measurements it is crucial to know the integrated 
luminosity of a given data set with good precision.
At CDF the luminosity measurement is performed by 
Cherenkov Luminosity Counters (CLC)~\cite{cdfLumi}. 
The detector consists of long conical, gaseous Cherenkov counters that 
are installed at small polar angles in the proton and antiproton 
directions and point to the collision region.
The Cherenkov counters measure the number of particles in the CLC
acceptance as well as their arrival time for each bunch crossing.
\par
The number of $p\overline{p}$ interactions per bunch crossing
follows a Poisson distribution with mean $\mu$.
The probability of empty crossings is given by 
$\mathcal{P}(0)=e^{-\mu}$. By measuring the fraction of empty
crossings with the CLC the average $\mu$ is determined~\cite{CDFJPsiXS2005}. 
The instantaneous luminosity is then given by
$\mathcal{L} = \bar{f}\cdot\mu/\sigma_\mathrm{inel}$, where $\bar{f}$ is the average 
bunch-crossing frequency and $\sigma_\mathrm{inel}$ the total inelastic 
$p\bar{p}$ cross section.
At the Tevatron in Run II we have $\bar{f}=1.7\,\mathrm{MHz}$.
To avoid a trigger bias the events for the luminosity measurement are
taken with a random beam-crossing trigger (zero bias trigger) running
at approximately 1~Hz.
The luminosity measurement has a systematic uncertainty of 6\% mainly due
to the uncertainty on the total inelastic cross section and the acceptance
of the CLC.

\subsection{The \DZero Experiment}
\label{sec:DZero}
\begin{figure}[!t]
  \begin{center}
    \epsfig{file=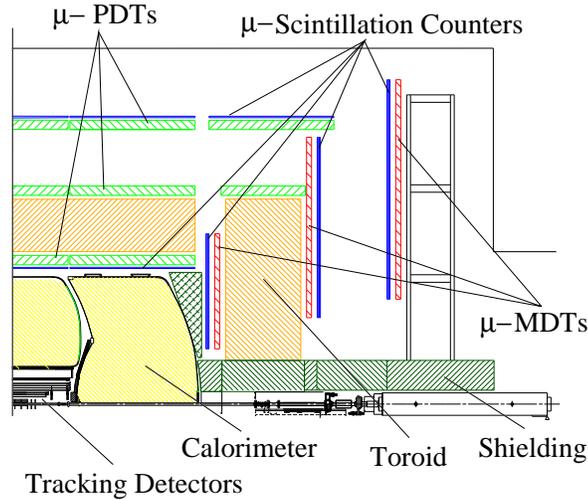, width=0.5\textwidth}
  \end{center}
  \caption{\label{fig:D0view} One quadrant of the \DZero detector.}
\end{figure}

The \DZero detector is a large general purpose detector primarily designed
to study high mass states and large transverse momentum phenomena.
The detector design was optimised with respect to three general goals:
(1) excellent identification and measurement of electrons and muons,
(2) good measurement of parton jets at large $p_T$ and
(3) a well-controlled measure of missing transverse energy ($\EtMiss$) as
a sign of the presence of neutrinos.
Figure~\ref{fig:D0view} shows one quadrant of the \DZero detector.
In the following paragraphs we discuss the subsystems of \DZero in more
details.

\paragraph{Tracking System}
The inner tracking system comprises two major subsystems:
a high precision silicon microstrip tracker (SMT)~\cite{D0SMT} and 
a scintillating fibre tracker~\cite{D0CFT}.
The inner tracker is encased in a superconducting solenoid, which provides
a magnetic field of 2 Tesla.
The SMT consists of six barrel modules and 16 disk modules.
Each barrel module has four radial layers of strip detector ladders
which are arranged parallel to the beam axis.
The layers are evenly spaced between radii of 2.5 and $10\,\mathrm{cm}$.
Single- and double-sided detectors provide measurements in the $r$-$\phi$ and
the $r$-$z$ views.
Layers 2 and 4 feature double-sided detectors which have axial strips on
each side with a small stereo angle of $2^\circ$ between the views.
There is one disk module with 12 wedge shaped double-sided detectors 
at one end of each barrel. 
These disks are called F-disks. 
The silicon wafers on a disk module are oriented perpendicular to the beam
axis and allow tracking at high pseudo-rapidities.
At each end of the barrel assembly there are additional three F-disk modules.
Thus, there is 12 F-disk modules in total.
Further along the beam pipe, on each side
of the detector, there are two larger disks of single-sided strip detectors (H-disks).

The Central Fibre Tracker (CFT) consists of scintillating fibres mounted on
eight concentric cylinders which extend radially from $19.5\,\mathrm{cm}$ to 
$51.5\,\mathrm{cm}$.
The detector is divided into 80 sectors in azimuth. Each sector contains
896 fibres, resulting in a total of 71680 channels.
The CFT provides full angular coverage of the central region up to
$|\eta|<1.7$.
Charged particles passing through a fibre produce scintillation light
that will travel along the fibre in both directions.
At one end, an aluminium mirror reflects the light back into the fibre.
At the other end, the fibre is joined to a wavelength shifting waveguide
that transmits the light to silicon avalanche photodiodes, 
which convert the light into an electrical signal.  
The momentum resolution of the \DZero tracking system can be parametrized
by $\Delta p_T / p_T = \sqrt{0.015^2+(0.0014\,p_T)^2}$, where
$p_T$ is measured in GeV.

\paragraph{Calorimetry}
\DZero has a sampling calorimeter based on depleted uranium, lead and copper
as absorber materials and liquid argon as sampling medium.
The calorimeter consists of three cryostats:
the Centre Calorimeter (CC), which covers the pseudorapidity region of
$|\eta|<1.2$ and two endcap calorimeters (EC) which extend up to $|\eta|\approx 4$.
The electromagnetic section of the calorimeter has a thickness of 
approximately 20 radiation lengths $X_0$. 
Calorimeter cells in the central region have a size of 
$\Delta \eta \times \Delta \phi = 0.1 \times 0.1$,
except for the third layer of the electromagnetic section, where the grid size
is $\Delta \eta \times \Delta \phi = 0.05 \times 0.05$ to provide better
spacial resolution in the maximum of electromagnetic showers.
At $\eta=0$, the CC has a total of 7.2 nuclear interaction lengths ($\lambda$),
at the smallest angle of the EC the total is $10.3\,\lambda$.
The energy resolution of the calorimeter is
$\Delta E/E=15\%/\sqrt{E}\oplus 0.4\%$ for electrons and photons.
For charged pions and jets the energy resolution is
$50\%/\sqrt{E}$ and $80\%/\sqrt{E}$, respectively ($E$ is measured in GeV).

To improve electron identification
\DZero features a preshower detector which consists of two subsystems:
the central system covering the region with $|\eta|<1.3$ (CPS) and the
forward systems covering $1.5 < |\eta| < 2.5$ (FPS)~\cite{D0ForwardPreshower}.
Both subdetectors are based on a combination of lead radiator and scintillator
layers. Electrons and photons shower in the radiator, while muons or charged
pions only deposit energy due to ionization.
In the central region the lead is mounted on the outer surface of the solenoid
amounting to a total of $2\,X_0$.
The scintillator is placed between the lead layer and the calorimeter.
In the forward region, which is not fully covered by the tracking system,
the radiator is sandwiched between two scintillator layers.
Photons do not deposit energy in the first scintillator layer and can thereby
be distinguished from electrons.

\paragraph{Muon System}
The muon detector of \DZero consists of three major components:
(1) a toroid magnet which provides a magnetic field of 
$|\vec{B}|=1.8\;\mathrm{T}$, (2) the Wide Angle Muon Spectrometer 
(WAMUS)~\cite{D0muon}
covering the central region of $|\eta|<1.0$, and (3) the 
Forward Angle Muon Spectrometer (FAMUS) covering the area of
$1.0<|\eta|<2.0$~\cite{D0muonForward}.
Both spectrometers are based on layers of proportional drift tubes and
scintillation counters.
Being immersed in a magnetic field allows the muon system to
provide a momentum measurement for muons independent of the inner tracker.
This feature of \DZero is a major difference to CDF where the muon
momentum is determined by matching a muon stub to a track in the central
tracker.
The total amount of material in the calorimeter and the iron toroids
varies between 13 and 19 nuclear interaction lengths, making the background from
hadronic punch-through negligible.

\paragraph{Trigger and Data Acquisition}
\DZero has a three-tier system of triggers to select events for recording.
The Level 1 trigger is a hardware based system with an output
rate of 10~kHz~\cite{D0level0}.
Data coming from the preshower detector and the fibre tracker (CFT) are
combined to form the Central Track Trigger (CTT)~\cite{D0CTT}.
In addition, there are conventional calorimeter and muon triggers.
Level 2 of the trigger system delivers an output rate to 1~kHz~\cite{Linnemann2001}.
At the second trigger level information of all detector components
can be combined at improved resolutions.
Level 3, running high level software algorithms on a PC farm, 
further reduces the rate to 50~Hz.

\paragraph{Luminosity Measurement}
At \DZero the luminosity is measured by two hodoscopes located on the inside
face of the calorimeters~\cite{D0lumi}. Each of these hodoscopes is made of 24
wedge shaped scintillators with fine mesh photo-multiplier tubes mounted
on the face of each wedge. 
The hodoscopes cover the pseudorapidity region of
$2.7 < |\eta| < 4.4$.
The luminosity counters are designed to 
distinguish between single and multiple interaction events. The measurement
of the rate of single inelastic interaction events is used to deduce the 
Poisson mean $\mu$ 
of the number of interactions per bunch-crossing.
The instantaneous luminosity is then calculated by using 
$\mathcal{L} = \bar{f}\cdot\mu/\sigma_{inel}$ as discussed in Sec.~\ref{sec:CDF}.
The luminosity measurement has an uncertainty of 7\%.

\subsection{Top quark signatures in $t\bar{t}$ events}
\label{sec:signatures}
In this section we discuss the experimental signature of top quark events.
We constrain the discussion to the decay mode which dominates in the SM: 
$t\rightarrow W+b$ with a branching ratio close to 100\%. 
We further concentrate on the signatures of $t\bar{t}$ events, 
since pair production is the main source of top quarks at the Tevatron
and at the LHC.
The signature of single top quark events and top quark decays via 
flavour changing neutral currents are presented in 
chapters~\ref{sec:singleTop} and \ref{sec:anotop}.

Once both top quarks have decayed, a $t\bar{t}$ event contains two $W$ 
bosons and two $b$ quarks: $W^+W^-b\bar{b}$.
Experimentally, $t\bar{t}$ events are classified according to the decay modes 
of the $W$ bosons.
There are three leptonic modes ($e\nu_e$, $\mu\nu_\mu$, $\tau\nu_\tau$) and
six decay modes into quarks of different flavour 
($u\bar{d}$, $u\bar{s}$, $u\bar{b}$, $c\bar{d}$, $c\bar{s}$, $c\bar{b}$).
We distinguish four $t\bar{t}$ event categories:
\begin{enumerate}
  \item Both $W$ bosons decay into light leptons 
    (either $e\nu_e$ or $\mu\nu_\mu$) which can be directly seen in the 
    detector. This category is called {\it dilepton} channel.
  \item One $W$ boson decays into $e\nu_e$ or $\mu\nu_\mu$. The second $W$
    decays into quarks. This channel is called {\it lepton-plus-jets}.
  \item Both $W$ bosons decay into quarks. We refer to this mode as the
    {\it all hadronic} channel.
  \item At least one $W$ boson decays into a tau lepton ($\tau\nu_\tau$)
    which itself can decay either leptonically (into $e\nu_e$ or $\mu\nu_\mu$)
    or hadronically into quarks.
\end{enumerate}
In good approximation we can neglect all lepton masses with respect to the 
$W$ mass and write:
\[ \Gamma^0_W \equiv \Gamma(W\rightarrow e\nu_e) = 
   \Gamma(W\rightarrow \mu\nu_\mu) =
   \Gamma(W\rightarrow \tau\nu_\tau) \]
At lowest order in perturbation theory the decay rate into a 
quark-antiquark pair, $q_1\bar{q}_2$, is given by the rate into leptons 
$\Gamma^0_W$ multiplied by
the square of the CKM matrix element $|V_{q_1q_2}|^2$ and enhanced by a 
colour factor of 3, which takes into account that quarks come in three 
different colours:
\[ \Gamma (W\rightarrow q_1 \bar{q}_2) = 3\,|V_{q_1q_2}|^2\;\Gamma^0_W \ .\]
The hadronic decay width $\Gamma_\mathrm{had}$ of the $W$ is summed over all six 
quark-antiquark modes 
\[ \Gamma_\mathrm{had} = 3\,\Gamma^0_W \cdot \sum_{q_1q_2} |V_{q_1q_2}|^2  
   \ \ \ \mathrm{with} \ \ \ 
   q_1q_2 \in \{ud, us, ub, cd, cs, cb \}.\]
The unitarity of the CKM matrix demands that the sum over the CKM elements 
squared is equal to 2 and 
one obtains $\Gamma_\mathrm{had}=6\,\Gamma^0_W$. 
As a result, each leptonic channel has a branching ratio of 1/9,
while the hadronic decay channel into two quarks has a branching ratio of 
6/9.
For the $t\bar{t}$ decay categories we get thus the probabilities as listed
in Tab.~\ref{tab:ttbarDecay}.
\begin{table}
  \caption{\label{tab:ttbarDecay} Categories of $t\bar{t}$ events and their 
    branching fractions.}
  \begin{center}
  \begin{tabular}{cccc}
    \br
    $W$ decays & $e/\mu\nu$ & $\tau\nu$ & $q\bar{q}$ \\ \mr
    $e/\mu\nu$ &   4/81     &  4/81     &   24/81    \\
    $\tau\nu$  &   --       &  1/81     &   12/81    \\
    $q\bar{q}$ &   --       &   --      &   36/81    \\ \br
  \end{tabular} 
  \end{center}
\end{table}
(i) Dilepton mode: 4/81, (ii) lepton-plus-jets: 24/81,
(iii) all-hadronic channel: 36/81, (iv) tau modes: 17/81. 
These four different types of $t\bar{t}$ events can be isolated by their
distinct event topologies.

\paragraph{Dilepton Channel}
This final state includes (1) two high $p_T$ leptons, electron or muon,
(2) a large imbalance in the total transverse momentum 
(missing transverse energy, $\EtMiss$) due to two neutrinos, 
and (3) two $b$ quark jets.
The dilepton event category has low backgrounds,
especially in the $e\mu$ channel, since $Z^0$ mediated events do not 
contribute. 
However, the drawback of the dilepton channel is its low branching
ratio of about 5\%.
There is a small contribution from tau events to the dilepton channel, if the
tau decays into $e$ or $\mu$. This cross feed has to be taken into account
when calculating acceptances for dilepton analyses.
Since dilepton events contain two neutrinos which contribute to the $\EtMiss$,
the top quarks cannot be fully reconstructed. 
This is a drawback if one wants to measure the top quark mass.
However, this disadvantage is partially compensated by the precisely measured
lepton momenta, in contrast to the only fair measurement of jet energies in
the lepton-plus-jets channel. 

\begin{figure}[!t]
  \begin{center}
  \includegraphics[width=0.4\textwidth]{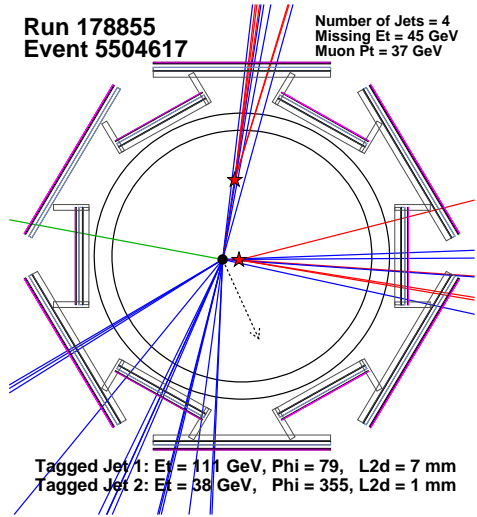}
  \end{center}
  \caption{\label{fig:ljetEvent}Event display of a CDF II muon-plus-jets
  event. The isolated line on the left hand side shows the muon
  trajectory. The arrow on the lower right indicated the direction of the 
  $\EtMissVec$. The event features four hadronic jets.}
\end{figure}

\paragraph{Lepton-Plus-Jets Channel}
The lepton-plus-jets channel is characterised by (1) exactly one
high-$p_T$ electron or muon, which is also called the primary lepton,
(2) missing transverse energy, (3)
two $b$ quark jets from the top decays, and (4) two additional jets
from one $W$ decay. 
An event display of a lepton-plus-jets candidate event measured at CDF II
is shown in figure~\ref{fig:ljetEvent}.
The big advantage of the lepton-plus-jets
channel is its high branching fraction of about 30\%. The backgrounds
are considerably higher than in the dilepton channel, but still manageable.
Several strategies of background suppression have been developed and
are discussed in section~\ref{sec:leptonPlusJets}. Again, as in the dilepton
channel, there is some cross feed from tau modes which has to be taken
into account for acceptances. 
In lepton-plus-jets events the momentum of the leptonically decaying 
$W$ boson can be reconstructed up to a twofold ambiguity. The
transverse momentum of the neutrino is assumed to be given by $\EtMiss$.
Two solutions for the $z$ component of the neutrino are obtained from the
requirement that the reconstructed invariant mass of the lepton and the 
neutrino be equal to the well known $W$ mass: $M_{\ell\nu}=M_W$.
To fully reconstruct the momenta of the top and antitop quark in the 
event, 
one has to assign the measured hadronic jets to the quarks.  
Without identification of $b$ quark jets
there are 24 possible combinations, including the ambiguity of the
neutrino reconstruction. If one jet is identified as likely to originate
from a $b$ quark 12 combinations remain. If there are two identified $b$
jets the ambiguity is down to 4 options. This illustrates the importance
of $b$ quark jet identification for top quark physics.

\paragraph{All Hadronic Channel}
The all hadronic channel has the largest branching ratio out of all
$t\bar{t}$ event categories, about 44\%. However, backgrounds are 
quite considerable. While there is the advantage that all final state
partons are measured, one has to deal with numerous combinations
when reconstructing the top quark momenta. Even if two jets are identified
as $b$ quark jets, 12 possible combinations remain. Another drawback is 
that 
jet energies as measured in the calorimeter have large uncertainties.
The combination of these disadvantages leads to the conclusion that the 
all hadronic 
channel, even though a clear $t\bar{t}$ signal has been established here,
proves not very useful to further investigate top quark properties.   

\paragraph{Tau Modes}
Top quark events containing the decay $W\rightarrow\tau\nu_\tau$ are
difficult to identify and have not been seen to date. The search for 
this decay mode is briefly described in section~\ref{sec:taumodes}. 

\renewcommand\arraystretch{1.0}

\subsection{Particle detection and identification}
In the previous section we have explained $t\bar{t}$ signatures in terms
of primary partons. In this section we will discuss how these partons
are detected and how their properties are measured.

\subsubsection{Electrons and muons}
\label{sec:eleMuoID}
Charged leptons emerging from a decay of a $W$ boson have high transverse
momenta that can be measured by the tracking system with good resolution.
Electrons interact within the first few segments of the calorimeter 
and form electromagnetic showers of photons and electron-positron
pairs. To select electrons from $W$ decays one typically requires
a transverse energy, $E_T = E\cdot\sin\theta$, of at least 20 GeV.
The energy resolution (at CDF for example) is 
$\sigma(E_T)/E_T=13.5\%/\sqrt{E_T}\oplus 2\%$, where $E_T$ is measured
in GeV.
The shower should be mainly contained in the electromagnetic part
of the calorimeter. A typical requirement is that 90\% of the total
measured energy is observed in the electromagnetic calorimeter.
Moreover, there must be a track that, if extrapolated to the 
calorimeter, matches the location of the electromagnetic shower
and has a momentum consistent with the shower energy ($E/p\sim1$).
Additional requirements include the transverse and longitudinal shapes
of the shower, which are known from test beam data. 
Energetic photons that interact with detector material prior 
to their entry into the tracking system and produce an electron-positron 
pair can mimic high-$p_T$ primary electrons and pose a serious background.
One can identify and veto these conversion pairs if both, the electron
and the positron track, are reconstructed and fulfil a matching 
criterion with respect to a common vertex. 
The major background to high-$p_T$ electrons arises from
hadronic showers developing early in the calorimeter and depositing 
most of their energy in the electromagnetic section.
Such a case can, e.g., be caused by photons from $\pi^0$ decays. 
The hadronic jet background can 
be considerably suppressed by exploiting the fact that leptons coming from 
a heavy boson decay are isolated from other jet activity in the event.
Thus, it is asked that the electron shower is isolated
from other energy deposits in the calorimeter. A typical requirement
is that the energy measured in an annulus of radius $\Delta R = 0.4$
around the electron shower is less than 10\% of the shower energy.

Muons can be reliably identified as tracks that penetrate the calorimeter
as well as additional shielding material and reach the outmost layers of the
collider experiment, the muon system. The track is typically required to
have a transverse momentum of at least 20 GeV/c and match, when 
extrapolated through the calorimeter, a short track segment measured
in the muon system. Uncertainties due to multiple scattering in the
calorimeter and the shielding material have to be considered accordingly
in the matching criteria. 
The transverse momentum of high-$p_T$ muons is measured with a typical resolution
of $\sigma(p_T)/p_T=0.1\%\,p_T$, where $p_T$ is given in GeV/$c$.
The energy measured in the calorimeter segment containing
the muon is required to be consistent with the deposition expected from
a minimum ionizing particle. 
Muons are required to be isolated from additional calorimeter activity 
as described above for electrons. 
Real muons from cosmic rays have to be 
distinguished from muons originating from hadronic collisions and removed
from the data sample. One requires that the muon track extrapolates to
the primary collision point of the event. Timing information of the 
tracker and the calorimeter are used to ensure that the passage of 
the muon through the primary interaction region falls into a narrow time
window around the beam crossing, that is known from the accelerator 
clock signal.
Other backgrounds to the muon signal arise from pion or kaon decays in
flight, or from hadrons that traverse the calorimeter and the absorber
material without producing a hadronic shower (punch-through). 

High-$p_T$ electrons and muons can be speedily reconstructed with good
precision and are
used in the trigger systems of collider experiments to select events
containing heavy gauge bosons and other high-$p_T$ events in real time.
The primary data samples for $t\bar{t}$ dilepton and lepton-plus-jets
events are defined in this way.

\subsubsection{Neutrinos}
Neutrinos interact only weakly with matter and therefore cannot be 
directly observed in a collider detector. Instead their presence is
inferred from an imbalance in the total transverse energy of the
event. As described in section~\ref{sec:factor} the hard scattering in 
hadronic collisions happens between two partons whose momentum
fraction is {\it a-priori} unknown. 
The remnants of the colliding hadrons, the spectator partons that do not
participate in the hard interaction, have little transverse
momentum and escape undetected down the beam pipe. 
In contrast to $e^+e^-$ collisions one can therefore not simply 
invoke energy and momentum conservation before and after the collisions.
However, the total transverse momentum is conserved and is known to be zero 
before the collision.
Any imbalance in the vector sum of transverse momenta can  
therefore be attributed to the presence of neutrinos that carry away momentum
undetected. In practice, one cannot determine the momenta of all particles
produced in the collision, but rather measures the imbalance of energy 
in the calorimeter. To each calorimeter cell $i$ one assigns a
transverse energy vector 
$\bi{E}_\mathrm{T}^i=(E_i\sin\theta_i\cos\phi_i,\;E_i\sin\theta_i\sin\phi_i)$ 
and calculates the sum over all cells: 
$\EtMissVec = - \sum_i\bi{E}_\mathrm{T}^i$.
Here $\theta_i$ and $\phi_i$ are the angular coordinates of calorimeter cell
$i$. Since muons loose only a minimal amount of their energy in the 
calorimeter, $\EtMissVec$ has to be corrected for identified muons.
The limiting factor on the resolution of $\EtMiss=|\EtMissVec|$ is the uncertainty 
in the measurement of hadronic jet energies in the calorimeter. Therefore,
one usually parametrizes the $\EtMiss$ resolution in terms of the
scalar sum of all transverse energies in the event, $\sum E_\mathrm{T}$, measured 
in GeV.
For $t\bar{t}$ dilepton candidate events
CDF has measured a resolution of 
$\sigma(\EtMiss)=0.7\cdot\sqrt{\sum E_\mathrm{T}}$~\cite{CDFtopEvidencePRD}.
D\O \ determined its $\EtMiss$ resolution for minimum bias data and quotes 
$\sigma(\EtMiss)=1.08\;\mathrm{GeV}+0.019\;\sum E_\mathrm{T}$~\cite{d0Evidence}.

\subsubsection{Jets of quarks and gluons}
\label{sec:jetReco}
The strength of the strong force increases with distance and prevents
elementary particles carrying colour charge, i.e. quarks and gluons, from 
existing 
as free objects. Quarks and gluons are confined to exist in hadrons, which are
colour singlets. In collider experiments quarks and gluons participate
in the hard interaction and subsequently form collimated jets of hadrons 
which can be observed in a detector. Jets tend to preserve the direction of 
motion of the original parton. In the calorimeter they are detected as an 
extended cluster of energy.
To compare measurements with theoretical predictions a precise mathematical
prescription is necessary how to form jets out of energy clusters 
measured in the calorimeter. Different jet algorithms are available,
but at hadron colliders calorimeter cells are conventionally combined 
within a cone of fixed radius $\Delta R$ in $\eta$-$\phi$ space, because jets
are approximately circular in the $\eta$-$\phi$ plane. Moreover, in this
parametrization the size of a jet of a particular $E_\mathrm{T}$ is independent of the 
jet rapidity. The cone size $\Delta R$ is chosen differently depending on the   
physics analysis. On one hand the cone must be big enough, such that most of the 
energy associated to the original quark is contained in the jet cone. 
On the other hand the cone size should be small enough, such that energy 
depositions 
corresponding to different primary partons are resolved individually rather
than merged to one jet.
This is especially a concern for $t\bar{t}$ events which have many jets
in their final state. At the Tevatron the optimal choice was found to be
$\Delta R = 0.4$ for CDF and $\Delta R = 0.5$ for D\O. The difference 
between the two experiments reflects a small dependence on the calorimeter
design and geometry.

Both Tevatron experiments feature an energy resolution for jets of
$\sigma(E_\mathrm{T})/E_\mathrm{T}\approx 100\%/\sqrt{E_\mathrm{T}}$, 
where $E_\mathrm{T}$ is measured in GeV. 
Several systematic effects compromise the jet energy measurement:
(1) intrinsic large fluctuations in the response of calorimeters to hadronic
showers, (2) nonlinear effects in calorimeter response, 
(3) calorimeter non-uniformities and energy loss in uninstrumented regions, 
such as cracks between modules, 
(4) increase in energy due to the overlap of multiple hard interactions in one 
bunch crossing,
(5) energy of the underlying event feeding into the jet cone
(The underlying event consists of particles coming from the fragmentation of 
 partons that do not participate in the hard scattering of the primary 
 $p\bar{p}$ or $pp$ interaction.), and
(6) energy loss due to the use of a finite cone size in jet reconstruction.
To understand these effects experimentally turns out to be one of the major
challenges in collider physics. Appropriate correction methods
have to be derived and the detector simulation has to be tuned to describe
the measurements.
  
Calorimeter calibration commonly starts with the electromagnetic section
for which an absolute energy scale can be derived by comparing electron
energies measured in the calorimeter to their momenta measured in the
tracking system or by reconstructing resonances with well known masses 
such as the $Z$ boson, the $\pi^0$ or the $J/\psi$. 
In a second step the hadronic part of the calorimeter can be calibrated
against the electromagnetic part by studying photon-plus-one-jet events, 
where the transverse energies should balance.
Dijet events are further used to calibrate one hadronic region of the 
detector relative to another better understood region.
Contributions from the underlying event are investigated by Monte Carlo 
simulations and comparison to data taken under special trigger conditions
and luminosities. 

Finally, it has to be noted that one cannot achieve
a one-to-one correspondence of primary quarks and gluons and the observed
jets in all kinematic regions and for all event topologies of a certain
physics process, and one cannot expect to do so. To define jets a minimum
of transverse energy is required. One may start counting soft jets with 
$8\,\mathrm{GeV}$.
However, this can complicate discrimination between signal and background.
In most $t\bar{t}$ analyses jets are typically defined with 
$E_\mathrm{T}>15\,\mathrm{GeV}$. This threshold causes, e.g., some $t\bar{t}$ 
lepton-plus-jets events to have only three instead of four jets.
Another source of inefficiency in jet reconstruction is the merging of
jets which are close together in $\eta$-$\phi$ space and cannot be resolved
separately. Conversely, if a parton radiates a gluon with large relative 
transverse momentum, that gluon can be reconstructed as an additional jet.

\subsubsection{Tagging of $b$ quark jets}
\label{sec:btagging}
As pointed out in section~\ref{sec:signatures}, $t\bar{t}$ events feature
two energetic $b$ quark jets, while the heavy flavour content in background
events is relatively low. Therefore, the identification, also called the 
tagging, of $b$ quark jets is an important tool to suppress background, enrich 
top quark data samples, and to facilitate the full reconstruction of top quark 
momenta by reducing the number of possible combinations of final state objects. 
In this section we briefly present three $b$ tagging methods employed
at hadron colliders. The first two methods are based on the relatively long 
lifetime of $b$ hadrons, which is about $1.5\,\mathrm{ps}$.
Since $b$ hadrons emerging from top quark decays have relatively high momenta, 
their long lifetime allows them to travel several mm before decaying.
Tracks from a $b$ hadron decay therefore typically originate from a secondary
vertex that is displaced from the primary interaction point.  
The first $b$ tagging method reconstructs the secondary vertex of
$b$ hadrons within a jet and bases the $b$ tag on the significance of the
displacement. The second method searches for tracks with large impact parameter
with respect to the primary vertex, but does not require the reconstruction
of the secondary vertex itself. 
The third method identifies leptons at intermediate momenta 
(typically: $2\,\mathrm{GeV}/c<p_\mathrm{T}<20\,\mathrm{GeV}/c$) from semileptonic 
$b$ decays within jets. Since momenta of these leptons are much smaller than
those of leptons coming from $W$ or $Z$ boson decays, this method is also 
called {\it soft lepton tag}.

\paragraph{Secondary vertex tag}
The secondary vertex method identifies $b$ jets by establishing 
a displaced secondary vertex in the jet from the decay of a long lived 
$b$ hadron. 
The following explanation goes along the lines of the algorithm 
employed by CDF, but the D\O \ version is quite similar.
In a first step the space point of the hard primary interaction is
precisely reconstructed. One starts by clustering the $z$ coordinates 
of all tracks at the point of their closest approach to the origin (perigee).
After applying several quality requirements this procedure yields an estimate
of the $z$ position of the primary vertex, $z_\mathrm{pv}$.
The trajectory of the proton and antiproton beams through the collision
region, also called the \emph{beamline}, is known with high precision,
$\mathcal{O}(1\,\mu \mathrm{m})$, 
from subsidiary measurements made on a run-by-run basis. 
Combining $z_\mathrm{pv}$ with the beamline yields
a first estimate of the primary vertex position $\bi{x}_\mathrm{pv}$. 
The precision of $\bi{x}_\mathrm{pv}$ is subsequently improved by including tracks
with low impact parameter significance into a full vertex fit. As a result, 
the uncertainty on the transverse
position of the primary vertex ranges from about 10 -- $30\,\mu\mathrm{m}$,
depending on the number of tracks and the event topology.

The subsequent steps of secondary vertex tagging are performed on a per-jet 
basis. Only tracks within the jet cone are considered. To ensure good track
quality cuts involving the transverse momentum, 
the significance of their impact parameter $d$ relative to $\bi{x}_\mathrm{pv}$,
$d/\sigma_{d}$, the number of silicon hits
attached to a track, the quality of those hits, and the quality of the final
track fit are applied. 
Tracks consistent with coming from the decays $K_s^0 \rightarrow \pi^+ \pi^-$ or 
$\Lambda \rightarrow \pi^- p$, or photon conversions are not accepted as good 
candidate tracks.
For a jet to be {\it taggable} at least two good tracks have to fall within 
the jet cone. 
One iterates over the set of good tracks and tries
to fit them to a common vertex with a certain minimum fit quality 
($\chi^2$ per degree of freedom).

\begin{figure}[!t]
  $\;\;\;$(a) \hspace*{80mm} (b) \\
  \includegraphics[width=0.57\textwidth]{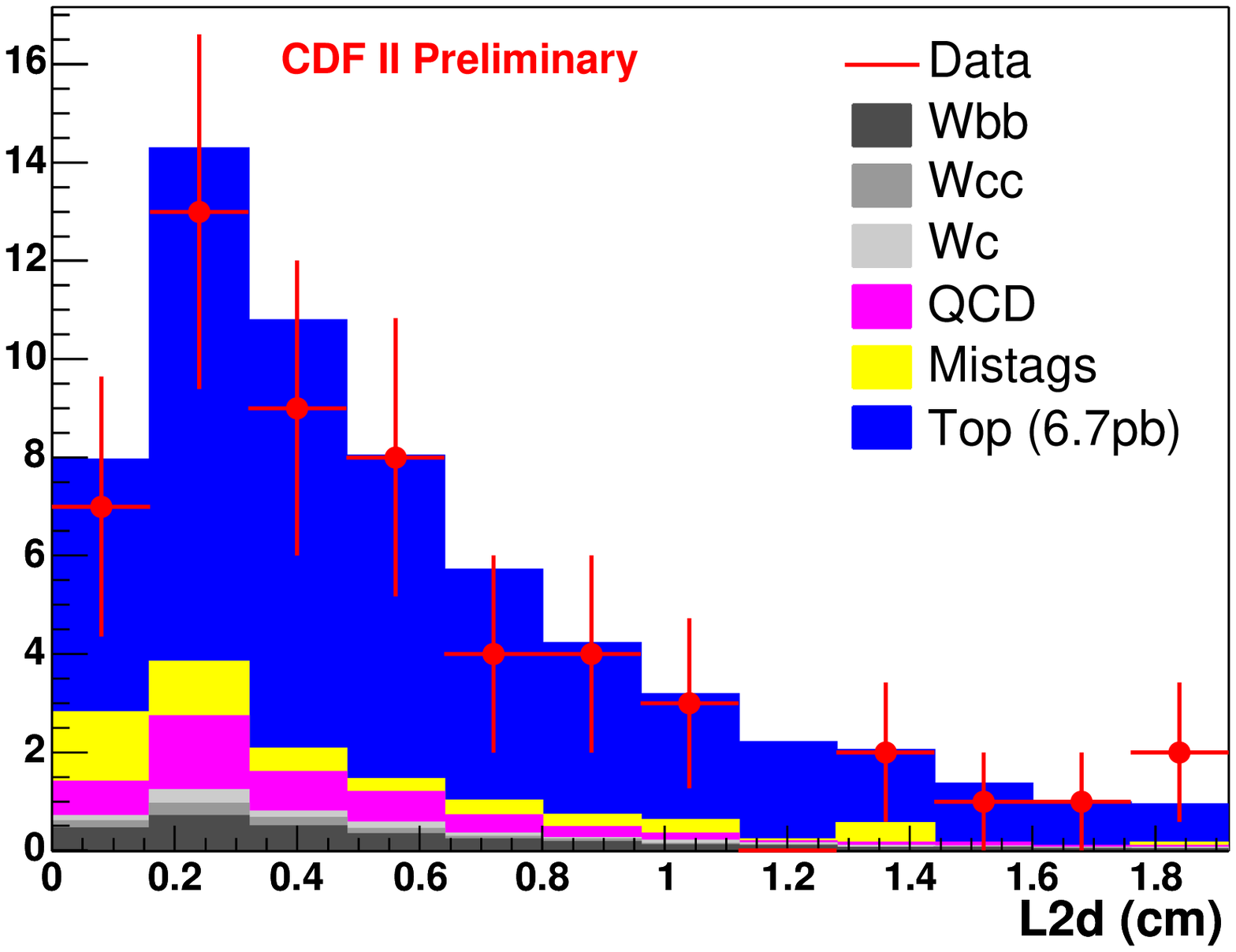}
  \resizebox{0.4\textwidth}{!}
  {\includegraphics[0pt,13pt][285pt,293pt]{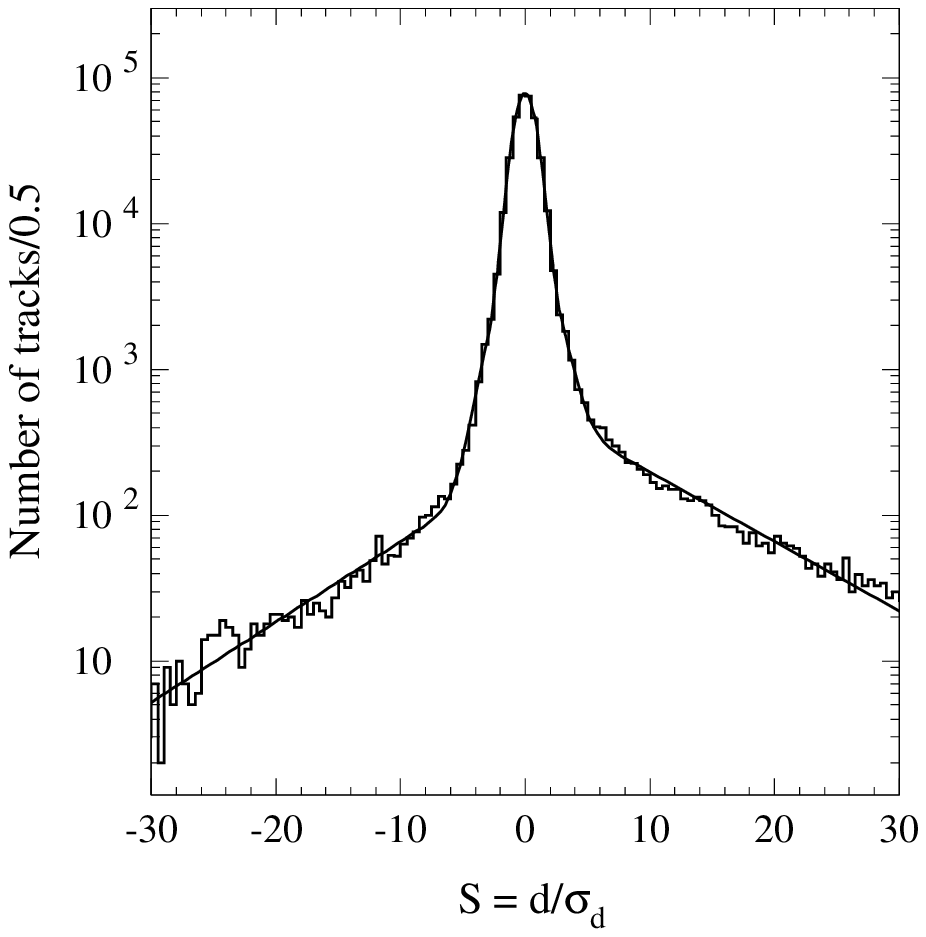}}
  \caption{\label{fig:btag}(a) Two-dimensional decay length $L_\mathrm{2D}$ 
  of secondary vertices
  as obtained in a CDF measurement of the $t\bar{t}$ cross 
  section~\cite{CDFsecvtxRun2}. (b) Distribution of the signed impact parameter
  significance of tracks in a jet data sample, where there is at least
  one jet with $E_\mathrm{T}>50\,\mathrm{GeV}$ per event. The data are fit with
  a resolution function consisting of two Gaussians plus an exponential
  function, separately for the positive and negative 
  sides~\cite{CDFttbarCrossSection2001LeptonPlusJetsRun1}.}
\end{figure}

%
Once a secondary vertex is found in a jet, the two-dimensional decay length
$L_\mathrm{2D}$ of the secondary vertex is calculated. $L_\mathrm{2D}$ is 
the projection
of the vector pointing from the primary to the secondary vertex onto 
the jet axis in the $xy$ plane. The sign of $L_\mathrm{2D}$ is determined by the 
angle between the jet axis and the secondary vertex vector in the $xy$ plane.
If the angle is $< 90^\circ$ the sign is positive, if the angle is 
$> 90^\circ$, the sign is negative. Large positive values of $L_\mathrm{2D}$ are 
predominantly attached to vertices of real $b$ hadron decays, while displaced 
vertices with negative $L_\mathrm{2D}$ are mainly due to track mismeasurements
or random combinations of tracks.
To reduce the background from false secondary vertices, called \emph{mistags},
a good secondary vertex is required to have 
a minimum decay length significance of typically
$L_\mathrm{2D}/\sigma_\mathrm{2D}>3$,
where $\sigma_\mathrm{2D}$ is the estimated uncertainty on $L_\mathrm{2D}$.
The uncertainty $\sigma_\mathrm{2D}$ is estimated for each vertex individually,
a typical value is about $200\,\mathrm{\mu m}$.
Figure~\ref{fig:btag}a shows the distribution of $L_\mathrm{2D}$ as obtained in a CDF 
measurement of the $t\bar{t}$ cross section~\cite{CDFsecvtxRun2}. 
The distribution has the form of a falling exponential, as expected from
a lifetime distribution. The content of the first bin is exceptionally lower 
due to the cut on the decay length significance, 
$L_\mathrm{2D}/\sigma_\mathrm{2D}>3$.
More detailed descriptions of secondary vertex $b$ tagging can be found
in references~\cite{CDFtopEvidencePRD,CDFttbarCrossSection2001LeptonPlusJetsRun1}
 (CDF Run I), reference~\cite{CDFsecvtxRun2} (CDF Run II) and
references~\cite{D0Zbjet,D0anoheavyflavor} (D\O \ Run II).
In the latest CDF analysis (Run II) the efficiency of tagging at least one 
$b$ quark jet in a $t\bar{t}$ event
is (53$\pm$3)\%~\cite{CDFsecvtxRun2}. This number includes the geometric 
acceptance of tracks from $b$ decays (taggability). 
The estimated uncertainty is purely systematic. 

\paragraph{Impact parameter tag}
The impact parameter $b$ tag algorithm compares track impact parameters
to measured resolution functions in order to calculate for each jet 
the probability that there are no long lived particles in the jet cone.
For light quark or gluon jets this probability is uniformly distributed 
between 0 and 1, while it is very small for jets containing tracks from
displaced vertices of heavy flavour decays.
The impact parameter significance $S_0$ is defined as the ratio of the 
impact parameter $d$ and its uncertainty $\sigma_d$.
The sign of $d$ is defined to be positive if the point of closest approach
to the primary vertex lies in the same hemisphere as the jet direction,
and negative otherwise.  
Figure~\ref{fig:btag}b shows the $S_0$ distribution of tracks in a jet data 
sample. The distribution is fit with a resolution function. 
The negative side of the resolution function is used
to compute the probability $P(S_0)$ that the $S_0$ of a particular track is 
due to detector resolution.
The probability that a jet is consistent with a zero lifetime hypothesis
is essentially given by the product of the individual probabilities $P(S_0)$
of tracks contained in the jet cone including an appropriate normalization.
Tracks used in the calculation of the jet probability are required to
satisfy certain quality criteria involving a cut on the impact parameter,
the transverse momentum and the number of hits in the silicon detector.
The impact parameter tagging method has the advantage of providing a 
continuous variable, the jet probability, to distinguish between light
and heavy flavour jets, which easily allows it to choose a certain value
of $b$ tagging efficiency by adjusting the cut value on the jet probability.  
In addition, the probability value can be used in subsequent steps of
an analysis, e.g. when applying multivariate techniques.

\paragraph{Soft Lepton Tag}
The Soft Lepton Tag (SLT) is based on the identification of leptons
originating from semileptonic $b$ decays $b \rightarrow \ell + X $,
where the lepton is either an electron or a muon.
Contributions come from the decay mode $b\rightarrow c\ell\nu$ or 
the sequential decay $b \rightarrow c \rightarrow s \ell \nu $.
The semileptonic branching ratio of $b$ and $c$ hadrons is measured to be
about 10\% per lepton species ($e$ or $\mu$)~\cite{PDG2004}. 
This means that about 80\% of $t\bar{t}$ events have at least one lepton
from a semileptonic decay. In contrast to leptons from $W$ decays
the leptons from $b$ decays are not isolated, but are rather contained
in the cone of the $b$ quark jet. Their $p_\mathrm{T}$ spectrum is much softer,
typically well below $20\,\mathrm{GeV}/c$. Therefore, the detection and identification
of these leptons is much more difficult.
In Run II, D\O \  and CDF have used only muons for the soft lepton tag so far.
D\O \ selects muons in the transverse momentum range of 
$4\,\mathrm{GeV}/c<p_\mathrm{T}<15\,\mathrm{GeV}/c$ and achieves a $b$ jet tagging 
efficiency of 11\%~\cite{D0anoheavyflavor}.

\subsection{Generation and simulation of Monte Carlo events}
\label{sec:mc}
Collider experiments are intricate devices and predicting the experimental
response to a certain physics process is non-trivial. 
To extract sensible information from data one needs first an accurate modelling
of the event kinematics and topology on parton and hadron level, and second
a detailed simulation of the detector response to particles interacting
with the detector material.  

There are two commonly used general purpose Monte Carlo event generators,
{\sc Pythia}~\cite{pythia} and {\sc Herwig}~\cite{herwig}, which provide an exclusive
description of individual events at hadron level. 
They are based on leading order matrix elements for the hard parton scattering 
convoluted with parametrizations of parton distribution functions and include 
approximate treatments of higher order perturbative effects, initial and final 
state gluon emission, parton shower, hadronization, secondary decays and the 
underlying event.
The strength of {\sc Pythia} and {\sc Herwig} is the modelling of the parton shower,
where outgoing partons are converted into a cascade of gluons and $q\bar{q}$
pairs with energy and angular distributions determined by the 
DGLAP equations, which describe the $Q^2$ evolution of quark and gluon 
densities~\cite{gribov1972,Dokshitzer1977,altarelli1977}. 
The parton shower is terminated when the virtual invariant mass of the
parton ($Q^2$) falls below a threshold value chosen such that
a perturbative treatment is valid above the threshold.
The evolution of the shower beyond this stage is determined by non-perturbative
physics. The partons are turned into colourless hadrons according to
phenomenological hadronization (or fragmentation) models.
Particles originating from interactions of the beam remnants are also
included in the model and are called the underlying event.
A set of parameters is tuned to reproduce hadron multiplicities and
transverse momentum spectra of the underlying event as measured in soft
$p\bar{p}$ collisions, so called minimum bias events.
To obtain a proper modelling of $b$ and $c$ hadron decays, heavy flavor 
jets are interfaced to the decay algorithm QQ which comprises
tables of the most up to date branching fractions.

While {\sc Pythia} and {\sc Herwig} provide a good description of $t\bar{t}$ 
events~\cite{frixione95},
they are insufficient to model multijet and $W$ + multijet background events.
Particularly critical is the generation of $W$ + heavy flavour jets.
In Run I of the Tevatron $W$ + jets events were generated with the
{\sc Vecbos} Monte Carlo program~\cite{vecbosDocu}, in Run II the {\sc Alpgen} 
program is
used~\cite{alpgen}, which generates high multiplicity partonic final states
based on exact leading order matrix elements. The parton level events are
passed to {\sc Herwig} for showering and hadronization and to QQ for the decay
of heavy flavour hadrons. To reproduce the entire $W$ + multijets spectrum
several {\sc Alpgen} samples have to be accurately merged together.

Monte Carlo event generators output a list of stable particles which can
be fed to a full detector simulation that reproduces the interaction of
those particles with the detector material. The simulation code is based on 
the {\sc Geant} package~\cite{geant3} which allows to implement a detailed geometry
description of the detector, track particles through the given detector 
volumes and compute their energy loss as well as multiple scattering.
In the so called digitization step the energy depositions obtained from 
{\sc Geant} 
are converted into raw detector data which are in the same format as physics
data recorded from real collisions. Most detectors are described by 
parametric models that convert energy deposits into detector signals.
The parameters of the models are tuned to correctly reproduce the detector
response.
The silicon detector response, for example, can be modelled by a simple
geometrical model based on the path length of the ionizing particle and
a Landau distribution measured from physics data. The raw data obtained from
Monte Carlo events are then subjected to the same reconstruction code as
data from collisions. These reconstructed Monte Carlo data can then be 
analysed in the same way as if they would contain physics events.
Monte Carlo samples generated in this way are used to compute acceptances
for certain physics processes and to compare kinematic distributions
between data and predictions.

\section{The quest for the top quark}
\label{sec:discover}
Immediately after the discovery of the $b$ quark in 1977 the existence of a 
weak isospin doublet partner, the top quark, was hypothesised.
The mass of the sixth quark was unknown and a wealth of predictions
appeared based on many different speculative ideas, see for example
references~\cite{pakvasa,preparata,sher1979}. Typical expectations were in
the mass range of about $20\,\mathrm{GeV}/c^2$, which became accessible two years later
with measurements at the PETRA $e^+e^-$ collider.

\subsection{Early searches}
In $e^+e^-$ annihilations $t\bar{t}$ pairs can be produced by the exchange
of a virtual photon or $Z$ boson. The mass hierarchy of Standard Model
quarks is beautifully reflected in a measurement of the ratio of the 
production rate of hadrons to that for muon pairs,
$R=\sigma(e^+e^-\rightarrow \mathrm{hadrons})/\sigma(e^+e^-\rightarrow \mu^+\mu^-)$, 
as a function of the centre-of-mass energy $\sqrt{s}$. 
The ratio $R$ is given by $R(s)=3\;\sum Q_i^2$, where the sum is over all
quark flavours with production thresholds greater than $\sqrt{s}$.
This corresponds to a step function, which is only altered near 
production thresholds where strong resonance effects occur.
If one compares $R$ above the $t\bar{t}$ production threshold to its
value below (but above the threshold for $b\bar{b}$ production) one 
expects to see an increase by $\Delta R = 4/3$.

Between 1979 and 1984 the five experiments at the PETRA collider,
CELLO, JADE, MARK J, PLUTO, and TASSO, measured $R$ at 
several centre-of-mass energies ranging from 12 to 46.8~GeV in steps
of about 20 to 30~MeV~\cite{CELLOtop,JADEtop,MARKJtop,PLUTOtop,TASSOtop}. 
No indication of top quark production was found. The final
lower limit of the top quark mass derived from PETRA measurements 
was $23\,\mathrm{GeV}/c^2$.

At the KEK laboratory in Japan a dedicated accelerator, TRISTAN, was built
to search for the top quark in $e^+e^-$ annihilations, with slightly increased
centre-of-mass energy than PETRA. The maximum energy reached was 61.4~GeV.
Between 1987 and 1990 several consecutive searches for top quark production
were presented by the two experiments operating at TRISTAN (AMY and VENUS).  
Again, no evidence for the top quark was found and the final lower limit on the
top quark mass was 30.2~GeV/$c^2$~\cite{AMYtop,VENUStop}.

In 1989 the Large Electron Position collider (LEP) at CERN and the Stanford
Linear Collider (SLC) started $e^+e^-$ collisions at the $Z^0$ pole,
$\sqrt{s}\approx 90\;\mathrm{GeV}$. The experiments at these accelerators
-- ALEPH, DELPHI, OPAL and L3 at LEP, and SLD at SLC -- searched for 
$Z^0\rightarrow t\bar{t}$ events of various 
topologies~\cite{ALEPHtop,DELPHItop,OPALtop,SLDtop}. 
The best lower limit on the top quark mass obtained from these direct searches 
was $45.8\,\mathrm{GeV}/c^2$. 

In the mid 1980s experiments at hadron colliders joined experiments at 
$e^+e^-$ machines in the quest for the top quark.
In contrast to $e^+e^-$ annihilations, searches in hadronic collisions have  
to deal with high backgrounds and thus model independent analyses are not 
feasible.
Instead the experiments had to concentrate on SM signatures.
Initially, one assumed the top quark mass to be below the mass of the $W$
boson. Top quark production via the electroweak interaction
was expected to be the dominating process at the CERN $Sp\bar{p}S$ collider
operating at $\sqrt{s}=546\;\mathrm{GeV}$, later at $\sqrt{s}=630\;\mathrm{GeV}$. 
The experiments UA1 and UA2 
searched for events were an on-shell $W$ boson is produced which decays into
a top and a bottom quark: $p\bar{p}\rightarrow W^\pm \rightarrow t\bar{b} / \bar{t}b$.
The top quark was reconstructed in its semileptonic decay mode into a 
$b$ quark, a charged lepton (electron or muon) and a neutrino. 
\par
First results reported by UA1 in 1984 seemed to be consistent with
the production of top quarks of mass 
$(40\pm10)\,\mathrm{GeV}/c^2$~\cite{UA1misdiscover}.
UA1 observed six events with one isolated lepton, 3 electron and 3 muon
events, and 2 hadronic jets. The effective invariant mass distribution of the lepton,
the two jets and the missing transverse energy shows a pronounced peak
at the $W$ mass, while the mass of the lepton, the missing transverse energy and
the jet with the lowest transverse energy clustered around 40~GeV/$\mathrm{c^2}$. 
The number of background events where the 
lepton was faked by a hadron or background events from heavy flavour production 
were believed to be negligible. 
Thus, these six events were interpreted as a first indication of
a top quark originating from a $W$ boson decay. 
However, subsequent analyses by UA1 were based on a more complete evaluation 
of the backgrounds and did not support this result~\cite{UA11988}.
Since UA1 did not apply a minimum cut on the missing transverse energy in the event,
backgrounds due the fake leptons, Drell-Yan production of lepton pairs, 
$\ell^+\ell^-$, as well as backgrounds from $b\bar{b}$ and $c\bar{c}$ 
production were found 
to be more significant than originally estimated.
\par
The final UA1 top quark analysis was published in 1990 and used data 
corresponding to an integrated luminosity of 5.4~$\mathrm{pb^{-1}}$.
The observed events were found to be consistent with the
non-top backgrounds, and a lower limit on the top quark mass of 
$60\,\mathrm{GeV}/c^2$ was set~\cite{UA1final}.
A slightly better limit was achieved by the second experiment operating at the 
CERN $Sp\bar{p}S$, UA2, based on an integrated luminosity of 
7.1~$\mathrm{pb^{-1}}$~\cite{UA2limit}. UA2 used only events containing an isolated 
central electron with $p_\mathrm{T}^e>12\;\mathrm{GeV}/c$ and $\EtMiss > 15\;\mathrm{GeV}$.
After all selection cuts 137 events were retained, in good agreement with
the expectation of $154.4\pm 14.6$ background events. To further enhance the
sensitivity of the analysis the transverse mass of the electron and the 
missing transverse energy  
\[M_\mathrm{T}^{e\nu} = 
  \sqrt{\;2\,p_\mathrm{T}^e\,\EtMiss\cdot(1 - \cos \Delta\phi_{e\nu})\; } \]
was formed, where $\Delta\phi_{e\nu}$ is the azimuthal angle between the 
electron and $\EtMissVec$ vectors.
The limit on the top quark mass were obtained by fitting the
observed $M_\mathrm{T}^{e\nu}$ distribution to template distributions of background events
alone or a combination of background and a top quark signal of certain mass.
As a result, top quark masses below $69\,\mathrm{GeV}/c^2$ were excluded.

It is important to note that experiments at hadron colliders do not provide
a direct limit on the top quark mass, but rather on the top quark production
cross section.
The mass limits are derived using the theoretically predicted production cross 
sections, that depend on the top quark mass, and the predicted branching ratios.
Systematic uncertainties on these predictions have thus to be properly accounted for.

\subsection{Searches and discovery at the Tevatron}
In 1988 the CDF experiment at the Tevatron joined the race for discovery of
the top quark. Due to the higher centre-of-mass energy at the Tevatron of 
$\sqrt{s}=1.8\;\mathrm{TeV}$  top quarks are predominantly produced 
as $t\bar{t}$ pairs. The first CDF top quark search uses a data sample with an
integrated luminosity of 4.4~$\mathrm{pb^{-1}}$ accumulated in Run 0 which 
lasted from 1988 to 1989~\cite{CDFtopSearch_90,CDFtopSearch_91}. 
CDF pursued a similar strategy 
as UA2 and used the $M_\mathrm{T}^{e\nu}$ distribution to discriminate between 
$W$+jets background events, where the $W$ boson decays on shell, and 
top quark decays where the electron and the neutrino originate from the exchange
of a virtual $W$ (this was still under the assumption that $M_\mathrm{top}<M_W+M_b$).  
The analysis is based on isolated central electron events with an 
electromagnetic transverse energy above 20~GeV and $\EtMiss>20\;\mathrm{GeV}$. 
In addition, at least two hadronic jets with $E_\mathrm{T}>10$~GeV and 
$|\eta|<2.0$ are required.
This analysis pushed the lower limit of the top quark mass to
$M_\mathrm{top}>77\,\mathrm{GeV}/c^2$ using
predictions of the cross section by Altarelli 
{\it et al.}~\cite{altarelli1988,nason1988}.

In addition to the search in the electron-plus-jets channel CDF performed an
analysis in the electron-plus-muon ($e\mu$) dilepton channel~\cite{CDFtopElMu90}. 
By requiring two leptons from different families, backgrounds from Drell-Yan
and $Z^0$ production as well as $W$+jets events are strongly suppressed.  
The remaining background is mainly due to $Z^0\rightarrow\tau^+\tau^-$ events.
Subsequently, CDF extended the dilepton analysis to include the $ee$ and 
$\mu\mu$ channels. Backgrounds from Drell-Yan and $Z^0$ events are 
reduced by additional requirements: (1) The dilepton
azimuthal opening angle $\Delta\phi_{\ell\ell}$ is required to be
below $160^\circ$. (2) $ee$ or $\mu\mu$ events with a dilepton invariant mass 
in the window $75<M_{\ell\ell}<105\;\mathrm{GeV}/c^2$ or with
$\EtMiss<20\;\mathrm{GeV}$ are rejected. The lepton+jets analysis was also
further improved. High-$p_T$ muons from the decay of a $W$ boson were 
included and a soft muon $b$ tag for at least one of the jets was 
required to reduce the $W$+multijet background. The dilepton and the 
lepton+jets analysis were combined to derive a common limit on the
$t\bar{t}$ cross section $\sigma_{t\bar{t}}$. Using theoretical expectations 
for $\sigma_{t\bar{t}}$~\cite{ellis1991}, and assuming SM decays for the 
top quark, the cross section limit was translated into a lower limit
on the top quark mass of $M_\mathrm{top}>91\,\mathrm{GeV}/c^2$ at the 95\% 
confidence level~\cite{CDFlimitFinalPRL,CDFlimitFinalPRD}. 
It is important to realize that with the top quark mass being above the sum of
the $W$ and the $b$ quark mass, the transverse mass $M_\mathrm{T}^{e\nu}$ 
is no longer a
suitable discriminant and a strategy change is necessary. Therefore, 
later CDF analyses were based on counting events only, rather than 
fitting the $M_\mathrm{T}^{e\nu}$ distribution. 

In 1992, with the start of Tevatron Run I, the D\O \ experiment joined the
hunt for the top quark. In April of 1994 D\O \ published its first
top quark analysis setting the last lower limit on the top quark mass
before its discovery~\cite{D0limitPRL}. The data sample was recorded in 
1992/93 and corresponds to an integrated luminosity of $15\;\mathrm{pb^{-1}}$.
In this analysis D\O \ uses the $e\mu$  and the $ee$ mode of the dilepton
channel and the lepton-plus-jets channels. In $e\mu$ events one electron
with $E_\mathrm{T}>15\;\mathrm{GeV}$ and one muon with $p_\mathrm{T} > 15\;\mathrm{GeV}/c$,
$\EtMiss>20$~GeV and at least one jet with $E_\mathrm{T}>15$~GeV are required.
For $ee$ candidates tighter cuts are applied, the two electrons must have
$E_\mathrm{T}>20$~GeV and $\EtMiss>25$~GeV. To suppress $Z^0\rightarrow e^+e^-$
background the cut on missing transverse energy is raised to 40 GeV
within a mass window of $|M_{ee}-M_Z|<12\;\mathrm{GeV}/c^2$, where $M_{ee}$
is the dielectron invariant mass. In the electron+jets mode one electron
with $E_\mathrm{T}>20$~GeV, at least four jets with $E_\mathrm{T}>15$~GeV, and 
$\EtMiss >30$~GeV are required. In muon+jets events the cut on the
missing transverse energy is slightly relaxed: $\EtMiss >20$~GeV.
To further reduce the $W$+jets background \DZero performs a topological
analysis of the events. The event shape is characterised by the aplanarity
$\mathcal{A}$ which is a quantity proportional to the lowest eigenvalue
of the momentum tensor of the observed objects. In $e$+jets events
a cut of $\mathcal{A}>0.08$ is applied, in $\mu$+jets events 
$\mathcal{A}>0.1$ is demanded. After all analysis cuts two dilepton and 
one $e$+jets event are observed, consistent with background estimates. 
The intersection of the derived upper limit on 
the $t\bar{t}$ cross section with the theoretical prediction~\cite{laenen1992}
yields a lower limit on the top quark mass of 
$131\,\mathrm{GeV}/c^2$~\cite{D0limitPRL}.  

In 1993 and 1994 CDF saw mounting evidence for a top quark signal. 
The detector upgrade for Run I, mainly the addition of a silicon 
vertex detector, was the keystone for the discovery of the top quark at CDF. 
The new silicon detector allowed for the reconstruction of secondary vertices 
of $b$ hadrons and a measurement of the transverse decay length $L_{xy}$ with a 
typical precision of $130\,\mu\mathrm{m}$.
Secondary vertex $b$ tagging proved to be a very powerful tool to
discriminate the top quark signal against the $W$+jets background and 
increase the sensitivity of the lepton-plus-jets $t\bar{t}$ analysis.
In July 1994 CDF published a paper announcing first evidence for $t\bar{t}$
production at the Tevatron based on events in the 
dilepton and the lepton-plus-jets channel~\cite{CDFtopEvidencePRL,CDFtopEvidencePRD}. 
The analysis uses a data sample with an integrated luminosity of 
$(19.3\pm0.7)\;\mathrm{pb^{-1}}$.
Two $e\mu$ dilepton events pass the selection cuts, but no $ee$ or $\mu\mu$
events. The expected background in the dilepton channel is 
$0.56^{+0.25}_{-0.13}$.
In the lepton-plus-jets channel three or more jets with 
$E_\mathrm{T}>15\;\mathrm{GeV}$
and $|\eta|<2.0$ are required. Secondary vertex and soft lepton $b$ tagging
reduce the $W$+jets background. Six events with a secondary
vertex tag are observed over a background of $2.3\pm0.3$ events.
Seven events have a soft lepton tag (muon or electron) over a background
of $3.1\pm0.3$ events. The soft lepton and the secondary vertex tagged samples
have an overlap of three events.
\par
Since the cross section of $W$+jets production cannot
be reliably predicted with sufficiently small uncertainties, special
techniques had to be developed for estimating backgrounds in the lepton-plus-jets
search directly from data. This technique assumes that the heavy quark content
($b$ and $c$) of jets in the $W$+jets sample is the same as in an inclusive
jet sample. This assumption is a conservative overestimate of the backgrounds, 
since the inclusive jet sample contains heavy quark contributions from direct
production (e.g. $gg\rightarrow b\bar{b}$), gluon splitting (where a final state
gluon branches into a heavy quark pair), and flavour excitation (where an
initial state gluon excites a heavy sea quark in the proton or antiproton),
while heavy quarks in $W$+jets events are expected to be produced entirely from
gluon splitting. Tag rates are measured in an inclusive jet sample and 
parametrized by the $E_\mathrm{T}$ and track multiplicity of each jet.
The parametrization is used to derive an estimate on the number of expected
background events based on the $W$+jets sample before applying the
$b$ tag algorithms. The probability that the estimated background has fluctuated
up to the total number of 12 candidate events, taking into account that
 three events have double tags,
is found to be 0.26\%. This corresponds to a $2.8\,\sigma$ excess for a
Gaussian probability function.
\par
\begin{figure}[!t]
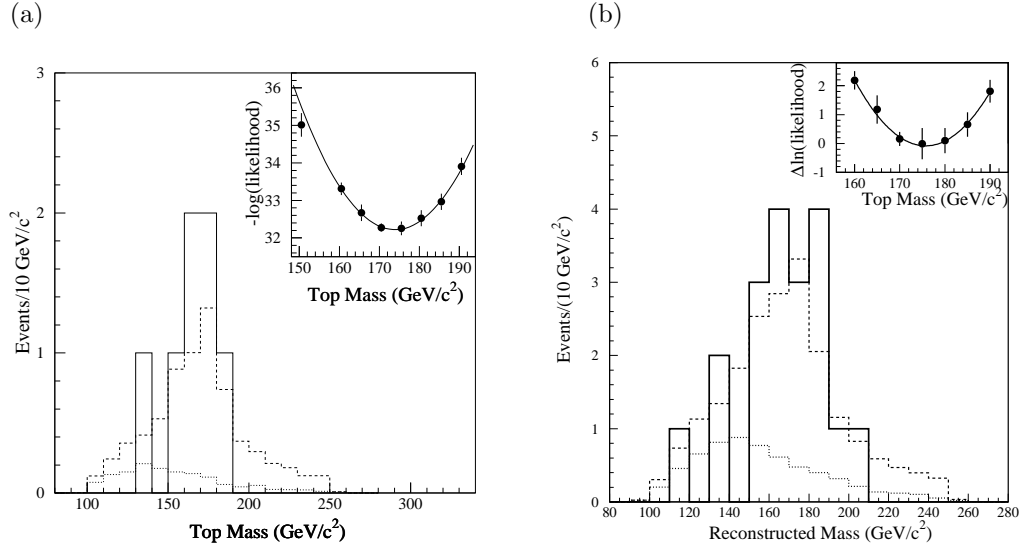

  \hspace*{0.1\textwidth} (a) \hspace*{70mm} (b) \\
   
  \hspace*{0.1\textwidth}
  \includegraphics[width=0.4\textwidth]{fig/datmc7_prl.epsi}
  \hspace*{0.05\textwidth}
  \includegraphics[width=0.4\textwidth]{fig/tag_cons_like.epsi}
\caption{\label{fig:mtopEvidence} 
  Reconstructed top mass distributions as published
  (a) in the CDF evidence paper of 1994~\cite{CDFtopEvidencePRL} and
  (b) in the CDF discovery paper~\cite{CDF_obsTop_95}. 
  The solid histogram shows CDF data. The dotted line shows the shape of the 
  expected background, the dashed line the sum of background plus $t\bar{t}$ 
  Monte Carlo events for $M_\mathrm{top}=175\;\mathrm{GeV}/c^2$.
  In both plots, the inset shows the 
  likelihood curve used to determine the top quark mass.}
\end{figure}

Assuming that the excess of $b$ tagged lepton-plus-jets events is due to
$t\bar{t}$ production a value for the top quark mass is estimated 
using a constrained kinematic fit. In the $W$+3 jets sample one additional
soft jet with $E_\mathrm{T}>8\;\mathrm{GeV}$ and $|\eta|<2.4$ is required. 
Seven events of the $b$ tagged lepton-plus-jets sample pass this requirement and
are fitted individually to the $t\bar{t}$ hypothesis. 
For each event the 12 
possible combinations of the jets, the lepton and the missing transverse
energy are considered. The combination with the best $\chi^2$ is chosen. 
The resulting top mass distribution, shown in figure~\ref{fig:mtopEvidence}a, 
is fitted to a sum of 
the expected distributions from $W$+jets and $t\bar{t}$ production
for different top quark masses.
The fit yields a value of
$M_\mathrm{top}=(174\pm 10 ^{+13}_{-12})\;\mathrm{GeV}/c^2$. The corresponding
log likelihood distribution is depicted in figure~\ref{fig:mtopEvidence}a. 
In November 1994 the D\O \ collaboration confirmed the evidence seen at
CDF. An update of the previous D\O \ analysis, now with an integrated
luminosity of $(13.5\pm 1.6)\;\mathrm{pb^{-1}}$, added soft muon $b$ 
tagging~\cite{d0EvidencePRL,d0Evidence}. In total, D\O \ observed nine events
over a background of $3.8\pm 0.9$.   

As Run I continued more data were accumulated and finally, in April 1995,
CDF and D\O \ were able to claim discovery of the top 
quark~\cite{CDF_obsTop_95,d0discovery}. 
CDF used a data sample corresponding to $67\;\mathrm{pb^{-1}}$ and significantly
improved its secondary vertex $b$ tagging techniques. The efficiency to
identify at least one $b$ quark jet in a $t\bar{t}$ event with 
more than three measured jets was found to be $(42\pm 5)\%$, almost double the
previous value of the 1994 analysis. 
In the new analysis the background estimate is also considerably improved. 
While the mistag rate due to track mismeasurements is 
again measured with samples of inclusive jets, the fractions of $W$+jets events 
that are $Wb\bar{b}$ or $Wc\bar{c}$ are disentangled from Monte Carlo samples, 
applying measured tagging efficiencies. There are 27 jets with a secondary 
vertex $b$ tag in 21 $W + \geq 3$ jets events. 
The estimated background is $6.7\pm 2.1$ $b$ tags.
The probability for this observation to be a background fluctuation is 
$2\cdot 10^{-5}$. 
The 1995 dilepton and soft lepton $b$ tag lepton-plus-jets analyses are only
slightly changed compared to those of 1994. Six dilepton events are observed
over a background of $1.3\pm0.3$. There are 23 soft lepton tags observed in
22 events, with $15.4\pm 2.0$ $b$ tags expected from background sources.
Six events contain both a jet with a secondary vertex and a soft lepton tag.      
The probability for all CDF data events to be due to a background fluctuation alone
is $1\cdot 10^{-6}$, which is equivalent to a $4.8\;\sigma$ deviation in a
Gaussian distribution. Again the top quark mass is kinematically reconstructed 
for $W + \geq 4$ jets events as described above. The mass distribution is
shown in figure~\ref{fig:mtopEvidence}b. The best fit is obtained for
$M_\mathrm{top}=(176\pm 8\pm 10)\;\mathrm{GeV}/c^2$.

\begin{figure}[!t]
\begin{minipage}{0.49\textwidth}
  \includegraphics[width=\textwidth]{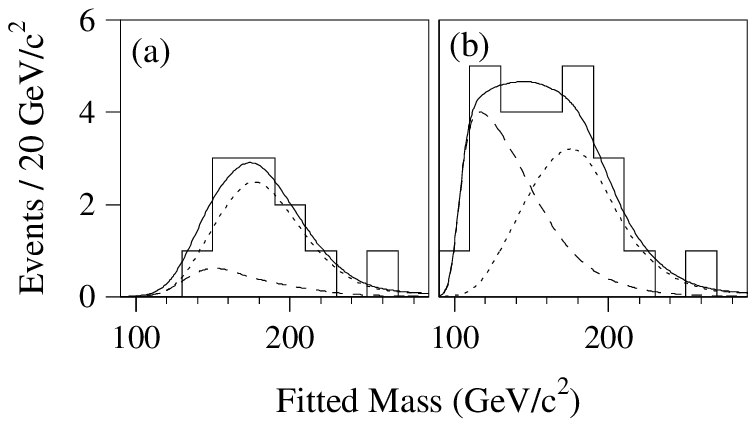}
\caption{\label{fig:mtopD0discover}Fitted mass distribution for candidate events
  (histogram) of the D\O \ discovery paper~\cite{d0discovery}. Overlaid is
  the expected mass distribution for top quark events with 
  $M_\mathrm{top}=199\;\mathrm{GeV}/c^2$ (dotted curve), the background (dashed curve), 
  and the sum
  of top and background (solid curve) for (a) the standard and (b) the loose
  event selection.}
\end{minipage} \ 
\begin{minipage}{0.49\textwidth}
  \includegraphics[width=\textwidth]{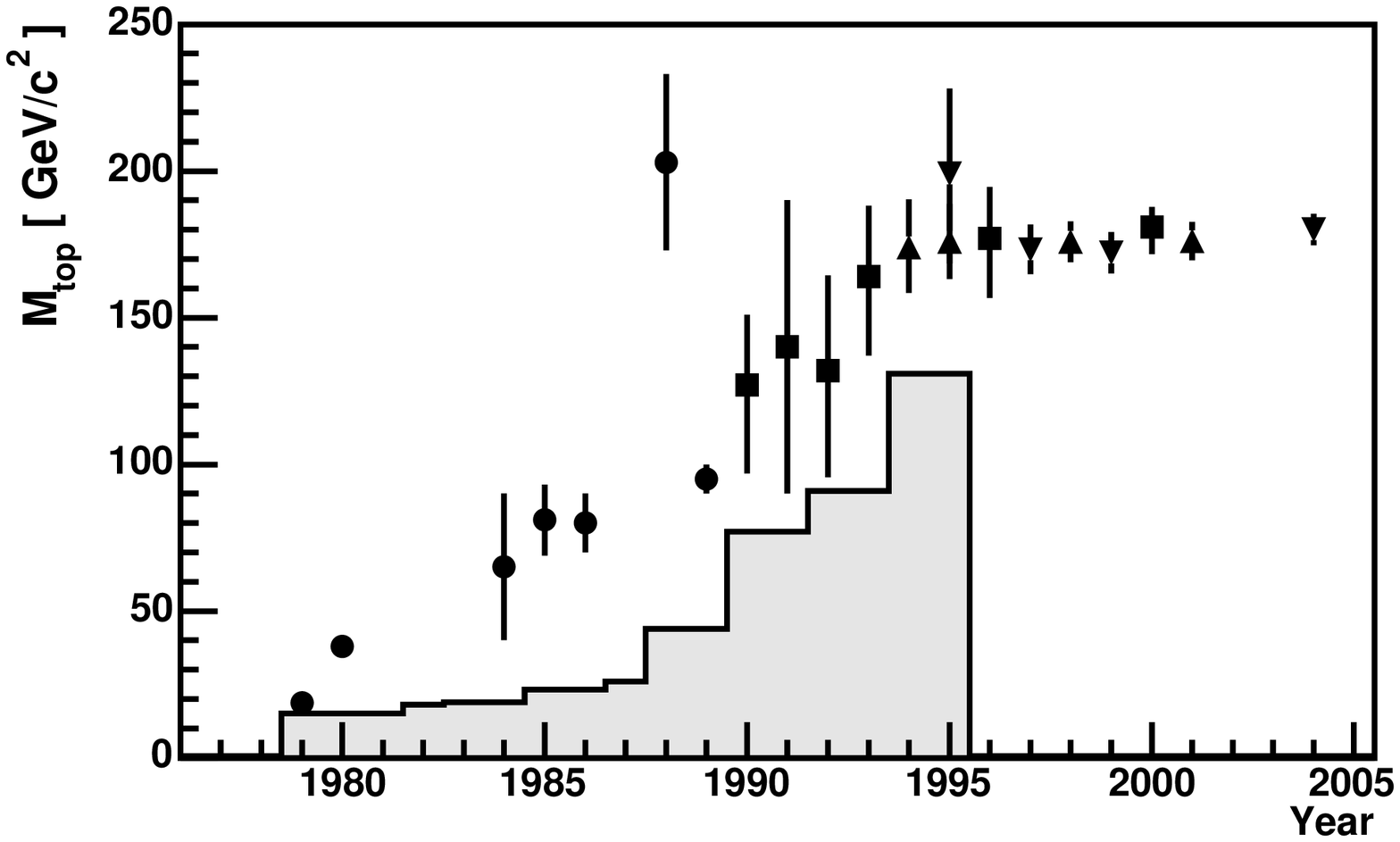}
\caption{\label{fig:mtopVersusTime}
History of the quest for the top quark. The shaded histogram
shows experimental lower limits on the top quark mass. 
Theoretical predictions based on flavour
symmetries are shown (dots) as well as predictions based on electroweak
precision measurements (squares). Direct measurements of $M_\mathrm{top}$ 
by CDF and D\O \ 
are represented by the triangle (triangles pointing up: CDF; triangles pointing 
down: D\O \ ).
}
\end{minipage} 
\end{figure}

Simultaneously to CDF the D\O \ collaboration updated its top quark analyses
based on data with an integrated luminosity of 
$50\;\mathrm{pb^{-1}}$~\cite{d0discovery}.
The updated analysis is very similar to the previous searches, involving
the dilepton channel, soft muon $b$ tagging and the topological analysis.
From all channels, D\O \ observes 17 events with an expected background
of $3.8\pm 0.6$ events. The probability for this measurement to be an
upward fluctuation of the background is $2\cdot 10^{-6}$, which corresponds
to 4.6 standard deviations for a Gaussian probability distribution.
To measure the top quark mass, lepton+4 jets events are subjected to a 
constrained kinematic fit. The resulting top quark mass distribution is 
shown in figure~\ref{fig:mtopD0discover} for the standard cuts (a) and looser
selection requirements (b). 
A likelihood fit to the observed mass distribution yields a central value for
the top quark mass of 
$M_{top}=199^{+19}_{-21}\;\mathrm{(stat.)}\pm 22\;\mathrm{(syst.)}\;\mathrm{GeV}/c^2$. 

In September 1995 CDF completed the series of publications establishing
the top quark discovery with a complementary kinematic analysis without
$b$ tagging, that used the $E_\mathrm{T}$ of the second and third highest 
$E_\mathrm{T}$ 
jets to calculate a relative likelihood for each event to be top quark like
or background like~\cite{CDFkinTop95,CDFkinTop95PRL}. The probability
for the observed data to be due to a background fluctuation was found
to be 0.26\%. 
  
Finally, 17 years after the discovery of the $b$ quark its weak isospin 
partner, the top quark, was firmly established. In 
figure~\ref{fig:mtopVersusTime} we summarise the long lasting 
quest for the top quark. The figure shows the lower experimental 
limits on the top quark mass 
(histogram)~\cite{PLUTOtop,TASSO82,MARKJ83,MARKJtop,VENUS87,UA11988,CDFtopSearch_90,CDFlimitFinalPRL,D0limitPRL}, theoretical predictions
mainly based on flavour symmetries within the quark mass matrix 
(dots)~\cite{sher1979,glashow_top,pinna84,kubo85,just86,alhendi88,marciano89} 
and predictions based on electroweak precision measurements
at LEP and SLC 
(squares)~\cite{ellis90,bernabeau91,arnaudon,dSchaile,ewwg96,ewwg2000} 
as discussed in section~\ref{sec:ewktopdetermine}. 
The direct measurements by 
CDF~\cite{CDFtopEvidencePRL,CDF_obsTop_95,CDFtopMass1998LJ,CDFtopMass2001LeptonPlusJetsRun1} and D\O \  
\cite{d0discovery,d0TopMass1997Run1,d0mtop99,d0mtopnature} 
are also included (triangles).
The good agreement between the measured top quark mass and the prediction
obtained from electroweak precision measurements constituted a major success
of the Standard Model.
\section{$\mathbf{t\bar{t}}$ cross section measurements}
\label{sec:ttbarxs}
Within the SM the $t\bar{t}$ cross section is calculated with
a precision of about 15\%~\cite{kidonakis2003,cacciari2004}, see also
section~\ref{sec:softgluonresum}. The SM further predicts that
the top quark decays to a $W$ boson and a $b$ quark with a branching ratio 
close to 100\%. Measuring the cross section in all possible channels
tests both, production and decay mechanisms of the top quark.
A significant deviation from the SM prediction would indicate either the
presence of a new production mechanism, e.g. a heavy resonance decaying into
$t\bar{t}$ pairs, or a novel decay mechanism, e.g. into supersymmetric
particles. The $t\bar{t}$ cross section depends sensitively on the top
quark mass. In the mass interval of 
$170 \leq M_\mathrm{top} \leq 190\,\mathrm{GeV}/c^2$ the cross section
drops by roughly 0.2~pb for an increase of 1~GeV/$c^2$ in $M_\mathrm{top}$.
This theoretically predicted dependence can be exploited
to turn a cross section measurement into an indirect determination of
the top quark mass. A 15\% measurement of the cross section is 
approximately equivalent to a 3\% measurement of $M_\mathrm{top}$.
One can also turn the argument around and use the measurements of the
cross section and the top quark mass and test their compatibility
with the theoretically predicted cross section and its mass dependence
as indicated in figure~\ref{fig:sigmattbar} in 
section~\ref{sec:softgluonresum}.

The $t\bar{t}$ cross section measurements are very fundamental to
top quark physics at the Tevatron, since these analyses
isolate data samples that are enriched in $t\bar{t}$ events and lay thereby the 
foundations for further investigations of top quark properties.
The $t\bar{t}$ cross section has been measured in the dilepton, the
lepton-plus-jets and the all hadronic channel. 
So far the $t\bar{t}$ tau modes evaded observation.
In this section we discuss the experimental methods applied in $t\bar{t}$ 
cross section measurements in more detail. We highlight representative 
Tevatron analyses for different channels and analyses methods, 
either from CDF or D\O. 
A comprehensive summary of all measurements and the combined cross section 
results is presented in section~\ref{sec:xs_summary}.

\subsection{Dilepton channel}
\label{sec:dilepton}
As mentioned in section~\ref{sec:signatures} the dilepton channel comprises
$t\bar{t}$ final states with two high $p_\mathrm{T}$ charged leptons 
(electrons or muons),
$\EtMiss$ and two $b$ quark jets. This clean signature allows to select
a $t\bar{t}$ sample with high purity, reaching a signal to background ratio
between 1.5 and 3 depending on the analysis.
In 2004 CDF published its first $t\bar{t}$ cross section measurement
at $\sqrt{s}=1.96\,\mathrm{TeV}$
in the dilepton channel based on a data sample with an integrated luminosity
of 
$\mathcal{L}_\mathrm{int}=(197\pm 12)\,\mathrm{pb^{-1}}$~\cite{CDFdileptonPRL04}, i.e.
about twice as much data as used in Run I.
Two complementary analyses are performed. The first one requires both 
leptons to be specifically identified as either electrons or muons, while 
the second technique allows one of the leptons to be identified only as
a high-$p_\mathrm{T}$ isolated track, thereby significantly increasing the lepton
detection efficiency at cost of a moderate increase in the expected 
backgrounds. In the following we will concentrate the discussion on the second, 
the lepton-plus-track analysis.

At trigger level the data samples used in dilepton analyses are selected
by finding events that either contain a central electron or muon candidate with 
$E_\mathrm{T}>18\;\mathrm{GeV}$, or an electron candidate in the forward calorimeter
with $E_\mathrm{T}>20\;\mathrm{GeV}$ and $\EtMiss > 15\;\mathrm{GeV}$.
In the offline analysis two oppositely charged leptons with
$E_\mathrm{T}>20\;\mathrm{GeV}$ are required. One lepton, the ``tight'' lepton has to 
pass strict lepton identification criteria and be isolated. Tight electrons 
have a well measured track pointing at an energy deposition in the calorimeter.
For electrons with $|\eta|>1.2$, this track association is made by a 
calorimeter-seeded silicon tracking algorithm.
Tight muons must have a well-measured track linked to a track segment in the 
muon chambers
and an energy deposition in the calorimeters consistent with that expected for
muons. The second lepton, the ``loose'' lepton, is defined as a well-measured,
isolated track with $p_\mathrm{T}>20\;\mathrm{GeV}/c$ and a pseudorapidity of 
$|\eta|<1.0$.

Candidate events must have a $\EtMiss > 25$~GeV. The value of $\EtMiss$ is 
corrected for all loose leptons if the associated calorimeter $E_\mathrm{T}$ 
is less 
than 70\% of the track $p_\mathrm{T}$. Events are rejected if the vector $\EtMissVec$ 
lies within $5^\circ$ of the loose lepton axis and $\EtMiss < 50$~GeV.
Jets are counted with a threshold of $E_\mathrm{T} > 20\,\mathrm{GeV}$ and 
within the 
pseudorapidity range of $|\eta|<2.0$. At least two jets defined in that way
are required.
Candidates being compatible with
cosmic ray muons or photon conversions are removed. To remove dilepton pairs
due to $Z^0$ boson production the cut on the missing transverse energy is 
tightened to $\EtMiss > 40$~GeV in a window of $\pm 15\;\mathrm{GeV}/c^2$ 
around the $Z^0$ mass. 

\begin{figure}[!t]
\begin{minipage}{0.49\textwidth}
  \includegraphics[width=\textwidth]{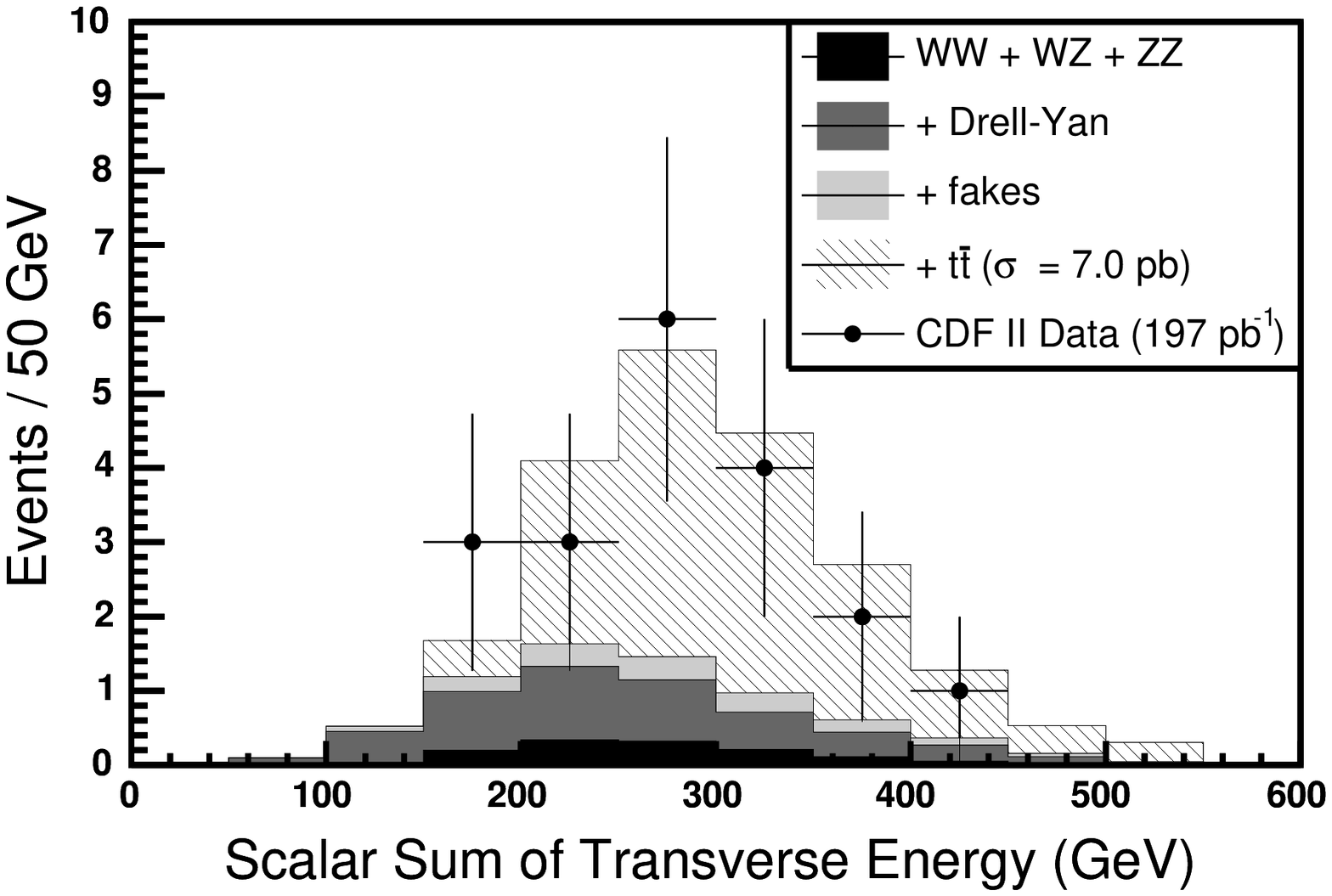}
\caption{\label{fig:htdilepton}The scalar sum of all transverse energies in the 
  event. The histograms are stacked. The shaded histograms are the backgrounds,
  the hatched histogram shows the $t\bar{t}$ signal distribution scaled to the 
  measured cross section. The dots plus error bars are data from the first
  CDF $t\bar{t}$ dilepton analysis in Run II~\cite{CDFdileptonPRL04}.}
\end{minipage} \ 
\begin{minipage}{0.49\textwidth}
  \includegraphics[width=\textwidth]{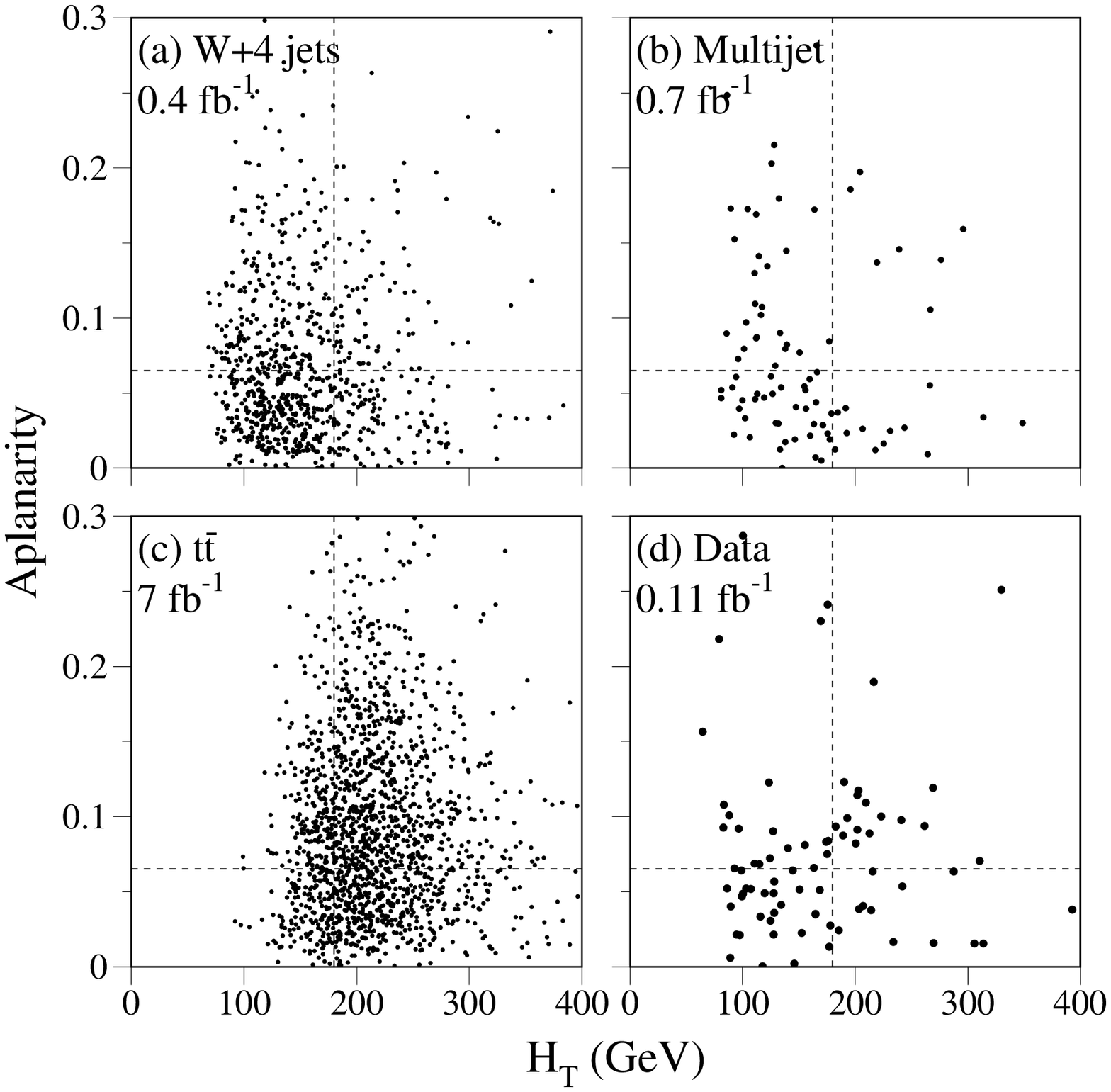}
\caption{\label{fig:aplanarityVsHT}Scatter plots of aplanarity versus $H_\mathrm{T}$ 
  for (a) the $W+4$ jets background sample, (b) QCD multijet background events,
  (c) $t\bar{t}$ Monte Carlo events from {\sc Herwig}, and (d) D\O \ lepton plus jets
  data. The dashed lines represent the threshold values used for the selection 
  in the D\O \ topological analysis~\cite{d0Sigmattbar2003Run1}.}
\end{minipage} 
\end{figure}

The dominant backgrounds to $t\bar{t}$ dilepton events are Drell-Yan
($q\bar{q}\rightarrow Z/\gamma^* \rightarrow \ell^+\ell^-$) production,
misidentified (fake) leptons in $W\rightarrow \ell\nu + \mathrm{jets}$
events where one jet is falsely reconstructed as a lepton candidate, and
diboson ($WW$, $WZ$, and $ZZ$) production. 
The Drell-Yan background is estimated by a combination of fully simulated
{\sc Pythia} Monte Carlo events and CDF data. A sample of $Z$ boson candidates
in the mass range of 76~--~106~GeV/$c^2$ is selected. The numbers of events
passing the nominal $t\bar{t}$ selection or a Drell-Yan specific selection
are obtained, respectively. These two numbers provide the normalization
for the expected Drell-Yan background contributions to the $t\bar{t}$
dilepton sample as obtained from Monte Carlo. 
The rate of lepton misidentification is obtained from an inclusive jet sample
after removing sources of real leptons such as $W$ and $Z$ boson decays.
The accuracy and robustness of the derived lepton fake rate is checked 
with several control samples and good agreement is found within the 
statistical uncertainties. The diboson background estimates are derived by
calculating the acceptances from Monte Carlo and applying the theoretical
cross sections which have a relatively small uncertainty. 
The total number of background events expected in the data sample
is $6.9\pm1.7$ of which 61\% are due to Drell-Yan processes, 22\% are due to 
fake leptons, and 17\% are due to diboson production.
CDF observes 18 events in data. 
The $t\bar{t}$ cross section is calculated according to the formula
\begin{equation}
  \sigma(t\bar{t}\,) = 
  \frac{N_\mathrm{obs}-N_\mathrm{bkg}}
  {\epsilon_\mathrm{evt}\cdot\mathcal{L}_\mathrm{int}} \ ,
  \label{eq:xs}
\end{equation}
where $N_\mathrm{obs}$ is the number of observed events, $N_\mathrm{bkg}$ is the
number of expected background events, and $\epsilon_\mathrm{evt}$ is the
event detection efficiency which includes the kinematic acceptance,
trigger and reconstruction efficiencies, and the branching ratio into
dilepton events. 
The event detection efficiency is 
$\epsilon_\mathrm{evt}=(0.88\pm0.12)\%$,
including a branching ratio of $\mathrm{BR}(W\rightarrow\ell\nu)=10.8\%$
for the $W$ boson decay into a charged lepton plus neutrino.
The kinematic acceptance is evaluated for a top mass of 
$M_\mathrm{top}=175\,\mathrm{GeV}/c^2$.
The resulting value of the cross section is 
$7.0^{+2.7}_{-2.3}\,(\mathrm{stat.})^{+1.5}_{-1.3}\,(\mathrm{syst.})\pm0.4\;(\mathrm{lumi.})\,\mathrm{pb}$, which is in very good agreement with the theoretical
prediction of $6.7^{+0.7}_{-0.9}\,\mathrm{pb}$~\cite{cacciari2004}. 
The distributions of various kinematic variables are found to be
consistent with the SM prediction as obtained from {\sc Pythia} Monte Carlo.
As an example, the distribution of the scalar sum of all transverse energies
in the event is shown in figure~\ref{fig:htdilepton}.
A Kolmogorov-Smirnov test of this distribution yields a $p$-value of 75\%.

\subsection{Lepton-plus-jets channel}
\label{sec:leptonPlusJets}
The signature of the $t\bar{t}$ lepton-plus-jets channel comprises a 
high-$p_\mathrm{T}$
electron or muon, missing transverse energy and four jets, see 
section~\ref{sec:signatures}. The branching fraction of this channel is
about 30\% which is one of the advantages over the dilepton channel.
However, $W+\mathrm{multijet}$ backgrounds are large and call for 
dedicated strategies to improve the signal to background ratio.
In this section we discuss four important methods to reduce backgrounds
in the lepton-plus-jets channel and obtain a measurement of the 
$t\bar{t}$ cross section. 

\subsubsection{Topological analyses}
The production of two very heavy objects in $t\bar{t}$ events is reflected in
several kinematic distributions that describe the general event topology.
Some analyses take advantage of the fact that the scalar sum of
transverse energies, $H_\mathrm{T}$, tends to be much higher in $t\bar{t}$ events than
in background events. This is due to the fact that jets originating from the
decay of a heavy object are typically much harder, i.e. higher in $E_\mathrm{T}$, 
than those from gluon radiation.   
Other analyses utilise the higher spherical symmetry of $t\bar{t}$ events.
More recent analyses combine several kinematic variables into one
kinematic discriminant using a neural network~\cite{CDFttbarXSkinANN2005}. 

We discuss here in more detail the final D\O \ topological analysis
used to measure the $t\bar{t}$ cross section in Run I at 
$\sqrt{s}=1.8\,\mathrm{TeV}$~\cite{d0Sigmattbar2003Run1}.
This analysis marks the final result of the technique used in D\O \ to discover
the top quark in the lepton-plus-jets channel in 1995~\cite{d0discovery}. 
The data were selected by a lepton-plus-jets trigger, requiring one loosely 
identified electron or muon and at least one jet.
The electron data set corresponds to a luminosity of 119.5~$\mathrm{pb^{-1}}$,
the muon data set to 107.7~$\mathrm{pb^{-1}}$.
The offline selection asks for an isolated, well identified electron with 
$E_\mathrm{T}>20\,\mathrm{GeV}$ and $|\eta|\leq2.0$ or an isolated, well identified
muon with $p_\mathrm{T}>20\,\mathrm{GeV}/c$ and $|\eta|\leq1.7$.
The missing transverse energy as calculated in the calorimeter has to be
above 25 GeV for electron-plus-jets events and above 20~GeV for 
muon-plus-jets events. The pseudorapidity of the reconstructed $W$ boson is 
required to be $|\eta(W)|\leq 2.0$. Jets are counted with a threshold of
$E_\mathrm{T}>15\,\mathrm{GeV}$ and $|\eta|<2.0$. Events with four or more jets
are accepted. To avoid an overlap with the soft muon $t\bar{t}$ analysis
events with jets having a soft muon $b$ tag are excluded.
\par
Three topological variables are used to further discriminate between signal
and background: (1) the $H_\mathrm{T}$ variable which is defined as the scalar sum
of all transverse jet energies using the jet definition as mentioned above,
(2) the scalar sum of $\EtMiss$ and the lepton $p_\mathrm{T}$ which 
is denoted $E_\mathrm{T}^\mathrm{L}$, and
(3) the aplanarity $\mathcal{A}$ which is an event shape variable quantifying
the ``flatness'' of an event and is defined as $\frac{3}{2}$ times the smallest
eigenvalue of the normalized laboratory momentum tensor $\mathcal{M}$
of the observed physics objects. The tensor $\mathcal{M}$ is defined
by:
\begin{equation}
  \mathcal{M}_{ij} = \frac{\sum_k p_{ki}\,p_{kj}}{\sum_k |\bi{p}_k|^2}
\label{eq:aplanarity}
\end{equation}
where $\bi{p}_k$ is the three momentum of the object $k$ and the indices
$i$ and $j$ are the $x$, $y$ and $z$ coordinates. The sum runs over all
objects under consideration, i.e. all jets defined by the cuts
$E_\mathrm{T}>15\,\mathrm{GeV}$ and $|\eta|<2.0$, and the $W$ boson reconstructed
from the lepton and $\EtMissVec$.
Large values of $\mathcal{A}$ are indicative of spherical events, whereas
small values correspond to more planar events.
The following cuts on topological variables are used:
$H_\mathrm{T}>180\,\mathrm{GeV}$, $E_\mathrm{T}^\mathrm{L}>60\,\mathrm{GeV}$ and 
$\mathcal{A}> 0.065$. 
These cuts result from an optimization procedure such that the smallest 
error on the measured cross section and the best signal to background ratio 
$S/B$ are achieved. 
   
The main backgrounds to the topological $t\bar{t}$ analysis arise from 
$W+\mathrm{jets}$ events and QCD multijet events that contain a misidentified electron or 
an isolated muon and mismeasured $\EtMiss$.
The background estimate proceeds in three major steps. In a first step the 
QCD multijet background is estimated as a function of jet multiplicity
from data samples where all selection cuts except the three topological cuts 
are applied. The electron and the muon samples are
treated separately because processes that give rise to a misidentified electron
or an isolated muon are significantly different. 

Jets that produce showers with 
a large fraction of the energy in the electromagnetic calorimeter can sometimes
pass the electron selection criteria and be falsely identified as an electron.
To determine the rate of multijet events containing such misidentified electrons
one examines the $\EtMiss$ spectrum of events that pass the electron trigger
but fail the full offline electron identification. These events correctly describe
the shape of the $\EtMiss$ distribution of QCD multijet events, while the 
normalization has to be found by matching the number of events at low 
$\EtMiss$ between this background sample and the sample where the full electron
identification has been applied. The number of events in the tail of the
normalized distribution above $\EtMiss>25\,\mathrm{GeV}$ provides an estimate
on the number of QCD multijet background events. 

Muons from semileptonic $b$ of $c$ quark decays are normally within a jet, i.e.
they are nonisolated. However, occasionally the decay kinematics is such, that
there is insufficient hadronic energy to produce a jet. In this case the muons
from heavy quark decays appear to be isolated and constitute the major source of 
QCD multijet background in the muon-plus-jets sample.
The rate of isolated muons from heavy quark decays is measured using multijet
samples with $\EtMiss<20\,\mathrm{GeV}$.

The multijet background is estimated for each inclusive $W+\mathrm{jet}$ 
multiplicity sample. 
In the $e+\geq4\;\mathrm{jets}$ sample the background
is estimated to be $7.2\pm2.2$ events, in the 
$\mu+\geq4\;\mathrm{jets}$ sample 
$13.9\pm4.4$ events.

In the second step of the background estimate the rate of $W+\mathrm{multijet}$
events is computed using a fit to the jet multiplicity spectrum that remains
after the subtraction of the QCD multijet background. This method exploits a
simple exponential relationship between the number of events and the jet
multiplicity
\begin{equation}
  \frac{\sigma(W+n \ \mathrm{jets})}{\sigma(W+(n-1)\ \mathrm{jets})} = \alpha \ ,
  \label{eq:berends}  
\end{equation}
where $\alpha$ is a constant that depends on the specific jet definition, i.e. 
the clustering algorithm as well as the $E_\mathrm{T}$ and $\eta$ requirements. 
Relation (\ref{eq:berends}) is known as ``Berends scaling''~\cite{ellis85,berends91}
and based on the assumption that the observed jets come from gluon radiation where
the emission of each additional jet is suppressed by a factor of the strong 
coupling constant $\alpha_s$. A fit to the observed D\O \ data yields a scaling
parameter of $\alpha=0.18\pm0.01$ for electron data and $\alpha=0.19\pm0.02$
for muon data. These fit values are used to predict the number of background
events in the $W+\geq 4$ jets sample and are found to be
$44.8\pm 8.6$ events for electron-plus-jets events and $25.8\pm 4.6$ events
for the muon-plus-jets sample. As a side remark it has to be noted that
$\alpha$ cannot simply be identified with $\alpha_s$, since the variation
of $\alpha_s$ in the PDFs has to be taken into account~\cite{d0alphas}.
If this is properly done the sensitivity to $\alpha_s$ is largely lost.

In the third step the cut efficiencies for the topological cuts are computed 
and then applied to the previously derived background estimate for the
$W+\geq 4$ jets sample. 
The efficiency on the QCD multijet background is again derived from data, 
while the efficiency on the $W+\mathrm{jets}$ background is taken from events 
generated with the {\sc Vecbos} Monte Carlo program~\cite{vecbosDocu}. 
After applying all cuts the total
background is predicted to be $4.51\pm0.91$ in the $e+\geq 4$ jets 
sample and $4.32\pm1.04$ events in the $\mu+\geq 4$ jets sample,
while 9 or 10 events are observed in data, respectively.

The event detection efficiency $\epsilon_\mathrm{evt}$ for the $t\bar{t}$ signal 
is determined from
{\sc Herwig} Monte Carlo events that were passed through the D\O \ detector simulation.
For a top quark mass of $M_\mathrm{top}=170\,\mathrm{GeV}/c^2$ a value
of $\epsilon_\mathrm{evt}=(1.29\pm0.23)\%$ is found for the electron sample,
and $\epsilon_\mathrm{evt}=(0.91\pm0.27)\%$ for the muon sample.
Using $\epsilon_\mathrm{evt}$ determined for a top mass of 
$M_\mathrm{top}=172.1\,\mathrm{GeV}/c^2$ as measured by D\O \ the cross
section is computed to be $(2.8\pm2.1)\,\mathrm{pb}$ in the 
electron-plus-jets channel and $(5.6\pm3.7)\,\mathrm{pb}$ in the 
muon-plus-jets channel.
Figure~\ref{fig:aplanarityVsHT} shows the aplanarity versus $H_\mathrm{T}$ 
plane for
the background samples, $t\bar{t}$ Monte Carlo and D\O \ data. The plots
illustrate the effectiveness of reducing the background via the cuts on
the topological variables $H_\mathrm{T}$ and $\mathcal{A}$. 

\subsubsection{Secondary vertex tag}
\label{sec:ttbarSecVtx}
While each $t\bar{t}$ event features two $b$ quark jets, only about 2\%
of the $W+\mathrm{jets}$ background contain a $b$ quark. Therefore,
the $t\bar{t}$ signal can be significantly enhanced by identifying 
$b$ quark jets. Three different $b$ jet identification methods are discussed
in section~\ref{sec:btagging}. In CDF the secondary vertex tagging method
proved to be the most effective one, yielding the cross section measurement
with the smallest total uncertainty.

We present here a summary of the first CDF Run II $t\bar{t}$ cross section
measurement using lepton-plus-jets events with secondary vertex $b$ 
tagging~\cite{CDFsecvtxRun2}. The $b$ tagging method has 
already been presented in section~\ref{sec:btagging}. That is why we
concentrate here on the event selection and the background estimate.
The analysis uses a data sample triggered by high momentum electrons
or muons corresponding to an integrated luminosity of 
162~$\mathrm{pb^{-1}}$. The electron selection requires an isolated 
cluster in the central calorimeter with $E_\mathrm{T}>20\,\mathrm{GeV}$ matched 
to a track 
with $p_\mathrm{T}>10\,\mathrm{GeV}/c$. Muon candidates have a track in the drift
chamber with $p_\mathrm{T}>20\,\mathrm{GeV}/c$ that is matched to a track segment
in the muon chambers. Events consistent with being photon conversions 
(electrons) or cosmic rays (muons) are rejected. The missing transverse
energy is required to be $\EtMiss>20\,\mathrm{GeV}$.
By requiring one and only one well identified lepton $t\bar{t}$ dilepton
events and $Z\rightarrow e^+e^-/\mu^+\mu^-$ events are suppressed. 
To improve the removal efficiency for $Z$ bosons,
events are also removed if a second, less stringently identified lepton 
is found that forms an invariant mass $M_{\ell\ell}$ with the primary lepton 
within the window of $76<M_{\ell\ell}<106\,\mathrm{GeV}/c^2$. 
Jets are defined as clusters in the hadronic calorimeter with 
$E_\mathrm{T}>15\,\mathrm{GeV}$ and $|\eta|<2.0$. At least three jets are 
required for an event to fall into the signal region.
One of these jets has to be identified as containing a $b$ quark
using the secondary vertex tag algorithm. 
The final cut is on the total transverse energy and demands
$H_\mathrm{T}>200\,\mathrm{GeV}$, which rejects approximately 40\% of the
background while retaining 95\% of $t\bar{t}$ signal events.

The secondary vertex tag algorithm is described in section~\ref{sec:btagging}.
Figure~\ref{fig:btag}a shows the distribution of the two-dimensional decay
length of secondary vertices in the CDF data sample compared to the SM 
prediction of $t\bar{t}$ signal and background. Good agreement is found.
The backgrounds to the secondary vertex tagged sample are (i) direct QCD production
of heavy flavour quarks without an associated $W$ boson (non-$W$ QCD), 
(ii) $W$ plus light quark jets events where one jet is falsely
identified as heavy flavour (mistags), (iii) $W$ plus heavy flavour jets,
(iv) diboson production, single top quark, and $Z\rightarrow \tau^+\tau^-$ 
production. 
The estimate of these backgrounds is partially derived from CDF data
and partially from Monte Carlo.
In particular, the number of $b$ tagged $W+\mathrm{jets}$ background events
is calibrated with the number of observed $W+\mathrm{jets}$ events before $b$ tagging.
Therefore, the first step in the background calculation is to estimate the number 
of background events that do not contain a $W$ boson in the pretag sample
and subtract that background from the observed number of events.

\emph{(i) Non-W QCD background:} The non-$W$ background is mainly due to events where 
a jet is misidentified as an electron and the $\EtMiss$ is mismeasured, or due to
muons from semileptonic $b$ decays which pass the isolation criterion.
Since the background sources in the electron and muon sample are different, one
has to treat these two samples separately. 
The non-$W$ background estimate uses the $\EtMiss$ variable and the isolation
variable $R_\mathrm{iso}$, that is defined as the ratio of calorimeter 
energy $E_\mathrm{iso}$ contained in an isolation annulus of $\Delta R = 0.4$ 
around the lepton (excluding the energy associated to the lepton) divided by the
lepton energy $E_\ell$.  
The $R_\mathrm{iso}$ versus $\EtMiss$ plane is divided into 
a signal region ($R_\mathrm{iso}< 0.1$ and $\EtMiss > 20\,\mathrm{GeV}$)
and three sideband regions. One assumes that the two variables are uncorrelated
for non-$W$ background events and calculates the number of background events in the 
signal region as a simple proportion of events in the sideband regions.
The contribution of true $W$ and $t\bar{t}$ events in the sideband regions
is subtracted using Monte Carlo predictions normalized to the observed
number of events in the signal region.
In the \emph{pretag} $e+\geq 3$ jets sample $(20\pm5)\%$ of the events are
estimated to be non-$W$ events, in the $\mu+\geq 3$ jets sample 
it is $(7.5\pm 2.3)\%$. 
The \emph{pretag} sample is defined as the data sample where all selection cuts
have been applied except requiring a $b$ tag. 
In the final sample, after requiring a secondary vertex
tag, about 18\% of the total background is due to non-$W$ events.

\emph{(ii) Mistags:} The mistag rate of the secondary vertex algorithm is measured
using inclusive jet samples. Mistags are caused mostly by a random combination
of tracks which are displaced from the primary vertex due to tracking errors.
The main idea is to use the rate of events
with negative two-dimensional decay length as an estimate of the mistag rate.
Corrections due to material interactions, long-lived light flavour particles
(e.g. $K_s^0$ and $\Lambda$), and negatively tagged heavy flavour jets are 
determined using fits to the effective lifetime spectra of tagged vertices.
The mistag rate is parameterized as a function of four jet variables:
$E_\mathrm{T}$, the good track multiplicity, $\eta$ and $\phi$ of the jet
as well as one event variable, i.e. the scalar sum of the $E_\mathrm{T}$ of all 
jets with 
$E_\mathrm{T}>8\,\mathrm{GeV}$. To estimate the mistag background in the 
$W+\mathrm{jets}$
 sample each jet in the pretag sample is weighted with its mistag rate. 
The sum of weights over all jets in the sample is then scaled down 
by the fraction of non-$W$ events in the pretag sample. Since the mistag rate
per jet is sufficiently low, this prediction of mistagged jets is a good
estimate on the number of events with a mistagged jet. It is found that
34\% of the background for the final $t\bar{t}$ selection are due to mistags.

\emph{(iii) W + heavy flavour:} The production of a $W$ boson in association with
heavy flavour quarks is the main background to the $t\bar{t}$ signal with 
a secondary vertex tag. Heavy quarks occur in the process 
$q_1\bar{q}_2\rightarrow W + g$ where the gluon splits into a $b\bar{b}$ or
a $c\bar{c}$ pair, and in the process $gq\rightarrow Wc$. 
As mentioned in section~\ref{sec:mc}, the {\sc Alpgen} Monte Carlo
program~\cite{alpgen} is used to generate several samples of exclusive 
$W+n\;\mathrm{jets}$ final states. This includes $W+b\bar{b}/c\bar{c} + n$ jets,
and $Wc+\mathrm{jets}$. The {\sc Alpgen} generation is followed by showering with
{\sc Herwig}. To reproduce the entire $W+\mathrm{multijets}$ spectrum the exclusive
samples are merged together taking the appropriate cross sections into account.
The combination procedure avoids double counting of jets produced at 
matrix element level with {\sc Alpgen} and hard jets produced by the showering
in {\sc Herwig}. A jet clustering algorithm is run at Monte Carlo particle level
after showering but before detector simulation. These Monte Carlo jets are 
matched to partons generated at matrix element level. If extra jets occur,
that do not match a parton, the event is rejected. 

\begin{figure}[!t]
  \hspace*{8mm} (a) \hspace*{63mm} (b) \\
  \includegraphics[width=0.50\textwidth]{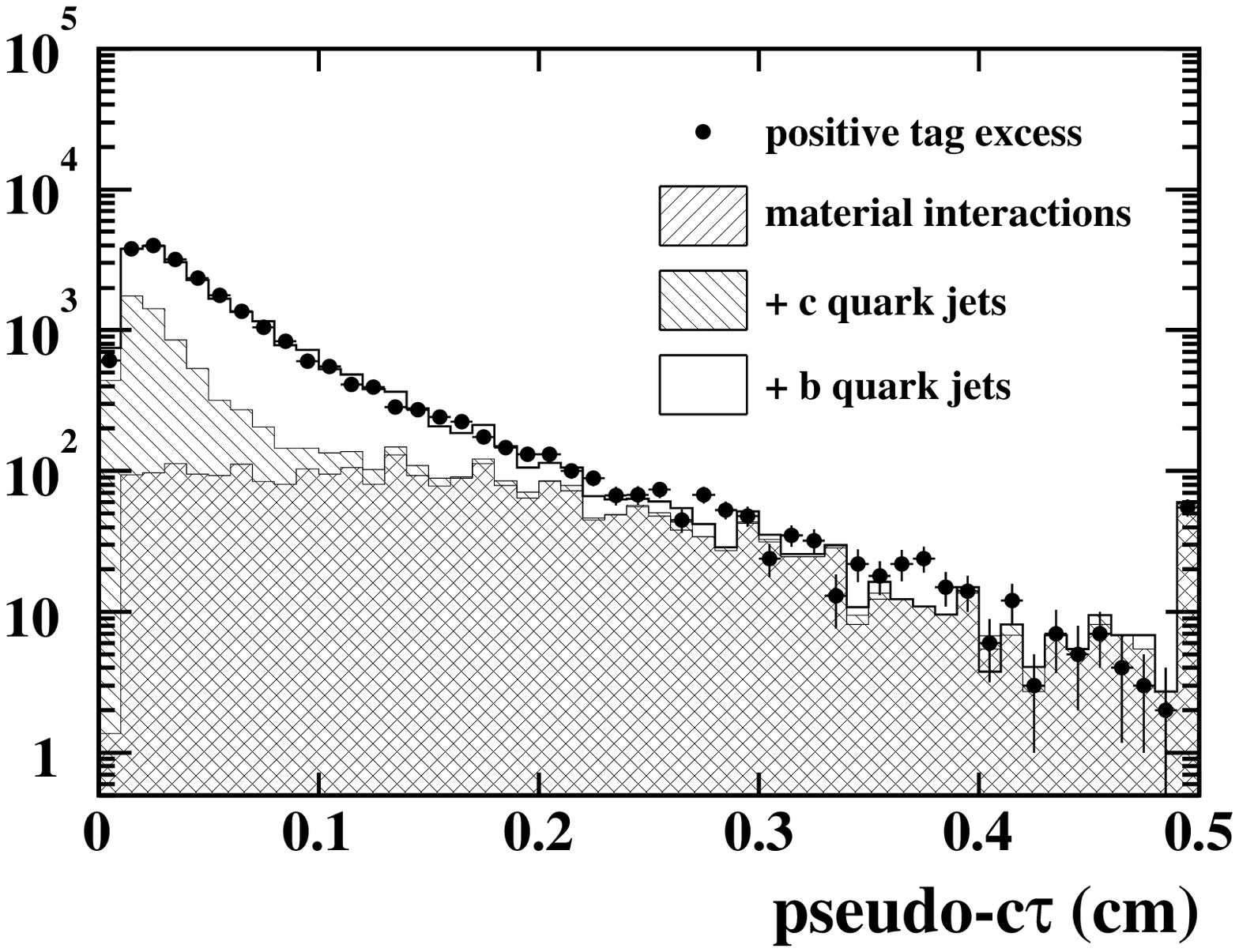}
  \includegraphics[width=0.48\textwidth]{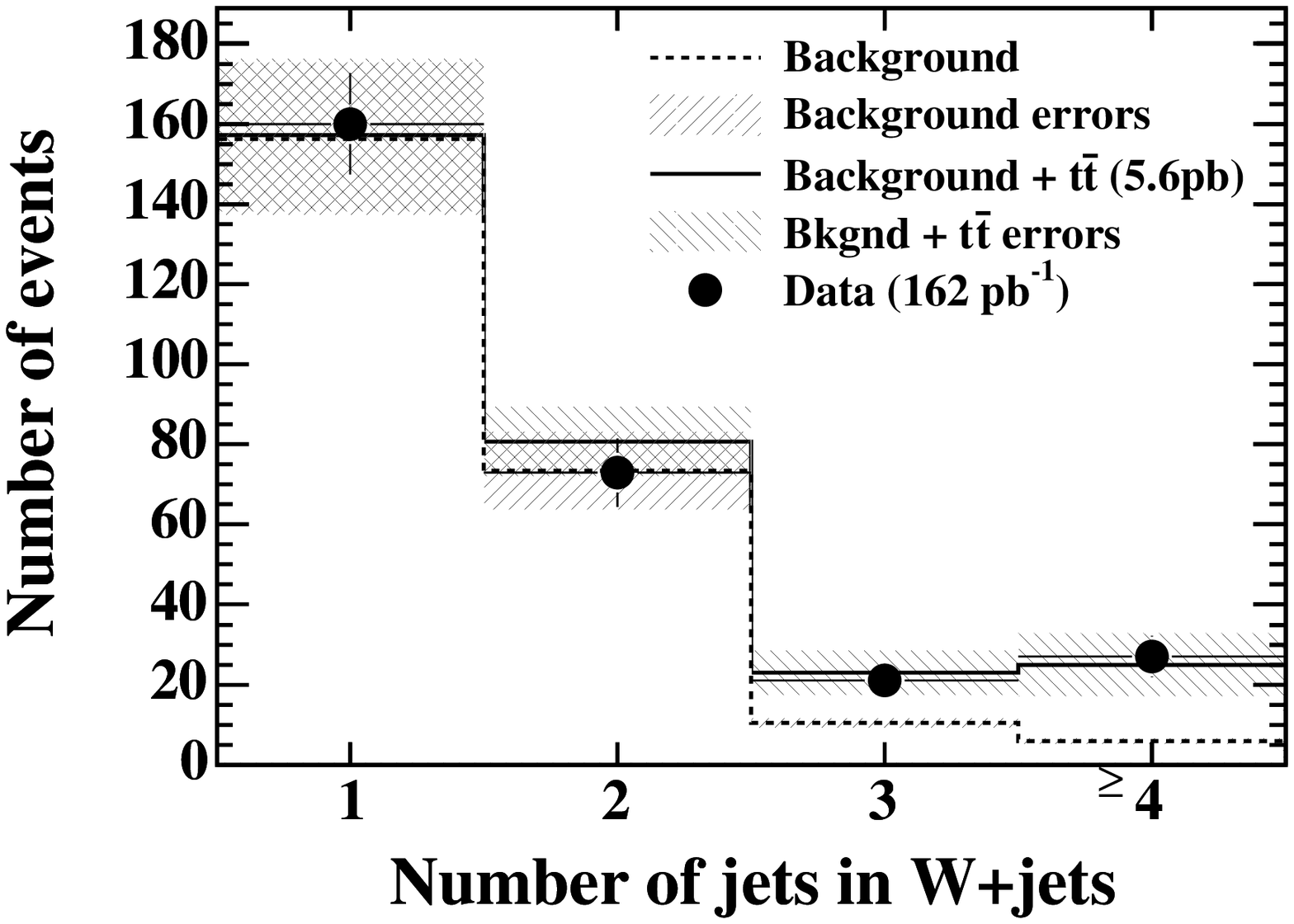}
\caption{\label{fig:ctauNjet} (a) Pseudo-$c\tau$ distribution for inclusive jet data.
  The distribution is used to measure the heavy flavour fractions in the jet sample.
  The fitted contributions for the different heavy flavour components and secondary
  interactions in light flavour jets are shown~\cite{CDFsecvtxRun2}.
  (b) Jet multiplicity spectrum for CDF data (dots), 
  the total background expectation (dashed histogram), and the sum of background 
  and the predicted $t\bar{t}$ signal ($\sigma=6.7\,\mathrm{pb}$, full line).}
\end{figure}

While the shape of the $W+n\;\mathrm{jets}$ spectrum can be reproduced well by
the Monte Carlo, the absolute normalization has large theoretical uncertainties
and is therefore taken from collider data. The heavy flavour fractions 
predicted by {\sc Alpgen} are found to be too small and are calibrated 
against fractions measured in inclusive jet data.
The composition of the inclusive jet sample is determined by fitting the
pseudo-$c\tau$ distribution for $b$ tagged jets. The pseudo-$c\tau$ is defined
as $L_\mathrm{2D}\cdot M_\mathrm{vtx}/p_\mathrm{T}^\mathrm{vtx}$, 
where $L_\mathrm{2D}$ is the 
two-dimensional decay length (see also section~\ref{sec:btagging}),
$M_\mathrm{vtx}$ is the invariant mass of all tracks associated to the secondary
vertex, and $p_\mathrm{T}^\mathrm{vtx}$ is the transverse momentum of the vertex 4-vector.
Figure~\ref{fig:ctauNjet}a shows the pseudo-$c\tau$ distribution for the 
inclusive jet data as well as the fitted heavy flavour components.
The template distributions for heavy flavour jets used in the fit are taken from
{\sc Alpgen} Monte Carlo samples.
The measured heavy flavour fractions in jet data are consistently higher by 50\% 
than the {\sc Alpgen} prediction. Therefore, a correction factor of $1.5\pm0.4$ 
is applied to the heavy flavour fractions in the $W+\mathrm{multijets}$ sample.
The assumption here is that this correction factor is universal and can be
transferred from the inclusive jets to the $W+\mathrm{multijets}$ sample.
As a result, it is found, that the fraction of $W+4$ jets events coming from
the $Wb\bar{b}$ process is $(4.8\pm1.3)\%$, the fraction with one or two $c$ 
jets from $Wc\bar{c}$ is $(7.3\pm2.0)\%$, and the fraction of $Wc$ events 
is $(6.1\pm1.3)\%$. It has to be noted that the $Wc$ fraction is directly
taken from the {\sc Alpgen} prediction and is not corrected, since it is due to
a different physics process.

The number of $W+$ heavy flavour background events for the $t\bar{t}$ analysis
is computed by multiplying the derived heavy quark fractions by the number of 
pretag
events, after subtracting the non-$W$ background. The heavy flavour contributions
to the total background in the $t\bar{t}$ candidate sample are as follows: 
25\% are due to $Wb\bar{b}$, 8\% due to $Wc\bar{c}$ and 6\% due to $Wc$ production.     

\emph{(iv) Diboson, single top and} $Z\rightarrow\tau^+\tau^-${\emph :}
There is a number of smaller backgrounds which can be reliably predicted 
by combining the event detection efficiency for these processes as determined 
from Monte Carlo events with the theoretically predicted cross section.
This method is feasible, since the cross section predictions for diboson
production processes, i.e. $WW$, $WZ$ and $ZZ$, and single top quark production
have relatively small uncertainties. Diboson events can mimic a $t\bar{t}$ 
signal if one boson decays leptonically and the other one decays into jets, 
where at least 
one jet is due to a $b$ or $c$ quark. In addition, $Z+\mathrm{jet}$ production
can mimic $t\bar{t}$ events if the $Z$ boson decays into $\tau^+\tau^-$ and 
one $\tau$ decays leptonically producing an isolated electron or muon, while the
second $\tau$ decays hadronically.  
The contribution of diboson, $Z\rightarrow\tau^+\tau^-$ and single top events 
to the total background is found to be 9\%.

The total background prediction is $13.5\pm1.8$ events, while 48 events 
are observed in the $W+ \geq 3$ jets sample. The clear excess of events is
attributed to $t\bar{t}$ production. 
Figure~\ref{fig:ctauNjet}b shows the jet multiplicity spectrum for CDF data, 
the total
background expectation, and the sum of background and the predicted
$t\bar{t}$ signal. A very good agreement between the data and the SM 
expectation is found.

To compute the $t\bar{t}$ cross 
section according to (\ref{eq:xs}), the event detection efficiency 
$\epsilon_\mathrm{evt}$ for $t\bar{t}$ events is needed.
It is obtained from a sample of {\sc Pythia} $t\bar{t}$ Monte Carlo events.
The trigger efficiency as well as several correction factors that account
for differences between data and Monte Carlo events are measured with 
control data samples. Particularly important is the correction factor
for the $b$ tagging efficiency.
A sample of inclusive electrons with $p_\mathrm{T}>7\,\mathrm{GeV}$ is used for 
this purpose. Many electrons in this momentum regime originate from 
semileptonic $b$ decays. Therefore, the sample is enriched in heavy flavour.
Using a double tag technique it is found that the average tagging efficiency
of a $b$ quark jet in data is $(24\pm1)\%$, while the Monte Carlo predicts
$(29\pm1)\%$, thus yielding a $b$ tagging correction factor of $0.82\pm0.06$,
where the error includes systematic uncertainties.
The efficiency to tag at least one jet in a $t\bar{t}$ event, after all
other cuts have been applied, is found to be $(53.4\pm3.2)\%$.
The uncertainty is almost purely systematic. The overall event detection 
efficiency is $\epsilon_\mathrm{evt}=(3.84\pm0.40)\%$. This yields a cross 
section measurement of 
$\sigma(t\bar{t})=5.6^{+1.2}_{-1.1}\,(\mathrm{stat.})\,^{+0.9}_{-0.6}\,(\mathrm{syst.})\,\mathrm{pb}$, which is in good agreement with the theoretical
expectation of $6.7^{+0.7}_{-0.9}\,\mathrm{pb}$. The statistical uncertainty
is larger than the systematic one, but very soon, using more data, the
systematic uncertainty will start to dominate the measurement. 
Although, having larger control samples at hand, it will be possible to also 
reduce the systematic uncertainty. 

\subsubsection{Impact parameter tag}
\label{sec:impactParameterTag}
The impact parameter $b$ tagging algorithm is an alternative method to 
identify $b$ quark jets. The method also relies on the long lifetime of $b$ 
hadrons, but the explicit reconstruction of a secondary vertex is not
required. The algorithm rather asks for tracks with high impact parameter
significance and is explained in section~\ref{sec:btagging}.
In the CDF Run II analysis presented here a jet is $b$ tagged if its 
probability to be consistent with a zero lifetime hypothesis is below 
0.01~\cite{CDFjetprob2004}.
The $b$ tagging efficiency of the algorithm is measured in inclusive electron
data using a double tag method. For heavy flavour jets with 
$E_\mathrm{T}>10\,\mathrm{GeV}$ the average tagging efficiency is found to be 
$(19.7\pm1.2)\%$. 

The event selection for the $t\bar{t}$ cross section measurement is essentially
the same as described in section~\ref{sec:ttbarSecVtx} for the 
analysis using a secondary vertex $b$ tag. There is one additional cut
which requires that the $\Delta \phi$ between the $\EtMissVec$ and the most
energetic jet is within $[0.5, 2.5]$ radians if $\EtMiss < 30\,\mathrm{GeV}$.
The cut on $H_\mathrm{T}$ is omitted.
The background is estimated based on the same techniques as the
measurement using the secondary vertex tag. The mistag rate of light
quark jets is measured in an inclusive jet data sample and found to be 
$(1.11\pm0.06)\%$. In a data sample corresponding to $162\,\mathrm{pb^{-1}}$
59 $W+\geq3$ jets events are observed after all cuts, while 
the total background is estimated to be $24.7^{+4.8}_{-4.6}$ events. 
The overall event detection efficiency is found to be
$\epsilon_\mathrm{evt}=(4.09\pm0.61)\%$ which includes the event $b$ tagging
efficiency of $(57.2\pm3.9)\%$. As a result, the $t\bar{t}$ cross section is 
calculated to be 
$\sigma(t\bar{t})=5.8^{+1.3}_{-1.2}\,(\mathrm{stat.})\pm1.3\,(\mathrm{syst.})\,\mathrm{pb}$, which is in good agreement with the result obtained with
the secondary vertex $b$ tagging algorithm.

\subsubsection{Soft lepton tag}
\label{sec:softLeptonTag}
The third technique to identify $b$ quark jets searches for electrons
or muons within jets originating from semileptonic $b$ decays. We present
here a CDF analysis using a soft muon tag~\cite{CDFsoftmuon2005}. 
A data sample with an integrated luminosity of 194~$\mathrm{pb^{-1}}$ is used.
The event selection before $b$ tagging is the same as for the secondary vertex tag 
analysis. After all cuts 337 events are retained.

The soft muon tag algorithm uses a global $\chi^2$ built from several 
characteristic distributions that separate muon candidates from background.
A jet is considered as $b$ tagged if it contains a muon with 
$p_\mathrm{T}>3\,\mathrm{GeV}/c$ and $\chi^2 < 3.5$ within $\Delta R<0.6$ of the jet 
axis. Events are rejected if the isolated high $p_\mathrm{T}$ lepton is a muon of 
opposite charge to the soft muon and the dimuon invariant mass is consistent
with a $J/\psi$, an $\Upsilon$ or a $Z^0$. The efficiency to tag a $t\bar{t}$
event is approximately 15\%, with small variations between the $W+3$ jets and 
$W+4$ jets samples. The rate of background events where a light quark jet is 
misidentified as containing a semileptonic $b$ hadron decay is estimated from
a sample of photon-plus-jets events. 
The misidentification probability of a track with
$p_\mathrm{T}>3\,\mathrm{GeV}/c$ and $\Delta R<0.6$ is found to be about 0.7\%. 

In the $W +\geq 3$ jets data sample 20 events with a soft muon tag are found,
while the total expected background is estimated to be $9.5\pm1.1$ events.
The total event detection efficiency is $\epsilon_\mathrm{evt}=(1.0\pm0.1)\%$ and 
the resulting $t\bar{t}$ cross section is found to be 
$\sigma(t\bar{t})=5.3\pm +3.3\,(\mathrm{stat.})\,^{+1.3}_{-1.0}\,(\mathrm{syst.})\,\mathrm{pb}$.

\subsection{All hadronic channel}
\label{sec:allhadronic}
While the all hadronic channel has the largest branching ratio of all $t\bar{t}$
event categories, the QCD multijet background is overwhelming and it is therefore
very challenging to isolate a $t\bar{t}$ signal.
We present here a D\O \ analysis based on a Run II dataset corresponding to 
162 $\mathrm{pb^{-1}}$~\cite{d0allJets}.
To enrich the signal content secondary vertex $b$ tagging
and a multivariate analysis with artificial neural networks is employed.

The data sample is collected with a dedicated jet trigger at an average trigger
efficiency of 74\%. The event pre-selection asks for six or more jets,
that are reconstructed with an algorithm using a fixed cone radius of 
$\Delta R = 0.5$. Jets are counted with $E_\mathrm{T}> 15\,\mathrm{GeV}$ and
$|\eta|<2.5$. Events that contain an isolated lepton are rejected to obtain
a dataset orthogonal to the lepton-plus-jets analysis. 
The reconstruction of secondary vertices is used to identify $b$ quark jets.
A jet is considered to be tagged if a vertex with signed decay length significance 
larger than 7.0 is found. Exactly one $b$ tagged jet is required. Double tagged
events are rejected to ease the background estimation.
The efficiency to tag exactly one jet in a hadronic $t\bar{t}$ event is 
determined to be 46\%.

The multivariate analysis combines 13 kinematic or event shape variables
using two neural networks. The background sample for training the 
networks is obtained from the pretag data sample.
A parametrization of the tagging rate is used to select events that have high
probability to be tagged, but do in fact not have a tag. This ensures on
one hand that the background sample is similar to the tagged events, but
on the other hand is disjoint to the candidate sample. Signal Monte Carlo
events are generated with {\sc Alpgen} in combination with {\sc Pythia}
for showering and hadronization.
After all cuts 220 candidate events are observed over a total background
of $186\pm5$ events. The event detection efficiency is
$\epsilon_\mathrm{evt}=(2.8\pm0.8)\%$ including a hadronic branching ratio
of 46.19\%. The $t\bar{t}$ production cross section is measured to be
$\sigma(t\bar{t})=7.7^{+3.4}_{-3.3}\,(\mathrm{stat.})^{+4.7}_{-3.8}\,(\mathrm{syst.})\pm 0.5\,(\mathrm{lumi.})\;\mathrm{pb}$.

\subsection{Cross section combination}
\label{sec:xs_summary}
\renewcommand\arraystretch{1.5}
In this section we summarise the final Run I $t\bar{t}$ cross section measurements
and give a brief overview on the status of the currently available Run II
measurements. The measured cross sections are presented in 
table~\ref{tab:xsSummary}. 
\begin{table}[t]
\caption{\label{tab:xsSummary}Measurements of the $t\bar{t}$ cross section at the
   Tevatron. The CDF measurements assume a top quark mass of 
   $M_\mathrm{top}=175\,\mathrm{GeV}/c^2$
   \cite{CDFttbarCrossSection2001LeptonPlusJetsRun1,CDFttbarCrossSection1998DileptonRun1,CDFttbarhadronic1997Run1}, the D\O \ measurements use 
   $M_\mathrm{top}=172.1\,\mathrm{GeV}/c^2$~\cite{d0Sigmattbar2003Run1}.
   In Run II a top mass of $M_\mathrm{top}=175\,\mathrm{GeV}/c^2$ is assumed
   for the measurements of both experiments.
   The D\O \ Run I lepton-plus-jets cross section is a combination of the 
   topological analysis
   and the soft lepton tag analysis.
   The given uncertainties include statistical and systematic contributions.
   } 
\begin{center}
\begin{tabular}{lllll}
  \br
   & \multicolumn{2}{c}{Run I at $\sqrt{s}=1.8\,\mathrm{TeV}$} & 
     \multicolumn{2}{c}{Run II at $\sqrt{s}=1.96\,\mathrm{TeV}$} \\ \mr
  Channel  & CDF & D\O & CDF & D\O \\ \mr
  Dilepton  & $8.4^{+4.5}_{-3.5}\;\mathrm{pb}$ & $6.0\pm3.2\;\mathrm{pb}$ & 
              $7.0^{+2.9}_{-2.4}\;\mathrm{pb}$ & $14.3^{+5.8}_{-4.8}\;\mathrm{pb}$ \\
  \hspace*{3mm} with secondary vertex tag & & & & $11.1^{+6.0}_{-4.6}\;\mathrm{pb}$ \\
  \hspace*{3mm} global kinematic fit & & & $8.6^{+2.7}_{-2.6}\;\mathrm{pb}$ & \\
  Lepton + jets & & $5.1\pm1.9\;\mathrm{pb}$ \\
  \hspace*{3mm} with secondary vertex tag & 
  $5.1\pm1.5\;\mathrm{pb}$ & & $5.6^{+1.5}_{-1.3}\;\mathrm{pb}$ &
  $8.6^{+1.7}_{-1.6}\;\mathrm{pb}$ \\ 
  \hspace*{3mm} with impact parameter tag & & & $5.8\pm1.8\;\mathrm{pb}$ &
  $7.6^{+1.8}_{-1.5}\;\mathrm{pb}$ \\
  \hspace*{3mm} with soft lepton tag & 
  $9.2\pm4.3\;\mathrm{pb}$ & & $5.2^{+3.2}_{-2.1}\;\mathrm{pb}$ & \\ 
  \hspace*{3mm} with $b$-tag and kinematic fit & & & $6.0\pm2.0\;\mathrm{pb}$ & \\
  \hspace*{3mm} with neural networks & & & $6.6\pm1.9\;\mathrm{pb}$ & \\
  All hadronic & $7.6^{+3.5}_{-2.7}\;\mathrm{pb}$ & $7.3\pm3.2\;\mathrm{pb}$ & 
                 $7.5^{+3.9}_{-3.1}\;\mathrm{pb}$ & 
                 $7.7^{+5.8}_{-5.1}\;\mathrm{pb}$ \\ \hline
  Combined & $6.5^{+1.7}_{-1.4}\,\mathrm{pb}$ & $5.7\pm1.6\;\mathrm{pb}$ & & \\ 
  \hline
  Predicted~\cite{cacciari2004} & 
  \multicolumn{2}{c}{$5.2^{+0.5}_{-0.7}\;\mathrm{pb}$} & 
  \multicolumn{2}{c}{$6.7^{+0.7}_{-0.9}\;\mathrm{pb}$} \\ \br
\end{tabular}
\end{center}
\end{table}
For Run I the cross sections obtained from combination of all $t\bar{t}$ event 
categories for CDF and D\O \ are also shown~\cite{CDFttbarCrossSection2001LeptonPlusJetsRun1,d0Sigmattbar2003Run1}. 
The combined values should be compared to the predicted cross section.
Within the uncertainties good agreement is found.

In Run II various new techniques to measure the $t\bar{t}$ cross section are 
introduced. While the traditional dilepton analysis uses two leptons that
meet strict identification criteria, CDF added an analysis where the 
identification of the second lepton is relaxed to increase the acceptance and
reduce the statistical uncertainty. This analysis is discussed in more detail
in section~\ref{sec:dilepton}. Table~\ref{tab:xsSummary} quotes the combined
value for both CDF dilepton analyses~\cite{CDFdileptonPRL04}. 
A third CDF dilepton analysis uses a 
kinematic fit to the $(\EtMiss, N_\mathrm{jet})$ phase space to disentangle
the major SM processes contributing to the dilepton sample 
(global kinematic fit)~\cite{CDFglobalDilepton}. 
D\O \ has made cross section measurements in a dilepton data sample without
$b$ tagging~\cite{D0dilepton2004} and in a $e\mu$ sample using a secondary vertex 
tag~\cite{D0emuBtag2004}.
\par
The CDF lepton-plus-jets analyses using three different $b$ tagging algorithms 
are presented in section~\ref{sec:leptonPlusJets}.
Based on a data set corresponding to $230\,\mathrm{pb^{-1}}$ the
D\O \ collaboration has also presented measurements using a secondary vertex 
$b$ tag and an impact parameter tag~\cite{Abazov:2005ey}.
CDF has performed two analyses that exploit kinematic properties of
$t\bar{t}$ lepton-plus-jets events. One analysis uses the sample with 
secondary vertex tags and adds a likelihood fit to the transverse energy
of the leading jet~\cite{CDFLjetsBtagKinematic}. The second analysis 
is based on the $W+\geq3$ jets sample before $b$ tagging and uses a neural
network to construct a powerful discriminant between the backgrounds and
the $t\bar{t}$ signal~\cite{CDFttbarXSkinANN2005}.
\par
The D\O \ measurement of the $t\bar{t}$ cross section in the all
hadronic channel~\cite{d0allJets} is discussed in section~\ref{sec:allhadronic}.
The CDF Run II analysis in the all hadronic channel is similar to the D\O \ one,
also requiring a secondary vertex $b$ tag and exploiting several kinematic
variables~\cite{Abulencia:2006se}. But CDF uses simple sequential cuts rather 
than a multivariate combination via a neural network.
At present, combined results on the $t\bar{t}$ cross section in Run II are
not yet available. The combination of the different $t\bar{t}$ event 
categories (dilepton, lepton-plus-jets, all hadronic) is relatively straight 
forward, since the data sets are 
essentially orthogonal. However, the different analyses within a particular 
category are correlated, sometimes even strongly correlated. The correlation
has to be determined and taken into account when a combination of the
different analyses is done.

\section{Top quark mass measurements}
\label{sec:mass}
As discussed in section~\ref{sec:ewktopdetermine}, the top quark mass is a very
important parameter for the description of electroweak processes and 
precision tests of the SM. 
The main implication is that the SM relates the masses of the top quark, the $W$
boson and the Higgs boson, such that precise measurements of $M_\mathrm{top}$ and
$M_W$ imply a prediction for $M_H$.
Therefore, the precise measurement of the top quark mass is one of the major goals
of Run II at the Tevatron.
However, since Run II results on the top quark mass are not yet published, we describe
in this chapter the two main methods used by CDF and D\O \ in Run I to determine 
$M_\mathrm{top}$: (1) the template method of CDF, which fits template distributions for 
different top quark masses to the distribution observed in data,
and (2) the matrix element method of D\O, which exploits the sensitivity of 
the leading order matrix element for $t\bar{t}$ production to $M_\mathrm{top}$.
Both methods are also used in Run II to measure $M_\mathrm{top}$.
In section~\ref{sec:MtopCombine} we discuss the combination of top quark mass measurements
in different $t\bar{t}$ event categories and among the two Tevatron experiments.

\subsection{Template method in $t\bar{t}$ lepton-plus-jets events}
\label{sec:template}
The best measurement of the top quark mass is achieved in the $t\bar{t}$ 
lepton-plus-jets channel, since it features a relatively high number of
candidate events, moderate background levels and allows for a full 
reconstruction of top quark momenta with reasonable accuracy.
That is why we present here the final top quark mass measurement in 
lepton-plus-jets events at CDF in Run I~\cite{CDFtopMass2001LeptonPlusJetsRun1}.
The data sample used in the analysis corresponds to an integrated luminosity of
106~$\mathrm{pb^{-1}}$. The event pre-selection is the same as the one in 
the Run I 
$t\bar{t}$ cross section analysis~\cite{CDFttbarCrossSection2001LeptonPlusJetsRun1},
which is essentially the same as the Run II selection described in 
section~\ref{sec:ttbarSecVtx}.
For the full reconstruction of $t\bar{t}$ candidate events a $W$ candidate
decaying into $e\nu$ or $\mu\nu$ and at least four jets are needed.
Therefore, events in the $W+3$ jets sample are used only if they feature 
at least one additional jet with $E_\mathrm{T}>8\,\mathrm{GeV}$ and $|\eta|<2.4$.
After pre-selection, but before requiring a $b$ tag the candidate sample 
consists of 163 events. 
Four subsamples are used to measure the top quark mass:
(1) The first subsample contains events with two secondary vertex $b$ tagged
 jets. (2) The second subsample consists of events with exactly one
secondary vertex tag. (3) The third subsample includes events with one or two
soft lepton tags, but no secondary vertex tag. (4) The fourth subsample 
contains events with no $b$ tag and at least four jets with 
$E_\mathrm{T}>15\,\mathrm{GeV}$ and $|\eta|<2.0$. One very important issue for the top
mass measurement is the reconstruction of jet energies, which is therefore 
detailed in the next section.

\subsubsection{Jet energy corrections}
The reconstruction of the top quark momenta in a $t\bar{t}$ candidate event
is based on the assumption that the four leading jets can be identified with
the primary partons from the top quark decay. To ensure the validity of this 
assumption several corrections have to be applied to the raw jet energies 
as measured in the calorimeter. While the general concept of jet reconstruction 
is briefly introduced in section~\ref{sec:jetReco}, we summarise here the 
jet corrections applied in the CDF top mass analysis.
\begin{enumerate}
  \item The {\it relative energy correction} accounts for non-uniformities 
  in the calorimeter response as a function of $\eta$ and is derived from 
  dijet data, where one jet is measured in the central calorimeter 
  ($0.2<|\eta|<0.7$) and the second jet is in the forward region
  ($1.1<|\eta|<4.2$). The forward calorimeter is thus calibrated against the
  response in the central region. The uncertainty on the relative correction
  varies between 0.2\% and 4\%.
  \item {\it Corrections for multiple interactions}. A fixed amount of 
  energy $\Delta E_\mathrm{mult}=0.297\,\mathrm{GeV}/c$ is subtracted from 
  the jet $E_\mathrm{T}$ for each additional reconstructed primary vertex in 
  the event.   
  \item The {\it absolute energy scale} is a multiplicative factor that
  converts the energy observed in the jet cone into the average true jet energy.
  To derive the absolute correction the calorimeter simulation is tuned to 
  agree with testbeam data of single electrons and pions. In a second step
  several parameters controlling the fragmentation process in the {\sc Isajet} 
  Monte Carlo generator are tuned such, that the multiplicity distribution, 
  the momentum spectrum, and the invariant mass distribution of charge particles, 
  as well as the ratio of charged to neutral energy agree between 
  {\sc Isajet} and dijet data.
  The absolute correction accounts for the nonlinearity of the calorimeter
  and energy losses near the boundaries of detector modules.
  The systematic uncertainty on the absolute energy scale is about 3\%.  
  \item {\it Underlying event corrections} account for extra energy contained
  in the jet cone due to particles coming from the fragmentation of partons that
  do not participate in the hard scattering of the primary $p\bar{p}$ interaction.
  A fixed amount of $0.65\,\mathrm{GeV}$ is subtracted from the $E_\mathrm{T}$
  of each jet .
  This correction is derived from minimum bias data.
  \item {\it Out-of-cone corrections} account for the energy that is physicswise
  associated to the jet but falling out of the jet cone. The out-of-cone 
  energy is related to low energy gluons emitted from initial partons,
  also referred to as soft gluon radiation. 
  The correction is derived from Monte Carlo events. 
  \item $t\bar{t}$ {\it specific corrections} are applied to the four leading
  jets and map their momenta to the momenta of the quarks from the $t\bar{t}$ 
  decay. The corrections account for three effects:
  (a) The difference in the jet $E_\mathrm{T}$ spectrum of $t\bar{t}$ induced jets and 
  the flat spectrum assumed in the previous corrections. 
  (b) The energy loss in semileptonic $b$ and $c$ hadron
  decays, where the undetected neutrinos or muons that loose only little of their
  energy in the calorimeter, carry away part of the energy of the primary parton.
  (c) The multijet structure of $t\bar{t}$ events as compared to dijet events
  used to derive the other corrections. While the jet corrections (i) through (v) are
  flavour independent the $t\bar{t}$ {\it specific corrections} treat $b$ jets
  differently than light quark jets. Four jet types are distinguished: 
  (1) jets used to reconstruct the hadronic $W$ decay, (2) jets assigned to the $b$ 
  quark from the top quark decay without $b$ tag or jets with secondary vertex tag, 
  (3) jets with a soft electron tag, (4) jets with a soft muon tag.
\end{enumerate}
Within the $E_\mathrm{T}$ range from 30 to 90 GeV the flavour independent corrections 
(number i through v) amount to an average correction factor of about 1.45.
The total systematic uncertainty including all corrections varies between
7\% for jets with corrected $E_\mathrm{T}$ of 20~GeV and 3.5\% for jets with 
$E_\mathrm{T}=150\,\mathrm{GeV}$.

\subsubsection{Event-by-event top mass fitting}
For top mass fitting the four-momenta of the particles of the $t\bar{t}$ decay 
chain are fully reconstructed. The four leading jets are assigned to the 
four quarks in the final state of the hard scattering. The mass of light
quark jets is assumed to be 0.5~GeV/$c^2$, the mass of $b$ quarks is set to
$5.0\,\mathrm{GeV}/c^2$. The jet-parton assignment bears some ambiguity.
If none of the jets is tagged as a $b$ jet candidate, there are 12 possible 
jet permutations. Including the two-fold ambiguity of the neutrino $p_z$
reconstruction, there are 24 combinations. With one $b$ tag that number is
reduced to 12, with two $b$ tags there are only four possible assignments.
For each event all of these possible combinations are tested.
A kinematic fit based on a $\chi^2$ criterion is used to find the best combination
and determine the top quark mass on an event-by-event basis. 
The $\chi^2$ expression implements six effective kinematic constraints:
(1,2) the two transverse momentum components of the $t\bar{t}+X$ system must be
zero, (3) the invariant mass of the lepton and neutrino, $M_{\ell\nu}$, 
must be equal to $M_W$,
(4) the invariant mass of the two light quarks, $M_{jj}$, must be equal to $M_W$,
(5,6) the two three-body invariant masses, $M_{\ell\nu j}$ and $M_{jjj}$, must be 
equal to $M_\mathrm{top}$.   
The $\chi^2$ expression is given by
\begin{eqnarray}
  \label{eq:topchi}
  \chi^2 & = & \sum_{\ell,\mathrm{jets}} 
  \frac{(\hat{p_\mathrm{T}}-p_\mathrm{T})^2}{\sigma^2_{p_\mathrm{T}}}+
  \sum_{i=x,y} \frac{(\hat{U_i}-U_i)^2}{\sigma^2_{U_i}}+
  \frac{(M_{\ell\nu}-M_W)^2}{\sigma^2_{M_W}}+ \nonumber \\
  &  & \frac{(M_{jj}-M_W)^2}{\sigma^2_{M_W}}+
  \frac{(M_{\ell\nu j}-M_\mathrm{top})^2}{\sigma^2_{M_\mathrm{top}}}+
  \frac{(M_{jjj}-M_\mathrm{top})^2}{\sigma_{M_\mathrm{top}}^2}\ .
\end{eqnarray}
The first sum in (\ref{eq:topchi}) is over the primary lepton $\ell$ and all jets
with raw $E_\mathrm{T}>8\;\mathrm{GeV}$ and $|\eta|<2.4$. The second sum is over the 
transverse components of the unclustered energy $\bi{U}_T$, which is defined
as the vector sum of the energies in the calorimeter towers after excluding the
primary lepton and all jets with raw $E_\mathrm{T}>8\;\mathrm{GeV}$ and $|\eta|<2.4$.   
The symbols with a hat represent quantities which are free to be altered in the 
fit procedure. The uncertainty on a quantity $X$ is denoted $\sigma_X$ and
occurs in the denominator of the respective $\chi^2$ terms.
$\sigma_{M_W}$ is set to $2.1\;\mathrm{GeV}/c^2$, $\sigma_{M_\mathrm{top}}$
is set to $2.5\;\mathrm{GeV}/c^2$.
For each combination of primary physics objects in an event the $\chi^2$ expression 
(\ref{eq:topchi}) is minimized. 
The combination with the lowest $\chi^2$ is chosen to be the best fit for that
event. Events with their lowest $\chi^2$ above 10 are rejected. The fit yields an estimate
of $M_\mathrm{top}$ for each event.

The mass fit performance is tested with $t\bar{t}$ Monte Carlo events. The best
results are obtained for events where all four leading jets are correctly assigned
to the appropriate quark. A mass resolution of 13~GeV/$c^2$ is reached for these 
events. For events with two secondary vertex tags the correct assignment is 
made in 49\% of the events, while for events without $b$ tag the 
assignment is correct only in 23\% of the cases. If the four leading partons
from the $t\bar{t}$ decay cannot be uniquely matched to the four leading jets
within $\Delta R < 0.4$, the mass resolution deteriorates to an average of
$34\;\mathrm{GeV}/c^2$.
   
\subsubsection{Final selection and backgrounds}
\renewcommand\arraystretch{1.5}
The final cut in the event selection requires that the $\chi^2$ of the best fit 
for an event is below 10. The background is estimated using the same methods as
discussed for the $t\bar{t}$ cross section measurement in 
section~\ref{sec:leptonPlusJets}. In addition, the background is renormalized 
using a maximum likelihood fit to the observed rates of events with secondary
vertex tag and soft lepton tag and their respective expectations.
In the fit the sum of $t\bar{t}$ signal and background events is 
constrained to be equal to the observed number of events. 
The number of expected and observed events in the four top mass subsamples
are given in table~\ref{tab:topMassEvents}.
\begin{table}[!t]
\caption{\label{tab:topMassEvents}Number of expected and observed number of
  events in the four top mass subsamples.}
\begin{center} 
\begin{tabular}{lccc}
 \br
 Data sample & Expected Background & Expected Signal & Observation \\ \mr
 SVX double tag & 0.2 & 6.1 & 5 \\  
 SVX single tag & 2.7 & 14.4 & 15 \\
 Soft Lepton Tag & 5.0 & 4.0 & 14 \\
 No Tag & 32.4 & 11.4 & 42 \\
 \hline 
 Total & 40.3 & 35.9 & 76 \\ \br  
\end{tabular}
\end{center}
\end{table}

\subsubsection{Top mass determination}
\begin{figure}[!t]
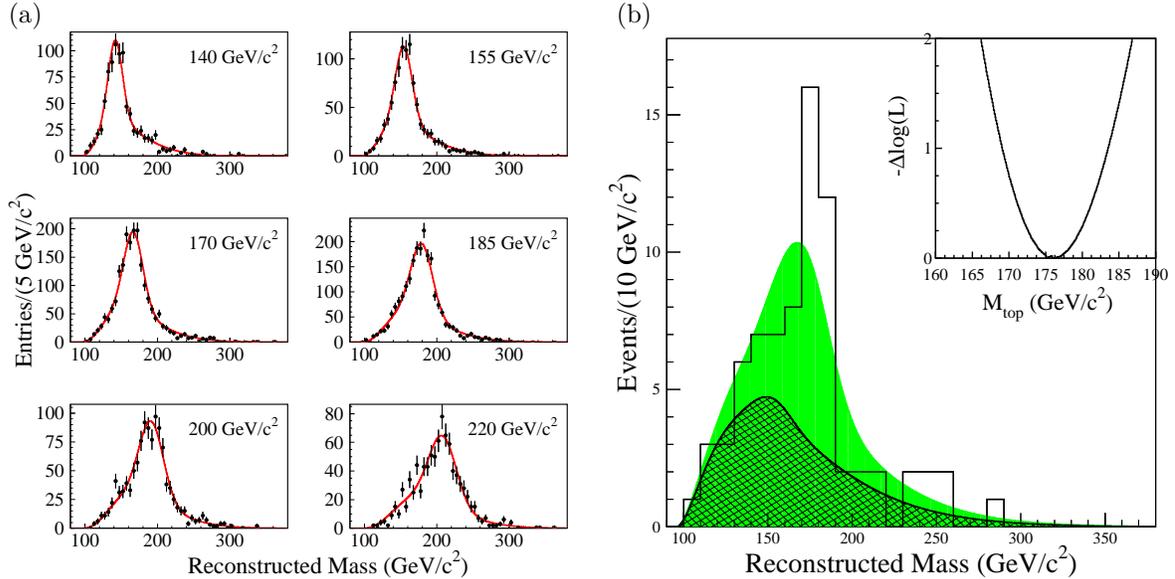

  (a) \hspace*{74mm} (b) \\
  \includegraphics[width=0.48\textwidth]{fig/herwig_templ_6_svx_sing.epsi}
  \hspace*{0.02\textwidth}
  \includegraphics[width=0.48\textwidth]{fig/opt1_bin_mod.epsi}
\caption{\label{fig:topmass} Top quark mass measurement with the template
   method at CDF in Run I~\cite{CDFtopMass2001LeptonPlusJetsRun1}.
   (a) Top mass distributions of $t\bar{t}$ signal 
   Monte Carlo for six different values of $M_\mathrm{top}$ (dots).
   The overlayed curves are the fitted template functions for that particular
   top quark mass.
   (b) The histogram shows the reconstructed top mass distribution for all 
   subsamples combined. The shaded region indicates the fitted 
   signal component, the hatched region the fitted background. The inset
   shows the log likelihood as a function of the top quark mass.}
\end{figure}

In a last step the best estimate of the top quark mass is determined by a 
maximum likelihood fit
to the distribution of reconstructed invariant masses. 
The shape of the reconstructed mass distribution for the $t\bar{t}$ signal
is obtained from {\sc Herwig} Monte Carlo events. 
Several samples for different top quark masses are generated.
The reconstructed mass distributions are referred to as templates,
which provides the name for the entire measurement method. 
The template distributions are parametrized using a function $f_s$ 
that depends on 12 parameters and the top quark mass. The parameters are
determined by a simultaneous $\chi^2$ fit to all template histograms.
For each of the four subsamples a different set of parameters is calculated. 
Figure~\ref{fig:topmass}a shows the top mass template histograms
and the fitted template function for six different values of $M_\mathrm{top}$.
The {\sc Vecbos}
Monte Carlo program is used to create templates for the background.
The background distribution is parametrized by a function $f_b$ with fewer 
parameters and no dependence on $M_\mathrm{top}$.
\par
Using the template functions and the observed mass distribution a likelihood
function is defined. The template parameters and the background fraction are
constrained to their central values within their uncertainties. The only parameter
which is entirely unconstrained is $M_\mathrm{top}$. The log likelihood function 
is minimized with respect to all parameters in a combined fit to all subsamples. 
The result of the minimization for the combined sample is shown in 
figure~\ref{fig:topmass}b. The minimum of the likelihood is reached for a
top quark mass of $M_\mathrm{top}=176.1^{+5.2}_{-5.0}\;\mathrm{GeV}/c^2$.
The given uncertainties are only statistical.
Systematic uncertainties due to various sources are evaluated. The biggest
uncertainty originates from the jet energy measurement 
($4.4\,\mathrm{GeV}/c^2$). The modelling of initial and final state 
radiation causes an uncertainty of $2.6\,\mathrm{GeV}/c^2$. The modelling
of the shape of the background spectrum induces an uncertainty of 
$1.3\,\mathrm{GeV}/c^2$. Smaller sources of uncertainties are:
the $b$ tagging efficiency ($0.4\,\mathrm{GeV}/c^2$), parton distribution
functions ($0.3\,\mathrm{GeV}/c^2$), and Monte Carlo generators 
($0.1\,\mathrm{GeV}/c^2$). The contributions of all effects are added in 
quadrature resulting in a total systematic uncertainty of 
$5.3\,\mathrm{GeV}/c^2$.

\subsection{Matrix element method}
\label{sec:mtopMatrix}
Originally D\O \ also used a template method to measure the top quark mass.
This method used a kinematically fitted mass, similar to CDF, and in addition
discriminants based on a likelihood ratio or a neural 
network~\cite{d0TopMass1998Run1LeptonPlusJets}.
In 2004 D\O \ reanalyzed its top mass data sample with a new technique based on a 
matrix element method, which we will describe below~\cite{d0mtopnature}.

The new measurement is based on the same data set of 91 lepton-plus-jets events
as the previous one. The basic selection cuts are very similar to those used
for the $t\bar{t}$ cross section measurement~\cite{d0ttbarXS1997}.
The data set corresponds to an integrated luminosity of $125\;\mathrm{pb^{-1}}$.
The result of the template analysis is 
$M_\mathrm{top}=173.3\pm5.6\;(\mathrm{stat.})\pm5.5\;(\mathrm{syst.})\;\mathrm{GeV}/c^2$.

The matrix element method is designed to extract more kinematic information
from the events than the previous template methods, and thus yield an 
improved precision. The basic idea of the new method is to exploit the fact
that the differential cross section for $t\bar{t}$ production depends 
sensitively on the top quark mass. The differential cross section can thus
be used to calculate a probability for a certain top mass hypothesis.
While the differential cross section consists of a phase space term
and the matrix element for $t\bar{t}$ production, the matrix element is the
more significant part, which motivates the name of the method.
Matrix element methods have already been used previously to analyse
$t\bar{t}$ dilepton events in CDF~\cite{kondoMatrix} and 
D\O~\cite{dalitzMatrix1992,dalitzMatrix1999,d0mtop99}. 
Since leading order matrix elements are used to calculate the event weights,
only events with exactly four jets are used. This jet cut minimizes the effect
of higher-order corrections and reduces the number of events from 91 to 71.

The production probability $P(x, M_\mathrm{top})$ for a $t\bar{t}$ event
with a measured set of variables $x$ at a certain top quark mass 
$M_\mathrm{top}$ is given as a convolution of the differential cross
section $d\sigma/dy$ with the parton distribution functions $f$ and
the transfer function $W(y,x)$ that maps the measured quantities $x$ into
the quantities $y$ at parton level:
\begin{equation}
 \label{eq:matrixFormel}
 P(x, M_\mathrm{top}) = \frac{1}{\sigma_\mathrm{total}} \;
 \int \rmd y\,\rmd q_1\,\rmd q_2 \; \frac{\rmd\sigma(y,M_\mathrm{top})}{\rmd y}\;
 f(q_1)f(q_2)\;W(y,x) \ .  
\end{equation}
The parton distribution functions $f(q_i)$ are evaluated for the incoming 
partons with momentum fraction $q_i$.
The integral in (\ref{eq:matrixFormel}) is properly normalized by dividing
by the total cross section $\sigma_\mathrm{total}$.
The integration runs over fifteen sharply measured variables, which are directly
assigned to the parton quantities without invoking a transfer function.
These variables are the eight jet angles, the three-momentum of the lepton, and
four equations of energy-momentum conservation. The jet energies are not well
measured and a transfer function 
$W_\mathrm{jets}(E_\mathrm{part}, E_\mathrm{jet})$
is needed to map jet energies $E_\mathrm{jet}$ measured in the detector
to parton level energies $E_\mathrm{part}$. 
The function $W_\mathrm{jets}(E_\mathrm{part}, E_\mathrm{jet})$ is a product of
four functions $F(E_\mathrm{part}^i, E_\mathrm{jet}^i)$, one for each jet in 
the event. The functional form of $F$ is the sum of two Gaussians.
The parameters of $F$ used for $b$ quarks are different from those for 
light quark
jets. After the first integration step there are five integrals left.
One integral runs over the energy of one of the quarks from the hadronic $W$ decay,
the other four are over the masses squared, $M_i^2, $ of the two $W$ bosons
and the two reconstructed top quarks in the event. This is an economical choice to safe
computing time.   
When computing $P(x, M_\mathrm{top})$ all possible 24 permutations of 
jet assignments and the neutrino $p_z$ solution are considered and the 
average is computed.

\begin{figure}[!t]
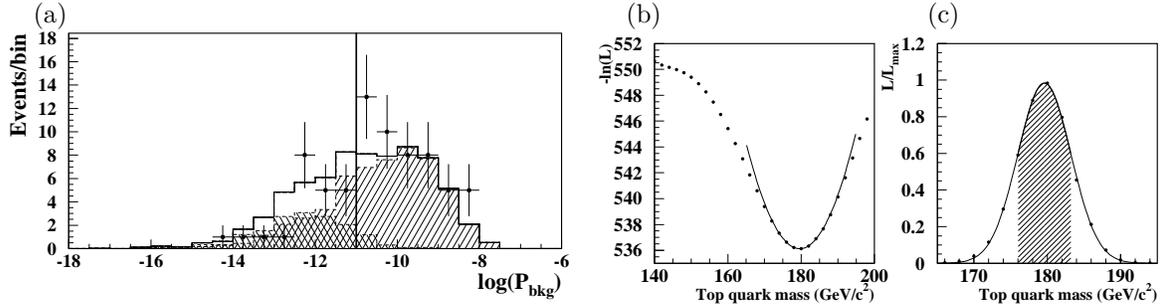

  \hspace*{2mm} (a) \hspace*{72mm} (b) \hspace*{33mm} (c) \\
  \includegraphics[width=0.48\textwidth]{fig/matrixbackg_prob.epsi}
  \hspace*{0.01\textwidth}
  \includegraphics[width=0.48\textwidth]{fig/toponly_bw2.epsi}
\caption{\label{fig:matrix} (a) Distribution of the logarithm of the 
  background probability $P_\mathrm{bkg}$ for 71 candidate events in the 
  D\O \ top mass sample~\cite{d0TopMass1998Run1LeptonPlusJets}. To increase 
  the purity of the sample a cut at
  $\log(P_\mathrm{bkg})<-11$ is performed, as indicated by the vertical line.
  The data are compared to $t\bar{t}$ signal Monte Carlo (left hatched)
  and $W+4$ jets background Monte Carlo events (right hatched).
  (b) Negative of the log likelihood as a function of $M_\mathrm{top}$. 
  (c) The likelihood normalized to its maximum value.
  The curve is fit to a Gaussian. The hatched area corresponds to the
  68.27\% probability interval.}
\end{figure}

Since the candidate sample still contains a considerable amount of background
events, it is useful to calculate a background probability 
$P_\mathrm{bkg}(x)$ based on the $W+4$ jets matrix element from {\sc Vecbos}.
Figure~\ref{fig:matrix}a shows the $P_\mathrm{bkg}$ distribution for the
sample of 71 candidate events and an overlay of $t\bar{t}$ and $W+4$ jets
Monte Carlo events.
To reduce the bias in the top mass measurement due to the background content
in the sample D\O \ employs a cut requiring $\log(P_\mathrm{bkg})<-11$.
After this cut the final top mass sample contains 22 events.
The final likelihood for the measurement of $M_\mathrm{top}$ is given by
\begin{eqnarray}
 \label{eq:mtopLL}
 -\ln L (M_\mathrm{top})&=&-\sum_{i=1}^{N}\ln[c_1P_{t\bar{t}}(x_i,M_\mathrm{top})
 +c_2P_\mathrm{bkg}(x_i)]+ \nonumber \\ 
 & & Nc_1 \int A(x)P_{t\bar{t}}(x,M_\mathrm{top})\,\rmd x 
 + Nc_2\int A(x)P_\mathrm{bkg}(x)\;\rmd x \ .
\end{eqnarray}
The integrals are calculated using Monte Carlo methods. The acceptance function
$A(x)$ is 1.0 or 0.0, depending on whether the event is accepted or rejected
by the analysis criteria. 
The sum runs over all $N=22$ candidate events.
The best value of $M_\mathrm{top}$ and the parameters $c_i$ are defined by 
minimizing $-\ln L(M_\mathrm{top})$, which is shown in figure~\ref{fig:matrix}b.
Figure~\ref{fig:matrix}c shows the likelihood normalized to its maximum value.
The Gaussian fit to the likelihood yields a top quark mass of
$M_\mathrm{top}=(179.6\pm3.6)\;\mathrm{GeV}/c^2$.
Monte Carlo studies show that the extracted top quark mass has to be corrected
by a shift of $\delta M_\mathrm{top}=+0.5\;\mathrm{GeV}/c^2$.
The total systematic uncertainty of the measurement is 3.9~GeV/$c^2$ which
is dominated by the uncertainty of the jet energy scale ($3.3\;\mathrm{GeV}/c^2$).
The final result is 
$M_\mathrm{top}=180.1\pm3.6\;(\mathrm{stat.})\pm3.9\;(\mathrm{syst.})\;\mathrm{GeV}/c^2$,
which is a substantial improvement over the previous template result.
The higher precision is mainly due to two differences in the analyses:
(a) Well measured events contribute more than poorly measured ones, since a probability 
is assigned to each event, (b) all possible permutations of the reconstruction
are included. Thus, the correct solution always contributes. 

\subsection{Top quark mass combination}
\label{sec:MtopCombine}

The best precision on the top quark mass is obtained if the individual measurements
in the different $t\bar{t}$ channels are combined for each experiment.
Further improvement is achieved if the results of the two Tevatron experiments,
CDF and D\O, are combined. In the all hadronic channel CDF uses a data sample
of 136 events with an estimated background of $108\pm 9$ 
events~\cite{CDFttbarhadronic1997Run1}.
A similar template method as described in section~\ref{sec:template} gives
a result of $186\pm10\,(\mathrm{stat.})\pm12\,(\mathrm{syst.})\;\mathrm{GeV}/c^2$.
Since the dilepton topology comprises two unobserved neutrinos a straightforward
full reconstruction of the event is not possible.
To get an estimate of $M_\mathrm{top}$ on an event-by-event basis a weighting
method is used. For each event a weight distribution is calculated as a function
of $M_\mathrm{top}$. The assigned weight depends on the agreement of the sum of the
assumed transverse neutrino momenta with the observed missing transverse energy.
The result in the dilepton channel at CDF is
$M_\mathrm{top}=167.4\pm10.3\;(\mathrm{stat.})\pm4.8\;(\mathrm{syst.})\;\mathrm{GeV}/c^2$~\cite{CDFtopMass1999DileptonRun1}.
The statistical uncertainties of the measurements in the three $t\bar{t}$ 
channels are uncorrelated, since the samples are statistically independent.
However, the systematic uncertainties are correlated. For simplicity, the 
correlation is either assumed to be 100\% or zero. The uncertainties concerning
the jet energy scale, the signal model (modelling of ISR and FSR, PDFs
and $b$ tagging), and the Monte Carlo generators are set to 100\%.
The correlation of uncertainties due to Monte Carlo statistics and the background model are
assumed to be zero. The combination procedure uses a generalized $\chi^2$ method
with full covariance matrix and yields a value of
$M_\mathrm{top}=(176.1\pm6.6)\;\mathrm{GeV}/c^2$~\cite{CDFtopMass2001LeptonPlusJetsRun1}.

In the all hadronic channel D\O \ has measured 
$M_\mathrm{top}=178.5\pm13.7\;(\mathrm{stat.})\pm7.7\,(\mathrm{syst.})\;\mathrm{GeV}/c^2$~\cite{d0allJetsMtop2005}. In the dilepton channel  D\O \ obtains
$M_\mathrm{top}=168.4\pm12.3\,(\mathrm{stat.})\pm3.6\,(\mathrm{syst.})\,\mathrm{GeV}/c^2$~\cite{d0TopMass1998Run1Dilepton}.
The combination of all $t\bar{t}$ topologies is entirely dominated by the 
lepton-plus-jets result described in detail in section~\ref{sec:mtopMatrix} and 
yields $M_\mathrm{top}=(179.0\pm5.1)\,\mathrm{GeV}/c^2$.
Combining all CDF and D\O \ Run I measurements the top quark mass is determined to be
\[ M_\mathrm{top}=\left(178.0\pm4.3\right)\;\mathrm{GeV}/c^2 \ \cite{mtopCombined2004}.\]
\subsection{Preliminary Run II results}
First preliminary Run II measurements of the top quark mass are now available
and prepared for publication.
CDF has been exploring several different techniques
for the measurement of $M_\mathrm{top}$. The best single measurement is obtained
from an extended mass template technique in the lepton-plus-jets channel. 
The analysis uses a data sample corresponding to $318\,\mathrm{pb^{-1}}$.
The measured invariant mass of the hadronic $W$ boson decay is used to reduce
the systematic uncertainty on the jet energy scale.
In a two-dimensional likelihood fit the top quark mass and the jet energy scale 
(JES) are obtained simultaneously. CDF measures 
$M_\mathrm{top}=173.5^{+2.7}_{-2.6}\,(\mathrm{stat.})\pm 2.5\,(\mathrm{JES})\pm 1.7\,(\mathrm{syst.})\,\mathrm{GeV}/c^2$.
\par
D\O \ has presented a preliminary result which is also based on a mass template
method. In $b$ tagged lepton-plus-jets events the top quark mass is found
to be $M_\mathrm{top}=170.6\pm4.2\,(\mathrm{stat.})\pm6.0\,(\mathrm{syst.})\,\mathrm{GeV}/c^2$. The analysis is based on a data sample of $229\,\mathrm{pb^{-1}}$ and uses
69 candidate events~\cite{d0mtopRun2}.

\subsection{Future prospects}
The aim of Run II at the Tevatron is to reach a total uncertainty of 2 to 
3 GeV/$c^2$ on $M_\mathrm{top}$.
At the LHC the top mass will be measured with negligible statistical uncertainty, while
the systematic uncertainty is predicted to be on the order of 1~GeV$/c^2$ 
towards the end of the running period~\cite{topLHC2000}. 
In this precision regime the concrete definition of 
the top quark mass becomes relevant. The experiments measure a kinematic top quark mass
which is approximately equal to the pole mass that appears in the 
perturbative top quark propagator. The ambiguity 
between pole and kinematic mass is on the order of 1 GeV and due to the fact that
at this level the top quark cannot be treated as a free quark anymore. 
At a future linear collider an energy scan in the threshold region for $t\bar{t}$
production will allow a very precise measurement of the $\msbar$ mass 
$\overline{M}_\mathrm{top}(\mu)$. An uncertainty of 20~MeV is envisaged.

\section{Top quark production and decay properties}
\label{sec:decayProperties}
While the $t\bar{t}$ cross section and top mass measurements have established
data samples that are compatible with SM top quark production, it is crucial
to check whether the candidate events are in agreement with other predictions
made for the SM top quark.
In this chapter we discuss the investigation of various production and decay
properties of top quarks:
the helicity of $W$ bosons
from the top quark decay, the measurement of $R_{tb}$, the ratio of branching 
ratios for $t\rightarrow W+b$ and $t\rightarrow W+q$, 
the search for the $t\bar{t}$ tau modes,
the analysis of spin correlations among the $t\bar{t}$ pair,
the measurement of the top quark $p_\mathrm{T}$ spectrum, and the search for
electroweak top quark production.

\subsection{W helicity in top quark decays}
\label{sec:Whelicity}
As discussed in chapter~\ref{sec:topdecay} the $V-A$ structure of the 
electroweak charged current interaction causes the $W$ bosons from the
top quark decay to be polarized. The fraction $\mathcal{F}_0$ 
of longitudinal $W$ bosons (helicity $h_W=0$) is predicted to be
\begin{equation}
\label{eq:Wlong} 
  \mathcal{F}_0 = \frac{M_\mathrm{top}^2/2M_W^2}{1+M_\mathrm{top}^2/2M_W^2}
   = (71 \pm 1)\% \ , 
\end{equation}
where we have used $M_\mathrm{top}=(178.0\pm4.3)\;\mathrm{GeV}/c^2$.
For $W^+$ bosons the remaining 29\% are left-handed ($h_W=-1$),
for $W^-$ bosons 29\% are right-handed ($h_W=+1$). 
In the SM right-handed $W^+$ and left-handed $W^-$ are strongly
suppressed (branching fraction: 0.04\%).
In the following discussion we refer only to the $W^+$,
but imply the $CP$-conjugate statement for the $W^-$.

\subsubsection{The lepton $p_\mathrm{T}$ spectrum}
\label{sec:ptspectrum}
\begin{figure}[!t]
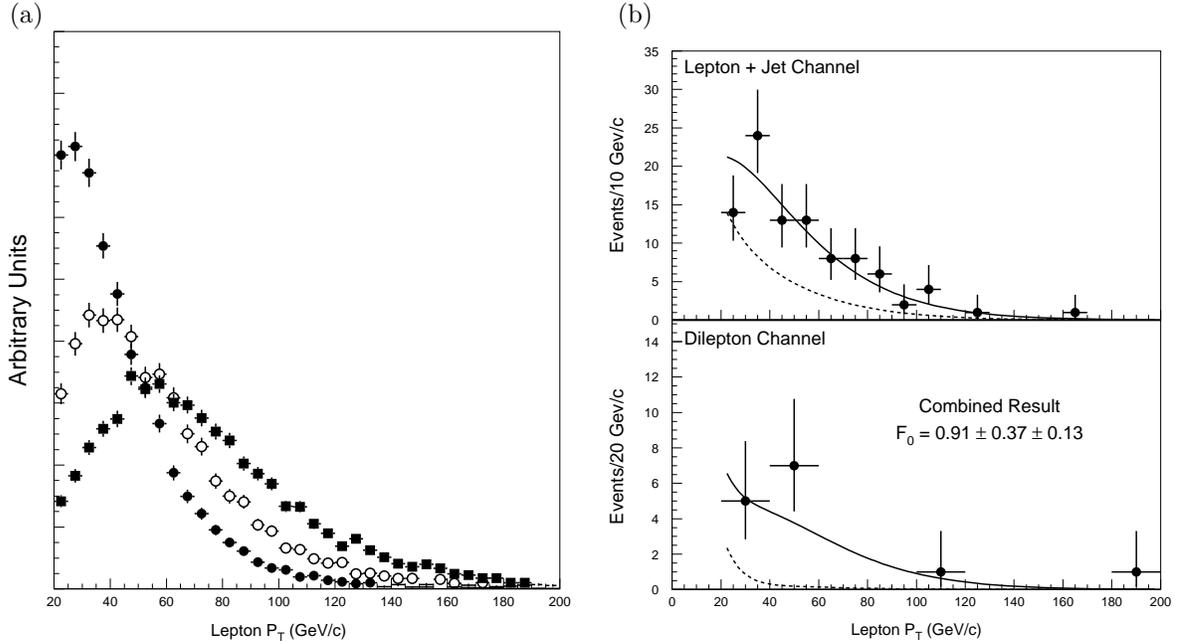

  (a) \hspace*{74mm} (b) \\
  \includegraphics[width=0.48\textwidth]{fig/WHEL_FIG1.epsi}
  \hspace*{0.02\textwidth} 
  \includegraphics[width=0.48\textwidth]{fig/WHEL_FIG2.epsi}
\caption{\label{fig:leptonPtWhelicity} 
  (a) Transverse momentum distributions of leptons from the $W$ decay for the 
  three different $W$ 
  helicities. The solid dots are from $W^+$ bosons with negative helicity
  or $W^-$ bosons with positive helicity. The open circles are from 
  longitudinally polarized $W$ bosons. The closed squares are for leptons 
  from right-handed
  $W^+$ bosons or from left-handed $W^-$ bosons. All three distributions are
  normalized to the same area. The figure is taken from 
  reference~\cite{CDFWhelicityLeptonPtRun1}.
  (b) Lepton $p_\mathrm{T}$ distributions of the 
  CDF Run I $W$ helicity analysis in the lepton-plus-jets and the dilepton
  sample~\cite{CDFWhelicityLeptonPtRun1}. The data (dots with error bars) are
  compared with the result of the combined fit (solid line) and with the 
  background component (dashed line).}
\end{figure}

We consider further the leptonic decay of the $W$ boson into
$e\nu_e$ or $\mu\nu_\mu$.
The $V-A$ structure of the $W$ boson decay causes a strong correlation
between the helicity of the $W$ boson and the lepton momentum.
Qualitatively, this can be understood as follows:
The $\nu_\ell$ from the $W^+$ decay is always left-handed, the 
$\ell^+$ is right-handed. In the case of a left-handed $W^+$ boson angular 
momentum conservation demands therefore that the $\ell^+$ is emitted
in the direction of the $W^+$ spin, that means anti-parallel to the 
$W^+$ momentum. That is why, charged leptons from the decay of 
left-handed $W$ boson are softer than charged leptons from longitudinal
$W$ bosons, which are mainly emitted in the direction transverse to the
$W$ boson momentum. 
The spectrum of leptons from right-handed $W$ bosons would be even 
more harder than  the one from longitudinal ones, 
since they would be emitted preferentially in the direction
of the $W$ momentum. Figure~\ref{fig:leptonPtWhelicity}a shows the 
$p_\mathrm{T}$ distributions of leptons for different $W$ helicities.
The significant differences for the three helicity states are apparent and
exploited by a CDF analysis which uses the dilepton and lepton-plus-jets 
data samples~\cite{CDFWhelicityLeptonPtRun1}. The threshold at 20 GeV/$c^2$
seen in figure~\ref{fig:leptonPtWhelicity}a is due to the event selection.
Out of the standard dilepton sample the helicity analysis uses only
those events that feature leptons of different flavour, one electron and 
one muon. This additional requirement removes the major part of the Drell-Yan
background. Seven dilepton events remain. The lepton-plus-jets sample is 
divided into three subsamples: (1) events with at least one secondary vertex
$b$ tag, (2) events with at least one soft lepton tag, but no secondary
vertex tag, and (3) events with no $b$ tags.
In the first two subsamples at least three jets with 
$E_\mathrm{T}>15\;\mathrm{GeV}$ 
and at least one jet with $E_\mathrm{T}>8\;\mathrm{GeV}$ are required. In the third 
subsample at least four jets with $E_\mathrm{T}>15\;\mathrm{GeV}$ are mandatory.
All jets are counted in the pseudorapidity region $|\eta|<2.4$.

An unbinned maximum likelihood fit to the observed lepton $p_\mathrm{T}$ spectrum
in each subsample is used to determine the fraction of top quark events
that decay to longitudinal $W$ bosons. The likelihood functions for each
subsample are added
and simultaneously minimized. In the fit function the lepton $p_\mathrm{T}$
spectrum for each subsample is parametrized as the product of an exponential
and a polynomial. Figure~\ref{fig:leptonPtWhelicity}b shows the observed lepton
$p_\mathrm{T}$ spectrum for the dilepton sample and the sum of all three 
lepton-plus-jets samples. The figure also shows the fit result (signal + background) and the
background component alone. The likelihood fit yields a result of
$\mathcal{F}_0=0.91\pm0.37\,(\mathrm{stat.})\pm0.13\,(\mathrm{syst.})$.
The largest contribution to the systematic uncertainty is due to the
uncertainty in the top quark mass, $\delta\mathcal{F}_0=0.07$.
The second largest source of uncertainty ($\delta\mathcal{F}_0=0.06$)
is due to the normalization
uncertainty of the non-$W$ background which peaks at low $p_\mathrm{T}$
and thereby mimics the shape of the lepton $p_\mathrm{T}$ distribution 
coming from negative
helicity $W$ bosons. Other sources of systematic uncertainties are
the $b$ tagging efficiency, Monte Carlo statistics, the modelling of 
gluon radiation, and uncertainties in the parton distribution functions.

CDF has applied the same method as described above to measure $\mathcal{F}_0$
in Run II data. The new analysis is based on data samples corresponding to 
$162\,\mathrm{pb}^{-1}$ or $193\,\mathrm{pb}^{-1}$ for the lepton-plus-jets
or the dilepton sample, respectively. 
The result for the combined sample is 
$\mathcal{F}_0=0.31^{+0.37}_{-0.23}\,(\mathrm{stat.})\pm 0.17\,(\mathrm{syst.})$
\cite{Abulencia:2005xf}.

\subsubsection{Measurement of $\mathcal{F}_0$ with the matrix element method}
D\O \ has used the matrix element method that is employed for the
top quark mass measurement, see section~\ref{sec:mtopMatrix}, to determine the 
longitudinal polarization fraction $\mathcal{F}_0$~\cite{Abazov:2004ym}.
The data sample consists of the same 22 events that are analyzed for the
mass measurement.
The $t\bar{t}$ probability density $P_\mathrm{t\bar{t}}(x, \mathcal{F}_0)$
includes contributions with longitudinal helicity ($\mathcal{F}_0$) and
negative helicity ($\mathcal{F}_{-}$), but only the ratio
$\mathcal{F}_0/\mathcal{F}_{-}$ is allowed to vary.
The likelihood maximization yields 
$\mathcal{F}_0=0.56\pm0.31\,(\mathrm{stat.})\pm0.07\,(\mathrm{syst.})$.  
The statistical error also contains a contribution from the top quark
mass uncertainty which is included in the measurement by integrating
over the mass from $M_\mathrm{top}=165\;\mathrm{GeV}/c^2$ to
$M_\mathrm{top}=190\;\mathrm{GeV}/c^2$.
The result for $\mathcal{F}_0$ is in good agreement with the SM prediction.

\subsubsection{The helicity angle $\cos\theta_{\ell W}$}
Another possibility to measure the polarization of $W$ bosons from 
top quark decays is to reconstruct the helicity angle $\theta_{\ell W}$, which
is defined as the angle between the lepton momentum in the $W$ rest 
frame and the $W$ momentum in the top quark rest frame.
Leptons from the decay of longitudinally polarized $W$ bosons have a
symmetric angular distribution of the form 
$1-\cos^2\theta_{\ell W}$. Leptons from left-handed $W$ bosons have an
asymmetric angular distribution of the form $(1-\cos\theta_{\ell W})^2$,
while right-handed $W$ bosons would have a distribution of the form
$(1+\cos\theta_{\ell W})^2$.

The direct reconstruction of $\cos\theta_{\ell W}$ requires information
on the top quark and $W$ rest frames and thus, the reconstruction of the 
respective momenta. Experimentally, the determination of $\bi{p}_\mathrm{top}$ 
and $\bi{p}_W$ is difficult, since it relies on the relatively poor 
measurement of jets and the $\EtMissVec$.
However, there is a very good approximation which relates 
$\cos\theta_{\ell W}$ to the invariant mass of the lepton and the $b$ quark jet
from the top quark decay:
\begin{equation}
  \label{eq:Mlb}
  M_{\ell b}^2 = \frac{1}{2} (M_\mathrm{top}^2 - M_W^2)\;
  (1+\cos\theta_{\ell W}) \ .
\end{equation} 
Using the quantity $M_{\ell b}^2$ instead of $\cos\theta_{\ell W}$ circumvents
the difficulties related to direct top quark reconstruction.

CDF has exploited relation (\ref{eq:Mlb}) to search for an anomalously large
right-handed fraction $\mathcal{F}_+$ of $W$ bosons in Run I 
data~\cite{CDFWHelicityFplusRun1}.
The large top quark mass has led to speculations that the top quark could
play an active role in electroweak symmetry breaking, which would lead to 
anomalous electroweak interactions of the top quark~\cite{peccei1991}.
One possibility are left-right symmetric models that lead to a significant
right-handed fraction of $W$ bosons in top quark decays.
An additional $V+A$ component would lead to a lower left-handed fraction, but leave
the longitudinal fraction $\mathcal{F}_0$ unchanged.
For the analysis CDF uses the same 7 dilepton events as for the lepton 
$p_\mathrm{T}$ analysis mentioned earlier in section~\ref{sec:ptspectrum}. 
Since the tagging of $b$ quark jets is not
employed in the dilepton sample, there are four $M_{\ell b}^2$ combinations
for each dilepton event. In the lepton-plus-jets sample at least one
$b$ tagged jet is required. There are 15 events with exactly one $b$ tag
and 5 events with two tags.
The $M_{\ell b}^2$ distributions are fit to a linear combination of 
template distributions for the three $W$ polarization states. Only the
shape information is used, the normalization is left floating.
The fit maximizes a binned likelihood as a function of $f_{V+A}$,
the right-handed fraction of the $Wtb$ vertex.   
If there are two possible $b$ jets that can be matched to the primary
lepton, the fit is performed to two-dimensional distributions taking
both solutions into account. The fit result is 
$f_{V+A}=-0.21^{+0.42}_{-0.24}\;(\mathrm{stat.})\pm 0.21 (\mathrm{syst.})$
which is an unphysical value, but gives more preference to the SM
$V-A$ interaction rather than a $V+A$ contribution. 
The dominating systematic uncertainty is due to the uncertainty in the 
top quark mass, which contributes $0.19$ to the total systematic
uncertainty of $0.21$.
The result can be converted into a one-sided upper limit
of  $f_{V+A}<0.80$ at the 95\% confidence level (C.L.).   
If one assumes the longitudinal fraction to have the SM value this limit
corresponds to $\mathcal{F}_+ < 0.24$ at the 95\% C.L..
This result can be combined with the measurement obtained from the
lepton $p_\mathrm{T}$ analysis. The correlation of the two results is about 40\%.
The combination yields an improved limit of $\mathcal{F}_+ < 0.18$ at the
95\% C.L., which is inconsistent with a pure $V+A$ theory at a confidence
level equivalent to the probability of a $2.7\;\sigma$ Gaussian statistical
fluctuation. 

In Run II CDF has provided a preliminary measurement of $\mathcal{F}_0$
using the $M_{\ell b}^2$ method. The analyzed data sample consists of
31 lepton-plus-jets events with one secondary vertex $b$ tag.
The resulting value is  
$\mathcal{F}_0=0.99^{+0.29}_{-0.35}\,(\mathrm{stat.})\pm0.19\,(\mathrm{syst.})$~\cite{Abulencia:2005xf}.

\subsection{Measurement of $R_{tb}$}
In the SM the top quark is predicted to decay to a $W$ boson and a $b$ quark
with a branching fraction of nearly 100\%. This is a consequence of the
CKM matrix element $|V_{tb}|$ being close to unity. Our knowledge on
$|V_{tb}|$ comes primarily from measurements of $b$ meson decays
using the unitarity condition of the CKM matrix 
($\mathbf{V V^\dag} = \mathbf{V^\dag V} = \mathbf{1}$). 
This indirect method yields
$0.9990 < |V_{tb}| < 0.9992$ with high precision~\cite{PDG2004}. 
However, if a fourth
generation of quarks was present, unitarity of the CKM matrix could be 
violated. Therefore, it is desirable to make a direct measurement of
the top quark branching fraction to $Wb$.
 
In Tevatron Run I CDF has measured the ratio 
$R_{tb} = \mathrm{BF}(t\rightarrow Wb)/\mathrm{BF}(t\rightarrow Wq)$, 
where $q$ can be any down-type quark~\cite{CDFBRtWbRun1}.
The analysis uses 9 events from the dilepton sample and 163 events from the 
lepton-plus-jets sample.
The ratio $R_{tb}$ is measured from the data by comparing the observed number
of events with secondary vertex $b$ tags or soft lepton $b$ tags to the number
of expected events based on the kinematic acceptances, tagging efficiencies and
background estimates.
Four event categories are considered:
(1) events with no $b$ tags, (2) events with one or more soft lepton tags,
(3) events with exactly one secondary vertex tag, and (4)
events with two secondary vertex tags. 
In the dilepton sample only secondary vertex $b$ tagging is used.
A maximum likelihood fit yields 
$R_{tb} = 0.94^{+0.26}_{-0.21}\,(\mathrm{stat.})\,^{+0.17}_{-0.12}\,(\mathrm{syst.})$,
consistent with the SM prediction. The result can also be expressed as a lower
limit: $R_{tb} > 0.56$ at the 95\% C.L..
\par
The CKM matrix element $|V_{tb}|$ is related to $R_{tb}$,
although in a model-dependent way. Under the assumption that the 
top quark decays only to $Wq$ final states and using
three generation unitarity one finds $R_{tb}=|V_{tb}|^2$.
As a result, CDF finds $|V_{tb}|=0.97^{+0.16}_{-0.12}$ or
$|V_{tb}|>0.75$ at the 95\% C.L.~\cite{CDFBRtWbRun1}.

In Run II CDF has obtained a preliminary result of this analysis using only the
secondary vertex $b$ tag algorithm and improved
the lower limit to $R_{tb} > 0.61$ at the 95\% C.L.~\cite{Acosta:2005hr}.
D\O \ has performed a similar analysis in the lepton-plus-jets sample 
using the secondary vertex and the impact
parameter $b$ tag method. The preliminary results are
$R_{tb}=0.70^{+0.27}_{-0.24}\,(\mathrm{stat.})\,^{+0.11}_{-0.10}\,(\mathrm{syst.})$
for the secondary vertex tagged sample and 
$R_{tb}=0.65^{+0.34}_{-0.30}\,(\mathrm{stat.})\,^{+0.17}_{-0.12}\,(\mathrm{syst.})$
for events with an impact parameter tag~\cite{d0Whelicity2005}.

\subsection{Search for $\mu \tau$ and $e\tau$ top decays in $t\tbar$ events}
\label{sec:taumodes}
Up to now the searches for the $t\bar{t}$ tau decay modes have not produced
enough evidence to claim the observation of this channel. We discuss in this paragraph
the latest search by CDF using Run II data corresponding to an integrated
luminosity of $194\,\mathrm{pb^{-1}}$~\cite{Abulencia:2005et}. 
The search is carried out in the $t\bar{t}$ 
tau dilepton channel where one $W$ boson decays into $e\nu_e$ or $\mu\nu_\mu$
and the second $W$ boson decays into $\tau\nu_\tau$ with a subsequent
hadronic decay of the tau lepton. The branching fraction of this $t\bar{t}$ mode
is about 5\%, the same size as the classical dilepton channel 
($ee$, $e\mu$, $\mu\mu$).
However, several factors lead to a reduced acceptance for the tau mode:
the hadronic branching ratio of the tau is 64\%, the kinematic acceptance is reduced
due to the undetected neutrino, and the tau selection is less efficient than
the electron or muon selection. In total, the event detection efficiency for the
tau modes is about five times smaller than the one for the standard dilepton 
analysis. This fact mainly explains why the tau mode so far evaded observation.

The data set for the CDF analysis was selected by a trigger on high-$p_\mathrm{T}$
electrons and muons. It is the same data set used for the dilepton cross section 
measurement discussed in section~\ref{sec:dilepton}. On analysis level 
electrons and muons are identified using standard cuts as mentioned in 
the description of the dilepton analysis. Hadronic tau decays have a distinct
signature of narrow isolated jets with low charged track multiplicity.
A tau candidate is defined by two components:
(1) A calorimeter cluster containing a seed tower with 
$E_\mathrm{T} > 6\,\mathrm{GeV}$ and a maximum of five adjacent towers
with $E_\mathrm{T}>1\,\mathrm{GeV}$. 
(2) At least one track with $p_\mathrm{T}>4.5\,\mathrm{GeV}/c$
is required to point to the tau calorimeter cluster.
Further cuts are employed to reduce background. A cone is defined around the 
seed track using a variable radius of
$\theta_\mathrm{cone} = \mathrm{min}\{0.17, \; (5\,\mathrm{GeV})/E_\mathrm{cluster}\}\,\mathrm{rad}$.
Within this cone there must be exactly one or three tracks with 
$p_\mathrm{T}>1\,\mathrm{GeV}/c$
including the seed track. In the case of three tracks, the charges must not all be
positive or negative. Within an isolation annulus extending from the outer edge of
the tau cone to $30^\circ$ no tracks or $\pi^0$ candidates are allowed.
The $\pi^0$ candidates are identified in the calorimeter by clusters of energy
in the shower maximum detector. 
The transverse momentum of the tau is estimated by the sum of the track 
$p_\mathrm{T}$ plus
the $E_\mathrm{T}/c$ of $\pi^0$ candidates identified within the tau cone. 
Additional requirements on the tau candidate are 
$p_\mathrm{T}>15\,\mathrm{GeV}/c$ 
and $m_\mathrm{candidate}<1.8\,\mathrm{GeV}/c^2$. 
The calorimeter towers within the isolation annulus
are required to have $E_\mathrm{T}$ less than 6\% of the tau candidate 
$E_\mathrm{T}$.
Electrons or muons that fake tau candidates are removed by asking that
the energy in the hadron calorimeter divided by the sum of the tau track momenta
is above 0.15, and that the $E_\mathrm{T}$ of the tau calorimeter cluster divided by the
tau seed track is above 0.5.
Additional requirements on candidate events are: $\EtMiss>20\,\mathrm{GeV}$, and
at least two jets within $|\eta|<2.0$. The first jet has to have 
$E_\mathrm{T}>25\,\mathrm{GeV}$,
the second jet $E_\mathrm{T}>15\,\mathrm{GeV}$. The scalar sum of the electron or muon
$p_\mathrm{T}$, the tau $p_\mathrm{T}$, the $\EtMiss$, and 
the $E_\mathrm{T}$ of the jets is defined as 
$H_\mathrm{T}$ and must exceed 205~GeV. 
Including all branching ratios, kinematic cuts and efficiencies the event detection 
efficiency for $t\bar{t}$ events is found to be 
$\epsilon_\mathrm{evt}=(0.080\pm0.015)\%$.  

The number of expected $t\bar{t}$ events is $1.00\pm0.17$. The estimated background
is $1.29\pm0.25$ events. The dominant contributions to the background are
$W+\mathrm{jets}$ events where a jet fakes a tau (58\%), 
$Z^0\rightarrow\tau^+\tau^- + \mathrm{jets}$ (19\%), $WW$ production (11\%),
and $Z^0\rightarrow e^+e^- + \mathrm{jets}$ where an electron fakes a tau (6\%).
In CDF II data two events are observed, compatible with the SM expectation.
The significance of the result is not high enough to claim the observation of 
the $t\bar{t}$ tau mode. The result can be converted into an upper limit on an 
anomalous enhancement of the $t\rightarrow \tau\nu_\tau q$ rate. 
An enhancement of more than a factor of
5.2 is excluded at the 95\% confidence level. The Run II analysis has improved the
expected signal to background ratio over the previous Run I $t\bar{t}$ tau 
analysis of CDF~\cite{CDFttbarTauRun1} from 0.5 to 0.78. 
With more data to come in Run II this improvement brings the observation 
of the $t\bar{t}$ tau mode within reach.
In the Run I analysis 4 candidate events
are observed over a background of 2 events and a signal expectation of 1 event.

\subsection{Spin correlations in $t\overline{t}$ events}
As discussed in detail in section~\ref{sec:topdecay}, top quarks decay as
quasi-free quarks due to their short lifetime. Thus, the spin information of
top quarks is transmitted to the decay products and is accessible experimentally.
In this way top quarks are a unique laboratory to study spin aspects of
heavy quark production, contrary to $b$ quarks where the hadronization to 
$b$ hadrons
removes the initial spin information.
At the Tevatron $t\bar{t}$ pairs are produced unpolarized.
However, the spins of the top and antitop quark are expected to have a 
strong correlation and point along the same axis in the $t\bar{t}$ rest frame
on an event-by-event basis.

An optimal spin quantization axis can be constructed using the velocity $\beta^*$
and the scattering angle $\theta^*$ of the top quark with respect to the
centre-of-mass frame of the incoming partons. The quantization axis forms
the angle $\psi$ with respect to the $p\bar{p}$ beam 
axis~\cite{mahlonParke1997}:
\begin{equation}
  \tan \psi = \frac{\beta^{*2}\sin\theta^*\cos\theta^*}{1-\beta^{*2}\sin^2\theta^*}
  \ .
\end{equation}
For the $q\bar{q}\rightarrow t\bar{t}$ process the spins of the top and the 
antitop quark are fully aligned along the same direction in this spin basis. 
If the gluon fusion process $gg\rightarrow t\bar{t}$ is included the 
correlation is reduced.
In the limit where the $t\bar{t}$ pair is produced at
rest ($\beta^*=0$) the spins are pointing along the beam axis.
The spin correlation of the $t\bar{t}$ pair is transmitted to the 
top quark decay products. Experimentally, it is particularly advantageous
to investigate correlations of the primary leptons in $t\bar{t}$ dilepton
events. The observables are the angles $\theta_{-}$ and $\theta_{+}$ between the 
momenta of the negatively or positively charged leptons in the rest frame
of their parent top quark and the spin quantization axis.
Using these angles the spin correlation can be expressed by the
differential cross section~\cite{mahlonParke1996}
\begin{equation}
  \frac{1}{\sigma_\mathrm{total}} \cdot 
  \frac{\rmd^2\sigma}{\rmd(\cos\theta_{+})\rmd(\cos\theta_{-})} = 
  \frac{1+\kappa \cos\theta_+ \, \cos\theta_{-}}{4} \ ,
\end{equation}
where $\kappa$ is the correlation coefficient which is predicted
to be $\kappa=0.88$ at the Tevatron (Run I).
The coefficient $\kappa$ can vary between $-1$, fully negative correlation,
and $\kappa=+1$ for a fully positive correlation.

The D\O \ collaboration has searched for evidence of $t\bar{t}$ spin
correlations in Run I data using dilepton events~\cite{d0SpinCorrRun1}.
Since the angles $\theta_{-}$ and $\theta_{+}$ are measured
in the top quark rest frame, the top quark has to be fully reconstructed.
In dilepton events this cannot be achieved unambiguously, since the 
two neutrinos are not reconstructed. Up to four different solutions per event
have to be considered. A weight is assigned to each solution 
based on how well the sum of the transverse momenta of the two neutrinos in
the solution agrees with the measured $\EtMiss$.
For each solution the leptons are boosted into the respective top quark
rest frame and $\cos\theta_{+}$ or $\cos\theta_{-}$ are computed.
\par
To deduce $\kappa$ the two-dimensional phase space of
$(\cos\theta_{+},\,\cos\theta_{-})$ is used in a binned likelihood analysis.
The bins are filled with the event weights obtained from the event fitter.
Since an event populates each bin with fractional probability, the
likelihood fit does not use a simple Poisson likelihood. The correlation between
different bins is obtained from Monte Carlo events and the bin contents 
of the $(\cos\theta_{+},\,\cos\theta_{-})$ space are rotated such, that they are
uncorrelated. The new variables are used to construct the likelihood.
D\O \ uses six dilepton data events, three $e\mu$, two $ee$ and one $\mu\mu$
event. The fit result from data is used to compute a lower limit of
$\kappa > -0.25$ at the 68\% confidence level, which is in agreement with the
SM expectation of $\kappa=0.88$.

\subsection{The top quark $p_\mathrm{T}$ spectrum}
\begin{figure}[!t]
  (a) \hspace*{73mm} (b) \\
  \includegraphics[width=0.49\textwidth]{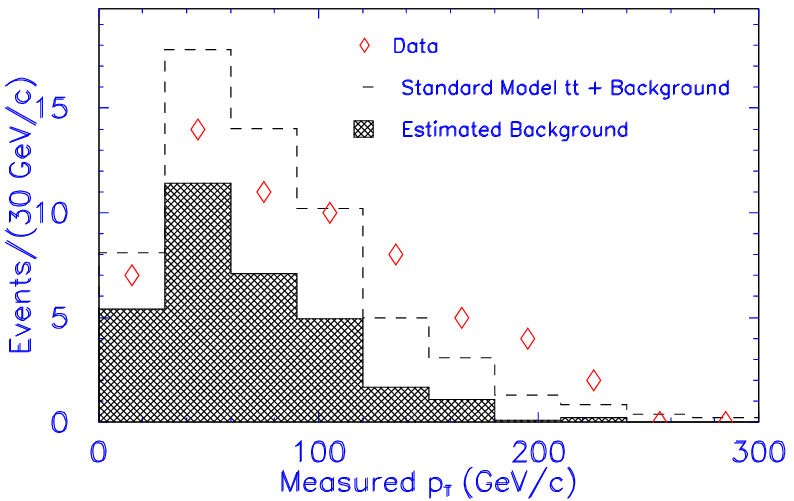}
  \hspace*{0.02\textwidth}
  \includegraphics[width=0.47\textwidth]{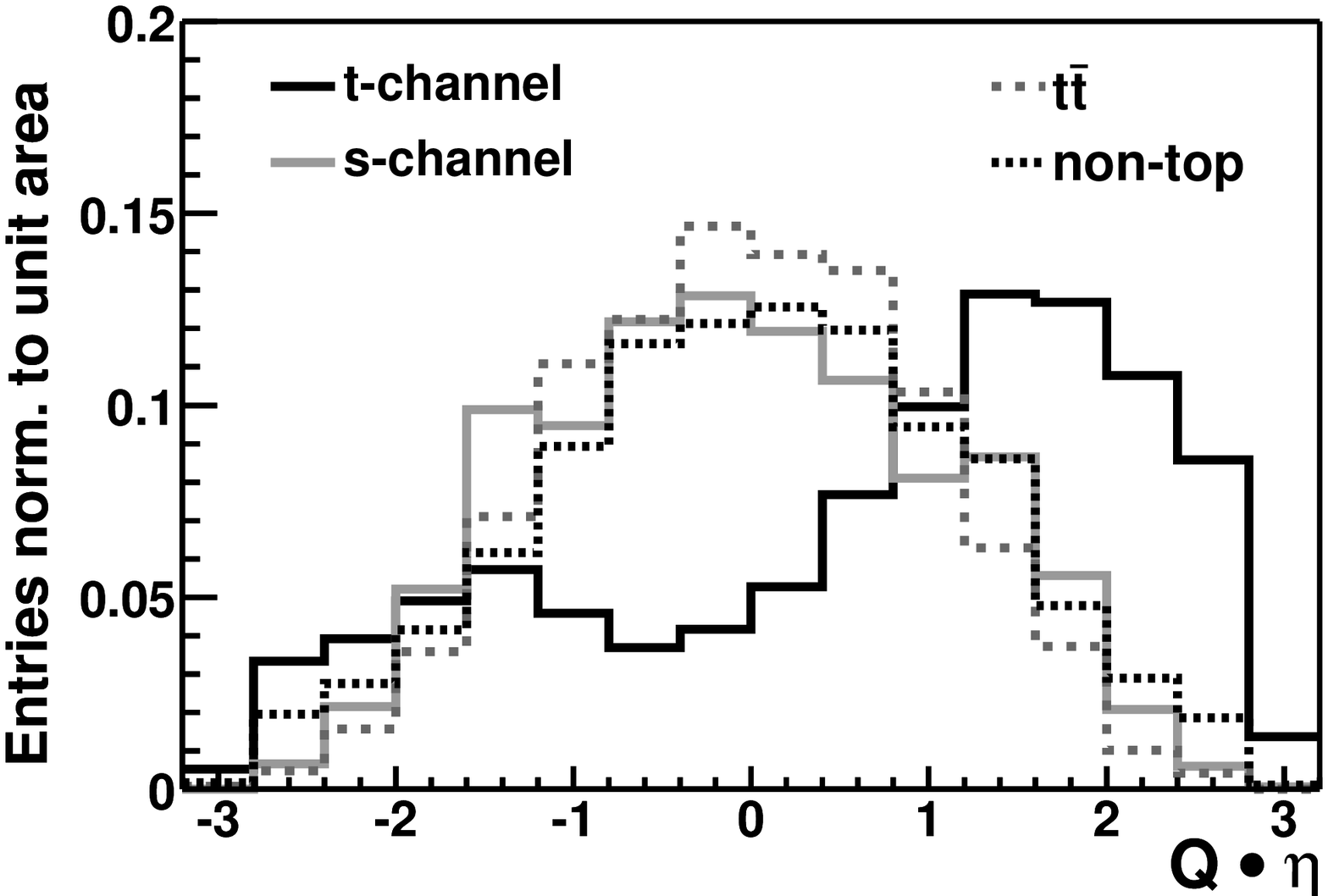}
\caption{\label{fig:topptSingleTop} 
  (a) Transverse momentum distribution of hadronically decaying top quarks
  from 61 candidate events of reference~\cite{CDFtopPtRun1}. 
  The hatched distribution shows the
  estimated background, the dashed distribution the SM prediction of
  background plus $t\bar{t}$ production.
  (b) $Q\cdot\eta$ distributions for 
  Monte Carlo templates normalized to unit area. The $Q\cdot\eta$ distribution
  is used in the CDF Run II single top analysis to distinguish between
  $t$-channel and $s$-channel single top events~\cite{CDF2005singleTopRun2}.}
\end{figure}

The CDF collaboration has further investigated the kinematics of $t\bar{t}$
events by measuring the top quark $p_\mathrm{T}$ spectrum~\cite{CDFtopPtRun1}.
This analysis is partially motivated by exotic models that predict 
alternative production mechanisms for top quarks at the Tevatron,
in particular an enhancement of top quarks with transverse momentum
above 200~GeV/$c$. The Run I lepton-plus-jets data set with secondary vertex $b$ 
tag or soft lepton tag are used in the analysis. The events are subjected to
a kinematic fit similar to the top quark mass analysis described in 
section~\ref{sec:template}. The invariant mass reconstructed from the top quark 
decay products is constrained to $175\,\mathrm{GeV}/c^2$.
Events with $\chi^2>10$ from the fit are rejected. The final data sample contains
61 candidate events. Figure~\ref{fig:topptSingleTop}a shows the 
$p_\mathrm{T}$ distribution of the hadronically decaying top quark. 
The differences in the spectrum do not constitute a significant deviation from
the SM prediction.
A maximum likelihood technique is used to estimate the fraction $R_4$ of
top quarks in the high transverse momentum region of
$225 < p_\mathrm{T}<425\,\mathrm{GeV}/c$. The measured fraction is 
$R_4=0.0^{+0.04}_{-0.0}$,
while the SM expectation is $R_4=0.025$. The data result can be converted into an
upper limit of $R_4<0.16$ at the 95\% confidence level. 

\subsection{Electroweak top quark production}
\label{sec:singleTop}
While $t\bar{t}$ pair production via the strong interaction is the dominant source 
of top quarks at the Tevatron (and also at the LHC), top quarks can also be 
produced as single quarks via the electroweak interaction. 
The theoretical aspects of single 
top production are discussed in section~\ref{sec:singleTopTheo} of this review. 
The relevant modes are $t$-channel and $s$-channel production,
see figure~\ref{fig:singleTop}.
Several analyses have searched for electroweak top quark production in 
Tevatron Run I 
data~\cite{d0SingleTopCutsRun1,d0SingleTopNNRun1,CDFsingleTopRun1,CDFsingleTopANNRun1}.
The best upper limits obtained on the production cross sections in Run I are 
$13\,\mathrm{pb}$ at the 95\% C.L. for the $t$-channel~\cite{CDFsingleTopRun1} 
and  
$17\,\mathrm{pb}$ at the 95\% C.L. for the $s$-channel~\cite{d0SingleTopNNRun1}.
\par
The CDF collaboration has published the first search for single top production
in Run II~\cite{CDF2005singleTopRun2}. In the following paragraphs we will briefly
review this analysis which uses a data sample corresponding to an integrated 
luminosity of $(162\pm10)\,\mathrm{pb^{-1}}$. The event selection is essentially
the same as the one used for the $t\bar{t}$ cross section measurement in
lepton-plus-jets events with secondary vertex $b$ tagging, see 
section~\ref{sec:ttbarSecVtx}. The jet definition differs in the pseudorapidity
range. While the $t\bar{t}$ cross section measurement counts jets up to 
$|\eta|<2.0$, the range is enlarged in the single top analysis to $|\eta|<2.8$.
This change is motivated by an increase in acceptance for the $t$-channel
process in the $W+2\,\mathrm{jets}$ sample by roughly 30\%.
Exactly two jets are required. At least one of these jets must be identified
as a $b$ quark jet by a secondary vertex tag.
To optimize the sensitivity, CDF applies 
a cut on the invariant mass $M_{\ell\nu b}$ of the charged lepton,
the neutrino and the $b$ tagged jet: 
$140\;\mathrm{GeV}/c^2 \leq M_{\ell\nu b} \leq 210\;\mathrm{GeV}/c^2$.
The transverse momentum of the neutrino is set equal to 
the missing transverse energy vector $\EtMissVec$; 
$p_z(\nu)$ is obtained up to a two-fold ambiguity from the 
constraint $M_{\ell\nu}=M_W$.
From the two solutions the one with the lower $|p_z(\nu)|$ is chosen.
If the $p_z(\nu)$ solution has non-zero imaginary part as a consequence
of resolution effects in measuring jet energies, 
only the real part of $p_z(\nu)$ is used.
Two analyses are performed on this data sample:
(1) a separate search, which measures the rates for the two single top
processes, $t$-channel and $s$-channel, individually, and
(2) a combined search where $t$- plus $s$-channel are treated as one
single top signal. 
For the separate search, the data sample is subdivided into events with 
exactly one $b$ tagged jet or exactly two $b$ tagged jets. 
For the 1-tag sample, at least one jet is required to have 
$E_\mathrm{T}\geq 30\;\mathrm{GeV}$.

The event detection efficiency for the signal is determined
from events generated by the matrix element event generator
{\sc MadEvent}~\cite{madevent1994,madevent2003}, followed by parton showering with
{\sc Pythia}~\cite{pythia}
and a full CDF II detector simulation.
{\sc MadEvent} features the correct spin polarization of the top quark and its
decay products.
For $t$-channel single top production two samples are generated,
one $b+q\rightarrow t+q^\prime$ and one $g+q\rightarrow t+\bar{b}+q^\prime$
which are merged together to reproduce the $p_\mathrm{T}$ spectrum of the 
$\bar{b}$ as expected from NLO differential cross section calculations.
This is an improved model compared to the {\sc Pythia} modelling used in 
the Run I analyses.
The event detection efficiency $\epsilon_\mathrm{evt}$ includes the 
kinematic and fiducial acceptance, branching ratios, 
lepton and $b$ jet identification as well as trigger efficiencies.
In the 1-tag sample one finds $\epsilon_\mathrm{evt}=(0.86\pm0.07)\%$ for 
the $t$-channel and  $\epsilon_\mathrm{evt}=(0.78\pm0.06)\%$ for the
$s$-channel. 

Two background components are distinguished:
$t\bar{t}$ and nontop background.
The nontop background is estimated using the same method as used
for the $t\bar{t}$ cross section analysis, see section~\ref{sec:ttbarSecVtx}.
The primary source (62\%) of the nontop background 
is the $W$+heavy flavour processes
$\bar{q}q^\prime\rightarrow Wg$ with $g\rightarrow b\bar{b}$ or 
$g\rightarrow c\bar{c}$, and $gq\rightarrow Wc$.
Additional sources are ``mistags'' (25\%), in which a light quark 
jet is erroneously identified as heavy flavour, 
``non-$W$''(10\%), e.g. direct $b\bar{b}$ production, and
diboson ($WW$, $WZ$, $ZZ$) production (3\%).
The non-$W$ and mistag fractions are estimated using CDF II data. 
The $W$+heavy flavour rates are extracted from {\sc Alpgen}~\cite{alpgen} 
Monte Carlo events normalized to data. 
The diboson rates are estimated from {\sc Pythia} events normalized 
to theory predictions~\cite{dibosonxsec}.
After all selection cuts CDF observes
33 events in the 1-tag sample, 6 events in the 2-tag sample, and
42 events for the combined search.
Within the uncertainties, the observations are in good agreement with 
predictions.

\begin{figure}[!t]
  (a) \hspace*{75mm} (b) \\
  \includegraphics[width=0.51\textwidth]{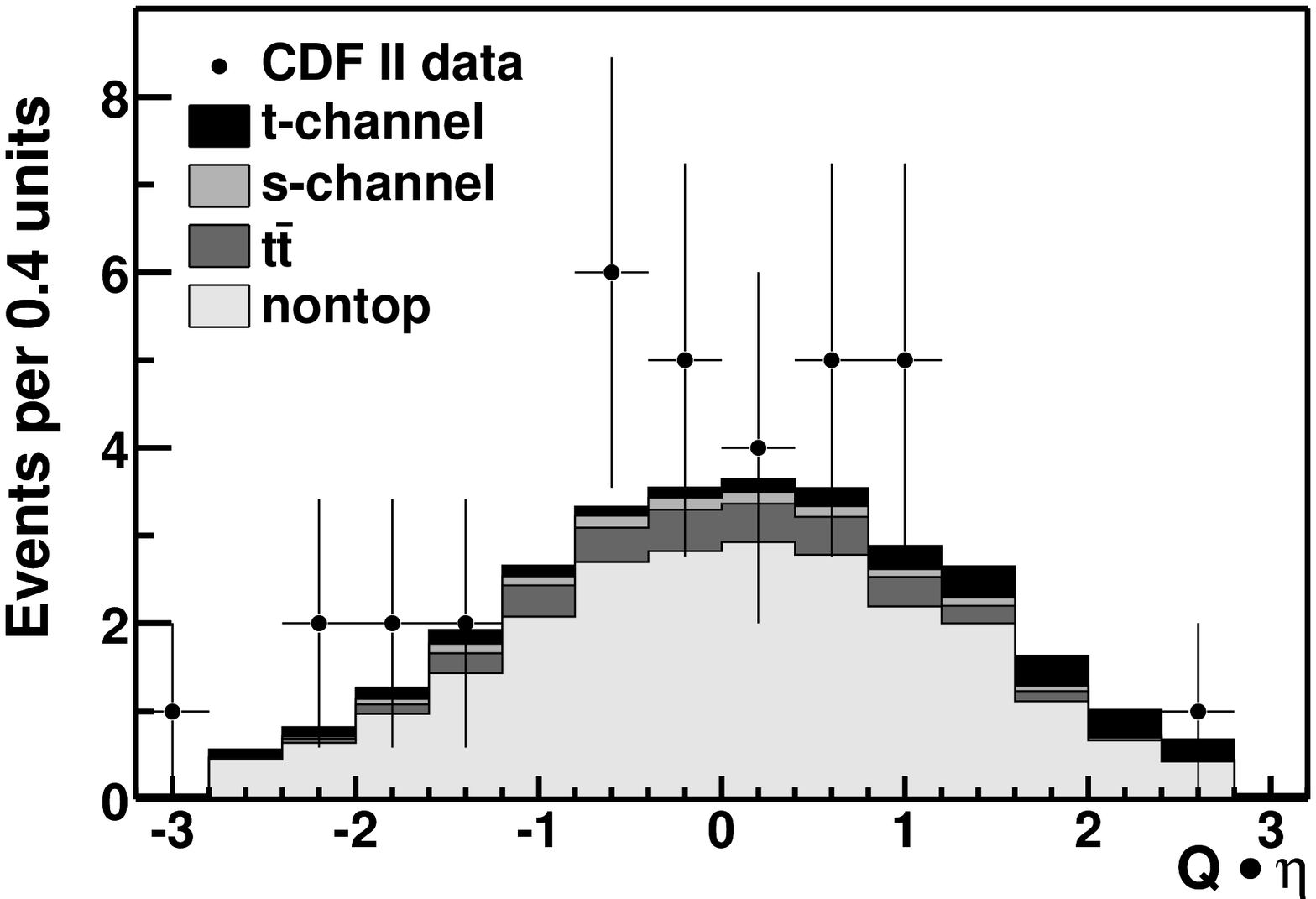}
  \hspace*{0.02\textwidth}
  \includegraphics[width=0.45\textwidth]{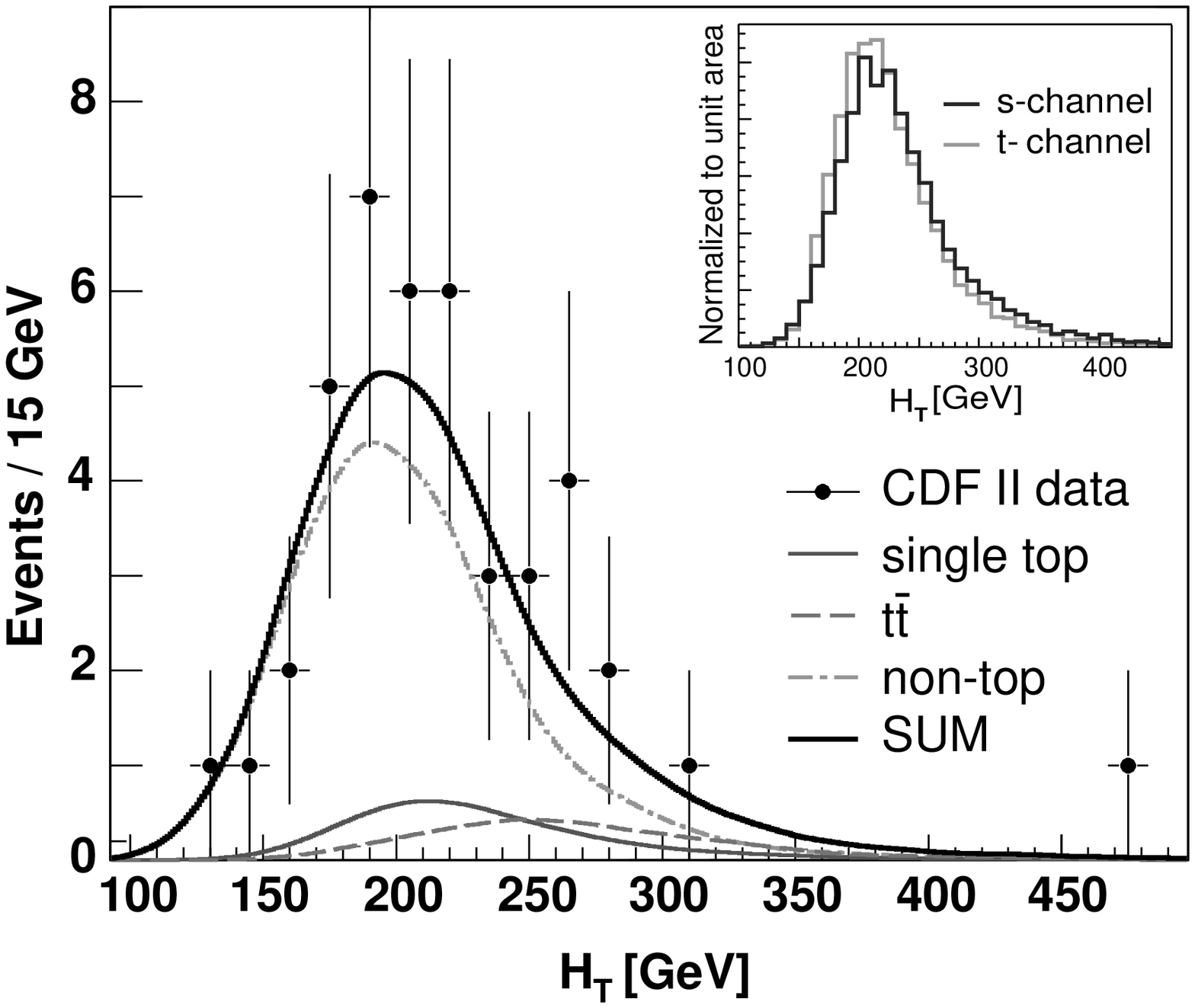}
\caption{\label{fig:singleTopCDF}
   Search for electroweak top quark production at CDF in 
   Run II~\cite{CDF2005singleTopRun2}. 
   (a) $Q\cdot\eta$ distribution for CDF data in the 
   1-tag sample (33 events) versus stacked Monte Carlo templates normalized to 
   the SM prediction.
   (b) $H_\mathrm{T}$ distribution for data (42 events) 
   in the combined single top search, 
   compared with smoothed Monte Carlo predictions for signal and background.
   The inset shows the $H_\mathrm{T}$ distributions for $t$-channel and 
   $s$-channel single top
   production to demonstrate that these distributions are very similar and
   may be combined.}
\end{figure}

To extract the signal content in data, a maximum likelihood technique
is employed. 
Separation of $t$- and $s$-channel events is achieved by using the 
$Q\cdot\eta$ distribution which exhibits a distinct asymmetry
for $t$-channel events, see figure~\ref{fig:topptSingleTop}b. 
$Q$ is the charge of the primary lepton and $\eta$ is the pseudorapidity 
of the untagged jet.
Figure~\ref{fig:singleTopCDF}a shows CDF data versus stacked Monte Carlo 
templates weighted by the expected number of events in the 1-tag sample.
The separate search defines a joint likelihood function for the
$Q\cdot\eta$ distribution in the 1-tag sample and for the number of events
in the 2-tag sample.
The background rates are constrained to their SM prediction by Gaussian 
priors. The systematic uncertainties are taken into account in the
likelihood definition.
Systematic shifts in the acceptance and in the shape of the 
$Q\cdot\eta$ template histograms, and their full correlation are considered.
The largest uncertainties on the acceptance are due to the 
uncertainty on the $b$ tagging efficiency (7\%),
the luminosity (6\%), the top quark mass (4\%), and the jet energy scale (4\%).
The likelihood procedure yields a most probable value of 
$0.0^{+4.7}_{-0.0}\,\mathrm{pb}$ for the $t$-channel cross section,
and  $(4.6\pm3.8)\,\mathrm{pb}$ for the $s$-channel cross section.
These results are translated into upper limits on the cross sections
which exclude an anomalous enhancement of single top quark production.
The upper limit for the $t$-channel is found to be 10.1~pb at the 95\% C.L.,
the upper limit for the $s$-channel is 13.6~pb at the 95\% C.L..

To measure the combined $t$-channel plus $s$-channel signal in data,
a kinematic variable is used whose distribution is very similar for the
two single top processes (see inset of figure~\ref{fig:singleTopCDF}b), 
but is different for background processes:
$H_\mathrm{T}$, which is the scalar sum  of $\EtMiss\;$ and the transverse 
energies of the lepton and all jets in the event.
In figure~\ref{fig:singleTopCDF}b the $H_\mathrm{T}$ distribution observed 
in data is compared with the SM prediction.
A likelihood function similar to that of the separate search
yields a most probable value of $7.7^{+5.1}_{-4.9}\,\mathrm{pb}$
for the single top cross section. The resulting upper limit
is 17.8~pb at the 95\% C.L.. 
The quoted result (combined search) excludes an anomalous enhancement 
of single top production which is more than 6.1 times larger than the 
SM production.
However, separate measurements of the $t$-channel and $s$-channel are important 
because the two processes are sensitive to different 
new physics contributions, see for example 
references~\cite{tait2001,Han_topFCNC,Hosch_topCollider,simmons1997,rizzo1996,malkawi95}.

D\O \ has presented results on a single top search using neural 
networks~\cite{Abazov:2005zz}. With a data sample corresponding to an integrated luminosity 
of $230\,\mathrm{pb^{-1}}$ D\O \ can set an upper limit
of 5.0 pb at the 95\% C.L. on the $t$-channel cross section and
6.4~pb at the 95\% C.L. on the $s$-channel cross section.
With more data at the Tevatron to come, a first observation
of electroweak top quark production is in reach within the next
years.
Preliminary studies indicate that data 
corresponding to about $1.5\,\mathrm{fb^{-1}}$ will allow to establish evidence
for electroweak top quark production at a probability level excluding a
background fluctuation that is equivalent to a $3\,\sigma$ deviation in a 
Gaussian distribution.  
Once electroweak top quark production is observed the measurement of the
production cross section will allow to directly extract the CKM matrix element
$|V_{tb}|$.

\section{Search for anomalous couplings}
\label{sec:anotop}

Mainly because of its large mass the top quark has fostered speculations
that it offers a unique window to search for physics beyond the SM.
One possibility to test this hypothesis is to measure top quark properties
and check whether the observation is in agreement with the SM
prediction. This approach is based on the measurements that are discussed
in previous chapters, involving properties such as the $t\bar{t}$ cross 
section, the top mass, or the $W$ helicity in top quark decays.
A second approach is to search directly for new particles coupling
to the top quark or for non-SM decays. 
In this section we present those topics which have led to specific
analyses at the Tevatron. In section~\ref{sec:anoProd} we will 
additionally discuss searches for anomalous single top quark production
at HERA and LEP. 

\subsection{Decays to a charged Higgs boson}
In the SM a single complex Higgs doublet scalar field is responsible for
breaking the electroweak symmetry and generating the masses of gauge
bosons and fermions. Many extensions of the SM include a Higgs sector
with two Higgs doublets and are therefore called Two Higgs Doublet 
Models (THDM). In a THDM electroweak symmetry breaking leads to five physical
Higgs bosons: two neutral $CP$-even scalars $h^0$ and $H^0$, one neutral $CP$-odd
pseudoscalar $A^0$, and a pair of charged scalars $H^\pm$.
The extended Higgs sector is described by two parameters:
the mass of the charged Higgs, $M_{H^+}$, and $\tan\beta=v_1/v_2$,
the ratio of the vacuum expectation values $v_1$ and $v_2$ of the 
two Higgs doublets. One distinguishes two types of THDMs.
In a type I THDM only one of the Higgs doublets couples to fermions,
in a type II THDM the first Higgs doublet couples to the up-type
quarks ($u$, $c$, $t$) and neutrinos, while the second doublet couples
to down-type quarks ($d$, $s$, $b$) and charged leptons.
The analyses we discuss in this section are concerned with type II models. 
A particular example for a type II THDM is the minimal supersymmetric
model (MSSM).

If the charged Higgs boson is lighter than the difference of top quark and 
$b$ quark mass, $M_{H^\pm}<M_\mathrm{top}-M_b$, the decay mode 
$t \rightarrow H^+ b$ is possible and competes with the SM decay
$t\rightarrow W^+b$. The branching fraction depends on $\tan\beta$ and
$M_{H^+}$. The MSSM predicts that the channel $t \rightarrow H^+ b$
dominates the top quark decay for $\tan\beta \lesssim 1$ and $\tan\beta \gtrsim 70$.
In most analyses it is assumed that 
$\mathrm{BF}(t\rightarrow W^+b)+\mathrm{BF}(t\rightarrow H^+b) = 1$.
At tree level the $H^\pm$ does not couple to vector bosons. Therefore,
the $H^\pm$ decays only to fermions. 
In the parameter region $\tan\beta < 1$ the dominant decay mode is
$H^+\rightarrow c\bar{s}$, while for $\tan\beta > 1$ the decay channel
$H^+\rightarrow \tau^+\nu_\tau$ is the most important one.
For $\tan\beta > 5$ the branching fraction to $\tau^+\nu_\tau$ is nearly
100\%. Thus, in this region of parameter space type II THDM models predict 
an excess of $t\bar{t}$ events with tau leptons over the SM expectation.

First searches for the $H^\pm$ in top quark events were already performed 
well before the top quark was discovered in 1994/95. The UA1 and UA2 
experiments at the CERN $Sp\bar{p}S$ excluded certain regions of the
$M_\mathrm{top}$ versus $M_{H^\pm}$ 
plane~\cite{chargedHiggsUA1,chargedHiggsUA2}. 
First searches of CDF improved these limits using events with a dilepton
signature~\cite{CDF_tHplus_94_2} or reconstructing tau leptons in their
hadronic decay mode~\cite{CDF_tHplus_94_1}.

At LEP the experiments searched for charged Higgs production in the
process $e^+e^-\rightarrow H^+H^-$. The analyses encompass the 
final states $c\bar{s}\bar{c}s$, $\tau^+\nu_\tau\tau^-\bar{\nu_\tau}$,
$c\bar{s}\tau^-\bar{\nu_\tau}$, and $ \tau^+\nu_\tau\bar{c}s$.
The combined result of all four experiments excludes charged Higgs
masses below $78.6\,\mathrm{GeV}/c^2$~\cite{lepCombinedChargedHiggs}.
The CLEO collaboration has used its measurement of the inclusive
$b\rightarrow s\gamma$ cross section to set an indirect limit
on the charged Higgs mass~\cite{cleoBsgamma}. 
The dependence of the cross section
on $M_{H^\pm}$ enters via quantum corrections. Since the process
$b\rightarrow s\gamma$ occurs only at loop level in the SM,
its sensitivity to new particles which might be exchanged in the 
loop is quite high. CLEO sets a limit of 
$M_{H^\pm} > (244 + 63/(\tan\beta)^{1.3})\,\mathrm{GeV}/c^2$ at the 95\%
C.L., under the assumption that a type II THDM is realized in nature.
If the Higgs sector has a richer structure, the CLEO limit can be 
circumvented. Therefore, direct searches for the $H^\pm$ which are less 
model dependent remain important.
   
After the top quark discovery the $H^\pm$ searches at the Tevatron 
looked for the decay modes $t\bar{t}\rightarrow H^\pm W^\mp b\bar{b}$
and $t\bar{t}\rightarrow H^\pm H^\mp b\bar{b}$. 
CDF published three analyses where the charged Higgs is assumed to 
decay into $\tau^+\nu_\tau$ and the tau lepton subsequently decays
semi-hadronically into a tau neutrino plus 
hadrons~\cite{CDF_tHplus_95,CDF_tHplus_97,CDFHiggsPlus2000Run1}.
The identification of hadronic tau decays in these CDF analyses is 
similar to the one described in detail in section~\ref{sec:taumodes}.
Two analyses (references~\cite{CDF_tHplus_95,CDF_tHplus_97}) used an 
inclusive approach, based on final states with missing transverse energy,
one identified hadronic tau decay, jets or high-$p_\mathrm{T}$ leptons.
One analysis (reference~\cite{CDFHiggsPlus2000Run1}) requires one 
high-$p_\mathrm{T}$ lepton ($e$ or $\mu$) in the event, which can originate 
from the decay $W\rightarrow e/\mu+\nu$ or 
$H^+\rightarrow \tau^+\nu_\tau \rightarrow \bar{\nu}_\tau + e^+/\mu^+ + \nu_{e/\mu} + \nu_\tau$.
The best exclusion limit is obtained from the inclusive 
analysis~\cite{CDF_tHplus_97}: For $M_{H^\pm}=80\,\mathrm{GeV}/c^2$
the region of $\tan\beta \gtrsim 25$ is excluded. For higher Higgs 
masses the exclusion becomes less stringent, reaching 
 $\tan\beta \gtrsim 200$ at $M_{H^\pm}=160\,\mathrm{GeV}/c^2$.  

The first D\O \ search for the $H^\pm$ is based on a disappearance strategy,
looking for a deficit of events in the SM lepton-plus-jets 
channel~\cite{d0ChargedHiggs1999}.
Since good agreement between the SM prediction and the observation is found,
certain regions of phase space for charged Higgs production can be excluded. 
This result is superseded by a direct search for hadronic tau decays in
top quark events~\cite{d0ChargedHiggs2002}. In the following paragraph
we will discuss this last Run I analysis on the charged Higgs in more
detail. 
\par
\begin{figure}[!t]
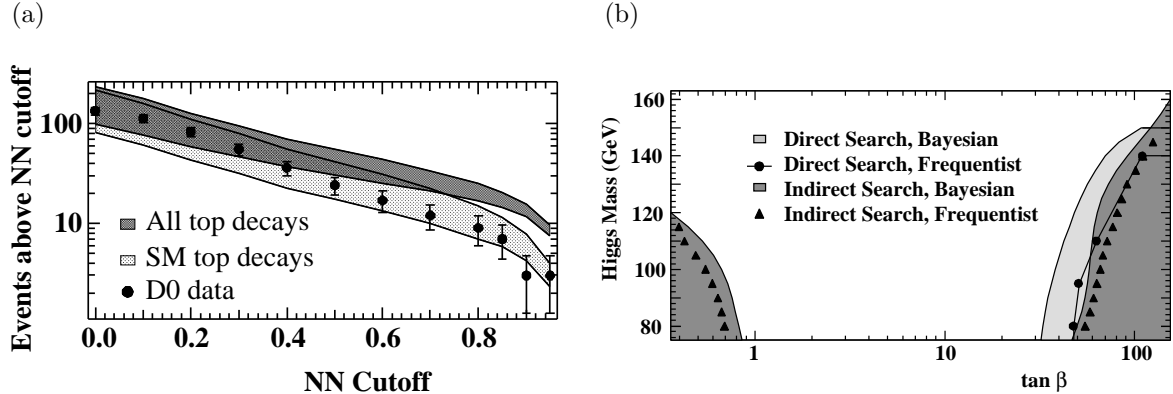

  (a) \hspace*{72mm} (b) \\
  \vspace*{1mm} \\
  \includegraphics[width=0.47\textwidth]{fig/HplusNNcutoff.epsi}
  \hspace*{0.02\textwidth}
  \includegraphics[width=0.49\textwidth]{fig/HplusExclusion.epsi}
\caption{\label{fig:HplusNN} Search for the $H^\pm$ at D\O \ in 
  Run I~\cite{d0ChargedHiggs2002}.
  (a) The number of observed events in data
  versus the cut on the neural network output (dots).
  Overlaid is the
  expectation from the SM (light shaded area) and a scenario with additional
  $H^\pm$ production in top quark decays for $\tan\beta = 150$ and
  $M_{H^\pm}=95\,\mathrm{GeV}/c^2$.
  (b) Exclusion region at the 95\% confidence level
  in $(M_{H^\pm}, \tan\beta)$ space.
  The top quark mass is assumed to be 175 GeV/$c^2$ and 
  $\sigma(t\bar{t})=5.5\,\mathrm{pb}$. The plot also contains the regions excluded
  by the indirect search method~\cite{d0ChargedHiggs1999}, where D\O \ looks for 
  a potential disappearance of the $t\bar{t}$ signal.}
\end{figure}

The D\O \ analysis uses data taken with a multijet + $\EtMiss$ trigger
corresponding to $(62.3\pm 3.1)\,\mathrm{pb^{-1}}$ of integrated luminosity.
Relatively loose preselection cuts are applied, followed by tighter cuts
involving an artificial neural network. The preselection requires that
$\EtMiss > 25\,\mathrm{GeV}$, at least four jets with 
$E_\mathrm{T}>20\,\mathrm{GeV}$ each,
but no more than eight jets with $E_\mathrm{T}>8\,\mathrm{GeV}$.
The selection is sensitive to $t\bar{t}\rightarrow H^\pm W^\mp b\bar{b}$ events
where the $W$ boson decays into quarks as well as 
$t\bar{t}\rightarrow H^\pm H^\mp b\bar{b}$ events.
The network has three input variables: the $\EtMiss$ and two of the three 
eigenvalues of the normalized momentum tensor. The training is done with
charged Higgs events generated by {\sc Isajet}~\cite{isajet} for the signal
and with 25000 multijet events from data for the background.
Figure~\ref{fig:HplusNN}a shows the number of remaining events in data
versus the cut on the neural network output. Overlaid is the number of events
expected from the SM as well as one scenario including a charged Higgs boson
with $M_{H^\pm}=95\,\mathrm{GeV}/c^2$.
The data show good agreement with the SM expectation, thereby falsifying the
chosen $H^\pm$ scenario. Based on a series of Monte Carlo experiments the
sensitivity for the $H^\pm$ search is optimized yielding on optimal cutoff 
at 0.91. 
\par
After the neural network cut hadronic tau decays are identified
in the data. A narrow jet with $\Delta R < 0.25$ is required, with 
1 to 7 tracks in the jet cone. Events with identified muons or electrons are
rejected. In addition, a $\chi^2$ requirement is constructed using 
calorimeter information from $W\rightarrow \tau\nu$ Monte Carlo events. 
After all cuts $3.2\pm 1.5$ multijet background events are 
expected, $1.1\pm0.3$ from $t\bar{t}$ and $0.9\pm0.3$ from $W+\mathrm{jets}$.
In data D\O \ observes three events. A Bayesian and a Frequentist statistical
analysis are used to deduce exclusion regions is the 
$(M_{H^\pm}, \tan\beta)$ phase space shown in figure~\ref{fig:HplusNN}b.
Under the assumption that $M_\mathrm{top}=175\,\mathrm{GeV}/c^2$ and 
$\sigma(t\bar{t})=5.5\,\mathrm{pb}$, the region of $\tan\beta > 32.0$ is
excluded at the 95\% C.L. for $M_{H^\pm}=75\,\mathrm{GeV}/c^2$.
For Higgs masses above 150 GeV/$c^2$ no limits can be set. This result can 
also be interpreted in terms of the branching fraction of 
$t\rightarrow H^+b$, setting a lower limit of 
$\mathrm{BF}(t\rightarrow H^+b)>0.36$ at the 95\% C.L..

\subsection{Search for $X^0\rightarrow t\bar{t}$ decays} 
Several extensions of the SM predict the existence of narrow resonances
that decay to $t\bar{t}$ pairs. One such model is, for example, a 
$Z^\prime$ predicted by top-colour-assisted technicolour~\cite{hillParke1994}.
This model speculates that the spontaneous breaking of electroweak symmetry
is related to the observed fermion masses, in particular the large top quark
mass, and can be accomplished by dynamical effects~\cite{bardeen1990}.

\begin{figure}[!t]
  (a) \hspace*{72mm} (b) \\
  \includegraphics[width=0.48\textwidth]{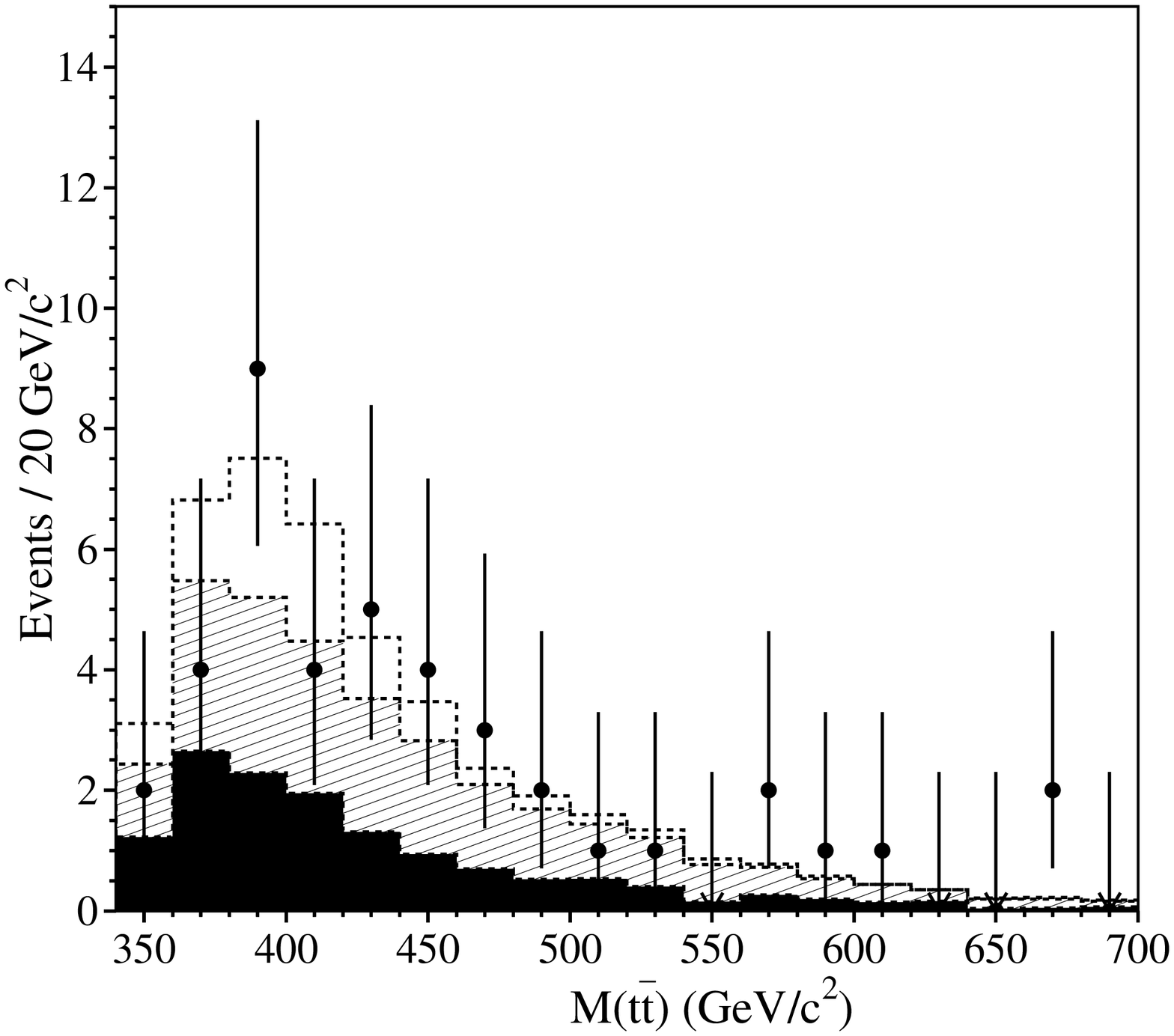}
  \hspace*{0.02\textwidth}
  \includegraphics[width=0.48\textwidth]{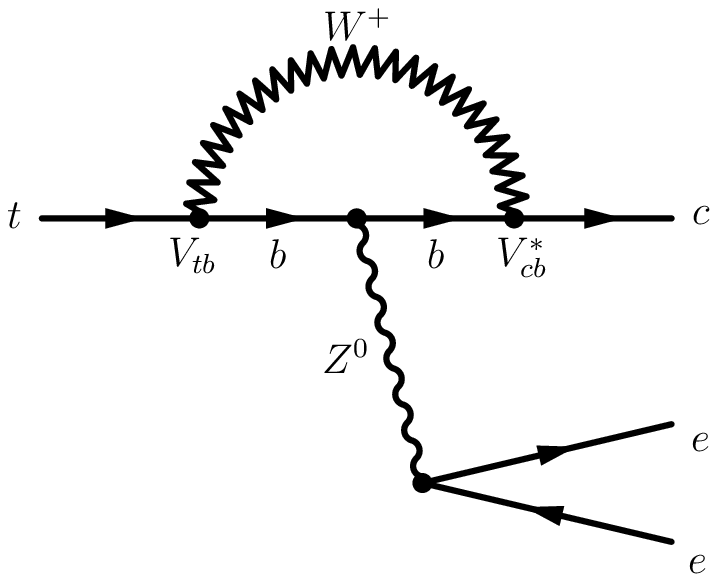}
\caption{\label{fig:x2ttbarANDtZc} (a) Invariant mass distribution of the 
  reconstructed $t\bar{t}$ pair of 41 events in the D\O \ lepton-plus-jets 
  data sample~\cite{d0X2ttbarRun1}. The solid histogram shows the shape of the
  $W +$ jets and QCD multijets background, the hatched histogram the sum of
  all SM backgrounds including $t\bar{t}$ production, and the open histogram
  shows the sum of signal ($M_X=400\,\mathrm{GeV}/c^2$) and background as 
  obtained from the fit to the data.
  (b) Feynman diagram for the FCNC decay
  $t\rightarrow cZ^0$ with $Z\rightarrow e^+e^-$. Other diagrams involve
  the $d$ and the $s$ quark within the loop instead of the $b$ quark.
  All diagrams together nearly cancel and cause the branching ratio
  of this decay to be very small, $\mathcal{O}(10^{-12})$, in the SM.
}
\end{figure}

CDF and D\O \ have performed model-independent searches for a narrow resonance
$X^0$ which decays into a $t\bar{t}$ pair~\cite{CDFX2ttbarRun1,d0X2ttbarRun1}.
Both analyses relate closely to the respective top quark mass measurements
of the two experiments~\cite{CDFtopMass1998LJ,d0TopMass1998Run1LeptonPlusJets}
and employ fits to the invariant mass spectrum of the $t\bar{t}$ pair.
D\O \ obtains a slightly better mass limit. The D\O \ analysis is based on
data with an integrated luminosity of $130\,\mathrm{pb^{-1}}$ and uses
events with a lepton-plus-jets signature passing the topological 
and the soft muon $b$ tag selection. The events are subjected to a kinematic
fit that comprises kinematic constraints on the reconstructed $W$ boson mass
and the top quark mass. Out of all possible permutations the option with 
the best $\chi^2$ is chosen. The $\chi^2$ is required to be below 10.
After all cuts 41 events are left in the data sample. 
Figure~\ref{fig:x2ttbarANDtZc}a shows the invariant mass distribution of the 
$t\bar{t}$ pair 
reconstructed from data.
The data are fit to distributions for the combined QCD multijet and 
$W+$ jets background, SM $t\bar{t}$ production, and a signal
for $X^0\rightarrow t\bar{t}$.
For comparison to the data figure~\ref{fig:x2ttbarANDtZc}a shows the 
backgrounds 
and the signal component ($M_X=400\,\mathrm{GeV}/c^2$) obtained from the fit. 
There is no significant deviation from the SM prediction.
The fit is repeated for several assumptions for the $X^0$ mass, again without
providing evidence for a signal. Therefore, upper limits on the
$X^0$ production cross section times branching ratio to $t\bar{t}$
are set, ranging from 5.0~pb at $M_X=400\,\mathrm{GeV}/c^2$ to 1.5~pb
at $M_X=850\,\mathrm{GeV}/c^2$. 
The confidence level is 95\%.
These limits are valid as long as the 
width of the resonance, $\Gamma_X$, is small compared to the D\O \ 
detector resolution. The limits quoted above assume a width of
$\Gamma_X = 0.012\,M_X$. 
If one compares the cross section limits with the predictions for a
leptophobic $Z^\prime$ boson, one can obtain a lower limit on the 
$Z^\prime$ mass: $M_{Z^\prime}>560\,\mathrm{GeV}/c^2$. 

\subsection{FCNC decays}
\label{sec:fcncDecays}
In the SM flavour changes at leading order (tree level) are induced by 
charged currents, the exchange of $W$ bosons. 
Flavour changing neutral currents (FCNC) are only possible at higher
orders in perturbation theory (loop level). FCNC decays of the 
top quark are strongly suppressed in the SM due to the 
Glashow-Iliopoulos-Maiani (GIM) mechanism~\cite{Glashow1970}.
Compared to the top quark mass the masses of down-type quarks
($d$, $s$, $b$) occurring in loop diagrams are small and degenerate.
Therefore, the sum over the respective amplitudes nearly cancels.
Fig.~\ref{fig:x2ttbarANDtZc}b shows one of the Feynman diagrams that describe
the decay $t\rightarrow c Z^0$. This type of diagram is also called
penguin diagram. The SM predicts the branching fractions for
$t\rightarrow c Z^0$ and $t\rightarrow c\gamma$ to be on the order of
$10^{-12}$, while 
$\mathrm{BF}(t\rightarrow cg) \simeq 10^{-10}$ and
$\mathrm{BF}(t\rightarrow cH^0) \simeq 10^{-13}$~\cite{eilam1991}.

In extensions of the SM FCNC top quark decays can be considerably
enhanced by several orders of magnitude if FCNC couplings
at tree level are allowed~\cite{fritzsch89,han96,han95}. 
In Two Higgs Doublet Models where the neutral scalar $h^0$ posses flavor
changing couplings the decay $t\rightarrow ch^0$ can be considerably
enhanced~\cite{hou1992}. A similar enhancement can be reached in 
supersymmetric models were $R$-parity is violated~\cite{eilam2001}. 
In topcolour-assisted technicolour theories the branching ratio
for $t\rightarrow c W^+W^-$ can reach values up to 
$10^{-3}$~\cite{yue2001}. Since the SM predictions for FCNC interactions
have much smaller rates, they are useful probes for new physics
beyond the SM. 

The CDF collaboration has searched for the FCNC decays
$t\rightarrow q Z^0$ and $t\rightarrow q\gamma$ in 
Run I data with an integrated luminosity of
$110\,\mathrm{pb^{-1}}$~\cite{CDF_FCNC_Run1}.
The $Z^0$ is measured in the decay channels $e^+e^-$ or $\mu^+\mu^-$.
Electron and muon identification is described in section~\ref{sec:eleMuoID}.
A Photon is identified as an energy cluster in the electromagnetic
calorimeter with no track pointing at it. Additionally, a cluster
with a single soft track pointing to it is accepted as a photon,
if the track carries less than 10\% of the cluster energy.
A typical photon cluster consists of two adjacent towers in the
electromagnetic calorimeter. The direction of the photon is 
defined by the line between the event vertex and the centroid of the
electromagnetic shower as measured in the shower maximum detector.
Background from hadronic jets is reduced by demanding that the
photon cluster is isolated from other energy depositions in the 
electromagnetic or the hadron calorimeter. In addition, the 
ratio of hadronic to electromagnetic energy is constrained to be less
than $0.055 + 0.00045\,E$, where $E$ is the sum of hadronic and 
electromagnetic energy of the cluster measured in the calorimeter.

The search strategy for the FCNC decays tries to isolate $t\bar{t}$ events
where one top quark decays via FCNC and the second top quark decays 
according
to the SM decay $t\rightarrow W^+b$. In the $t\rightarrow q\gamma$ search
two signatures are considered, where the $W$ boson decays either leptonically
into $e\nu_e$ or $\mu\nu_\mu$ or hadronically into quarks.
In the first subsample a well identified charged lepton ($e$ or $\mu$)
with $p_\mathrm{T}>20\,\mathrm{GeV}/c$, $\EtMiss > 20\,\mathrm{GeV}$,
a photon with $E_\mathrm{T} > 20\,\mathrm{GeV}$, and at least 
two jets with $E_\mathrm{T}>15\,\mathrm{GeV}$
are required. The second subsample contains events with at least four jets and
one photon with $E_\mathrm{T}>50\,\mathrm{GeV}$. 
At least one jet must be identified
as a $b$ quark jet with a secondary vertex tag. 
In both subsamples,
there must be a photon-plus-jet combination with an invariant mass
in the top quark mass window $140 < M_{j\gamma} < 210\,\mathrm{GeV}/c^2$. 
In the hadronic subsample, the remaining jets must have
$\sum E_\mathrm{T}\geq 140\,\mathrm{GeV}$.
One event passes all selection criteria. The event features a 
$72\,\mathrm{GeV}$ muon,
$\EtMiss=24\,\mathrm{GeV}$, a $88\,\mathrm{GeV}$ photon, and three jets.
The background is expected to be about 0.5 events, mainly from $W\gamma$ 
production with two additional jets.   

In the $t\rightarrow q Z^0$ search, the $W$ from the second top quark decay
is considered to decay hadronically, while the $Z^0$ is reconstructed in its
leptonic decay modes to $e^+e^-$ or $\mu^+\mu^-$.
This yields an event signature of four jets and two leptons with an invariant
mass close to the $Z^0$ mass ($75 < M_{\ell\ell} < 105\,\mathrm{GeV}/c^2$). 
Since the branching ratio of $Z^0\rightarrow \ell^+\ell^-$ is small, 
the total detection efficiency of the $t\rightarrow q Z^0$ mode is much 
smaller 
than the one for the $t\rightarrow q\gamma$ search. Each of the four jets
must have $E_\mathrm{T}>20\,\mathrm{GeV}$ and $|\eta|<2.4$. 
One $Z\rightarrow \mu^+\mu^-$ event passes all selection criteria. The expected
background is 0.6 events.

Since no statistically significant excess is observed in both searches,
the data are used to set upper limits on the top quark branching ratios
into FCNC decays
\begin{equation}
\label{tcgammaLimit}
  \BR \left( t \rightarrow u/c + \gamma \right) < 3.2\%
  \ \ \ \ \ \ \ \ \ \ \ \ 
  \BR \left( t \rightarrow u/c + Z^0 \right) < 33\%
\end{equation}
at the $95\%$ confidence level.
These limits are still far from the interesting region. 
At the LHC experiments will reach a sensitivity level of 
$10^{-4}$~\cite{topLHC2000},
and thereby reach a region where some models predicting very large
enhancements can be excluded~\cite{yang1998}.

\subsection{Anomalous single top production}
\label{sec:anoProd}

\begin{figure}[!t]
\begin{minipage}{0.47\textwidth}
  \includegraphics[width=\textwidth]{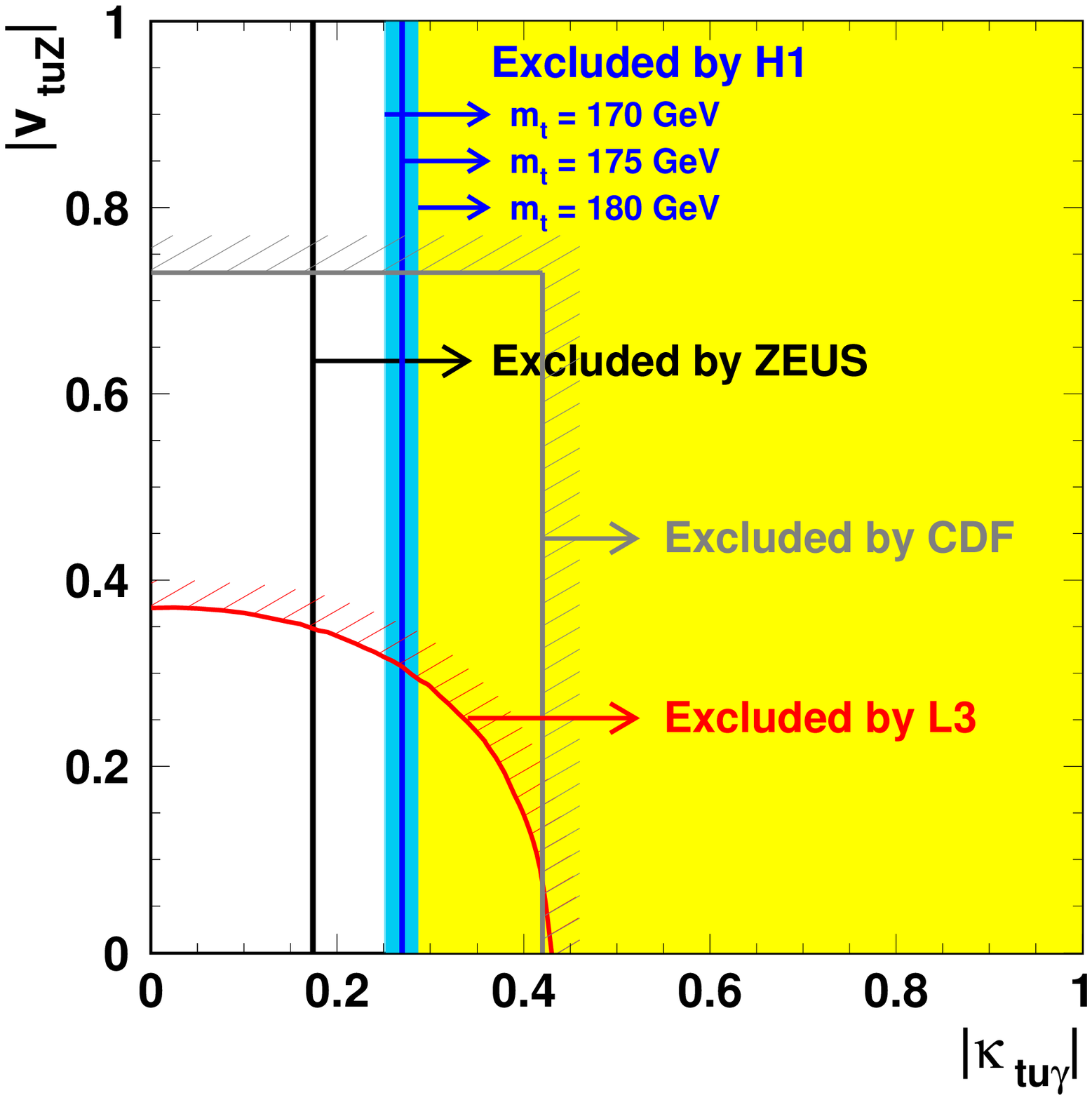}
\caption{\label{fig:limitAnoTop} Limits at the 95\% confidence level
  on anomalous couplings $\kappa_{tu\gamma}$ and 
  $\kappa_{tuZ}$ (in this plot noted as $v_{tuZ}$) 
  of the top quark to the $\gamma$ or the $Z$ boson 
  from reference~\cite{H1singleTop}.
  The plot shows the limits obtained by the HERA experiments
  H1 and ZEUS~\cite{H1singleTop,ZEUSsingleTop}, 
  by L3~\cite{l3singleTop} and by CDF~\cite{CDF_FCNC_Run1}
  (see section~\ref{sec:fcncDecays}).} 
\end{minipage} \hspace*{0.05\textwidth} 
\begin{minipage}{0.44\textwidth}
   \includegraphics[width=0.42\textwidth]{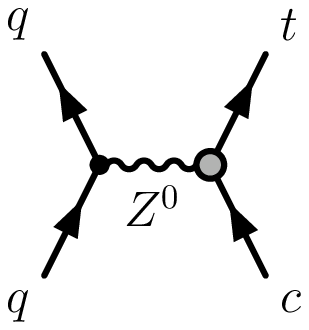}
   \includegraphics[width=0.42\textwidth]{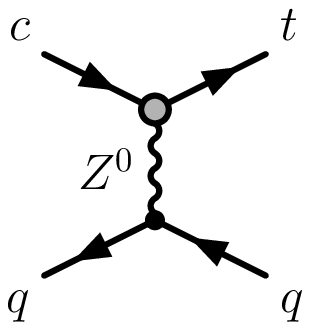}
 
   \vspace*{3mm}

   \includegraphics[width=0.42\textwidth]{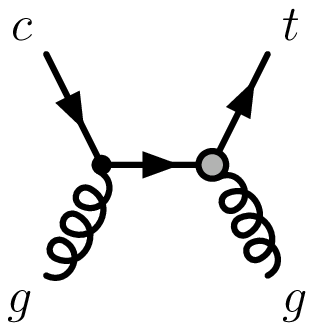}
   \includegraphics[width=0.42\textwidth]{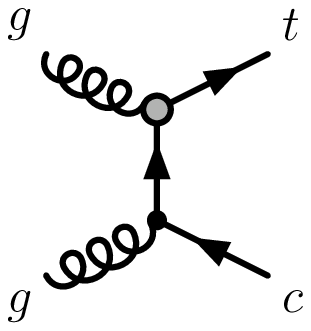}
 \caption{\label{fig:anoProd} Feynman diagrams for single top production via 
   $2 \rightarrow 2$ FCNC processes in hadron collisions.}
\end{minipage} 
\end{figure}

At the Tevatron the search for anomalously enhanced FCNC in top quark decays
is statistically restricted by the number of $t\bar{t}$ pairs produced.
More stringent limits on anomalous $tq\gamma$ of $tqZ$ couplings at tree 
level are set by searching for the production of single top quarks via
FCNC at LEP and HERA. At LEP all four experiments have searched for the
reactions 
$e^+e^-\rightarrow t\bar{c} / t\bar{u}$
and presented upper limits on anomalous 
couplings~\cite{delphiSingleTop,l3singleTop,alephSingleTop,opalSingleTop}.
The limits obtained by the L3 collaboration are shown in 
figure~\ref{fig:limitAnoTop}.
For $\kappa_\gamma = 0$ the coupling to the $tqZ$ coupling is 
constrained by $\kappa_Z < 0.41$, for $\kappa_Z = 0$ the $tq\gamma$ coupling
is confined to $\kappa_\gamma < 0.49$. Both limits are taken at the 95\%
confidence level. The other LEP collaborations obtain very similar results.
Figure~\ref{fig:limitAnoTop} also visualises the limits that result from the 
CDF analysis discussed in section~\ref{sec:fcncDecays}.
The LEP results can be translated into limits on the branching ratios:
$\mathrm{BR}(t\rightarrow u/c+\gamma)<4.2\%$ and
$\mathrm{BR}(t\rightarrow u/c+Z^0)<14\%$~\cite{alephSingleTop}.

At HERA the two experiments H1 and ZEUS searched for top quarks
produced in the inclusive FCNC reaction 
$ep\rightarrow etX$~\cite{H1singleTop,ZEUSsingleTop}.
The SM process for single top production at HERA is 
the charged current reaction
$ep\rightarrow \nu t\bar{b}X$~\cite{schuler1988} with a predicted cross 
section of less than 1~fb~\cite{stelzerSullivan1997}, well below the
sensitivity of the experiments. An observed excess in this channel can 
therefore signal physics beyond the SM.
ZEUS has analyzed a data sample corresponding to $130\,\mathrm{pb^{-1}}$ 
of integrated luminosity. The top quark is assumed to decay according to 
the SM decay $t\rightarrow W^+ b$. The $W$ boson is detected in its leptonic
and the hadronic decays channels. The observed data are in good agreement
with the background expectation. Thus, ZEUS is able to compute an upper
limit on the anomalous $tu\gamma$ coupling: 
$\kappa_{tu\gamma}<0.174$ at the 95\% confidence level, which is the best
limit on this quantity to date. The ZEUS result is depicted in 
figure~\ref{fig:limitAnoTop}. 

The H1 analysis uses data with an integrated luminosity of 
$118.3\,\mathrm{pb^{-1}}$~\cite{H1singleTop}. 
In the leptonic channel five events are observed
over a SM background of $1.31\pm0.22$. If this excess is attributed to
FCNC top quark production a total cross section of
$\sigma(ep\rightarrow etX)=0.29^{+0.15}_{-0.14}\,\mathrm{pb}$ is found.
In the hadronic channel there is no excess in the data. Only more data will
help to decide whether the observation is a real signal or just a
statistical fluctuation. 

At the Tevatron there are no searches for anomalous single top
quark production yet. Possible FCNC processes at tree level could 
be $q_1\bar{q}_1\rightarrow t \bar{q}_2$,  
$q_1 \bar{q}_2 \rightarrow t \bar{q}_2$, 
$q g \rightarrow t g$, and $g g \rightarrow t \overline{q}$.
Example Feynman diagrams for these processes are shown in 
figure~\ref{fig:anoProd}.
Another option for FCNC top quark production is the $2 \rightarrow 1$ quark-gluon 
fusion process $g + u/c \rightarrow t$.
Studies at parton level suggest that the anomalous up quark
coupling parameter $\kappa_{tug}/\Lambda$ can be measured down to 
$0.02~\mathrm{TeV^{-1}}$ in Run II of the Tevatron~\cite{Hosch_topCollider}. 
For the anomalous charm quark coupling parameter 
$\kappa_{tcg}/\Lambda$ a bound of $0.06~\mathrm{TeV^{-1}}$ is in reach.
Studies with a fast detector simulation for ATLAS indicate a similar
reach at the LHC~\cite{cakir2005}.

\section{Summary and outlook}

Ten years after the discovery of the top quark many of its properties have been
thoroughly investigated. All measurements are in good agreement with the
predictions made by the Standard Model (SM).
At the Tevatron and at LHC the production of $t\bar{t}$ pairs is the dominating
source of top quarks. The experiments CDF and D\O \ have established
$t\bar{t}$ signals in the dilepton, the lepton-plus-jets and the all hadronic
channels and measured the production cross section. 
The $t\bar{t}$ channels involving tau leptons have not been 
observed yet, but they are well within reach in Run II at the Tevatron. 
\par
The most important parameter of the top quark is its mass. The large value
of $M_\mathrm{top}$ distinguishes the top quark strongly from the other
quarks, since it decays essentially as a free quark. Thus, the top quark
is an ideal laboratory to study polarization effects in heavy quark production
and decay. 
Via higher order perturbative corrections the top quark mass is strongly correlated 
with the $W$ boson mass and the Higgs boson mass. Therefore, electroweak precision tests
and predictions of $M_H$ are very sensitive to $M_\mathrm{top}$.
The final combined Run I result on the top quark mass yields:
$M_\mathrm{top}=(178.0\pm4.3)\,\mathrm{GeV}/c^2$.
Run II measurements are underway. First preliminary measurements indicate a
slightly lower mass value. 
CDF finds 
$M_\mathrm{top}=173.5^{+2.7}_{-2.6}\,(\mathrm{stat.})\pm 2.5\,(\mathrm{JES})\pm 1.7\,(\mathrm{syst.})\,\mathrm{GeV}/c^2$, D\O \ 
$M_\mathrm{top}=170.6\pm4.2\,(\mathrm{stat.})\pm6.0\,(\mathrm{syst.})\,\mathrm{GeV}/c^2$.
At the end of Run II the top quark mass will be measured with an
uncertainty of $2\;\mathrm{to}\;3\,\mathrm{GeV}/c^2$.
At the LHC the precision of $M_\mathrm{top}$ will reach about 1~GeV$/c^2$.
A limiting factor will be the theoretical understanding of the relation between
the kinematic quark mass measured by experiments and the theoretically well
defined pole mass or $\msbar$ mass.
The final word on $M_\mathrm{top}$ will come from an $e^+e^-$ linear collider
where a scan of the $t\bar{t}$ production threshold will allow for a precision
of about $20\,\mathrm{MeV}/c^2$.  
\par
After the discovery of the top quark the focus of research by the CDF and 
D\O \ collaborations 
shifted to a detailed investigation of its production and decay properties.
The helicity of $W$ bosons from the top quark decay was measured and found
to agree with the SM $V-A$ theory of the electroweak interaction.
The ratio of branching ratios 
$R_{tb} = \mathrm{BF}(t\rightarrow Wb)/\mathrm{BF}(t\rightarrow Wq)$
was determined and found to agree with a value close to 1 as predicted
by the SM. D\O \ has investigated $t\bar{t}$ dilepton events for
spin correlations among the two charged leptons, CDF has measured the 
top quark $p_\mathrm{T}$ spectrum. Both analyses show no significant
deviations from the SM.
The search for electroweak single top quark production is about to reach
a critical phase in Run II. A discovery of this top quark production
mode is around the corner. 
\par
Its large mass is motivation to view the top quark as an ideal probe for new
physics beyond the SM.
Already before the top quark discovery the search for the decay
$t\rightarrow H^+b$ started. CDF and D\O \ searched also for a heavy
resonance which decays into a $t\bar{t}$ pair. A CDF analysis looked
for FCNC decays of the top quark. No indication of these phenomena beyond 
the SM is found.
\par 
The LHC will be a top quark factory where several millions of top quarks 
will be produced per year. This wealth of data will allow to enter a 
precision era of top quark physics where top quark properties will be
very accurately investigated. The Tevatron analyses provide an excellent
basis for further studies at the LHC. But also new topics 
where huge amounts of data are needed will become accessible at the LHC,
for example, the search for like sign top pair production 
($tt$ or $\bar{t}\bar{t}$), production of single top quarks via FCNC,
the search for $CP$ violation in $t\tbar$ production, or the direct 
measurement of the electric charge of the top quark. The abundance of analysis topics
and open questions guarantees that top quark physics will remain a
lively and thrilling field of elementary particle research in the coming
years which will see the end of Run II at the Tevatron and the turn on
of the LHC.
\ack
I am indebted to my colleagues from the Tevatron experiments CDF and D\O \ 
who have devoted their energy and intellect to build and operate these
superb telescopes to investigate the microcosm. Without their
dedication to thoroughly analyze the data and produce first class physics
results this review would not have been possible.
I also want to thank Prof. Dr. Thomas M\"uller for his encouragement to write
this review article and reading the manuscript.
%
\section*{References}
\begin{footnotesize}
\bibliography{abbrevi,toprefs,cdfpubs,d0pubs,leprefs,lehrbuch,reviews,theorie}
\bibliographystyle{toprep}
\end{footnotesize}  
\end{document}